\documentclass[12pt]{report}

\usepackage[12pt]{ee-report}

\usepackage[toc,page]{appendix}
\usepackage{cite}
\usepackage{epsfig}
\usepackage{bm}
\usepackage{amssymb,amsmath}
\usepackage{soul,color}
\usepackage{psfrag}
\usepackage[english]{babel}
\usepackage{blindtext}
\usepackage{titlepage}
\usepackage{nomencl}
\makenomenclature

\usepackage{pdflscape}

\usepackage{amsmath,epsfig,bm,amssymb,amsthm}
\usepackage{cite}
\usepackage{multirow}

\usepackage{amsmath,epsfig,bm,amssymb,amsthm,balance}

\newcommand{\comment}[1]{}

\newcommand{\bt}{\mathrm{b}}
\newcommand{\dd}{{\mathrm d}}
\newcommand{\Et}{\mathrm{E}}
\newcommand{\SRi}{{\mathrm{SR}_i}}
\newcommand{\sr}{{\mathrm{sr}}}
\newcommand{\SR}{{\mathrm{SR}}}
\newcommand{\RD}{{\mathrm{RD}}}
\newcommand{\rd}{{\mathrm{rd}}}
\newcommand{\sri}{{\mathrm{sr}_i}}
\newcommand{\RDi}{\mathrm{R}_i\mathrm{D}}
\newcommand{\rdi}{\mathrm{r}_i\mathrm{d}}
\newcommand{\sd}{\mathrm {sd}}

\newcommand{\SNR}{\mathrm {SNR}}
\newcommand{\opt}{\mathrm {opt}}
\newcommand{\toep}{\mathrm {toeplitz}}

\newcommand{\Var}{\mathrm {Var}}
\newcommand{\dgy}{\mathrm {diag}\{\yb\}}
\newcommand{\E}{\mathrm {E}}

\newcommand{\CN}{\mathcal{CN}}
\newcommand{\CoN}{\mathcal{C}^N}
\newcommand{\Z}{\mathbb{Z}}
\newcommand{\Vc}{\mathcal{V}}

\newcommand{\s}{\mathbf{s}}
\newcommand{\V}{\mathbf{V}}
\newcommand{\w}{\mathbf{w}}
\newcommand{\yb}{\mathbf{y}}
\newcommand{\rb}{\mathbf{r}}
\newcommand{\I}{\mathbf{I}}
\newcommand{\0}{\mathbf{0}}
\newcommand{\A}{\mathbf{A}}
\newcommand{\B}{\mathbf{B}}

\newcommand{\h}{\mathbf{h}}
\newcommand{\xb}{\mathbf{x}}
\newcommand{\e}{\mathbf{e}}

\newcommand{\ub}{\mathbf{u}}
\newcommand{\z}{\mathbf{z}}

\newcommand{\q}{\mathbf{q}}
\newcommand{\Sb}{\mathbf{S}}
\newcommand{\G}{\mathbf{G}}
\newcommand{\C}{\mathbf{C}}
\newcommand{\Lb}{\mathbf{L}}
\newcommand{\U}{\mathbf{U}}

\newcommand{\R}{\mathbf{R}}

\newcommand{\Ib}{\mathbf{I}}
\newcommand{\dgs}{\mbox{diag}\{\mathbf{s}\}}
\newcommand{\dgsc}{\mbox{diag}\{\mathbf{s}^*\}}

\newcommand{\MSD}{\mathrm{msd}}
\newcommand{\CDD}{\mathrm {cdd}}

\newcommand{\lpn}{\lim \limits_{P_0/N_0 \rightarrow \infty}}

\newcommand{\bb}{{\mathrm b}}
\newcommand{\CpN}{\mathbf{\mathcal{C}}^N}

\newcommand{\hb}{\mathbf{h}}
\newcommand{\Rb}{\mathbf{R}}
\newcommand{\Sgw}{\mathbf{\Sigma}_{\w}}
\newcommand{\gth}{\gamma_{\mathrm{th}}}
\newcommand{\dgh}{\mbox{diag}\{\mathbf{h}_2\}}
\newcommand{\dghc}{\mbox{diag}\{\mathbf{h}_2^*\}}
\newcommand{\Sgy}{\mathbf{\Sigma}_{\yb}}

\newcommand{\Sg}{\mathbf{\Sigma}}
\newcommand{\tb}{\tilde{b}}

\newcommand{\bDelta}{\mathbf{\Delta}}

\newcommand{\diag}{\mathrm {{diag}}}
\newcommand{\Omg}{\mathbf{\Omega}}

\newcommand{\rt}{\mathrm {r}}
\newcommand{\ovy}{\overline{\yb}}
\newcommand{\ovS}{\overline{\Sb}}
\newcommand{\ovV}{\overline{\V}}
\newcommand{\ovG}{\overline{\G}}
\newcommand{\ovq}{\overline{\q}}
\newcommand{\ovw}{\overline{\w}}

\newcommand{\ovh}{\overline{\h}}
\newcommand{\Sig}{\mathbf{\Sigma}}
\newcommand\norm[1]{\left\lVert#1\right\rVert}
\newcommand{\Vcb}{\bm{\mathcal{V}}}

\newcommand{\Sgwb}{\mathbf{\Sigma}_{\ovw}}
\newcommand{\X}{\mathbf{X}}
\newcommand{\Q}{\mathbf{Q}}
\newcommand{\bSigy}{\overline{\Sgb}_{\yb}}
\newcommand{\Sgb}{\mathbf{\Sigma}}
\newcommand{\et} {\mathrm{e}}

\newcommand\blfootnote[1]{%
  \begingroup
  \renewcommand\thefootnote{}\footnote{#1}%
  \addtocounter{footnote}{-1}%
  \endgroup
}

\begin{document}
\setcounter{page}{1}
\pagenumbering{roman}

\begin{titlepage}
\begin{center}
\vspace*{0in} {\Large \bf Differential Modulation and Non-Coherent Detection in Wireless Relay Networks \vspace{.5in} \\
}
{ A Thesis Submitted \\
to the College of Graduate Studies and Research \\
in Partial Fulfilment of the Requirements \\
for the Degree of Doctor of Philosophy \\
in the Department of Electrical and Computer Engineering \\
University of Saskatchewan\\
\vspace{.5in}
by} \\
\vspace{.5in}
{\bf M. R. Avendi \\
}

%\vspace{1.0cm}
%\begin{figure}[htb!]
%\centering
%\mbox{\scalebox{.5}{\includegraphics*{figure/uofs_logo.eps}}}
%\end{figure}
%\vspace{.8cm}
%{{\large Ph. D. Dissertation Advisor:}\\
%\vspace{0.2cm} {\Large \bf Prof. Ha H. Nguyen}
%}

\vspace{2.6in}
%Department of Electrical and Computer Engineering\\
%University of Saskatchewan\\
Saskatoon, Saskatchewan, Canada \\
%December 2004 \\
%\vspace{.8in}
\copyright\ Copyright M. R. Avendi, January, 2014. All rights reserved. \\

\end{center}
\end{titlepage}

%\centerline{\epsfig{figure={figure/blk_coop.eps},width=8.5cm}} 

\addcontentsline{toc}{chapter}{Permission To Use}
\chapter*{Permission To Use}
In presenting this thesis in partial fulfilment of the requirements for a Postgraduate
degree from the University of Saskatchewan, it is agreed that the Libraries of this
University may make it freely available for inspection. Permission for copying of this
thesis in any manner, in whole or in part, for scholarly purposes may be granted by
the professors who supervised this thesis work or, in their absence, by the Head of
the Department of Electrical and Computer Engineering or the Dean of the College
of Graduate Studies and Research at the University of Saskatchewan. Any copying,
publication, or use of this thesis, or parts thereof, for financial gain without the
written permission of the author is strictly prohibited. Proper recognition shall be
given to the author and to the University of Saskatchewan in any scholarly use which
may be made of any material in this thesis.
Request for permission to copy or to make any other use of material in this thesis
in whole or in part should be addressed to:

\noindent
Head of the Department of Electrical and Computer Engineering \\
57 Campus Drive \\
University of Saskatchewan \\
Saskatoon SK S7N 5A9 \\
Canada

%\addcontentsline{toc}{chapter}{Acknowledgements}
%\include{acknowledge}

\addcontentsline{toc}{chapter}{abstract}
\chapter*{Abstract} 

The technique of cooperative communications is finding its way in the next generations
of many wireless communication applications. Due to the distributed nature of cooperative networks, acquiring fading channels information for coherent detection is more challenging than in the traditional point-to-point communications. To bypass the requirement of channel information, differential modulation together with non-coherent detection can be deployed. This thesis is concerned with various issues related to differential modulation and non-coherent detection in cooperative networks. Specifically, the thesis examines the behaviour and robustness of non-coherent detection in mobile environments (i.e., time-varying channels). The amount of channel variation is related to the normalized Doppler shift which is a function of user's mobility. The Doppler shift is used to distinguish between slow time-varying
(slow-fading) and rapid time-varying (fast-fading) channels. The performance of
several important relay topologies, including single-branch and multi-branch dual-hop relaying with/without a direct link that employ amplify-and-forward relaying and two-symbol non-coherent detection, is analyzed. For this purpose, a time-series model is developed for characterizing the time-varying nature of the cascaded channel encountered in amplify-and-forward relaying. Also, for single-branch and multi-branch dual-hop relaying without a direct link, multiple-symbol differential detection is developed. 

First, for a single-branch dual-hop relaying without a direct link, the performance of
two-symbol differential detection in time-varying Rayleigh fading channels is evaluated. It is seen that the performance degrades in rapid time-varying channels. Then, a multiple-symbol differential detection is developed and analyzed to improve the system performance in fast-fading channels. Next, a multi-branch dual-hop relaying with a direct link is considered. The performance of this relay topology using a linear combining method and two-symbol differential detection is examined in time-varying Rayleigh fading channels. New combining weights are proposed and shown to improve the system performance in fast-fading channels. The performance of the simpler selection combining at the destination is also investigated in general time-varying channels. It is illustrated that the selection combining method performs very close to that of the linear combining method. Finally, differential distributed space-time coding is studied for a multi-branch dual-hop relaying network without a direct link. The performance of this network using two-symbol differential detection in terms of diversity over time-varying channels is evaluated. It is seen that the achieved diversity is severely affected by the channel variation. Moreover, a multiple-symbol differential detection is designed to improve the performance of the differential distributed space-time coding in fast-fading channels.

\addcontentsline{toc}{chapter}{Table of Contents}
\tableofcontents

\addcontentsline{toc}{chapter}{List of Abbreviations}

\nomenclature{RF}{Radio Frequency}
\nomenclature{MIMO}{Multiple-Input Multiple-Output}
\nomenclature{3 GPP}{3rd Generation Partnership Project}
\nomenclature{3 GPP}{3rd Generation Partnership Project}
\nomenclature{OFDM}{Orthogonal Frequency Division Multiplexing}
\nomenclature{LTE}{Long Term Evolution}
\nomenclature{LAN}{Local Area Network}
\nomenclature{SD}{Source-Destination}
\nomenclature{SR}{Source-Relay}
\nomenclature{RD}{Relay-Destination}
\nomenclature{AF}{Amplify and Forward}
\nomenclature{DSTC}{Distributed Space-Time Code}
\nomenclature{TDD}{Time Division Duplex}
\nomenclature{CSI}{Channel State Information}
\nomenclature{D-AF}{Differential Amplify and Forward}
\nomenclature{MRC}{Maximum-Ratio Combining}
\nomenclature{SC}{Selection Combining}
\nomenclature{MSD}{Multiple Symbol Differential}
\nomenclature{VAA}{Virtual Antenna Array}
\nomenclature{AR}{Auto Regressive}
\nomenclature{AR(1)}{First-Order Auto Regressive}
\nomenclature{PSK}{Phase Shift Keying}
\nomenclature{WLAN}{Wireless LAN}
\nomenclature{RFID}{Radio Frequency Identification}
\nomenclature{$M$-DPSK}{Differential $M$-PSK}
\nomenclature{AWGN}{Additive White Gaussian Noise}
\nomenclature{ML}{Maximum Likelihood}
\nomenclature{STC}{Space-Time Code}
\nomenclature{STBC}{Space-Time Block Code}
\nomenclature{PAM}{Pulse Amplitude Modulation}
\nomenclature{QAM}{Quadrature Amplitude Modulation}
\nomenclature{DPSK}{Differential Phase Shift Keying}
\nomenclature{USTC}{Unitary Space-Time Code}
\nomenclature{D-USTC}{Differential Unitary Space-Time Code}
\nomenclature{RFID}{Radio Frequency Identification}
\nomenclature{SNR}{Signal to Noise Ratio}
\nomenclature{CDD}{Conventional Differential Detection}
\nomenclature{PEP}{Pairwise Error Probability}
\nomenclature{BER}{Bit Error Rate}
\nomenclature{BEP}{Bit Error Probability}
\nomenclature{cdf}{Cumulative Distribution Function}
\nomenclature{pdf}{probability density function}
\nomenclature{MSDSD}{Multiple-Symbol Differential Sphere Decoding}
\nomenclature{CCI}{Co-Channel Interference}
\nomenclature{ISI}{Inter-Symbol Interference}

\printnomenclature[5em]

\addcontentsline{toc}{chapter}{List of Figures}
\listoffigures
%\listoftables
\pagenumbering{arabic}
\chapter{Introduction and Thesis Outline}
\label{ch:intro}

\setcounter{page}{1}
\pagenumbering{arabic}

\section{Introduction}
Perhaps when Heinrich R. Hertz mentioned ``I do not think that the wireless waves I have discovered will have any practical application", from his modesty, he did not really believe in great advances in this field. Soon, Nicola Tesla increased the distance of electromagnetic transmission and Guglielmo Marconi made a breakthrough in wireless communications with the discovery of short waves. Nowadays, wireless communications are non-detachable parts of our life. From small cordless gadgets and cellular phones to radars and satellites communications, they all have one thing in common, an antenna and an RF transceiver which wirelessly connects them to the world.

In wireless communications, long distances, natural or artificial barriers and mobility of users introduce a notorious effect, known as fading, which can be divided into large-scale and small-scale fading. Large-scale fading is due to path loss of signals as a function of distance and shadowing by large objects such as buildings and hills \cite{fundamental-digital,gold_wireless,DigComFad-Simon}. On the other hand, small-scale fading is due to the constructive and destructive interference of the multiple signal copies received over multiple paths between the transmitter and receiver. It is the later case that causes rapid fluctuation in the signal strength which limits the transmission reliability substantially over wireless channels.

The increasing demand for better quality and higher data rate in wireless communication systems motivated the use of \emph{diversity} techniques to mitigate the destructive effect of fading. The basic idea of all diversity techniques is to provide different replicas of the same information over multiple independently-faded paths in order to decrease the probability that the received signal is in deep fade (i.e., when the channel gain is dropped dramatically in magnitude), thus increase the reliability and the probability of successful transmission. Common diversity techniques that have been studied intensively in the literature and applied in practice include time diversity (e.g., channel coding, interleaving), spatial diversity (e.g., multiple-input multiple-output (MIMO) systems), combination of multi-path and frequency diversity (e.g., orthogonal frequency division multiplexing (OFDM) together with channel coding or pre-coding).

Among different diversity techniques, spatial diversity using multiple antennas has been shown to be a very effective technique both in the literature and practice because of its better spectral efficiency. However, using multiple antennas is not always feasible in many applications. The obvious example is in personal mobile units in which there is insufficient space to make wireless channels corresponding to multiple antennas uncorrelated. This limitation was however addressed by the technique of cooperative communications \cite{user-coop1,user-coop2}.

Today, cooperative communications has become a mature research topic in the literature. Currently, a special type of cooperative communication (with the help of one relay) has been standardised in the 3 GPP LTE technology to leverage the coverage problem of cellular networks and it is envisaged that LTE-advanced version will include cooperative relay features to overcome other limitations such as capacity and interference \cite{coop-dohler}. There are also applications for cooperative relay networks in wireless LAN, vehicle-to-vehicle communications \cite{coop-V2V} and wireless sensor networks that have been discussed in \cite{coop-LTE,coop-WiMAX,coop-deploy,coop-dohler,coop-sensor} and references therein.

\begin{figure}[t]
\psfrag {Source} [] [] [1.0] {Source}
\psfrag {Relay} [] [] [1.0] {Relay}
\psfrag {Destination} [] [] [.9] {Destination}
\psfrag {Base Station} [] [] [.9] {Destination}
\psfrag {hsrd} [] [] [1.0] {Cascaded channel}
\psfrag {h0} [] [] [1.0] {Direct channel}
\centerline{\epsfig{figure={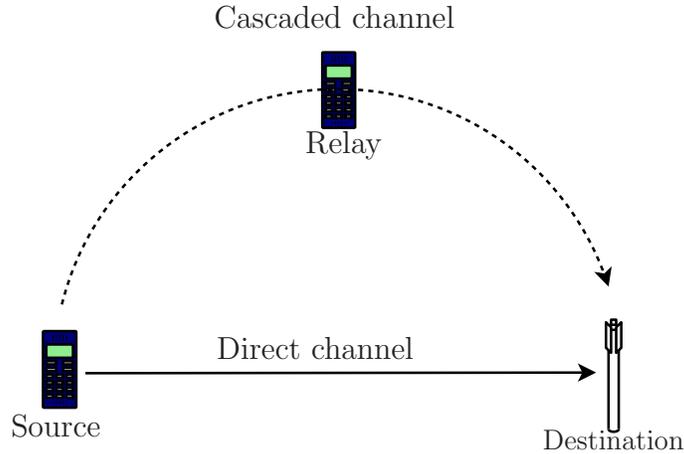},width=8.5cm}}
\caption{A simple cooperative network with three nodes.}
\label{fig:blk_coop}
\end{figure}

An illustration of a simple cooperative network with three nodes is given in Figure~\ref{fig:blk_coop}. As can be seen, there are two links from Source to Destination. The first link is the direct channel from Source to Destination (SD), which is similar to the channel encountered in conventional point-to-point communication. On the other hand, there is a channel from Source to Relay (SR) and a channel from Relay to Destination (RD). Since Relay can also listen to Source from SR channel, it would be able to re-broadcast the received data to Destination through RD channel. In this way, the second link is established through Source-Relay-Destination path. For convenience, the overall channel of Source-Relay-Destination is called the cascaded or the equivalent channel. Therefore, the overall diversity and performance of the network would benefit from the extra antenna which is constructed using the help of Relay. Similarly, multiple relays can be used to achieve higher diversity.

Depending on the protocol that relays utilize to process and re-transmit the received signal to the destination, the relay networks have been generally classified as decode-and-forward or amplify-and-forward \cite{coop-laneman}. Among these two protocols, amplify-and-forward (AF) has been the focus of many studies because of its simplicity in the relay's function. Specifically, the relay's function is to multiply the received signal with an amplification factor and forward the result to the destination.

Moreover, depending on the strategy that relays utilize to cooperate, relay networks are categorized as repetition-based and distributed space-time coding (DSTC)-based \cite{DSTC-Laneman}. In the repetition-based strategy, relays forward the received signals to the destination in time-division duplex (TDD) fashion, whereas in the DSTC-based strategy, the relays simultaneously transmit the received symbols such that a space-time code can be constructed at the destination. The later strategy has a better spectral efficiency than the former but it is more complicated to design and build \cite{DSTC-Laneman}.

At the destination, based on the type of modulation, either coherent or non-coherent detection would be applied. In coherent detection, it is required that the instantaneous channel state information (CSI) of all transmission links are known at the destination. Although this requirement can be accomplished by sending pilot (training) signals and using channel estimation techniques in \emph{slow-fading} environments, it is not feasible in fast \emph{time-varying} channels. Moreover, collecting the CSI of SR channels at the destination is questionable due to noise amplification at relays. Furthermore, the computational complexity and overhead of channel estimation increase proportionally with the number of relays. In addition, in fast \emph{time-varying} channels a more frequent channel estimation is needed, which reduces the effective transmission rate and spectral efficiency. Also, all channel estimation techniques are subject to impairments that would directly translate to  performance degradation.

To circumvent these limitations, in repetition-based strategy, differential modulation with two-symbol non-coherent detection has been considered in \cite{DAF-Liu,DAF-DDF-QZ,DAF-General,DAF-MN-Himsoon,DAF-GLAT,DAF-ML,DAF-FSK} for AF relay networks. This technique is referred as differential AF (D-AF) transmission. In D-AF transmission, information bits are differentially encoded at the source. Only the second-order statistics of the SR channels (no instantaneous CSI) are needed at the relays to determine the amplification factor \cite{DAF-Liu,DAF-DDF-QZ,DAF-General,DAF-MN-Himsoon}. Then, the decision variables, computed from the received signals in different links, are weighted and summed at the destination. Computing the optimum weights for Maximum-Ratio-Combining (MRC) require the instantaneous CSI of Relay-Destination (RD) channels, which are unknown, and the amplification factors of relays. Thus, the second-order statistics of the RD channels have been used to define a set of fixed weights in \cite{DAF-Liu,DAF-DDF-QZ,DAF-General,DAF-MN-Himsoon}. For further reference, this method is referred as semi-MRC. %The output of the combiner is used in two-symbol non-coherent detection of the transmitted symbols with diversity.

Distributed space-time coding (DSTC) is another strategy that has been considered in cooperative networks to provide a better spectral efficiency than the repetition-based strategy \cite{DSTC-Laneman}. In the DSTC-based strategy \cite{DSTC-Y,DSTC-Kaveh,DSTC-Laneman}, the relays cooperate to process and forward the received signals to the destination so that a space-time code can be constructed at the destination and therefore allow the system to enjoy the higher spectral efficiency of space-time codes \cite{stc-jafar}. Moreover, the constructed virtual antenna array (VAA) enables one to extend established techniques of traditional MIMO systems to relay networks. For instance, most of the designed space-time codes for MIMO systems can be utilized in cooperative networks \cite{DSTC-Y,DSTC-Kaveh,DSTC-Laneman}. Also, differential space-time codes \cite{D-SPT-Hughes,D-USTC-HOCH} can be adapted for relay networks as in \cite{D-DSTC-Amin,D-DSTC-Giannakis,D-DSTC-Y,D-DSTC-Usayl} so that non-coherent detection can be done without any requirement of the CSI.

{
\section{Research Objectives}}

Motivated from the previous discussion, this thesis is concerned with differential modulation and non-coherent detection in wireless relay networks. The main objectives are outlined as follows.

$\bullet$
{
Most of the existing literature on differential amplify-and-forward transmission assumes a \emph{slow-fading} environment and shows that a 3-4 dB loss is observed between coherent and non-coherent detection. However, with the increase of vehicles' speed (e.g., when a mobile user travels in a high-speed train) \cite{train-speed,train-speed2}, the wireless channels become more time-varying. This faster variation thus leads to a higher degradation in the performance. Hence, it is important to analyse the performance and examine the robustness of non-coherent detection in \emph{fast-fading} channels. The first objective of our research is therefore to study differential amplify-and-forward relay networks in \emph{time-varying} Rayleigh fading channels. In point-to-point communications, to study the performance of differential modulation over time-varying channels, a time-series model is often used for modelling the direct channel. In relay networks, the cascaded channels have a more complex distribution than that of the direct channel. To the best of our knowledge, to date, there is no study on the time-series model of cascaded channels. Hence, a time-series model is developed to characterize the evolution of the cascaded channel in time. Based on this model, the performance of several important topologies are investigated. This investigation would be a useful tool to design a robust system and prevent the network to fall into regions that additional power transmission will not improve the performance (error floor regions).}

$\bullet$
{
In the conventional non-coherent detection, the decision variable is computed from the latest two received symbols. However, two-symbol differential detection would not perform well in fast-fading channels. Hence, it would be useful to improve the performance of non-coherent detection in \emph{time-varying} channels using other techniques. One of the techniques that has been used in point-to-point links to improve the performance of non-coherent detection in time-varying channels is multiple-symbol differential (MSD) detection \cite{msdd-div2}. In this technique, a larger window of the received symbols are jointly processed for detection. Here, we consider the application of MSD decoding and investigate its effectiveness in the context of relay networks for two relay topologies.}

$\bullet$
{
The Maximum-Ratio Combining (MRC) technique needs at least the second-order statistics of all transmission links. However, collecting the second-order statistics of all channels at the destination might be a challenge (if not impossible) for some applications. On the other hand, selection combining does not need any kind of CSI. Although the SC method has been considered for point-to-point communications with diversity reception, this technique has not been considered for relay networks employing differential amplify-and-forward strategy. Hence, our objective is to develop selection combining for differential amplify-and-forward relay networks. Performance analysis of the SC method shall also be considered and compared with that of the MRC method to determine a trade-off between simplicity and performance.
}

{
\section{Research Methodology}}
{
The main methodology in our research is summarized below.}

\begin{itemize}

\item
{
Communication theory \cite{com_th_craig} is the main tool to develop new signal processing algorithms and conduct performance analysis of the networks under consideration. Probability theory, random variables and processes, linear algebra and matrix analysis are also extensively used in our research.}

\item
{
Channel modelling in time-varying scenarios for relay networks is identified as an important and crucial task. This will be done by applying and extending the modelling techniques in point-to-point channels.}

\item
{
As any design starts with the system modelling stage and it is not possible to build the complete physical system at the beginning, computer simulation using MATLAB is a common and important tool in communications research (and perhaps in many other related areas). Here, MATLAB is used to simulate various elements of communication links such as Source, wireless channels, Relays and Destination. The accuracy of channel modelling and any important or major approximations made in the theoretical development will be verified with computer simulation. In addition, the performance of the developed signal processing algorithms will be checked and verified with the theoretical analysis and the results will be interpreted.}

\end{itemize}

\section{Organisation of the Thesis}
This dissertation is organized in a manuscript-based style. The first two chapters of the thesis discuss relevant background of point-to-point and relay wireless communications. The published or submitted manuscripts are included as the contributions of the thesis.

Chapter~\ref{ch:p2p} contains the background on point-to-point wireless communications which will be extended to relay networks in Chapter~\ref{ch:coop}. In Chapter~\ref{ch:p2p}, first, single-antenna communication systems, the block diagram of transmitter and receiver, the wireless channel model and modulation and demodulation techniques are described. Next, multiple-antenna communication systems, receive and transmit diversity, combining methods and space-time coding are presented. Chapter~\ref{ch:coop} covers the essential background knowledge on cooperative networks. The major relay topologies, relay protocols and cooperative strategies, that are relevant to this thesis, are introduced in this chapter.

The manuscript in Chapter~\ref{ch:dh} studies a dual-hop relaying system without direct link that employs differential $M$-PSK together with two-symbol and multiple-symbol differential detection. The performance of this system in time-varying channels is analysed. A multiple-symbol detection is also developed and theoretically analysed for this system. The manuscript in Chapter~\ref{ch:mnode} considers multi-branch dual-hop relaying with direct link using semi-MRC at the destination. Differential $M$-PSK and two-symbol non-coherent detection are employed in this system and its performance is evaluated in time-varying channels. The manuscript in Chapter~\ref{ch:sc} studies selection combining (SC) at the destination of relay networks. The performance of this system in slow-fading channels is analysed and compared with the system using the semi-MRC method. The manuscript in Chapter~\ref{ch:sc_tv} examines the SC method in general time-varying Rayleigh fading channels.
While Chapters~\ref{ch:dh}-\ref{ch:sc_tv} are concerned with the repetition-based strategy, Chapter~\ref{ch:dstc} considers a multi-branch dual-hop relaying without direct link and with the use of distributed space-time coding (DSTC) strategy. The performance of this system using two-symbol differential detection in terms of diversity over time-varying channels is analysed. Moreover, a multiple-symbol differential detection is developed for this system to improve its performance in fast-fading channels. Finally, Chapter~\ref{ch:sum} concludes this thesis by summarizing the contributions and suggesting potential research problems for future studies.

%\clearpage

\emph{Notation}: Bold upper-case and lower-case letters denote matrices and vectors, respectively. $(\cdot)^t$, $(\cdot)^*$, $(\cdot)^H$ denote transpose, complex conjugate and Hermitian transpose of a complex vector or matrix, respectively. $|\cdot|$  denotes the absolute value of a complex number and $\|\cdot \|$ denotes the Euclidean norm of a vector. $\mathcal{CN}(0,N_0)$ stands for complex Gaussian distribution with zero mean and variance $N_0$. $\mbox{E}\{\cdot\}$ denotes expectation operation. Both ${e}^{(\cdot)}$ and $\exp(\cdot)$ show the exponential function. $\dgs$ is the diagonal matrix with components of $\s$ on the
main diagonal and $\Ib_N$ is the $N \times N$ identity matrix. A symmetric $N\times N$ Toeplitz matrix is defined by $\mbox{toeplitz}\{x_1,\cdots,x_N\}$. $\mbox{det}\{\cdot\}$ denotes determinant of a matrix. $\CoN$ is the set of complex vectors with length $N$.
$\Z$ is the set of integer numbers.

%\addcontentsline{toc}{chapter}{References}
%\bibliographystyle{IEEEbib}
%\bibliography{H:/latex/references}

\chapter{Background on Point-to-Point Communications}
\label{ch:p2p}
This chapter discusses point-to-point communication systems using single antenna and multiple antennas in wireless fading channels. First, for a single antenna system, the block diagram of the transmitter and receiver, the channel model and modulation and demodulation techniques are described. Specifically, the focus is on differential encoding and non-coherent detection techniques which do not require channel estimation. Next, diversity systems using multiple antennas, different combining methods and space-time codes are described. The background given in this chapter will be extended and applied to the context of relay networks in the next chapters.

\section{Single-Antenna Wireless Communication}
\label{sec:single-antenna}
\begin{figure}[t]
\psfrag {Source} [] [] [1.0] {Source}
\psfrag {Base Station} [] [] [1] {Destination}
\psfrag {h0} [] [] [1.0] {Wireless channel}
\centerline{\epsfig{figure={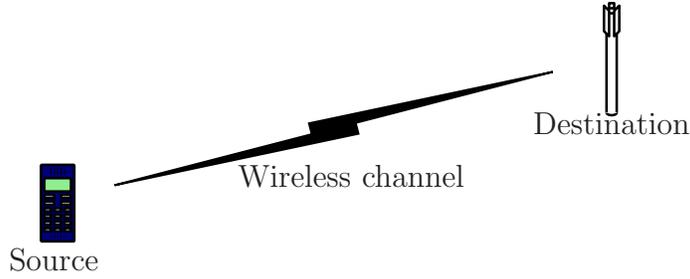},width=8.5cm}}
\caption{A point-to-point wireless communication link.}
\label{fig:blk_p2p}
\end{figure}

Figure~\ref{fig:blk_p2p} depicts a point-to-point communication link in which a source transmits signals to a destination using a single antenna over a wireless channel. The transmitted signal is an electromagnetic wave in the radio frequency band (RF). This signal is generated by the transmitter, whose detailed operation is described next.%, typically around 2 GHz for cellular networks. %Such a transmitted signal would be affected by the fading phenomena of the wireless channel.%Hence, it is important to look at mathematical model of a fading channel.

\subsection{Transmission}
\label{subsec:modulation}
\begin{figure}[t]
\psfrag {bits} [] [] [1.0] {bits}
\psfrag {Modulation} [] [] [1.0] {Modulation}
\psfrag {Differential} [] [] [1] {Differential}
\psfrag {Encoder} [] [] [1] {Encoder}
\psfrag {D} [] [] [1.0] {D}
\psfrag {vm} [] [] [1.0] {$v[k]$}
\psfrag {sk} [] [] [1.0] {$s[k]$}
\psfrag {Pulse} [] [] [1.0] {Pulse}
\psfrag {Shaping} [] [] [1.0] {Shaping}
\psfrag {baseband} [] [] [1] {$s_b(t)$}
\psfrag {passband} [] [] [1] {$\tilde{s}(t)$}
\psfrag {carrier} [] [] [1.0] {$e^{j2\pi f_ct}$}
\psfrag {antenna} [] [] [1.0] {Tx}

\psfrag {channel} [] [] [1.0] {Channel}
\psfrag {rx antenna} [] [] [1.0] {Rx}
\psfrag {wt} [] [] [1.0] {AWGN}
\psfrag {ytilde} [] [] [1.0] {$\tilde{y}(t)$}
\psfrag {carrier rx} [] [] [1.0] {$e^{-j2\pi f_ct}$}
\psfrag {Match} [] [] [1.0] {Match}
\psfrag {Filter} [] [] [1] {Filter}
\psfrag {ybt} [] [] [1.0] {$y_b(t)$}
\psfrag {yk} [] [] [1.0] {$y[k]$}
\psfrag {Detection} [] [] [1.0] {Detection}
\psfrag {vhm} [] [] [1.0] {$\hat{v}[k]$}
\psfrag {Channel} [] [] [1.0] {Channel}
\psfrag {Estimator} [] [] [1.0] {Estimator}
\centerline{\epsfig{figure={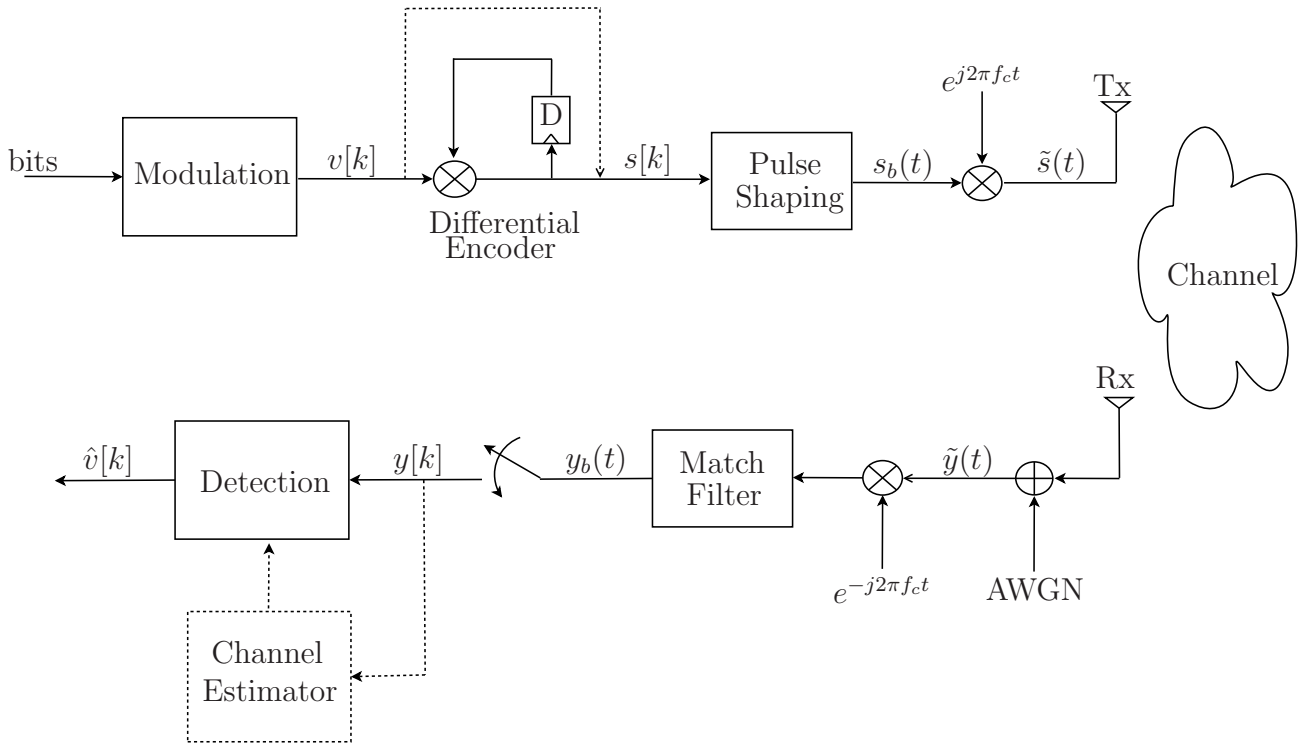},width=17cm}}
\caption{A communication system operating over a wireless fading channel.}
\label{fig:txrx-p2p}
\end{figure}

The structure of the transmitter of a point-to-point communication system over a wireless channel is illustrated in Figure~\ref{fig:txrx-p2p}. For simplicity, all signals and signal processing blocks are shown in the complex form. In reality, the complex form is divided into real and imaginary parts, which correspond to the in-phase and quadrature components of the RF signal. Moreover, due to Digital Signal Processing (DSP) implementation, the discrete-time equivalent is used for signal representation before the pulse-shaping block. Information to be transmitted, either analog signals such as audio or video, or digital signals such as text or multi-media, are converted to binary (bit) sequence by previous stages (e.g., source coding). Then, the binary sequence is given to the modulation block.

\subsubsection{$M$-PSK Modulation}
\begin{figure}[t]
\psfrag {v1} [] [] [1.0] {$v_1$}
\psfrag {v2} [] [] [1.0] {$v_2$}
\psfrag {v3} [] [] [1.0] {$v_M$}
\psfrag {phi1} [b] [] [1.0] {in-phase}
\psfrag {phi2} [] [] [1.0] {quadrature}
\psfrag {0} [br] [] [1.0] {$0\;\;$}
\psfrag {t1} [] [] [.8] {$\qquad2\pi/M$}
\psfrag {t2} [] [] [.8] {$\qquad -2\pi/M$}
\centerline{\epsfig{figure={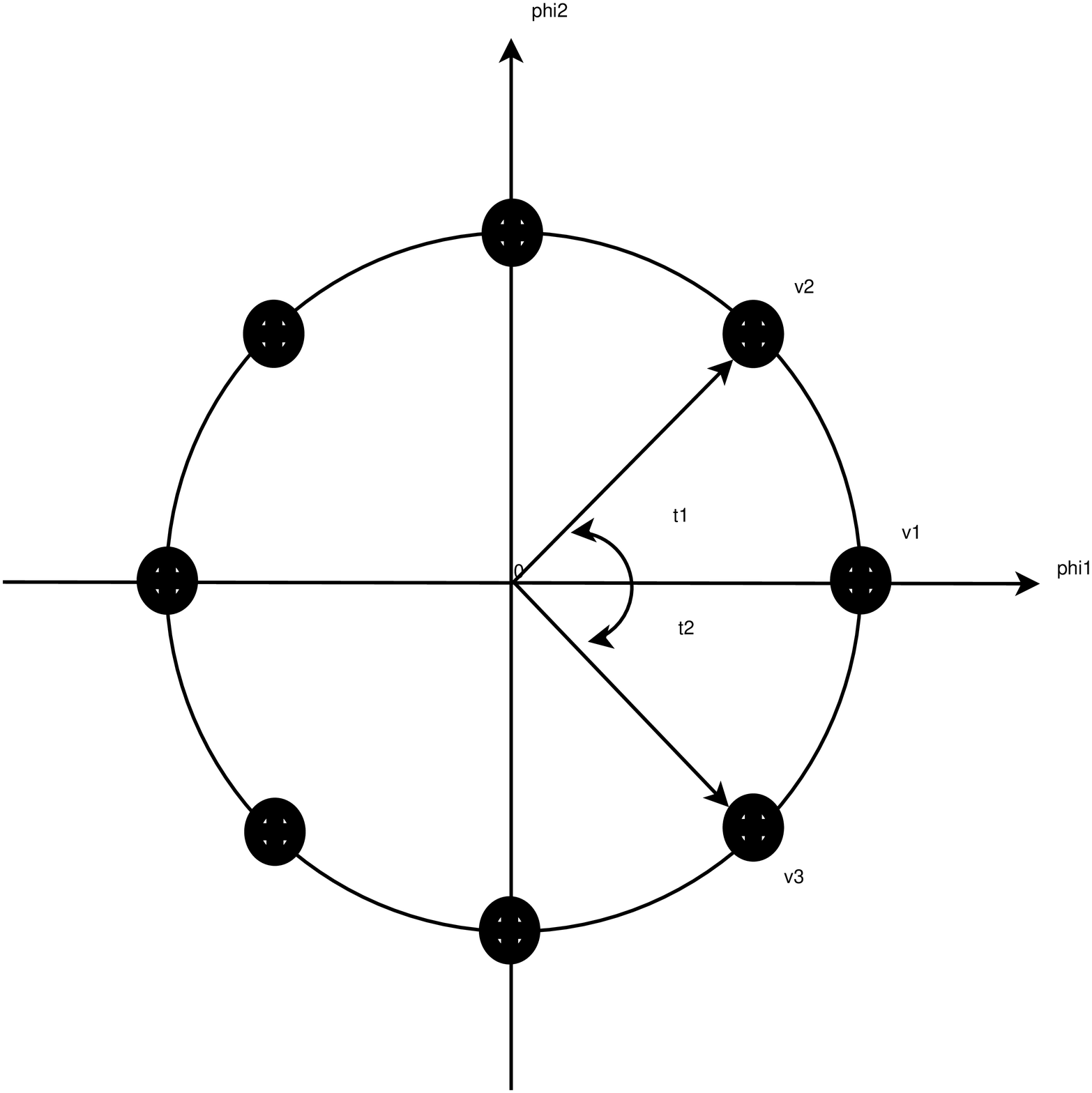},width=8.5cm}}
\caption{Signal space plot for $M$-PSK. Here $M=8$.}
\label{fig:mpsk}
\end{figure}

Information bits are mapped to symbols using a signalling (or modulation) scheme. Among different signalling techniques, the $M$-ary phase-shift keying ($M$-PSK) is widely used in existing technologies such as WLANs, RFID standards, Bluetooth, satellite communications etc., owing to its constant envelope property and good bandwidth efficiency. In $M$-PSK, a group of $\log_2 M$ information bits are encoded into the phase of symbol $v[k]\in \Vc$ where $\Vc=\{e^{j2\pi m/M},\; m=0,\dots, M-1\}$ and $k$ is the discrete-time index. For a $M$-PSK there are $M$ signal points equally spaced on the circle. Signal space plot of $8$-PSK is depicted in Figure~\ref{fig:mpsk}, in which eight signal points are shown.

\subsubsection{Differential Encoding}
The second block can be bypassed for coherent detection such that $s[k]=v[k]$.
However, for applications that channel estimation is not feasible, information symbols can be differentially encoded, so that the receiver does not need the CSI. Such a scheme is called differential $M$-PSK ($M$-DPSK). Given $v[k]\in \Vc$ and $s[k-1]\in \Vc$, $M$-PSK and $M$-DPSK symbols at time indices $k$ and $k-1$, respectively, the $M$-DPSK symbol at time $k$ is obtained as
\begin{equation}
\label{eq:DPSK}
s[k]=v[k] s[k-1], \quad  k\in \Z, \quad s[0]=1.
\end{equation}

\subsubsection{Pulse Shaping and Up conversion}
The resulting discrete-time symbol is then converted to a continues-time signal using the pulse shaping block in the baseband. The baseband signal, $s_b(t)$, is then up-converted by a local oscillator to produce an RF signal $\tilde{s}(t),~ 0<t<T_s$, where $T_s$ is the symbol time (or symbol duration)\footnote{To simplify the notation, signal equations in the passband are avoided and only the baseband representations are used.}. Next, the RF signal is propagated through the wireless channel and would be affected by large scale fading (pathloss and shadowing) and small scale fading. Small scale fading or simply fading is the main cause of rapid fluctuation in the signal strength. Hence, it is important to look at the mathematical model of a fading channel.

\subsection{Wireless Channel}
\label{subsec:ch-model-p2p}
In wireless communications, to avoid dealing with the complexity of electromagnetic equations, wireless channels are modelled with a linear time-varying system \cite{fundamental-digital}. Depending on the propagation delay, channels are divided to frequency selective and flat fading channels. In frequency selective channels, the propagation delay is larger than the symbol time whereas in flat-fading channels the delay spread is much less than the symbol time. The focus of this thesis is on \emph{flat-fading} channels, which are applicable for narrowband communication systems. For flat-fading channels, the channel impulse response is represented by one filter tap (or coefficient). In addition, since most of the processing is actually done at the baseband, the baseband representation of the channel coefficient is used. This coefficient is modelled as a complex random variable whose distribution depends on the nature of the radio propagation environment.

\subsubsection{Statistical Model}
Typical distributions for the baseband channel coefficient are Rayleigh, Rician, Nakagami, etc. In this thesis, the Rayleigh flat-fading model is adopted since it is a popular model for many applications such as mobile networks. Let $h[k]$ represent the channel coefficient in a Rayleigh flat-fading model at time index $k$. Then, $h[k]$ is modelled as a complex Gaussian random variable with zero mean and variance $\sigma_h^2$. In fact, the name ``Rayleigh" fading comes from the distribution of the envelope $\eta=|h[k]|$, which is a Rayleigh distribution:
\begin{equation}
\label{eq:Rayleigh-pdf}
f_{\eta}(\eta)= \frac{\eta}{\sigma_h^2} \exp\left( -\frac{\eta}{2\sigma_h^2}\right),\quad \eta\geq 0.
\end{equation}
Also, the related random variable $\lambda=|h[k]|^2$ is exponentially distributed with density
\begin{equation}
\label{eq:exp-pdf}
f_{\lambda}(\lambda)=\frac{1}{\sigma_h^2} \exp\left(-\frac{\lambda}{\sigma_h^2}\right),\quad \lambda \geq 0.
\end{equation}
The exponential random variable is mostly encountered in the performance analysis.

\subsubsection{Auto-Correlation}
Due to mobility of users, the fading channel also changes over time and thus the rate of channel variation has a significant impact on many aspects of a wireless communication system. A statistical quantity that models this variation in time is known as the channel auto-correlation function, $\varphi(n)$, defined as
\begin{equation}
\label{eq:ch-auto-corr}
\varphi(n)= \E \{ h[k] h^*[k+n] \}
\end{equation}
where $n$ is the time-distance between two channel coefficients at time indices $k$ and $k+n$, respectively. If the symbol duration is smaller than the period of time over which the fading process is correlated, the channel is called slow-fading. Otherwise it is fast-fading.

A popular model for the auto-correlation function of a flat-fading channel is Clark's or Jakes' model \cite{microwave-jake}, given as
\begin{equation}
\label{eq:clark-Jakes}
\varphi(n)=\sigma_h^2 J_0(2\pi f_D T_s n),
\end{equation}
where $J_0(\cdot)$ is the zeroth-order Bessel function of the first kind:
\begin{equation}
\label{eq:Bessel0}
J_0(x)=\frac{1}{\pi} \int \limits_{0}^{\pi} e^{j x \cos(\theta)} \dd \theta
\end{equation}
and $f_D$ and $T_s$ are the Doppler frequency and symbol duration, respectively. The Doppler frequency is caused by the mobility of users and determined as
$
\label{eq:doppler-f}
f_D=f_c v/c
$
where $f_c$ is the carrier frequency used by the communication system, $v$ is the velocity of the user and $c=3\times 10^{8}$ is the speed of light. Usually, the rate of the channel variation is shown with the product $f_DT_s$ and it is called the normalized Doppler frequency.

\begin{figure}[ht]
\centering
\begin{minipage}[b]{0.45\linewidth}
\psfrag {|h[k]|} [] [] [1.0] {$|h[k]|$}
\psfrag {time} [] [] [1.0] {time index, $k$}
\epsfig{figure={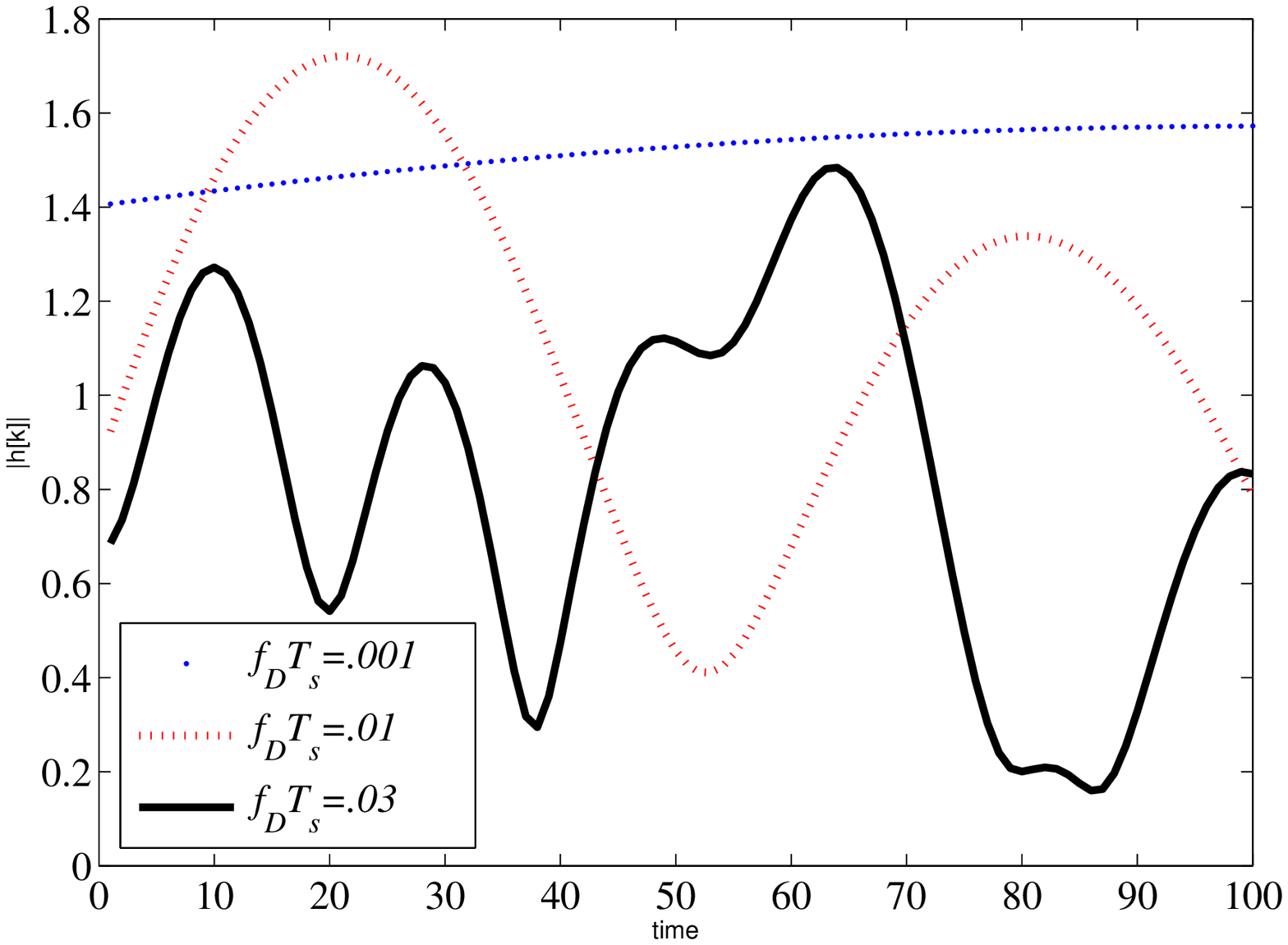},height=6cm,width=8cm}
%\label{fig:minipage1}
\end{minipage}
\quad
\begin{minipage}[b]{0.45\linewidth}
\psfrag {auto-correlation} [] [] [1.0] {$\varphi(n)$}
\psfrag {time} [] [] [1.0] {$n$}
\epsfig{figure={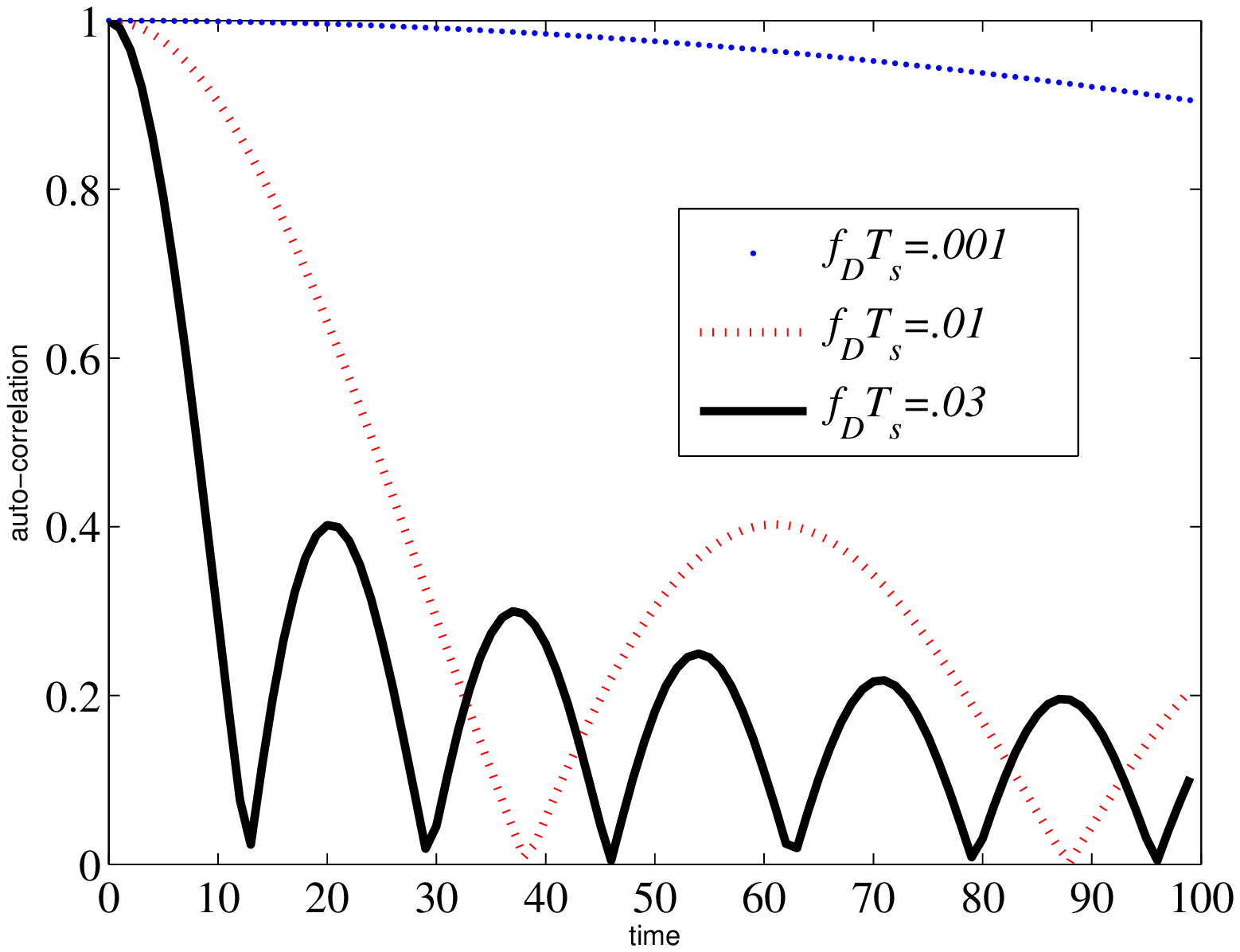},height=6cm,width=8cm}
%\label{fig:minipage2}
\end{minipage}
\caption{Evolution of channel coefficients and corresponding auto-correlation values of a Rayleigh flat-fading channel under different fading rates.}
\label{fig:ch-auto}
\end{figure}

To get more insights about the relationship between the channel variation and the normalized Doppler frequency, the evolution of the amplitude of a Rayleigh flat-fading channel coefficient over time, under different fade-rates, and also its auto-correlation function is plotted in Figure.~\ref{fig:ch-auto}. As can be seen from the figure, the variation rate of the channel is directly related to the normalized Doppler value and the auto-correlation function value. For $f_DT_s=.001$, the channel coefficients are approximately fixed over time, which is also related to the small slope of the corresponding auto-correlation plot. Such a channel is called slow-fading. For $f_DT_s=0.01$ and $0.03$ the auto-correlation values change faster and the channel variations are larger. Such channels would be called fast-fading. {The above values of the normalized Doppler frequency can be translated to different vehicle speeds of communication nodes in a typical wireless system. For example, in a system with carrier frequency $f_c=2$ GHz and symbol duration $T_s=0.1$ ms, the corresponding Doppler shifts would be around $f_D=f_DT_s/T_s=10,100,300$ Hz, which would correspond to the speeds of $v=cf_D/f_c=5,54,162$ km/hr, respectively ($c=3\times 10^{-8}$ m/s is the speed of light).}
%\begin{figure}[t]
%\psfrag {|h[k]|} [r] [] [1.0] {$|h[k]|$}
%\psfrag {time} [t] [] [1.0] {time index, $k$}
%\centerline{\epsfig{figure={figure/ch_fdts.eps},width=8.5cm}}
%\caption{Evolution of Rayleigh flat-fading channel coefficients over time under different fading rates.}
%\label{fig:ch-fdts}
%\end{figure}
%\begin{figure}[ht]
%\psfrag {auto-correlation} [l] [] [1.0] {$\varphi(n)$}
%\psfrag {time} [l] [] [1.0] {$n$}
%\centerline{\epsfig{figure={figure/ac_fdts.eps},width=8.5cm}}
%\caption{Auto-correlation function of a Rayleigh flat-fading channel under different fading rates.}
%\label{fig:ac-fdts}
%\end{figure}

\subsubsection{Auto-Regressive Models}
It is also important to mathematically model the evolution of channel coefficients in time using a time-series model. Such a model would be useful for simulation purpose or in performance analysis of a communication system operating over time-varying channels. The first-order AR model, AR(1), has been widely used and verified as a simple yet effective time-series model for Rayleigh flat-fading channels \cite{AR2-ch,AR1-ch}. It is given as
\begin{equation}
\label{eq:AR1}
h[k]= \alpha h[k-1]+\sqrt{1-\alpha^2} \; e[k]
\end{equation}
where $\alpha\leq 1$ is the auto-correlation value of the channel and $e[k]$ is distributed as $\CN(0,\sigma_h^2)$ and independent of $h[k-1]$. It is easy to see that with the above model, the distribution of $h[k]$ is complex Gaussian with zero mean and variance $\sigma_h^2$.

\subsubsection{Channel Simulation}
There are several methods to generate time-correlated channel coefficients in computer simulation. In this thesis, the simulation algorithm of \cite{ch-sim} is used. This simulation algorithm utilizes a sum-of-sinusoids method to generate time-correlated Rayleigh-faded channel coefficients. For instance, to generate $h[k]$:
\begin{gather}
\label{eq:qi_gen}
\nonumber
h[k]=\Re\{h[k]\}+j \Im\{h[k]\}, \\
\Re\{h[k]\}=\sqrt{\frac{2}{N_1}} \sum \limits_{n=1}^{N_1} \cos(2\pi f_{D} k \cos(a_n)+\phi_n) \\
\Im\{h[k]\}=\sqrt{\frac{2}{N_1}} \sum \limits_{n=1}^{N_1} \cos(2\pi f_{D} k \sin(a_n)+\psi_n)\\ \nonumber
a_n=\frac{2\pi n-\pi +\theta}{4N_1},\quad n=1,2,\cdots,N_1
\end{gather}
where $\phi_n,\psi_n,$ and $\theta$ are statistically independent and uniformly distributed on $[-\pi,\pi)$ for all $n$ and $N_1$ is the number of multipaths chosen arbitrarily large enough for an accurate model \cite{ch-sim}. The input to the simulation algorithm is the normalized Doppler frequency of the channels, which is a function of the velocity of users and the symbol duration (i.e., the transmission rte). For the same transmission rate, a higher velocity causes a higher fade-rate and thus less correlation between channel coefficients. Therefore, by changing the Doppler values, various fading scenarios from slow-fading to fast-fading channels can be simulated.

\subsection{Receiver and Detection}
The structure of the receiver is illustrated in Figure~\ref{fig:txrx-p2p}. The signal from the receive (Rx) antenna is added by AWGN noise in the passband. Then, the received RF signal $\tilde{y}(t)$ is down-converted by a local oscillator\footnote{It is assumed that the transmitter and receiver are synchronized. In practice, propagation over a long distance leads to a delay in the received signal, which can be estimated and compensated for by a synchronization algorithm.} and passed through the match filter to obtain the continuous-time baseband signal $y_b(t)$. The signal $y_b(t)$ is then sampled at the symbol rate $1/T_s$ to obtain the discrete-time baseband signal as
\begin{equation}
\label{eq:y-p2p}
y[k]=\sqrt{P} h[k] s[k]+w[k],
\end{equation}
where $P$ is the transmit power per symbol, $h[k]\sim \CN(0,\sigma_h^2)$ is the channel coefficient and $w[k]\sim \CN(0,N_0)$ is the discrete-time white Gaussian noise component at the receiver. The signal-to-noise ratio (SNR), defined as the ratio of the average received signal power per (complex) symbol time to noise power per (complex) symbol time, is given as
\begin{equation}
\label{eq:SNR}
\SNR=\frac{P\sigma_h^2}{N_0}.
\end{equation}

\subsubsection{Coherent Detection}
In the case of coherent detection, the channel coefficient is obtained by the channel estimator block. Assuming perfect channel estimate, coherent detection of $M$-PSK can be performed on a symbol by symbol basis by multiplying $y[k]$ and $h^*[k]$ as
\begin{equation}
\label{eq:coh-p2p}
h^*[k] y[k]= |h[k]|^2 v[k] + h^*[k] w[k]
\end{equation}
where $s[k]=v[k]$ when the differential encoder is bypassed in a coherent system. Then the maximum likelihood (ML) detection of the transmitted symbol can be obtained as
\begin{equation}
\label{eq:coh-det-p2p}
\hat{v}[k]= \arg \min \limits_{v[k]\in \Vc} |h^*[k] y[k]- v[k]|.
\end{equation}
The error probability of such a system at high SNR can be shown to behave as \cite{fundamental-digital}
\begin{equation}
\label{eq:Pe_coh-p2p}
P_e\propto \frac{1}{\SNR}.
\end{equation}
The above expression shows that the achieved diversity order, which is defined as \cite{fundamental-digital},
\begin{equation}
\label{eq:diversity}
G_d=-\lim \limits_{\SNR \rightarrow \infty} \frac{\log(P_e(\SNR))}{\log(\SNR)}
\end{equation}
is equal one for this system.

\subsubsection{Two-Symbol Non-Coherent Detection}
When differential encoding is performed at the transmitter, non-coherent detection can be applied without requiring the knowledge of $h[k]$ (i.e., no channel estimation is required). Assuming the channel coefficients stay approximately constant for two successive symbols i.e., $h[k]\approx h[k-1]$, one has
\begin{equation}
\label{eq:yk-p2p}
y[k]=v[k] y[k-1]+ \tilde{w}[k]
\end{equation}
where $\tilde{w}[k]=w[k]-s[k] w[k-1]$ is the equivalent noise at the output of the detector, which is also a complex Gaussian random variable with zero mean and variance $2N_0$.

Hence, based on the observations $y[k-1]$ and $y[k]$, differential non-coherent detection of $M$-DPSK can be performed by computing the following decision variable
\begin{equation}
\label{eq:decision}
\zeta=y^*[k-1] y[k]=v[k] |y[k-1]|^2 +y^*[k-1]\tilde{w}[k].
\end{equation}
From the above expression it can be seen that the ML detection of the transmitted symbol at time $k$, can be obtained as \cite{Dig-ITC-porakis}:
\begin{equation}
\label{eq:detection}
\hat{v}[k]=\arg \min \limits_{v[k]\in \Vc} |\zeta-v[k]|
\end{equation}
which shows that no channel information is needed for the detection. It is also well known that, because the noise variance is doubled, about 3 dB performance loss exists between coherent and non-coherent detections in slow-fading environment \cite{Dig-ITC-porakis,DigComFad-Simon}. However, the two-symbol non-coherent detection suffers a larger performance loss in time-varying channels \cite{DigComFad-Simon}.

\subsubsection{Multiple-Symbol Differential Detection}
To overcome the performance limitations of two-symbol detection, multiple-symbol differential (MSD) detection \cite{msdd-div2,msdd_fung,MSDSD-L} has been proposed for point-to-point communications. In MSD detection, blocks of $N>2$ received symbols are jointly processed to decide on $N-1$ data symbols. Let's collect $N$ received symbols into vector
$\yb=[\;y[1],\cdots,y[N]\;]^t $, which can be written as
\begin{equation}
\label{eq:MSD-p2p}
\yb=\sqrt{P} \dgs \h +\w
\end{equation}
where $\s=[\; s[1],\cdots,s[N] \; ]^t$, $\h=[\;h[1],\cdots,h[N] \;]^t$ and $\w=[\;w[1],\cdots,w[N] \;]^t$. The maximum likelihood MSD detection decision rule reads \cite{MSDSD-L}
\begin{equation}
\label{eq:ML-MSDD}
\hat{\s}=\arg \min \limits_{\s \in \CoN} \left\lbrace \left( \dgy \s^* \right)^{H} \C^{-1} \dgy \s^* \right\rbrace
\end{equation}
where $\C=\E\{\h \h^H\}+N_0 \I_N$ and $\E\{\h\h^H\}=\toep\{1,\varphi(1),\cdots,\varphi(N-1)\}$ and $\varphi(\cdot)$ is the auto-correlation function of the channel. Using the Choleskey decomposition $\C^{-1}=\Lb \Lb^H$, the decision rule can be further simplified to
\begin{equation}
\label{eq:ML-MSDD2}
\hat{\s}=\arg\min\limits_{\s \in \CoN}\left\lbrace \| \U \s \|^2 \right\rbrace
\end{equation}
where $\U=\left( \Lb^H \dgy \right)^*$ is an upper triangular matrix. The above decision rule can be solved by the sphere decoding algorithm \cite{MSDSD-L} with low complexity.

\section{Multiple Antenna and Spatial Diversity}
As described in the previous section, the diversity order of a single-antenna communication system over fading channels is equal to one. To improve the performance of communication systems over fading channels, diversity techniques such as time, frequency or spatial diversity can be used. Here, spatial diversity is considered as it provides better bandwidth efficiency. Spatial diversity is an effective method to combat detrimental effects in wireless fading channels by using multiple antennas at the transmitter and/or the receiver. Spatial diversity can be classified as receive diversity, transmit diversity and transmit and receive diversity.

\subsection{Receive Diversity}
\begin{figure}[t]
\psfrag {h1} [l] [] [1.0] {$h_1[k]$}
\psfrag {h2} [l] [] [1.0] {$h_2[k]$}
\psfrag {hr} [l] [] [1.0] {$h_R[k]$}
\psfrag {y1k} [] [] [1.0] {$y_1[k]$}
\psfrag {y2k} [] [] [1.0] {$y_2[k]$}
%\psfrag {yik} [] [] [1.0] {$\cdots$}
\psfrag {yRk} [] [] [1.0] {$y_R[k]$}
\psfrag {sk} [] [] [1.0] {$s[k]$}
\psfrag {antenna1} [] [] [1.0] {Rx 1}
\psfrag {antenna2} [] [] [1.0] {Rx 2}
\psfrag {antennaR} [] [] [1.0] {Rx $R$}
\psfrag {Tx} [] [] [1.0] {Tx}
%\psfrag {time} [] [] [1.0] {}
\centerline{\epsfig{figure={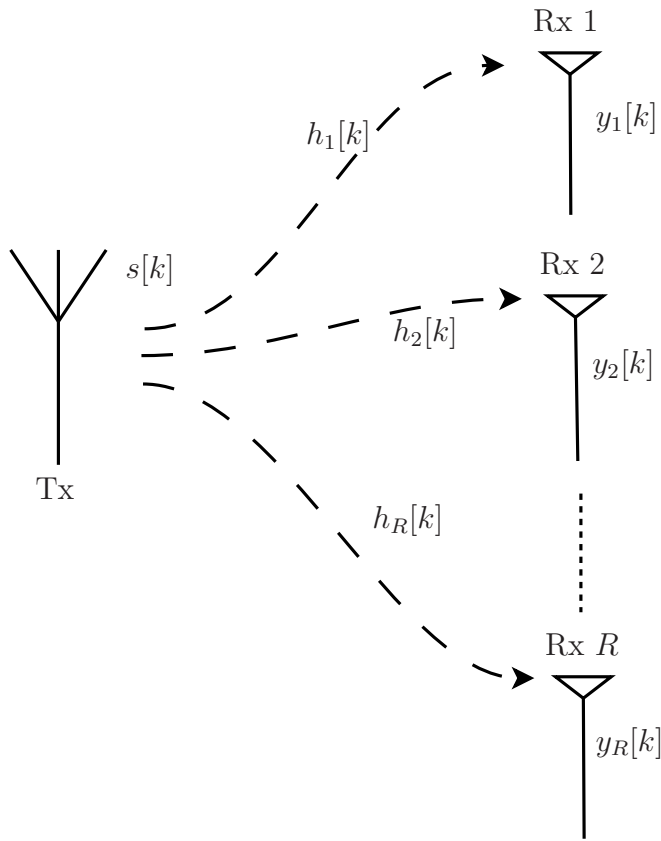},width=8.5cm}}
\caption{Spatial diversity using multiple receive antennas.}
\label{fig:rx_div}
\end{figure}

Receive diversity is simply achieved by employing multiple antennas at the receiver as depicted in Figure~\ref{fig:rx_div}. The distance between the receiver antennas is such that the transmitted symbol experiences an independent fading value in each link. Assume that symbol $s[k]$ is transmitted from the transmit (Tx) antenna at time $k$. The received signals at Rx antennas are given as
\begin{equation}
\label{eq:Rx-div2}
y_i[k]=h_i[k] s[k]+w_i[k], \quad i=1,\cdots, R,
\end{equation}
where the $i$-th channel $h_i[k]\sim \CN(0,\sigma_h^2)$ and the noise $w_i[k]\sim \CN(0,N_0)$ are independent across the antennas. To achieve a diversity, the received signals need to be combined using some combining technique. There are two main combining methods that are considered  as follows.

\subsubsection{Maximum Ratio Combining (MRC)}
The MRC method weights the received signal in each branch in proportion to the signal strength and also aligns the phases of the signals to maximize the output SNR \cite{fundamental-digital}. The sufficient statistic is given as
\begin{equation}
\label{MRC1-p2p}
\zeta= \sum \limits_{i=1}^{R} a_i \zeta_i
\end{equation}
where $a_i$ are the combining weights. For coherent detection, $v[k]=s[k]$ and the quantities $\zeta_i$ are
\begin{equation}
\label{MRC2-p2p}
\zeta_i= h_i^*[k]y_i[k]=|h_i[k]|^2 v[k]+h_i^*[k]w_i[k], \quad i=1,\cdots,R.
\end{equation}
On the other hand, for non-coherent detection one has
\begin{equation}
\label{MRC-diff-p2p}
\zeta_i= y_i^*[k-1]y_i[k]= |y_i[k-1]|^2 v[k]+ y_i^*[k-1] \tilde{w}_i[k],
\end{equation}
where $\tilde{w}_i[k]=w_i[k]-s[k]w_i[k-1]$.
If all links have the same average signal strength, one has $a_i=1,~i=1,\cdots,R.$
The output of the combiner is then used to detect the transmitted signal as
\begin{equation}
\label{eq:mrc-detect}
\hat{v}[k]=\arg \min \limits_{v[k]\in \Vc} |\zeta-v[k]|.
\end{equation}

It has been shown that the error probability of the MRC scheme behaves as \cite{fundamental-digital}
\begin{equation}
\label{eq:Pe-MRC}
P_e \propto \frac{1}{(\mathrm{SNR})^R}
\end{equation}
which reveals that the achieved diversity order is $R$. Also, non-coherent detection performs around 3 dB worse than its coherent version in slow-fading channels \cite{DigComFad-Simon}.

\subsubsection{Selection Combining}
Another important combining method is selection combining (SC). In the SC method, the decision statistics of each link is computed and compared to choose the link with the highest SNR. For coherent detection, this requires estimation of the channel coefficients of all links. In practice, the link with the highest amplitude of the received signal is chosen instead and then only the channel in the chosen link is estimated for coherent detection. Similarly, for non-coherent detection, the link with the highest amplitude of the decision variable is chosen. The output of the combiner is therefore
\begin{equation}
\label{eq:zeta-sc-p2p}
\zeta = \arg \max \limits_{\zeta_{i,i=1,\cdots,R}} \{ |\zeta_i| \}
\end{equation}
where
\begin{equation}
\label{eq:zi}
\zeta_i=
\left\lbrace
\begin{matrix}
y_i[k], & \mbox{for coherent detection} \\
y_i^*[k-1] y_i[k], & \mbox{for non-coherent detection}.
\end{matrix}
\right.
\end{equation}
The output of the combiner can be used to detect the transmitted signal using \eqref{eq:coh-det-p2p} and \eqref{eq:mrc-detect}, for coherent and non-coherent, respectively. The SC method is simpler than the MRC method and can also provide a diversity of $R$.  However, its performance is inferior to that of the MRC method \cite{stc-jafar}.

\subsection{Transmit Diversity}
\begin{figure}[t]
\psfrag {h1} [l] [] [1.0] {$h_1$}
\psfrag {h2} [l] [] [1.0] {$h_2$}
\psfrag {hr} [l] [] [1.0] {$\qquad h_R$}
\psfrag {y1k} [] [] [1.0] {$y[k]$}
\psfrag {y2k} [] [] [.9] {$y[k+1]$}
\psfrag {yik} [] [] [1.0] {$\cdots$}
\psfrag {yRk} [] [] [.9] {$y[k+R]$}
\psfrag {sk} [] [] [1.0] {$x$}
\psfrag {antenna1} [] [] [1.0] {Tx 1}
\psfrag {antenna2} [] [] [1.0] {Tx 2}
\psfrag {antennaR} [] [] [1.0] {Tx $R$}
\psfrag {rx antenna} [] [] [1.0] {Rx}
\psfrag {time} [] [] [1.0] {}
\centerline{\epsfig{figure={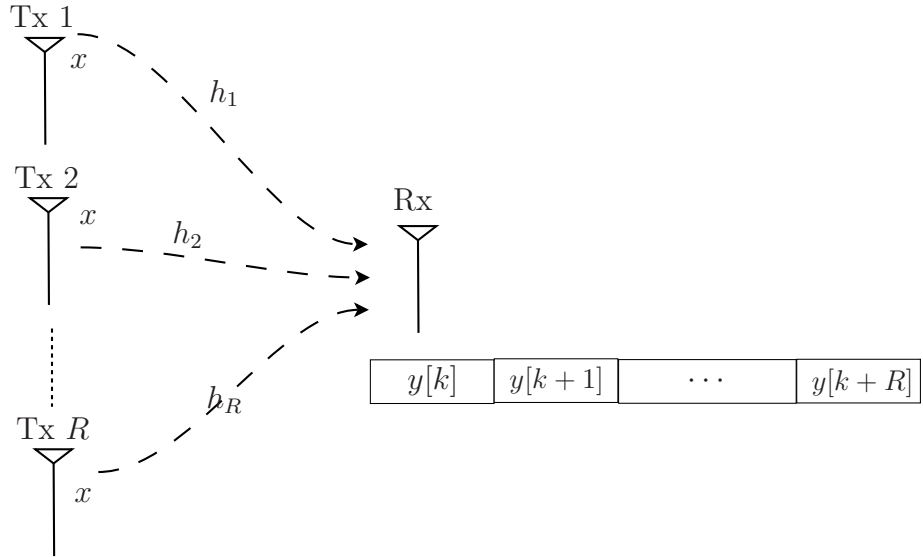},width=12cm}}
\caption{Spatial diversity using multiple transmit antennas and repetition-based code.}
\label{fig:tx_div}
\end{figure}

Providing a spatial diversity by using multiple antennas at the transmitter is depicted in Figure~\ref{fig:tx_div}. There are $R$ transmit antennas and one receive antenna. There are two main methods to send signals from Tx antennas to Rx antenna that achieve a diversity of $R$. They are discussed next.

\subsubsection{Repetition-Based Transmission}
First, consider a time division duplex (TDD) transmission where all antennas send the same symbol in different time slots to the receiver. In any time, only one antenna is turned on and the rest are silent. This is similar to repetition code \cite{fundamental-digital} and hence not efficient in terms of spectral efficiency. However, this method is similar to one of the main cooperative strategies that employed in this thesis.

Let, $x$ be the transmitted symbol and $y[k+i-1]$ be the received symbol from the $i$th transmit antenna. It can be written as
\begin{equation}
\label{eq:y-div-p2p}
y[k+i-1]= h_i x + w[k+i-1], \quad i=1,\cdots, R
\end{equation}
where $h_i \sim \CN(0,\sigma_h^2)$ is the channel gain from the $i$th transmit antenna to the receiver and $w[k+i-1]\sim \CN (0,N_0)$ is the noise component in each time slot and $R$ is the number of transmit antennas. As can be seen, there are $R$ copies of the transmitted symbol perturbed by independent fading values and noises. Similar to receive diversity, the received signals in different time slots can be combined using the MRC or SC method to achieve diversity.

On the other hand, instead of the repetition-based transmission, space-time codes can be utilized in multiple-antenna systems to achieve a better spectral efficiency.

\subsubsection{Space-Time Coding}
A space-time code (STC) is essentially a rule that maps the input bits to the transmitted symbols for a multiple antennas system. With space-time coding, symbols can be transmitted simultaneously from different antennas and hence a higher data rate can be achieved.

\begin{figure}[t]
\psfrag {h1} [] [] [1.0] {$h_1$}
\psfrag {h2} [] [] [1.0] {$h_2$}
\psfrag {antenna1} [] [] [1.0] {Tx 1}
\psfrag {antenna2} [] [] [1.0] {Tx 2}
\psfrag {rx antenna} [] [] [1.0] {RX}
\psfrag {s1} [] [] [1.0] {$s_1$}
\psfrag {s2} [] [] [1.0] {$-s_2^*$}
\psfrag {s3} [] [] [1.0] {$s_2$}
\psfrag {s4} [] [] [1.0] {$s_1^*$}
\psfrag {y1} [] [] [1.0] {$y_1$}
\psfrag {y2} [] [] [1.0] {$y_2$}
\centerline{\epsfig{figure={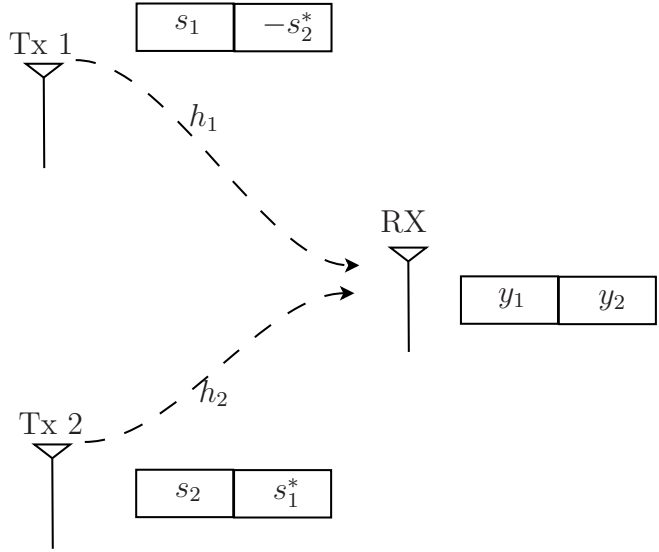},width=8.5cm}}
\caption{Block diagram of the Alamouti orthogonal space-time block coding scheme.}
\label{fig:alamouti}
\end{figure}

As an example, consider a system with two transmit and one receive antennas. For such a system, there is a very famous STC known as Alamouti orthogonal space-time block code \cite{STC-Alam}. The code is described in Figure~\ref{fig:alamouti}. A modulation scheme such as PAM, PSK, etc., with $M$ symbols is used to map $\log_2 M$ bits to a symbol. Let $s_1$ and $s_2$ be two symbols to be transmitted. In the first time slot, the transmitter sends $s_1$ from antenna 1 and $s_2$ from antenna 2. Then, in the second time slot, it transmits $-s_2^*$ and $s_1^*$ from antenna 1 and antenna 2, respectively \cite{stc-jafar}. The transmitted codeword is expressed as
\begin{equation}
\label{eq:C-Alam}
\Sb=
\begin{pmatrix}
s_1 & -s_2^* \\
s_2 & s_1^*
\end{pmatrix}.
\end{equation}
Here, it is assumed that the channel gains are quasi-static (i.e., they are constant during two time slots) and hence for simplicity the time index is dropped for all symbols. Also, transmission of one codeword during two time slots is called one block transmission.

Then, the received signal at the receiver over two time slots is
\begin{equation}
\label{eq:y-Alam}
\yb=[y_1,y_2]^t=\sqrt{P} \Sb \h +\w
\end{equation}
where $\h=[h_1,h_2]^t\sim \CN(\0,\I_2)$ is the channel vector, $\w=[w_1,w_2]^t\sim \CN(\0,N_0\I_2)$ is the noise vector and $P$ is the average received power per symbol.

After some simple manipulations, \eqref{eq:y-Alam} can be re-written as
\begin{equation}
\label{almaouti-p2p-manu}
\begin{bmatrix}
y_1 \\
y_2^*
\end{bmatrix}=
\begin{bmatrix}
h_1 & h_2 \\
h_2^* & -h_1^*
\end{bmatrix}
\begin{bmatrix}
s_1 \\
s_2
\end{bmatrix}
+
\begin{bmatrix}
w_1 \\
w_2^*
\end{bmatrix}.
\end{equation}
The very important property that follows from the specific structure of the Alamouti STC is that the columns of the above square channel matrix are orthogonal, regardless of the actual values of the channel coefficients. Hence, with known channel information at the receiver, the transmitted symbols can be separately detected by projecting $[y_1,y_2^*]^t$ onto each of the two columns to obtain the sufficient statistics as follows:
\begin{align}
\label{eq:suff-stat-alam}
h_1^*y_1+h_2y_2^*=(|h_1|^2+|h_2|^2)s_1+\tilde{w}_1 \\
h_2^*y_1-h_1y_2^*=(|h_1|^2+|h_2|^2)s_2+\tilde{w}_2
\end{align}
where $\tilde{w}_1=h_1^*w_1+h_2w_2^*$ and $\tilde{w}_2=h_2^*w_1-h_1w_2^*$ are effective noise components. Based on the above expressions, the information symbols $s_1$ and $s_2$ can be detected in the same way as for single-antenna point-to-point system, albeit with more favourable effective channel gain of $|h_1|^2+|h_2|^2$. In fact, such an effective channel gain yields a diversity order of 2.

The above discussion also shows that the Alamouti code can send one symbol per one time slot, while the TDD scheme sends only one symbol per two time slots for $R=2$ transmit antennas. Thus, the Alamouti code provides twice the data rate over the TDD scheme while a full diversity of two can be achieved by both schemes.

\subsubsection{Differential Space-Time Coding}
To avoid channel estimation at the receiver, similar to the differential PSK described for single-antenna system, differential unitary space-time codes (D-USTC) have been investigated in \cite{D-USTC-HOCH,D-SPT-Hughes} for multiple-antenna systems.
In D-USTC, information symbols at block index $k$ are encoded as an unitary matrix $\V[k] \in \Vc$ where $\Vc=\{\V_l| \V_l^* \V_l=\V_l\V_l^*=\I_R,\; l=1,\cdots,L \}$. Here $R$ and $L$ are the number of transmit antennas and the total number of codewords, respectively. Designing these unitary matrices has been studied in \cite{D-USTC-HOCH,D-SPT-Hughes}. The codeword is then differentially encoded as
\begin{equation}
\label{eq:S-D-USTC}
\Sb[k]=\V[k] \Sb[k-1],\quad \Sb[0]=\I_R,
\end{equation}
which is also a unitary matrix. The encoded codeword is then transmitted from multiple antennas to the receiver. The vector form of the received signal at the receiver at
block index $k$ is given as
\begin{equation}
\label{eq:y-ustc}
\yb[k]= \sqrt{P} \Sb[k] \h[k]+\w[k].
\end{equation}

\subsubsection*{Two-Codeword Non-Coherent Detection}
Also, similar to DPSK, two-symbol non-coherent detection can be applied to D-USTC used in the multiple-antenna system. In the case of slow-fading, it is assumed that the channel vector is approximately constant for two block transmissions, i.e.,
\begin{equation}
\label{eq:hk_hk-1_MISO}
\h[k]\approx \h[k-1].
\end{equation}
Substituting \eqref{eq:S-D-USTC} and \eqref{eq:hk_hk-1_MISO} into \eqref{eq:y-ustc} gives
\begin{equation}
\label{eq:cdd-ustc}
\yb[k]=\V[k] \yb[k-1]+ \tilde{\w}[k]
\end{equation}
where
\begin{equation}
\label{eq:w_tilde-ustc}
\tilde{\w}[k]=\w[k]-\V[k] \w[k-1]
\end{equation}
is the equivalent noise vector at the detector output, which is $\CN(\0,2N_0\I_R)$.
Based on \eqref{eq:cdd-ustc}, non-coherent detection of the transmitted codeword is as folows \cite{D-USTC-HOCH,D-SPT-Hughes}:
\begin{equation}
\label{eq:D-USTC-detect}
\hat{\V}[k]= \arg \min \limits_{\V[k]\in \Vc} \|\yb[k]-\V[k] \yb[k-1] \|.
\end{equation}
Again, around 3 dB performance degradation exists between coherent and non-coherent detections \cite{D-USTC-HOCH,D-SPT-Hughes} over slow-fading channels. The performance degradation is, however, larger for time-varying channels \cite{ustc-tv-peel}.

\subsubsection*{Multiple-Symbol Non-Coherent Detection}
Similar to multiple-symbol differential (MSD) decoding \cite{msdd-div2,MSDSD-L,msdd_fung} of DPSK signals in single-antenna systems, the MSD detection has been investigated for D-USTC in MIMO channels in \cite{MSDUSTC-P}. Let the $N$ received symbols be collected in vector $\bar{\yb}=[\yb^t[1],\cdots,\yb^t[N]]^t$, which can be expressed as
\begin{equation}
\label{eq:D-USTC-MS}
\bar{\yb}= \sqrt{P} \bar{\Sb} \bar{\h}+\bar{\w}
\end{equation}
where $\bar{\Sb}=\diag\{ \Sb[1],\cdots,\Sb[N]\; \}$ is a $NR\times NR$ block-diagonal matrix and $\bar{\h}=[\; \h^t[1],\cdots,\h^t[N]\;]^t$ and $\bar{\w}=[\; \w^t[1],\cdots,\w^t[N]\;]^t$.

The maximum likelihood MSD detection processes blocks of $N$ consecutively received symbols to find estimates for $(N-1)$ codewords $\bar{\V}=\diag\{ \V[1],\cdots,\V[N-1] \}$ which correspond to $N$ transmit codewords in $\bar{\Sb}$. The ML MSD detection rule can be written as \cite{MSDUSTC-P}
\begin{equation}
\label{eq:ML-MSD-USTC}
\hat{\bar{\V}}=\arg \min \limits_{\bar{\V}\in \Vc^{N-1}} \{  \bar{\yb}^H \bar{\Sb} (\C^{-1}\otimes \I_R) \bar{\Sb}^H \bar{\yb} \},
\end{equation}
where $\C=\R_{\h}+N_0 \I_N$ and $\R_{\h}$ is given as
\begin{equation}
\label{eq:Rh}
\R_{\h}=\toep\{\varphi(0),\varphi(R),\cdots,\varphi((N-1)R) \}
\end{equation}
and $\varphi(\cdot)$ is the auto-correlation function. Since the complexity of the ML MSD detection grows exponentially with $N$, a tree-search decoding algorithm (i.e., sphere decoding) has been developed in \cite{MSDUSTC-P} to solve the minimization of \eqref{eq:ML-MSD-USTC} with low complexity.

\section{Summary}
In this chapter, point-to-point communication systems using single antenna and multiple antennas were introduced. For single-antenna systems, the structures of the transmitter, channel fading model and receiver were described. For multiple-antenna systems, receive diversity using two important combining techniques (MRC and SC)  and also transmit diversity using repetition coding and space-time coding were presented. Also, differential modulation and non-coherent detection were discussed for both cases. It was pointed out that 3 dB performance loss exists between coherent and non-coherent detections in slow-fading channels and that the performance degradation would be larger in fast time-varying channels. Multiple-symbol differential detection is also introduced for both scenarios to mitigate performance degradation in fast time-varying channels.

In the next chapter, these fundamental concepts will be extended and applied to relay networks. Important topologies, relay protocols and strategies will be introduced. Specifically, amplify-and-forward relaying together with repetition-based relaying and distributed space-time coding will be elaborated.

%\addcontentsline{toc}{chapter}{References}
%\bibliographystyle{IEEEbib}
%\bibliography{H:/latex/references}

\chapter{Cooperative Communications}
\label{ch:coop}
In the previous chapter, point-to-point communications was introduced. It was explained that the effect of multipath fading can be mitigated using diversity techniques. Specifically, spatial diversity using multiple antennas at either transmitter or receiver was elaborated. However, in many wireless applications, such as cellular networks, WLAN, and ad-hoc sensor networks, deploying multiple antennas is not feasible. This is mainly due to the size and weight restrictions of these applications. In addition, due to power limitation of wireless systems, users at far locations or on the borders of wireless cells, suffer from coverage limit and interference from other neighbour cells. Fortunately, providing spatial diversity became feasible in these applications thanks to the work of Sendonaris et al. on cooperative communications\cite{user-coop1,user-coop2}.

In this chapter, cooperative or relay communications are presented as a mean to overcome the above mentioned limitations of deploying multiple antennas at one communication terminal. The canonical relay network topologies, relay protocols and cooperative strategies are introduced. The background in this chapter is useful for better understanding of the main contributions in subsequent chapters.

%In this chapter, we may interchangeably use Source or the source and Destination or the destination as the origin and the end of a signal, respectively.

\section{Overview and Topologies of Relay Networks}
\label{sec:archs}
Due to non-directional propagation of electromagnetic waves, all users in a network are able to receive signals from other users. In cooperative communications, users act as relays that receive and process signals from Source and forward the results to Destination. Depending on the availability of the direct channel from the source to the destination or the number of relays in the network, various topologies can be constructed. Moreover, relay networks can be distinguished by the protocol that the relays use to process the received signals from Source or by the strategy of cooperation. In the following, the main topologies, protocols and strategies, that will be employed in the next chapters, are introduced.

\begin{figure}
\label{fig:dh}
\psfrag {h1} {$h_{\sr}$}
\psfrag {h2} {$h_{\rd}$}
\psfrag {Source}[][][1] {Source}
\psfrag {Relay}[][][1] {Relay}
\psfrag {Destination}[][][1] {Destination}
\centerline{\epsfig{figure={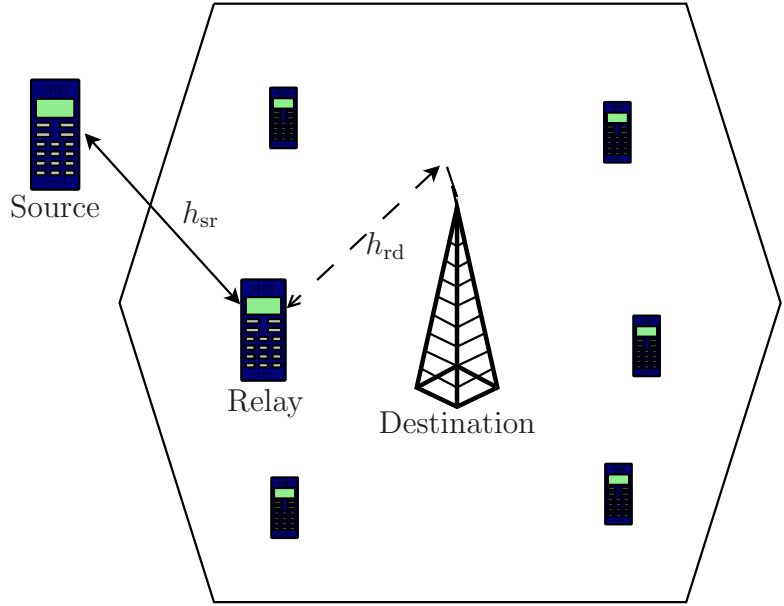},height=8cm,width=10cm}}
\caption{Single-branch dual-hop relaying without direct link for coverage extension.}
\end{figure}
\subsubsection*{{Single-Branch Dual-Hop Relaying Without Direct Link}}
Depending on the propagation conditions, the direct link between the source and the destination may not be sufficiently strong to facilitate data transmission. For instance, a mobile user at the cell edge of a cellular network would receive a weak signal of interest from the base station and experience a coverage limit. In this situation, a neighbour user can act as a relay to bridge the source to the destination, creating a canonical relaying topology, namely single branch dual-hop (DH) relaying without direct link. Figure~\ref{fig:dh} depicts a dual-hop relaying system in a cellular network. As can be seen, a mobile user (Source) communicates with the base station (Destination) via another mobile user (Relay). Dual-hop relaying has been studied in the literature as a solution to overcome coverage limits of many wireless applications such as cellular networks, WLAN, wireless sensor networks, etc \cite{coop-dohler}. In particular, dual-hop relaying has been developed in WLAN IEEE 802.11 technology and WiMAX IEEE 802.16j and has been standardized in 3GPP LTE \cite{coop-dohler}. Moreover, in addition to its own importance, dual-hop relaying is the backbone of other topologies with multi-branches to be studied in the next chapters. In Chapter~\ref{ch:dh}, a dual-hop relay network will be studied in details.
\begin{figure}
\psfrag {Source} [][][1] {Source}
\psfrag {Relay1} [][][1]{Relay 1}
\psfrag {Relay2} [][][1]{Relay 2}
\psfrag {RelayR} [][][1]{Relay R}
\psfrag {Destination} [][][1]{Destination}
\psfrag {f1} {$h_{\sr_1}$}
\psfrag {g1} {$h_{\rd_1}$}
\psfrag {f2} {$h_{\sr_2}$}
\psfrag {g2} {$h_{\rd_2}$}
\psfrag {f3} {$h_{\sr_R}$}
\psfrag {g3} {$h_{\rd_R}$}
\centerline{\epsfig{figure={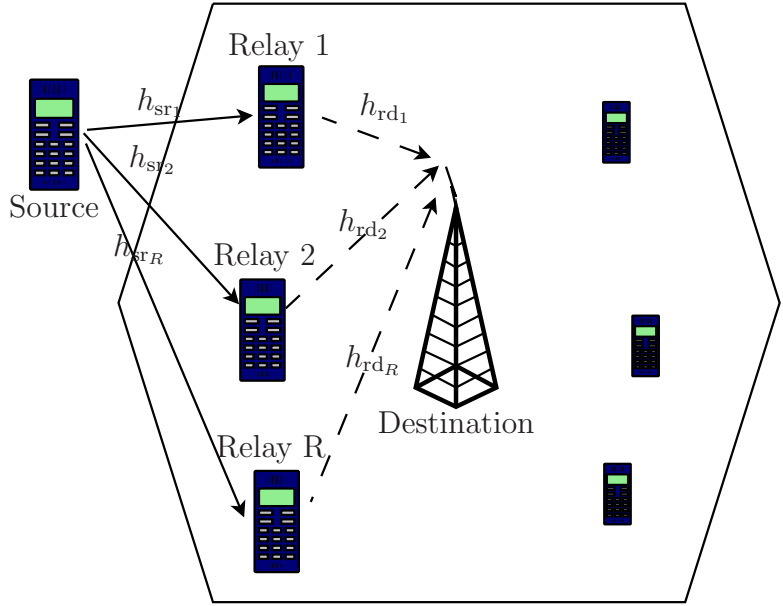},height=8cm,width=10cm}}
\caption{Multi-branch dual-hop relaying without direct link for coverage extension and diversity improvement.}
\label{fig:mbdh}
\end{figure}
\subsubsection*{{Multi-Branch Dual-Hop Relaying Without Direct Link}}
Single-branch dual-hop relaying can be extended to multi-branch dual-hop relaying if there are more relays willing to cooperate. This is depicted in Figure~\ref{fig:mbdh}. In this topology, the user experiencing coverage limit can benefit from both coverage extension and diversity improvement with the help of other users. The maximum achievable diversity in this topology equals $R$, the number of relays.
In Chapter~\ref{ch:dstc} a multi-branch dual-hop relaying without direct link is studied.

\begin{figure}
\psfrag {Source} [][][1] {Source}
\psfrag {Relay1} [][][1]{Relay}
\psfrag {Destination} [][][1]{Destination}
\psfrag {h0} {$h_{\sd}$}
\psfrag {f1} {$h_{\sr}$}
\psfrag {g1} {$h_{\rd}$}
\centerline{\epsfig{figure={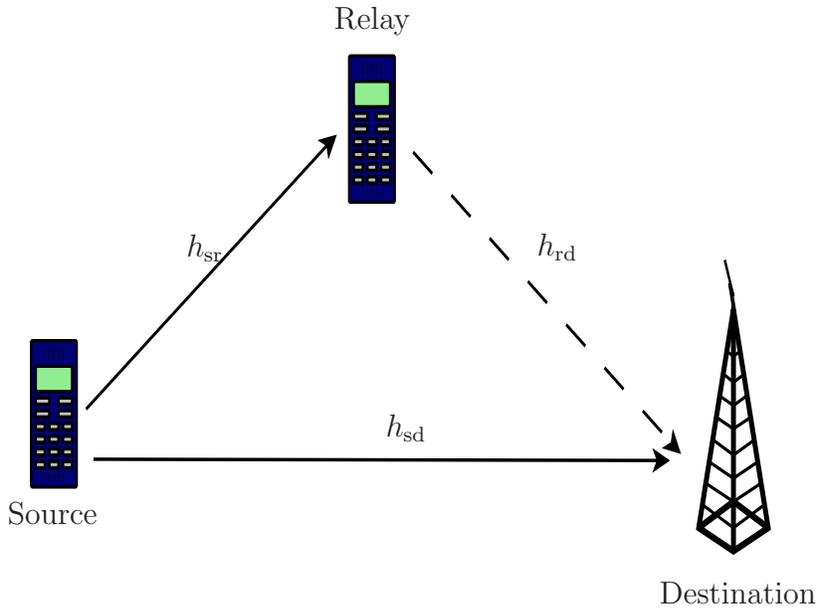},height=8cm,width=10cm}}
\caption{Single-branch dual-hop relaying with direct link.}
\label{fig:dhpd}
\end{figure}
\subsubsection*{{Single-Branch Dual-Hop Relaying with Direct Link}}
In case that the received signal from the direct link is sufficiently strong, it can be used to improve the diversity gain of the network. Figure~\ref{fig:dhpd} depicts a relay network with a direct link and a single branch dual-hop relaying link. The maximum achievable spatial diversity for this network is two. This architecture has been examined in several studies as one of the important relay topologies. This topology is also considered in Chapters~\ref{ch:mnode},\ref{ch:sc},\ref{ch:sc_tv} of this thesis.

\subsubsection*{{Multi-Branch Dual-Hop Relaying with Direct Link}}
\begin{figure}
\psfrag {Source}[][][1] {Source}
\psfrag {Relay} [][][1] {Relay}
\psfrag {Destination} [][][1] {Destination}
\psfrag {Relay1} [][][1] {Relay 1}
\psfrag {Relay2} [][][1] {Relay 2}
\psfrag {RelayR} [][][1] {Relay R}
\psfrag {f1} {$h_{{\sr}_1}$}
\psfrag {f2} {$h_{{\sr}_2}$}
\psfrag {fR} {$h_{{\sr}_R}$}
\psfrag {g1} {$h_{{\rd}_1}$}
\psfrag {g2} {$h_{{\rd}_2}$}
\psfrag {gR} {$h_{{\rd}_R}$}
\psfrag {h0} {$h_{\sd}$}
\centerline{\epsfig{figure={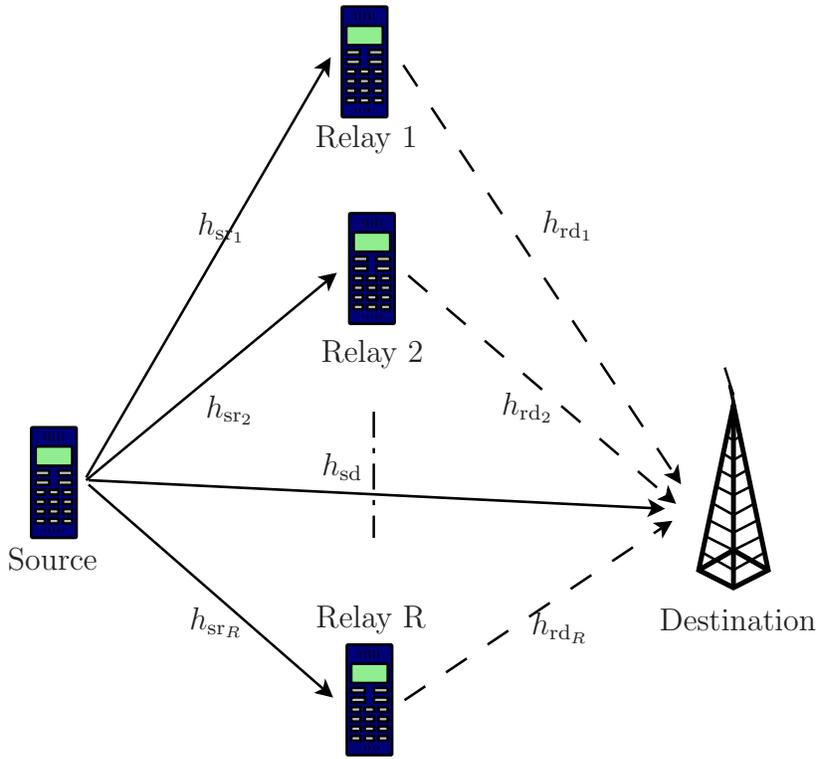},height=10cm,width=10cm}}
\caption{Multi-branch dual-hop relaying with direct link.}
\label{fig:mbpd}
\end{figure}

Single-branch relaying with direct link can be extended to multi-branch relaying with direct link if there are more relays in the network willing to help. In this way, additional dual-hop branches can be constructed to get a higher cooperative diversity, as depicted in Figure~\ref{fig:mbpd}. The maximum achievable spatial diversity of multi-branch relay networks with direct link is $R+1$, where $R$ is the number of relays. In Chapter~\ref{ch:mnode}, a multi-branch relay network with direct link will be studied.

\section{Relaying Protocols}
\label{sec:protocols}
The received signals from Source are processed by the relays before forwarding to the destination. Depending on the type of processing, the relay networks are classified as decode-and-forward or amplify-and-forward.

\subsubsection*{{Decode-and-Forward Relaying}}
In decode-and-forward relaying (DF), the relays decode the received signal from the source, re-encode it and then re-broadcast the result to the destination. The process of decoding and re-encoding at the relays introduces additional computational burden to the relays. Moreover, the decoding process cannot be free of error and leads to a error propagation problem.

\subsubsection*{{Amplify-and-Forward Relaying}}
In amplify-and-forward (AF) relaying, the decoding process is avoided and the received signal is simply multiplied by a multiplication factor before being re-transmitted to the destination. The main drawback in AF relaying is that the noise at the relay is also amplified together with the signal. It should be also mentioned that the term ``amplify" does not necessarily mean that the signal amplitude will be enhanced by the amplification factor, but it could be either ways. This thesis, only focuses on amplify-and-forward relaying in the considered systems.

\section{Relay Channel Model}
By employing AF relaying, a cascaded channel is constructed between a source and a destination. This cascaded channel has different properties than a direct channel. Assume that symbol $v$ is transmitted from Source to Relay (Figure~\ref{fig:dh}). The received signal at Relay is
\begin{equation}
\label{eq:x-dh}
x=\sqrt{P_0} h_{\sr} v + w_1, \quad w_1 \sim \CN(0,N_0),
\end{equation}
where $P_0$ is the average transmitted power per symbol at Source. The received signal at Relay is multiplied by amplification factor $A$ and re-transmitted to Destination. The received signal at Destination is
\begin{equation}
\label{eq:y-dh}
y= A h_{\rd} x + w_2, \quad w_2 \sim \CN(0,N_0).
\end{equation}
Substituting \eqref{eq:x-dh} into \eqref{eq:y-dh}, gives
\begin{equation}
\label{eq:y-dh2}
y= A \sqrt{P_0} h v + w
\end{equation}
where
$
h= h_{\sr} h_{\rd}
$
and
$
w=A h_{\rd} w_1+w_2
$
are the equivalent channel and noise, respectively.
As can be seen, from Destination's point of view, the transmitted symbol $v$ is distorted by $h$, the product of two channels. This channel would be interchangeably called the cascaded, the equivalent or double Rayleigh channel.

With Rayleigh faded assumption for individual channels, i.e., $h_{\sr}\sim \CN(0,\sigma_{\sr}^2),$ $h_{\rd}\sim \CN(0,\sigma_{\rd}^2),$ the real and imaginary parts of $h=X+jY$ are identically distributed Laplacian random variables with pdfs \cite{SPAF-P}
\begin{gather}
\label{eq:f_X}
f_X(x)=\frac{1}{\sigma_{\sr}\sigma_{\rd}} \exp\left( \frac{-2|x|}{\sigma_{\sr}\sigma_{\rd}}\right),\\
\label{eq:f_Y}
f_Y(y)=\frac{1}{\sigma_{\sr}\sigma_{\rd}} \exp\left( \frac{-2|y|}{\sigma_{\sr}\sigma_{\rd}}\right).
\end{gather}
Also, the pdf of the envelope $\eta=|h[k]|$ is
\begin{equation}
\label{eq:f_h}
f_{\eta}(\eta)=\frac{4\eta}{\sigma_{\sr}^2\sigma_{\rd}^2} K_0\left(\frac{2\eta}{\sigma_{\sr}\sigma_{\rd}} \right),
\end{equation}
where $K_0(\cdot)$ is the zero-order modified Bessel function of the second kind \cite{SPAF-P}, \cite{DGC-M}. In addition, the time-series model of the cascaded channel is important for performance study of relay networks in time-varying channels.
To derive this model, depending on the mobility of the nodes with respect to each other, three cases are considered as follows. For simplicity, let set $\sigma_{\sr}^2=\sigma_{\rd}^2=1$ in all three cases.

\begin{figure*}[t]
\psfrag {h} [] [] [.8] {$|h[k]|$}
\psfrag {delta} [] [] [.8] {\eqref{eq:Delta}}
\psfrag {mdelta} [] [] [.8] {$|\Delta[k]|$ or $|\hat{\Delta}[k]|$}
\psfrag {deltahat} [] [] [.8] {\eqref{eq:delta_h_hat}}
\psfrag {PDF} [] [] [1] {pdf}
\psfrag {exact} [] [c] [.8] {\eqref{eq:hsr_hrd}}
\psfrag {approx} [] [c] [.8] {\eqref{eq:AR2-model-approx}}
\psfrag {theory} [] [c] [.8] {\eqref{eq:pdf-envelope}}
\centerline{\epsfig{figure={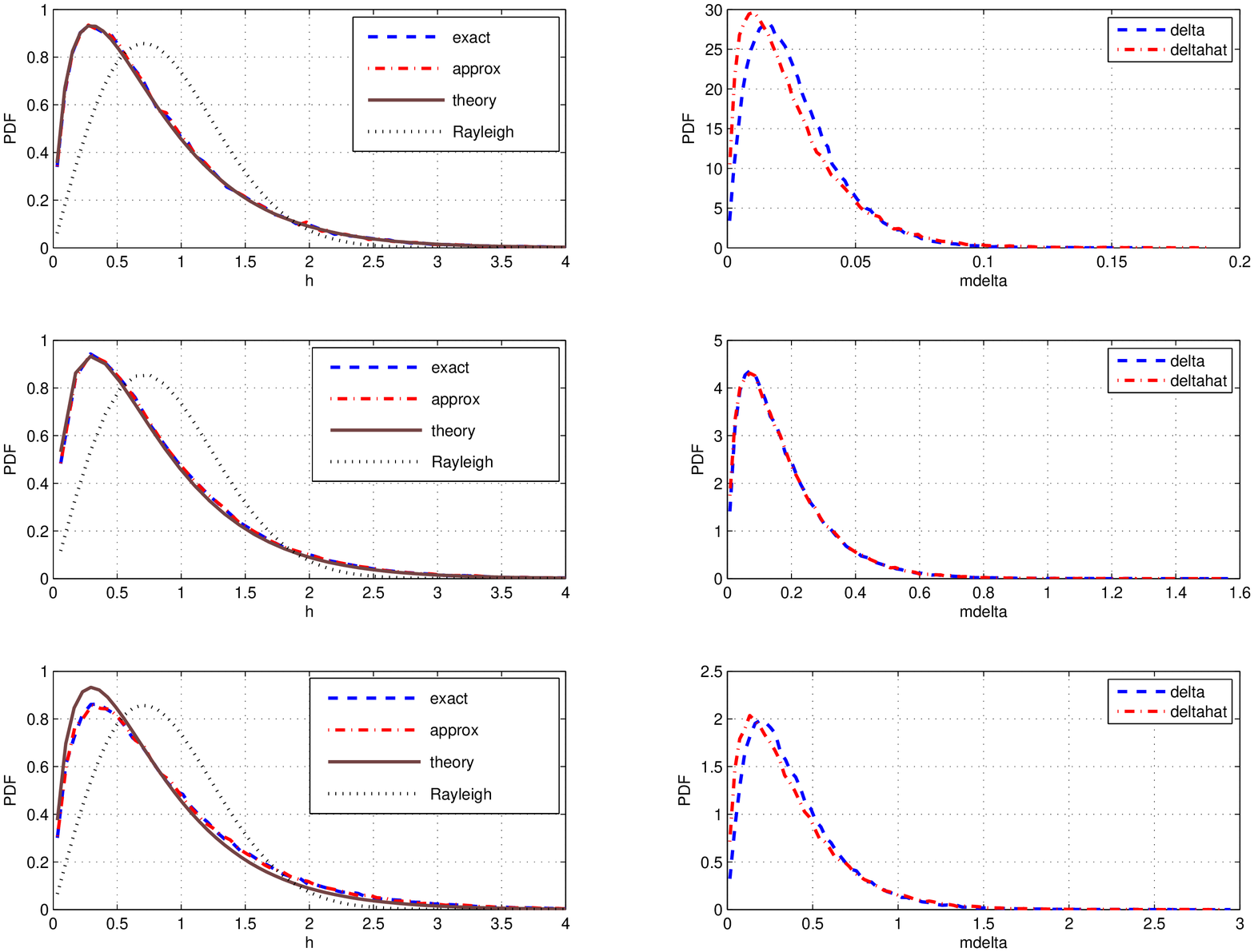},width=19cm}}
\caption{Theoretical pdf of $|h[k]|$ and obtained distributions of $|\Delta[k]|$, $|\hat{\Delta}[k]|$ and $|h[k]|$ for various cases.}
\label{fig:pdf}
\end{figure*}

\subsubsection*{Mobile Source, Fixed Relay and Destination}
\label{subsubsec:mobile-source}
When Source is moving but Relay and Destination are fixed, the SR channel becomes time-varying and their statistical properties follow the fixed-to-mobile 2-D isotropic scattering channels \cite{microwave-jake}. However, the RD channel remains static.
The $\SR$ channel can be described by an AR(1) model as
\begin{equation}
\label{eq:AR_hsri}
h_{\sr}[k]=\alpha_{\sr} h_{\sr}[k-1]+\sqrt{1-\alpha_{\sr}^2}e_{\sr}[k]
\end{equation}
where $\alpha_{\sr}=J_0(2\pi f_{\sr} n)\leq 1$ is the auto-correlation of the $\SR$ channel, $f_{\sr}$ is the normalized Doppler frequency of the SR channel and $e_{\sr}[k]\sim \mathcal{CN}(0,1)$ is independent of $h_{\sr}[k-1]$.  Also, under the scenario of fixed relays and destination, two consecutive $\RD$ channel uses are approximately equal, i.e.,
\begin{equation}
\label{eq:hrdi-fixed}
h_{\rd}[k]\approx h_{\rd}[k-1].
\end{equation}

Thus, for the cascaded channel, multiplying \eqref{eq:AR_hsri} by \eqref{eq:hrdi-fixed} gives
\begin{equation}
\label{eq:AR2-model}
h[k]=\alpha_{\sr} h[k-1] +\sqrt{1-\alpha_{\sr}^2} h_{\rd}[k-1] e_{\sr}[k]
\end{equation}
which is an AR(1) model with the parameter $\alpha_{\sr}$ and $h_{\rd}[k-1] e_{\sr}[k]$ as the input white noise.

\subsubsection*{Mobile Source and Destination, Fixed Relay}
When both Source and Destination are moving, but Relay is fixed, the SR and RD channels become time-varying and again follow the fixed-to-mobile scattering model \cite{microwave-jake}. Therefore, the AR(1) model in \eqref{eq:AR_hsri} is used for modelling the SR channel.

Similarly, for $\RD$ channel, the AR(1) model is
\begin{equation}
\label{eq:AR_hrdi}
h_{\rd}[k]=\alpha_{\rd} h_{\rd}[k-1]+\sqrt{1-\alpha_{\rd}^2}e_{\rd}[k]
\end{equation}
where $\alpha_{\rd}=J_0(2\pi f_{\rd} n)\leq1$ is the auto-correlation of the $\RD$ channel, $f_{\rd}$ is the normalized Doppler frequency of the RD channel and $e_{\rd}[k]\sim \mathcal{CN}(0,1)$ is independent of $h_{\rd}[k-1]$.

Then, for the cascaded channel, multiplying \eqref{eq:AR_hsri} by \eqref{eq:AR_hrdi} gives
\begin{equation}
\label{eq:hsr_hrd}
h [k]=\alpha  h [k-1]+\Delta[k],
\end{equation}
where $\alpha=\alpha_{\sr}\alpha_{\rd}\leq 1$ is the equivalent auto-correlation of the cascaded channel and
\begin{multline}
\label{eq:Delta}
\Delta[k]  =\alpha_{\sr} \sqrt{1-\alpha_{\rd}^2} h_{\sr}[k-1] e_{\rd}[k]+\alpha_{\rd} \sqrt{1-\alpha_{\sr}^2}\\
h_{\rd}[k-1] e_{\sr}[k]
+\sqrt{(1-\alpha_{\sr}^2)(1-\alpha_{\rd}^2)}e_{\sr}[k]e_{\rd}[k]
\end{multline}
represents the time-varying part of the equivalent channel, which is a combination of three uncorrelated complex-double Gaussian distributions \cite{DGC-M} and uncorrelated to $h[k-1]$. Since $\Delta[k]$ has a zero mean, its auto-correlation function is computed as
\begin{equation}
\label{eq:var_Delta}
E\{\Delta[k]  \Delta^*[k+m]\}=
\begin{cases}
1-\alpha^2, & \text{if}\;\;  m=0,\\
0, & \text{if} \;\; m\neq 0.
\end{cases}
\end{equation}
Therefore, $\Delta[k]$ is a white noise process with variance $\mbox{E}\{\Delta[k]\Delta^*[k]\}=1-\alpha^2$.

However, using $\Delta[k]$ in the way defined in \eqref{eq:Delta} is not feasible for the performance analysis. Thus, to make the analysis feasible, $\Delta[k]$ shall be approximated with an adjusted version of one of its terms as
\begin{equation}
\label{eq:delta_h_hat}
\hat{\Delta}[k]= \sqrt{1-\alpha^2} {h}_{\rd}[k-1] {e}_{\sr}[k]
\end{equation}
The above approximation of $\Delta[k]$ is also a white noise process with first and second order statistical properties identical to that of $\Delta[k]$ and uncorrelated to $h[k-1]$.

By substituting \eqref{eq:delta_h_hat} into \eqref{eq:hsr_hrd}, the time-series model of the equivalent channel can be described as
\begin{equation}
\label{eq:AR2-model-approx}
h[k]=\alpha h[k-1]+\sqrt{1-\alpha^2} h_{\rd}[k-1]e_{\sr}[k].
\end{equation}
The above approximation of $\Delta[k]$ is again an AR(1) with parameter $\alpha$ and $h_{\rd}[k-1]e_{\sr}[k]$ as the input white noise.

Comparing the AR(1) models in \eqref{eq:AR2-model} and \eqref{eq:AR2-model-approx} shows that, in essence, they are only different in the model parameter, i.e., $\alpha_{\sr}$ and $\alpha$,: the parameter contains the effect of the $\SR$ channel in the former model, while the effects of both the $\SR$ and $\RD$ channels are included in the later model. This means that the model in \eqref{eq:AR2-model-approx} can be used as the time-series model of the cascaded channel for the analysis in both cases. Specifically, for static $\RD$ channels  $\alpha_{\rd}=1$ and hence \eqref{eq:AR2-model-approx} turns to \eqref{eq:AR2-model}.

To validate the model in \eqref{eq:AR2-model-approx}, its statistical properties are verified with the theoretical counterparts. Theoretical mean and variance of $h[k]$ are shown to be equal to zero and one, respectively \cite{SPAF-P,DGC-M}. This can be seen by taking expectation and variance operations over \eqref{eq:AR2-model-approx} so that $\mbox{E}\{h[k]\}=0$, $\mbox{Var}\{h[k]\}=1$. Also, the theoretical auto-correlation of $h[k]$ is obtained as the product of the auto-correlation of the $\SR$ and $\RD$ channels in \cite{SPAF-P}. By multiplying both sides of \eqref{eq:AR2-model-approx} with $h^*[k-1]$ and taking expectation, one has
\begin{equation}
\label{eq:hi-auto}
\mbox{E}\{h[k]h^*[k-1]\}=\alpha\mbox{E} \{h[k-1]h^*[k-1]\}\\+\mbox{E}\{\hat{\Delta}[k]h^*[k-1]\}.
\end{equation}
Since $\hat{\Delta}[k]$ is uncorrelated to $h[k-1]$ then $\mbox{E}\{\hat{\Delta}[k]h^*[k-1] \}=0$ and it can be seen that
\begin{equation}
\label{eq:h[k]}
\mbox{E}\{h[k]h^*[k-1]\}=\alpha=\alpha_{\sr} \alpha_{\rd}.
\end{equation}
In addition, the theoretical pdf of the envelope $\lambda=|h[k]|$ is
\begin{equation}
\label{eq:pdf-envelope}
f_{\lambda}(\lambda)=4\lambda K_0\left( 2 \lambda \right)
\end{equation}
where $K_0(\cdot)$ is the zero-order modified Bessel function of the second kind \cite{SPAF-P}, \cite{DGC-M}.
To verify this, using Monte-Carlo simulation the histograms of $|h[k]|$, $|\Delta[k]|$ and $|\hat{\Delta}[k]|$, for different values of $\alpha$, are obtained for both models in \eqref{eq:hsr_hrd} and \eqref{eq:AR2-model-approx}. The values of $\alpha$ are computed from a wide range of the normalized Doppler frequencies. These histograms along with the theoretical pdf of $|h[k]|$ are illustrated in Figure~\ref{fig:pdf}.
Although, theoretically, the distributions of $\Delta[k]$ and $\hat{\Delta}[k]$ are not exactly the same, it is seen that for practical values of $\alpha$ they are very close. Moreover, the resultant distributions of $h[k]$, regardless of $\Delta[k]$ or $\hat{\Delta}[k]$, are similar and close to the theoretical distribution.
The Rayleigh pdf is depicted in the figure only to show the difference between the distributions of an individual and the cascaded channels.

\subsubsection*{All Nodes are Mobile}
In this case, all links follow the mobile-to-mobile channel model \cite{m2m-Akki}. However, they are all individually Rayleigh faded and the only difference is that the auto-correlation of the channel should be replaced according to this model. Thus, the channel model \eqref{eq:AR2-model-approx} again can be used as the time-series model of the cascaded channel in this case, albeit with appropriate auto-correlation value. The reader is referred to the discussion in \cite{SPAF-P} and \cite{m2m-Akki} for more details on computing these auto-correlations as well as the tutorial survey on various fading models for mobile-to-mobile cooperative communication systems in \cite{M2M-Talha}. For the analysis in this thesis, it is assumed that the equivalent maximum Doppler frequency of each link, regardless of fixed-to-mobile or mobile-to-mobile, is given and then the auto-correlation of each link is computed based on Jakes' model \cite{microwave-jake}.

\section{Cooperative Strategies}
\label{sec:strategy}
The transmission process in cooperative networks is usually divided into two phases. In the first phase, relays are silent and listening to the transmitted signal from the source. In the second phase, the relays cooperate to deliver the received signals to the destination. There are two main cooperative strategies: Repetition-based and distributed space-time coding based.

\subsection{Repetition-Based Cooperation}
\label{subsec:repet}
\begin{figure}[t]
\psfrag {source} [] [] [0.8] {Source broadcasts}
\psfrag {Relay1} [] [] [.8] {Relay 1 forwards\;\;}
\psfrag {Relay2} [] [] [.8] {Relay 2 forwards\;\;}
\psfrag {Relayi} [] [] [.8] {$\cdots$}
\psfrag {RelayR} [] [] [.8] {Relay $R$ forwards\;\;\;}
\psfrag {phaseI} [] [] [1.0] {Phase I}
\psfrag {phaseII} [] [] [1.0] {Phase II}
\psfrag {Time} [] [] [1.0] {Time $\rightarrow$ }
\centerline{\epsfig{figure={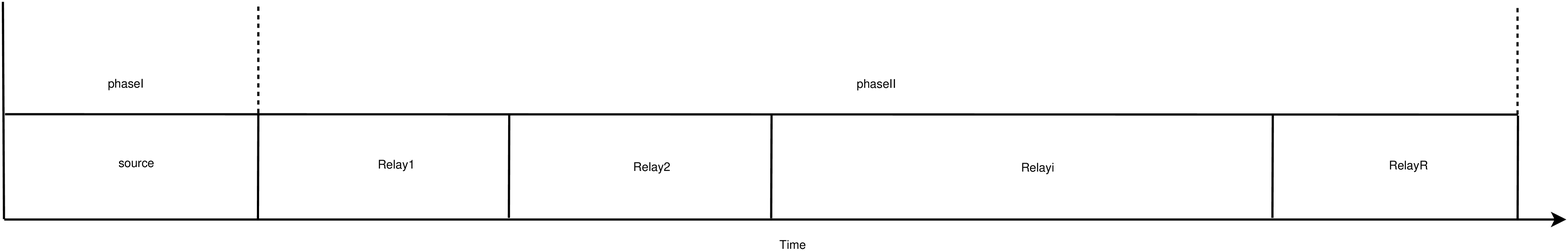},width=17cm}}
\caption{Transmission process in repetition-based relaying strategy.}
\label{fig:tx_rep}
\end{figure}

Repetition-based strategy is actually a combination of receiver diversity and transmit diversity using the repetition code as described in Chapter~\ref{ch:p2p}. All relays receive different copies of the same signal from Source in Phase I, which is similar to receiver diversity in point-to-point communications (Chapter~\ref{ch:p2p}). In Phase II, the relays amplify and re-broadcast the received symbols to Destination sequentially in time (TDD manner) as depicted in Figure~\ref{fig:tx_rep}. This is similar to transmit diversity using repetition coding. Repetition-based strategy is simple to implement. It does not require a complicated encoding and decoding at Source or Destination. Moreover, during the transmission of each relay, other relays remain silent and hence the relays do not have to be synchronized in the symbol level. However, the repetition-based strategy has a low spectral efficiency.

In the repetition-based strategy, the received signals from multiple links at Destination need to be combined using a combining technique to achieve the cooperative diversity. The maximum-ratio-combining scheme is considered in Chapter~\ref{ch:mnode}, while the selection combining method will be used in Chapters~\ref{ch:sc}-\ref{ch:sc_tv}.

\subsection{Distributed Space-Time Coding-Based}
\label{subsec:DSTC}
\begin{figure}[t]
\psfrag {source} [] [] [1] {Source broadcasts}
\psfrag {relays} [] [] [1] {All relays forward}
\psfrag {phase1} [] [] [1] {Phase I}
\psfrag {phase2} [] [] [1] {Phase II}
\psfrag {time}   [] [] [1] {Time $\rightarrow$}
\centerline{\epsfig{figure={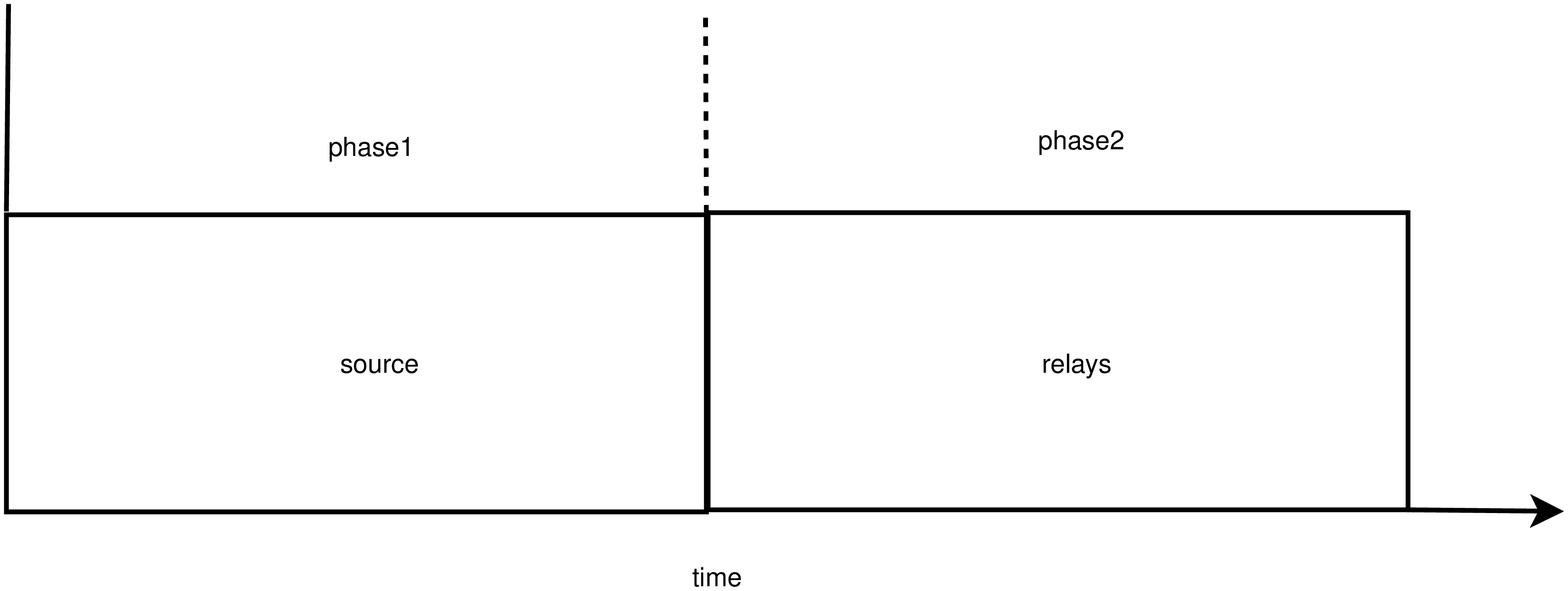},width=12cm}}
\caption{Transmission process in DSTC-based strategy.}
\label{fig:dstc_tx}
\end{figure}

The repetition-based strategy, suffers from a low data rate inherent in TDD transmission. A logical solution would be to deploy space-time coding in a distributed way using the help of relays. In distributed space-time coding (DSTC), different relays receive different copies of the same information symbols in Phase I. The relays process these received signals and simultaneously forward them to the destination in Phase II. The transmission process in DSTC-based strategy is depicted in Figure~\ref{fig:dstc_tx}. The distributed processing at different relay nodes forms a virtual antenna array (VAA). Therefore, conventional space-time block coding schemes can be applied to relay networks to achieve the cooperative diversity and coding gain \cite{coop-dohler,DSTC-Y}.

\section{Summary}
This chapter presented the main relay topologies such as single-branch and multi-branch dual-hop relaying with and without direct link. Also, relay protocols and cooperative strategies were described. More importantly, a time-series model was developed for the cascaded channel. The background in this chapter will be necessary for better understanding of the systems and networks considered in the subsequent chapters. In the next chapter, a dual-hop relay network employing differential encoding and decoding will be studied first.

%\addcontentsline{toc}{chapter}{References}
%\bibliographystyle{IEEEbib}
%\bibliography{H:/latex/references} 

\chapter{Performance of Differential Amplify-and-Forward Dual-Hop Relaying}
\label{ch:dh}
In wireless communications, service coverage limit happens due to power limitation of wireless applications and attenuations of transmitted signals in long distance. In this case, the received signal power at the destination is weak and hence the direct link between the source and the destination cannot be practically used. As described in the previous chapter, an important relay topology to overcome coverage limit is dual-hop relaying. In dual-hop relaying, depicted in Figure~\ref{fig:dh}, a neighbour user in the coverage area of the wireless network acts as a relay to bridge the communications from a far user to its destination.

The manuscript in this chapter studies a single-branch dual-hop relaying system without direct link that employs differential encoding and decoding (see Figure~\ref{fig:dh}). To circumvent channel estimation, differential $M$-PSK is utilized at Source to encode the data symbols. Amplify-and-forward relaying protocol is used at Relay. At the destination, first, two-symbol differential detection is considered. Different from the conventional slow-fading assumption, commonly made in the literature, here, the users can be highly mobile and hence the channels become fast time-varying. The first goal in this manuscript is to examine the performance and robustness of two-symbol differential detection in general time-varying Rayleigh fading channels. For this purpose, a time-series model has been developed to characterize the time-varying nature of the cascaded channel. An exact BER probability expression of two-symbol differential detection is derived and shown to approach an error floor at high SNR region in fast-fading environments. This analysis is useful in optimizing the system parameters in order to design more robust relaying systems.

Next, as a solution to overcome the error floor, a multiple-symbol differential detection scheme is developed for the dual-hop relaying system. The main disadvantage of multiple-symbol differential detection is its complexity, which increases exponentially with the number of processed symbols. In addition, the decision metric of multiple-symbol differential detection is more complicated when applied to relay networks due to the complicated distribution of the received signal at the destination. Hence, it is important to obtain a decision metric that can be computed with low complexity and without sacrificing much the performance. Moreover, to approach the optimal performance promised by multiple-symbol detection, it is important to determine the system parameters accurately based on the channel information. These objectives are accomplished in the rest of the manuscript, in which the multiple-symbol differential sphere decoding in point-to-point communications \cite{MSDSD-L} is adapted for dual-hop relay networks. Furthermore, theoretical error performance of the multiple-symbol detection (MSD) scheme is also obtained. This analysis is useful to investigate a trade-off between the MSD window size and the desired performance. Simulation results in various fading and channel scenarios are provided to verify the analysis of two-symbol and multiple-symbol differential detection algorithms and also performance improvements gained by multiple-symbol detection.

The results of our study on dual-hop relaying systems are reported in manuscripts [Ch4-1] and [Ch4-2], listed below. Manuscript [Ch4-1] considers the general case of dual-hop relaying and is included in this chapter.

[Ch4-1] M. R. Avendi, Ha H. Nguyen,`` Differential Dual-Hop Relaying under User Mobility", submitted to \textit{IET Communications}.

[Ch4-2] M. R. Avendi, Ha H. Nguyen,`` Differential Dual-Hop Relaying over Time-Varying
Rayleigh-Fading Channels", \textit{IEEE 13th Canadian Workshop on Information Theory, Toronto, Canada, June 2013.}.

\begin{center}
{\bf{\Large
Differential Dual-Hop Relaying under User Mobility}}
\end{center}
\begin{center}
M. R. Avendi, Ha H. Nguyen
\end{center}
%\thanks{The authors are with the Department of Electrical and Computer Engineering,
%University of Saskatchewan, Saskatoon, Canada, S7N5A9.
%Email: m.avendi@usask.ca, ha.nguyen@usask.ca.}

\begin{center}
\bf Abstract
\end{center}
\label{dh:abs}
This paper studies dual-hop amplify-and-forward relaying system employing differential encoding and decoding over time-varying Rayleigh fading channels. First, the convectional ``two-symbol'' differential detection (CDD) is theoretically analysed in terms of the bit-error-rate (BER). It is seen that the performance of two-symbol differential detection severely degrades in fast-fading channels and reaches an irreducible error floor at high transmit power. Next, to overcome the error floor experienced with fast-fading, a nearly optimal ``multiple-symbol'' detection (MSD) is designed and theoretically analysed. The analysis of CDD and MSD are verified and illustrated with simulation results under different fading scenarios.

\begin{center}
\bf Index terms
\end{center}
Dual-hop relaying, amplify-and-forward, differential M-PSK, non-coherent detection, time-varying fading channels, multiple-symbol detection.

\section{Introduction}
\label{se:intro}
Dual-hop relaying without a direct link has been considered in the literature as a technique to leverage coverage problems of many wireless applications such as 3GPP LTE, WiMAX, WLAN, Vehicle-to-Vehicle communication and wireless sensor networks \cite{coop-deploy,coop-V2V,coop-WiMAX,coop-dohler,coop-LTE}. Such a technique can be seen as a type of cooperative communication in which one node in the network helps another node to communicate with (for example) the base station when the direct link is very poor or the user is out of the coverage area.

A two-phases transmission process is usually utilized in such a network. Here, Source transmits data to Relay in the first phase, while in the second phase Relay performs amplify-and-forward (AF) strategy to send the received data to Destination \cite{coop-laneman}. Error performance of dual-hop relaying without direct link employing AF strategy has been studied in \cite{dual-hop-Hasna,dh-smith09,dh-hasna03}. Also, the statistical properties of the cascaded channel between Source and Destination in a dual-hop AF relaying have been examined in \cite{SPAF-P}.

In the existing literature, either coherent detection or \emph{slow-fading} environment is assumed. In coherent detection, instantaneous channel state information (CSI) of both Source-Relay and Relay-Destination links are required at Destination. This requirement, and specifically Source-Relay channel estimation, would be challenging to meet for some applications. Also, when Source and/or Relay are mobile, the constructed channels become time-varying. On one hand, the channel variation makes coherent detection inefficient or sometimes impossible due to the requirement of fast channel estimation or tracking. On the other hand, by employing differential modulation and two-symbol non-coherent detection, this variation can be somewhat tolerated as long as the channel does not change significantly over two consecutive symbols. For the case of conventional differential detection (CDD) using two symbols, it is well known that over slow-fading channels, around 3 dB loss is seen between coherent and non-coherent detections. However, in practical time-varying channels, the effect of channel variation can lead to a much larger degradation.

Motivated from the above discussion, the first goal in this article is to analyse the performance of single-branch dual-hop relaying employing differential $M$-PSK and two-symbol non-coherent detection in time-varying Rayleigh fading channels. We refer to this system as differential dual-hop (D-DH) relaying. In \cite{DAF-ITVT}, we studied a multi-branch differential AF relaying with direct link in time-varying channels, however only a lower bound of the bit-error rate (BER) was derived for the multi-branch system. Although, the dual-hop relaying considered in this article is a special case of multi-branch relaying, the analysis here is different than that of \cite{DAF-ITVT}. Specifically, for the case of two-symbol non-coherent detection, an exact bit error rate (BER) expression is obtained. It is also shown that the system performance quickly degrades in fast-fading channels and reaches an error floor at high transmit power. The theoretical value of the error floor is also derived and used to investigate the fading rate threshold. Interestingly, it is seen that the error floor only depends on the auto-correlation value of the cascaded channel and modulation parameters.

%Counterpart to point-to-point communications \cite{msdd-div,msdd-div2,MSDSD-L,msdd_fung,MSDUSTC-P,msd-Rick},
The second goal in this article is to design a multiple-symbol detection (MSD) for the D-DH relaying to improve its performance in fast-fading channels. Multiple-symbol detection has been considered for point-to-point communications in \cite{msdd-div,msdd-div2,MSDSD-L,msdd_fung,MSDUSTC-P,msd-Rick}. The challenge in developing multiple-symbol detection for AF relay networks is that, due to the complexity of the distribution of the received signal at Destination, the optimum decision metric does not yield a closed form solution. To circumvent this problem, here, the optimum decision rule is replaced with an alternative decision rule and further simplified to be solved with low complexity. Furthermore, theoretical error performance of MSD is obtained. This analysis is useful to investigate a trade-off between the MSD window size and the desired performance. The error analysis of both  CDD and MSD are thoroughly verified with simulation results in various fading scenarios.

The outline of the paper is as follows. Section \ref{sec:system} describes the system model. In Section \ref{sec:two-symbol}, two-symbol differential detection and its performance over time-varying channels are studied. Section \ref{sec:MSDSD} develops the MSD algorithm and analyses its performance. Simulation results are given in Section \ref{sec:sim}. Section \ref{sec:con} concludes the paper.

\emph{Notation}: Bold upper-case and lower-case letters denote matrices and vectors, respectively. $(\cdot)^t$, $(\cdot)^*$, $(\cdot)^H$ denote transpose, complex conjugate and Hermitian transpose of a complex vector or matrix, respectively. $|\cdot|$  denotes the absolute value of a complex number and $\|\cdot \|$ denotes the Euclidean norm of a vector. $\mathcal{CN}(0,N_0)$ stands for complex Gaussian distribution with zero mean and variance $N_0$. $\mbox{E}\{\cdot\}$ denotes expectation operation. Both ${e}^{(\cdot)}$ and $\exp(\cdot)$ show the exponential function. $\dgs$ is the diagonal matrix with components of $\s$ on the main diagonal and $\Ib_N$ is the $N \times N$ identity matrix. A symmetric $N\times N$ Toeplitz matrix is defined by $\mbox{toeplitz}\{x_1,\cdots,x_N\}$. $\mbox{det}\{\cdot\}$ denotes determinant of a matrix. $\CpN$ is the set of complex vectors with length $N$. $\Re\{\cdot\}$ and $\Im\{\cdot\}$ denote the real and imaginary parts of a complex number.

\section{System Model}
\label{sec:system}
\begin{figure}[t]
\psfrag {Source} [] [] [1.0] {Source}
\psfrag {Relay} [] [] [1.0] {Relay}
\psfrag {Destination} [] [] [1.0] {Destination}
\psfrag {h1} [] [] [1.0] {$\qquad h_1[k]$\;\;\;}
\psfrag {h2} [] [] [1.0] {$\qquad h_2[k]$}
\centerline{\epsfig{figure={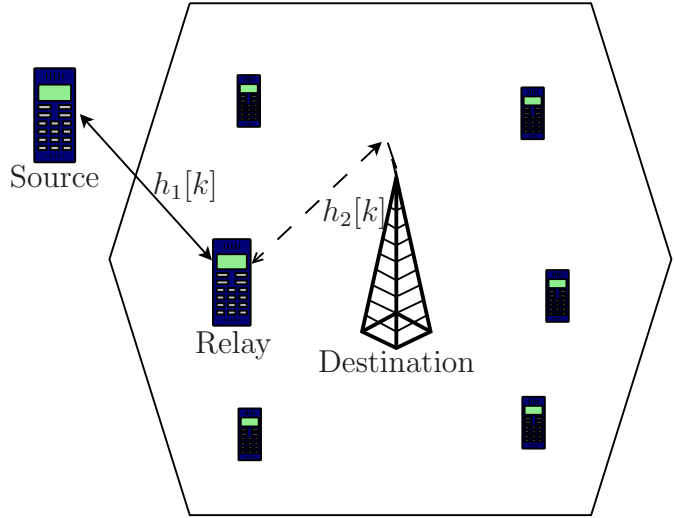},width=8.5cm}}
\caption{Illustration of single-branch dual-hop relaying system without direct link.}
\label{fig:sysmodel}
\end{figure}
The wireless relay model under consideration, depicted in Figure~\ref{fig:sysmodel}, has one Source, one Relay and one Destination. Source is out of the cell coverage and hence the received signal in the direct link is not sufficiently strong to facilitate data transmission. Therefore, with the help of another user (Relay), a dual-hop relaying system without direct link is constructed to connect Source to Destination. Each node has a single antenna, and the communication between nodes is half duplex (i.e., each node is able to only send or receive in any given time). The channels from Source to Relay (SR) and from Relay to Destination (RD) are denoted by $h_1[k]$ and $h_2[k]$, respectively, where $k$ is the symbol time. A Rayleigh flat-fading model is assumed for each channel, i.e., $h_1[k]\sim \CN(0,\sigma_1^2)$ and $h_2[k]\sim \CN(0,\sigma_2^2)$. The channels are spatially uncorrelated and changing continuously in time. The time-correlation between two channel coefficients, $n$ symbols apart, follows the Jakes' model \cite{microwave-jake}:
\begin{equation}
\varphi_i(n)=\E \{h_i[k]h_i^*[k+n]\}=\sigma_i^2 J_0(2\pi f_i n),\quad i=1,2
\end{equation}
where $J_0(\cdot)$ is the zeroth-order Bessel function of the first kind and $f_i$ is the maximum normalized Doppler frequency of the $i$th channel.

At time $k$, a group of $\log_2M$ information bits is mapped to a $M$-PSK symbol as $v[k]\in \mathcal{V}$ where $\mathcal{V}=\{e^{j2\pi m/M},\; m=0,\dots, M-1\}$. Before transmission, the symbols are encoded differentially as
\begin{equation}
\label{eq:s-source}
s[k]=v[k] s[k-1],\quad s[0]=1.
\end{equation}
The transmission process is divided into two phases. Block-by-block transmission protocol is utilized to transmit a frame of symbols in each phase as symbol-by-symbol transmission causes frequent switching between reception and transmission, which is not practical. However, the analysis is the same for both cases and only the channel auto-correlation value is different ($n=1$ for block-by-block and $n=2$ for symbol-by-symbol).

In phase I, the symbol $\sqrt{P_0}s[k]$ is transmitted by Source to Relay, where $P_0$ is the average source power per symbol. The received signal at Relay is
\begin{equation}
\label{eq:relay_rx}
x[k]=\sqrt{P_0}h_1[k]s[k]+w_1[k]
\end{equation}
where $w_1[k]\sim \CN(0,N_0)$ is the noise component at Relay. Also, the average received SNR per symbol at Relay is defined as
\begin{equation}
\label{eq:rho1}
\rho_1=\frac{P_0\sigma_1^2}{N_0}.
\end{equation}

The received signal at Relay is then multiplied by an amplification factor $A$, and re-transmitted to Destination. The amplification factor, based on the variance of SR channel, is commonly used in the literature as
\begin{equation}
\label{eq:A}
A =\sqrt{\frac{P_1}{P_0\sigma_1^2+N_0}},
\end{equation}
to normalize the average power per symbol at Relay to $P_1$. {Typically, the total power $P=P_0+P_1,$ is allocated between Source and Relay such that the average BER of the system is minimized.} The corresponding received signal at Destination is
\begin{equation}
\label{eq:dest-rx1}
y[k]=Ah_2[k]x[k]+w_2[k],
\end{equation}
where $w_2[k]\sim \CN(0,N_0)$ is the noise component at Destination. Substituting (\ref{eq:relay_rx}) into (\ref{eq:dest-rx1}) yields
\begin{equation}
\label{eq:Destination-rx}
y[k]= A \sqrt{P_0}h[k]s[k]+w[k],
\end{equation}
where $h[k]=h_1[k]h_2[k]$ is the cascaded channel with zero mean and variance $\sigma_1^2\sigma_2^2$ \cite{SPAF-P}, and
\begin{equation}
\label{eq:w[k]}
w[k]=A h_2[k]w_1[k]+w_2[k]
\end{equation}
is the equivalent noise at Destination.
It should be noted that for a given $h_2[k]$, $w[k]$ is a complex Gaussian random variable with zero mean and variance
\begin{equation}
\label{eq:sig_wk}
\sigma_{w}^2=N_0 \left(1+ A^2 |h_2[k]|^2\right).
\end{equation}
Thus $y[k]$, conditioned on $s[k]$ and $h_2[k]$, is a complex Gaussian random variable as well.

In the following section, the conventional two-symbol differential detection (CDD) of the received signals at Destination and its performance are considered.

\section{Two-Symbol Differential Detection}
\label{sec:two-symbol}

\subsection{Detection Process}
\label{subsec:channel-model}
Given two consecutive received symbols at a time, non-coherent detection of the transmitted symbol is obtained by the following minimization:
\begin{equation}
\label{eq:ml-detection}
\hat{v}[k]= \arg \min \limits_{v[k]\in \mathcal{V}} |y[k]- v[k] y[k-1]|^2.
\end{equation}
As can be seen, no channel information is needed for detection. Two-symbol detection is simple to implement and it is mainly based on the assumption that the channel coefficients are approximately constant during two adjacent symbols. Although this assumption would be true for slow-fading channels, it would be violated when users are fast moving. In the next section, the performance of two-symbol non-coherent detection in time-varying Rayleigh fading channel is analysed.

\subsection{Performance Analysis}
\label{sec:symbol_error_probability}
Using the unified approach in \cite[eq.25]{unified-app}, it follows that the conditional BER for two-symbol differential detection can be written as
\begin{equation}
\label{eq:Pb-gama-hrd}
P_{\bb}^{\CDD}(E|\gamma,h_2)=\frac{1}{4\pi} \int \limits_{-\pi}^{\pi} g(\theta) e^{-q(\theta)\gamma} \dd \theta
\end{equation}
where $g(\theta)=(1-\beta^2)/(1+2\beta\sin(\theta)+\beta^2)$, $q(\theta)=(b^2/\log_2 M) (1+2\beta\sin(\theta)+\beta^2)$, and $\beta=a/b$. The values of $a$ and $b$ depend on the modulation size \cite{unified-app}. Also, $\gamma$ is the instantaneous effective SNR at the output of the differential detector which needs to be determined for time-varying channels.

To proceed with the performance analysis of two-symbol differential detection in time-varying channels, it is required to model the time-varying nature of the channels. For this purpose, individual Rayleigh-faded channels, i.e., Source-Relay and Relay-Destination channels, are expressed by a first-order auto-regressive (AR(1)) model as
\begin{gather}
\label{eq:ARi}
h_i[k]=\alpha_i h_i[k-1]+\sqrt{1-\alpha_i^2} e_i[k],\quad i=1, 2
\end{gather}
where $\alpha_i=\varphi_i(1)/\sigma_i^2$ is the auto-correlation of the $i$th channel and $e_i[k]\sim \mathcal{CN}(0,\sigma_i^2)$ is independent of $h_i[k-1]$. Based on these expressions, a first-order time-series model has been derived in \cite{DAF-ITVT} to characterise the evolution of the cascaded channel in time. The time-series model of the cascaded channel is given as (the reader is referred to \cite{DAF-ITVT} for the detailed derivations/verification)
\begin{equation}
\label{eq:ARmodel}
h[k]=\alpha h[k-1]+\sqrt{1-\alpha^2}\ h_2[k-1]e_1[k]
\end{equation}
where $\alpha=\alpha_1 \alpha_2 \leq 1$ is the equivalent auto-correlation of the cascaded channel, which is equal to the product of the auto-correlations of individual channels, and $e_1[k]\sim \mathcal{CN}(0,\sigma_1^2)$ is independent of $h[k-1]$.

By substituting (\ref{eq:ARmodel}) into (\ref{eq:Destination-rx}) one has
\begin{equation}
\label{eq:ykk-1}
y[k]=\alpha v[k]y[k-1]+n[k],
\end{equation}
where
\begin{equation}
\label{eq:n[k]}
n[k]=w[k]- \alpha v[k]w[k-1]
+ \sqrt{1-\alpha^2}A\sqrt{P_0} s[k] h_2[k-1] e_1[k].
\end{equation}
From expression \eqref{eq:ykk-1}, with given $y[k]$ and $y[k-1]$, the non-coherent detection process can be interpreted as coherent detection of data symbol $v[k]$ distorted by a fading channel equivalent to $y[k-1]$ and in the presence of the equivalent noise $n[k]$.
%As can be seen from the model in \eqref{eq:n[k]}, the exact distribution of $n[k]$ is difficult to find. However, by substituting $h_2[k]$ from \eqref{eq:ARi} into $w[k]$ one has
%\begin{multline}
%\label{eq:w[k]-hat}
%w[k]=w_2[k]+A \alpha_2 w_1[k]  h_2[k-1]\\+A\sqrt{1-\alpha_2^2}w_1[k] e_2[k]
%\approx w_2[k]+A \alpha_2 w_1[k]  h_2[k-1]
%\end{multline}
%where the approximation comes from the observation that even for very fast-fading channels the term $A\sqrt{1-\alpha_2^2}$ is very small. Hence, using the approximated value of $w[k]$ into \eqref{eq:n[k]},
%\begin{multline}
%\label{eq:n[k]-app}
%n[k]\approx w_2[k]+A\alpha_2 w_1[k] h_2[k-1]\\
%-\alpha v[k] \left( w_2[k-1]+Ah_2[k-1]w_1[k-1] \right)\\
%+\sqrt{1-\alpha^2} A \sqrt{P_0} s[k] h_2[k-1] e_1[k]
%\end{multline}
%which shows that, conditioned on $h_2[k-1]$, $n[k]$ is a linear combination of complex Gaussian random variables and hence it is also complex Gaussian random variable.
%From now on, the time index $[k-1]$ is omitted in this section to simplify the notation.
Hence, for time-varying channels, based on \eqref{eq:ykk-1} and \eqref{eq:n[k]}, $\gamma$ is computed as
\begin{equation}
\label{eq:gama_d}
\gamma=\bar{\gamma} |h_1|^2
\end{equation}
where
\begin{equation}
\label{eq:gama_b}
\overline{\gamma}=\frac{\alpha^2 A^2 (P_0/N_0) |h_2|^2}{1+\alpha^2+\left[1+\alpha^2+(1-\alpha^2)\rho_1 \right] A^2|h_2|^2}.
\end{equation}
In the above, the time index $[k-1]$ is omitted to simplify the notation. Clearly, for slow-fading channels ($\alpha=1$), the equivalent noise power is only enhanced by a factor of two and $\gamma$ is half of the received SNR in coherent detection $A^2 (P_0/N_0) |h_2|^2 |h_1|^2/(1+A^2|h_2|^2)$ \cite{dual-hop-Hasna,SPAF-P}, as expected. However, for fast-fading channels, $\alpha<1$, the noise power is dominated by the last term in \eqref{eq:n[k]} and then significantly increases with increasing transmit power. This leads to a larger degradation in the effective SNR and poor performance of two-symbol non-coherent detection in fast-fading channels.

Since, $\lambda_1=|h_1|^2$ is exponentially distributed, i.e., $f_{\lambda_1}(\lambda)=\exp(-\lambda/\sigma_1^2)/\sigma_1^2$, the variable $\gamma$, conditioned on $|h_2|$, follows an exponential distribution with the following pdf and cdf:
\begin{equation}
\label{eq:pdf-gama_b}
f_{\gamma|h_2}(\gamma)=\frac{1}{\overline{\gamma}\sigma_1^2} \exp\left(-\frac{\gamma}{\overline{\gamma}\sigma_1^2}\right)
\end{equation}
\begin{equation}
\label{eq:cdf-gama_b}
F_{\gamma|h_2}(\gamma)=1- \exp\left(-\frac{\gamma}{\overline{\gamma}\sigma_1^2}\right).
\end{equation}

By substituting $\gamma$ into \eqref{eq:Pb-gama-hrd} and taking the average over the distribution of $\gamma$, one has
\begin{equation}
\label{eq:Pb-hrd}
P_{\bb}^{\CDD}(E|h_2)=\frac{1}{4\pi} \int \limits_{-\pi}^{\pi} g(\theta) I(\theta) \dd \theta
\end{equation}
where
\begin{equation}
\label{eq:I_theta}
I(\theta)=\int \limits_{0}^{\infty} e^{-q(\theta)\gamma} \frac{1}{\overline{\gamma}\sigma_1^2} e^{-\frac{\gamma}{\overline{\gamma}\sigma_1^2}} \dd \gamma=\frac{1}{\overline{\gamma}\sigma_1^2 q(\theta)+1}
=b_3(\theta) \frac{\lambda_2+b_1}{\lambda_2+b_2(\theta)}
\end{equation}
with $\lambda_2=|h_2|^2$, $b_3(\theta)=b_2(\theta)/b_1$ and $b_1$, $b_2(\theta)$ defined as
\begin{align*}
\label{eq:b1b2b3}
b_1=\frac{1+\alpha^2}{(1+\alpha^2)A^2+(1-\alpha^2)A^2\rho_1}\\
b_2(\theta)=\frac{1+\alpha^2}{(1+\alpha^2)/b_1+\alpha^2 q(\theta)A^2\rho_1}
\end{align*}
%The later expression in \eqref{eq:I_theta} is obtained by substituting $\bar{\gamma}$.

Now, by taking the final average over the distribution of $\lambda_2=|h_2|^2$, $f_{\lambda_2}(\lambda)=\exp\left(-\lambda/\sigma_2^2\right)/\sigma_2^2$, it follows that
\begin{equation}
\label{eq:Pb}
P_{\bb}^{\CDD}(E)=\frac{1}{4\pi} \int \limits_{-\pi}^{\pi} g(\theta) J(\theta) \dd\theta
\end{equation}
where
\begin{equation}
\label{eq:J_theta}
J(\theta)=\int \limits_{0}^{\infty} b_3(\theta) \frac{\lambda+b_1}{\lambda+b_2(\theta)} \frac{1}{\sigma_2^2} e^{\left(-\frac{\lambda}{\sigma_2^2}\right)} \dd\lambda
=b_3(\theta) \left( 1+\frac{1}{\sigma_2^2}(b_1-b_2(\theta)) e^{\left(\frac{b_2(\theta)}{\sigma_2^2}\right)} E_1\left(\frac{b_2(\theta)}{\sigma_2^2}\right) \right)
\end{equation}
and $E_1(x)=\int \limits_x^{\infty} (e^{-t}/t) \dd t$ is the exponential integral function. The definite integral in \eqref{eq:Pb} can be easily computed using numerical methods and it gives an exact value of the BER of the D-DH system under consideration in time-varying Rayleigh fading channels.

It is also informative to examine the expression of $P_{\bb}(E)$ at high transmit power. In this case,
\begin{equation}
\label{eq:gb_limit}
\lim \limits_{(P_0/N_0)\rightarrow \infty} E[\bar{\gamma}]= \frac{\alpha^2}{(1-\alpha^2)\sigma_1^2},
\end{equation}
which is independent of $|h_2|^2$ and $(P_0/N_0)$. Therefore, by substituting the converged value into (\ref{eq:Pb-hrd}), the error floor appears as
\begin{equation}
\label{eq:PEP-floor}
\lim \limits_{(P_0/N_0) \rightarrow \infty} P_{\bb}^{\CDD}(E)= \frac{1}{4\pi} \int \limits_{-\pi}^{\pi} g(\theta) \frac{1-\alpha^2}{\alpha^2q(\theta)+1-\alpha^2} \dd \theta.
\end{equation}
It is seen that the error floor is determined based on the amount of the equivalent channel auto-correlation and also the parameters of the signal constellation. Thus, one way to control this error floor would be to keep the normalized Doppler frequency as low as possible by reducing the symbol duration of the system. {Moreover, the BER expression can be used to optimize the power allocation between Source and Relay in the network. This is explained further in Appendix \ref{ddh:Appen1}.}
%In addition, the outage probability, i.e., the probability that the instantaneous effective SNR at the output of the differential detector drops below a SNR threshold $\gth$ is
%\begin{multline}
%\label{eq:Pout}
%P_{\mathrm{out}}=\mathrm{Pr}(\gamma \leq \gth)=\int \limits_{0}^{\infty} F_{\gamma|h_2}(\gth) \frac{1}{\sigma_2^2} e^{\left( -\frac{\lambda}{\sigma_2^2} \right)} \dd \lambda\\=
%1- e^{\left(- \frac{c_2\gth}{c_1}\right)} \int \limits_{0}^{\infty} \frac{1}{\sigma^2_2} e^{\left(- \frac{\lambda}{\sigma_2^2}-\frac{c_3\gth}{c_1}\frac{1}{\lambda} \right)} \dd \lambda\\=1- e^{\left(- \frac{c_2\gth}{c_1}\right)} \sqrt{\frac{4c_3 \gth}{c_1 \sigma_2^2}} K_1 \left( \sqrt{\frac{4c_3\gth}{c_1\sigma_2^2}} \right)
%\end{multline}
%where $K_1(\cdot)$ is the first-order modified Bessel function of the second kind and
%$$c_1=\alpha^2 A^2 \rho_1$$
%$$c_2=\left[1+\alpha^2+(1-\alpha^2)\rho_1\right] A^2$$
%$$c_3=1+\alpha^2.$$
%
%Similarly, by substituting \eqref{eq:gb_limit} into \eqref{eq:cdf-gama_b}, the outage probability reaches an floor, determined as
%\begin{equation}
%\label{eq:pout_floor}
%\lim \limits_{(P_0/N_0 \rightarrow \infty)} \mathrm{Pr}(\gamma \leq \gth)= 1- \exp{\left(-\frac{\gth (1-\alpha^2)}{\alpha^2} \right)}
%\end{equation}

\section{Multiple-Symbol Detection}
\label{sec:MSDSD}
As discussed in the previous section, two-symbol non-coherent detection suffers from a high error floor in fast-fading channels. To overcome such a limitation, this section designs and analyses a multiple-symbol detection scheme that takes a window of the received symbols at Destination for detecting the transmitted signals. %The detection process is still non-coherent, i.e., no instantaneous CSI is needed, though, the second order statistics of the channels (variance and auto-correlation) are required for the detection process.

\subsection{Detection Process}
Let the $N>2$ received symbols be collected in vector $\yb=\left[\; y[1],y[2],\dots, y[N]\; \right]^t$, which can be written as
\begin{equation}
\label{eq:Y}
\yb=A\sqrt{P_0} \dgs \diag\{\hb_2\} \hb_1 +\w
\end{equation}
where
\begin{gather}
\s= \left[\; s[1],\cdots, s[N]\; \right]^t \\
\hb_2=\left[\; h_2[1],\cdots, h_2[N] \;\right]^t \\
\hb_1=\left[\; h_1[1],\cdots, h_1[N]\; \right]^t \\
\w=\left[\; w[1],\cdots, w[N]\; \right]^t.
\end{gather}
Therefore, conditioned on both $\s$ and $\hb_2$, $\yb$ is a circularly symmetric complex Gaussian vector with the following pdf:
\begin{equation}
\label{eq:pdfY}
P(\yb|\s,\hb_2)=\frac{1}{\pi^N \mathrm{det}\{\Sgy\}} \exp\left( -\yb^H \Sgy^{-1} \yb \right).
\end{equation}
In \eqref{eq:pdfY}, the matrix $\Sgy$ is the conditional covariance matrix of $\yb$, defined as
\begin{equation}
\label{eq:RY}
%\begin{split}
\Sgy=\E \{ \yb \yb^H | \s,\hb_2 \}=
A^2 P_0 \dgs \diag\{\hb_2\} \Sg_{\hb_1} \diag\{\hb_2^*\} \dgsc+\Sgw
%\end{split}
\end{equation}
with
\begin{gather}
\Sg_{\hb_1}=\E\{ \hb_1 \hb_1^H \}=\mathrm{toeplitz}\left\lbrace \varphi_1(0),\dots,\varphi_1(N-1)\right\rbrace, \\
\Sgw= N_0 \mbox{diag}\left\lbrace (1+A^2|h_2[1]|^2),\cdots,(1+A^2|h_2[N]|^2) \right\rbrace
\end{gather}
as the covariance matrices of $\hb_1$ and $\w$, respectively.

Based on \eqref{eq:pdfY}, the maximum likelihood (ML) detection would be given as
\begin{equation}
\label{eq:ML}
\hat{\s}=\arg \max \limits_{\s \in \CpN} \left\lbrace \underset{\hb_2}{\E} \left\lbrace
\frac{1}{\pi^N \mathrm{det}\{\Sgy\}} \exp\left( -\yb^H \Sgy^{-1} \yb \right)
\right\rbrace \right\rbrace.
\end{equation}
where $\hat{\s}= \left[\; \hat{s}[1],\cdots, \hat{s}[N] \; \right]^t$.
As it can be seen, the ML metric needs the expectation over the distribution of $\hb_2$, which does not yield a closed-form expression. %Certain assumptions can be made to simplify the decision metric. For instance, if the RD link is very strong, the effect of the RD channel can be ignored to give a decision metric without any expectation. However, because of the dependence of $\Sgy$ to $\s$, one still needs an exhaustive search over all possible data sequences.
As an alternative, it is proposed to use the following modified decision metric:
\begin{equation}
\label{eq:ML-Modified}
\hat{\s}=\arg \max \limits_{\s \in \CpN} \left\lbrace
\frac{1}{\pi^N \mathrm{det}\{\overline{\Sg}_{\yb}\}} \exp\left( -\yb^H \overline{\Sg}_{\yb}^{-1} \yb \right)
\right\rbrace
\end{equation}
where
\begin{equation}
\label{eq:bSig_y}
\overline{\Sg}_{\yb}=\underset{\hb_2}{\E} \{ \Sgy \}=A^2P_0 \dgs \Sg_{\hb} \dgsc+(1+A^2\sigma_2^2)N_0 \Ib_N
%A^2P_0 \dgs \Rb_{\hb} \dgsc+(1+A^2\sigma_2^2)N_0 \dgs \dgsc \\
=\dgs\; \C \; \dgsc
\end{equation}
with
\begin{equation}
\label{eq:C}
\C=A^2P_0  \Sg_{\hb} +(1+A^2\sigma_2^2)N_0\Ib_N
\end{equation}
\begin{equation}
\label{eq:R_h}
\Sg_{\hb}=\E\left\lbrace \dgh \Sg_{\hb_1} \dghc \right\rbrace=
\mathrm{toeplitz} \{ \varphi_1(0)\varphi_2(0),\cdots,\varphi_1(N-1)\varphi_2(N-1) \}.
\end{equation}
Although the simplified decision metric is not optimal in the ML sense, it will be shown by simulation results that nearly identical performance to that obtained with the optimal metric can be achieved.

Using the rule $\det \{\A\B\}=\det\{\B\A\}$, the determinant in \eqref{eq:ML-Modified} is no longer dependent to $\s$ and the modified decision metric can be further simplified as
\begin{equation}
\label{eq:ML-simp}
\hat{\s}=\arg \min \limits_{\s\in \CpN} \left\lbrace \yb^H \overline{\Sg}_{\yb}^{-1} \yb \right\rbrace
=\arg \min \limits_{\s\in \CpN} \{\yb^H \dgs \C^{-1} \dgsc \yb \}.
%\\
%=\arg \min \limits_{\s \in \CpN} \left\lbrace (\diag\{\yb\} \s^*)^H \Lb\Lb^H \diag\{\yb\} \s^* \right\rbrace
%=\arg \min \limits_{\s\in \CpN} \left\lbrace  \|\U \s \|^2 \right\rbrace.
\end{equation}
Next using the Cholesky decomposition of $\C^{-1}=\Lb \Lb^H$ gives
\begin{equation}
\label{eq:ML-simp2}
\hat{\s}=\arg \min \limits_{\s \in \CpN} \left\lbrace (\diag\{\yb\} \s^*)^H \Lb\Lb^H \diag\{\yb\} \s^* \right\rbrace
=\arg \min \limits_{\s\in \CpN} \left\lbrace  \|\U \s \|^2 \right\rbrace,
\end{equation}
where $\U=(\Lb^H \diag\{\yb\})^*$.

The minimization in \eqref{eq:ML-simp2} can then be solved using sphere decoding described in \cite{MSDSD-L} to find $N-1$ information symbols. The MSD algorithm adapted for D-DH relaying is summarized in \emph{Algorithm I}. The complexity of the detection process is similar to that of point-to-point (P2P) communications, considered in \cite{MSDSD-L,msd-hassibi}. Although, here, the second-order statistics of two channels are required, in contrast to one channel in P2P communications. It should be mentioned that steps 1 to 3 are performed once, whereas steps 4 to 6 will be repeated for every $N$ consecutive received symbols. Also, the processed windows overlap by one symbol, i.e., the observation window of length $N$ moves forward by $N-1$ symbols at a time. Since, one symbol is taken as the reference, the output of the detection process contains $N-1$ detected symbols.

In the next section, the error analysis of the developed multiple-symbol detection is presented.

\begin{table}[thb!]
\renewcommand{\arraystretch}{2}
\label{tb:msdsd}
\centering
\begin{tabular}{l}
\hline
\bf Algorithm 1: MSD-DH \\
\hline
{\bf Input:} $\sigma_1^2,\sigma_2^2, f_1, f_2, A, P_0,N_0, M, N, \yb$ \\
{\bf Output:} $\hat{v}[k],\quad k=1,\cdots,N-1$ \\
\hline
1: Find $\Rb_{\hb}$ from \eqref{eq:R_h} \\
2: Find $\C$ from \eqref{eq:C} \\
3: Find $\Lb$ from $\C^{-1}=\Lb\Lb^H$  \\
4: Find $\U=(\Lb^H \diag\{\yb\})^*$ \\
5: Call function $\hat{\s}$=MSDSD ($\U$,$M$) \cite{MSDSD-L}\\
6: $\hat{v}[k]=\hat{s}^*[k] \hat{s}[k+1],\quad k=1,\cdots,N-1$\\
\hline
\end{tabular}
\end{table}

\subsection{Performance Analysis}
\label{subsec:msd-analysis}
Assume that vector $\s$ is transmitted and it is decoded as vector $\hat{\s}$. Based on the decision rule \eqref{eq:ML-simp}, an error occurs if
\begin{equation}
\label{eq:err1}
\yb^H \diag\{\hat{\s}\} \C^{-1} \diag\{\hat{\s}^*\} \yb \leq \yb^H \diag\{{\s}\} \C^{-1} \diag\{{\s}^*\} \yb,
\end{equation}
which can be simplified as
\begin{equation}
\label{eq:pep1}
\Delta=\yb^H \Q \yb \leq 0,
\end{equation}
with
\begin{equation}
\label{eq:Q}
\Q=\diag\{\hat{\s}-\s\} \C^{-1} \diag\{\hat{\s}^*-\s^*\}.
\end{equation}
Therefore, the PEP is defined as
\begin{equation}
\label{eq:msdPEP1}
P(\s \rightarrow \hat{\s})=P(\Delta\leq 0|\s,\hat{\s}).
\end{equation}
The above probability can be solved using the method of \cite{ber-biglieri} as
\begin{equation}
P(\Delta\leq 0|\s,\hat{\s})
\approx
 -\frac{1}{q} \sum \limits_{k=1}^{q/2} \left\lbrace c \Re[\Phi_{\Delta}(c+jc\tau_k)]+\tau_k \Im[\Phi_{\Delta}(c+jc\tau_k)] \right\rbrace
\end{equation}
where $\Phi_{\Delta}(\cdot)$ is the characteristic function of the random variable $\Delta$ and $\tau_k=\tan((2k-1)\pi/(2q))$. Also, the constant $c$ can be set equal to one half the smallest real part of the poles of $\Phi_{\Delta}(t)$ and $q=64$ gives enough accuracy for the approximation \cite{ber-biglieri}.

To proceed with computing \eqref{eq:msdPEP1}, the characteristic function of random variable $\Delta$ should be determined. Based on the modified decision rule \eqref{eq:ML-Modified}, conditioned on $\s$, $\yb$ is circularly symmetric complex Gaussian vector with zero mean and covariance matrix $\bSigy$. Thus the characteristic function of the quadratic form $\Delta=\yb^H \Q \yb$ can be shown to be \cite{ch_quad}
\begin{equation}
\label{eq:phi_d}
\Phi_{\Delta}(t)=\frac{1}{\det\{\Ib_N+t\bSigy\Q\}}.
\end{equation}
By substituting \eqref{eq:phi_d} into \eqref{eq:msdPEP1}, the PEP of multiple-symbol detection is obtained. The BER can be obtained by the union bound of the PEP over the set of dominant errors as \cite{msdd-div}
\begin{equation}
\label{eq:msd-Pb}
P_{\bb}^{\MSD}\approx \frac{w}{\log_2(M)(N-1)} P(\s \rightarrow \hat{\s}),
\end{equation}
where $w$ is the sum of Hamming distances between the bit streams corresponding to the dominant errors. In the dominant errors, the detected vector $\hat{\s}$ is different than the transmitted vector $\s$ by only one closest symbol. For $M$-PSK, without loss of generality one can set
\begin{gather}
\s=[\;1,\cdots,1,1\;], \\
\hat{\s}=[\; 1,\cdots,1,\exp(j2\pi/M)\;]
\end{gather}
to find $P(\s\rightarrow \hat{\s})$. Also, with Gray-mapping \cite{msdd-div}
\begin{equation}
\label{eq:w}
w=\left\lbrace
\begin{matrix}
2(N-1), & \mbox{if} \; M=2 \\
4(N-1), & \mbox{if} \; M>2.
\end{matrix} \right.
\end{equation}

\section{Simulation Results}
\label{sec:sim}
In this section the dual-hop relay network under consideration is simulated for various channel qualities using both CDD and MSD to verify the analysis.

Information bits are differentially encoded with either BPSK ($M=2$) or QPSK ($M=4$) constellations. Note that, $\left\lbrace a=0,\; b=\sqrt{2}\right\rbrace$ and $\left\lbrace a=\sqrt{2-\sqrt{2}},\; b=\sqrt{2+\sqrt{2}}\right\rbrace$ are obtained for DBPSK and DQPSK, respectively \cite{unified-app}. The amplification factor at Relay is fixed to $A=\sqrt{P_1/(P_0\sigma_1^2+N_0)}$ to normalize the average relay power to $P_1$. Also, $N_0=1$ is assumed.

Based on the fading powers of the channels, three scenarios would be considered as: symmetric channels with $\sigma_1^2=1,\sigma_2^2=1$, non-symmetric channels with strong SR channel $\sigma_1^2=10,\sigma_2^2=1,$ and non-symmetric channels with strong RD channel $\sigma_1^2=1,\sigma_2^2=10$. The fading powers are listed in Table~\ref{table:sig12}.

\begin{figure}[htb!]
\psfrag {PAF} [t][] [1]{$\varrho=\frac{P_0}{P}$}
\psfrag {BER} [] [] [1] {BER}
\psfrag {sig12=110} [] [] [1] {$\quad \qquad \sigma_1^2=1,\sigma_2^2=10$}
\psfrag {sig12=101} [] [] [1] {$\quad \qquad\sigma_1^2=10,\sigma_2^2=1$}
\psfrag {sig12=11} [] [] [1] {$\quad \qquad\sigma_1^2=1,\sigma_2^2=1$}
\psfrag {P/N0} [] [] [1] {$P/N_0=[30,35,40]$ dB}
\centerline{\epsfig{figure={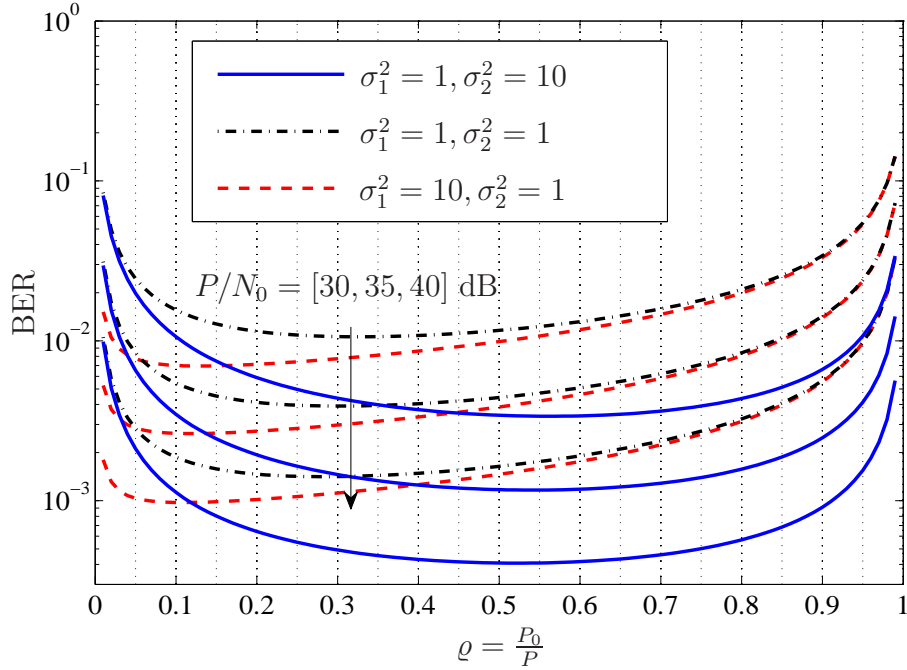},height=8.5cm,width=12cm}}
\caption{BER as a function of power allocation factor for various channel variances and $P/N_0=[ 30,35,40 ]$ dB from top to bottom, using DQPSK.}
\label{fig:pw_m4_all}
\end{figure}

\begin{table}[!ht]
\begin{center}
\caption{Fading powers and corresponding optimum power allocation factors.}
\label{table:sig12}
\vspace*{.1in}
\begin{tabular}{cc|c|c|}
\cline{2-3}
&\multicolumn{1}{|c|}  {$[\sigma_1^2, \sigma_2^2]$} & $\varrho_{\opt}$   \\ \cline{1-3}
\multicolumn{1}{ |c| } {Symmetric} & {$[1,1]$} & 0.30   \\
\cline{1-3}
\multicolumn{1}{ |c| }{Strong SR} & {$[10,1]$} & 0.12   \\
\cline{1-3}
\multicolumn{1}{ |c| }{Strong RD} & {$[1,10]$} & 0.54  \\
\cline{1-3}
\end{tabular}
\end{center}
\end{table}
First, the BER expression of CDD is used to optimize the power allocation between Source and Relay in the network. The optimization problem aims to minimize the BER for a given total power $P=P_0+P_1$. Let $P_0=\varrho P$, where $\varrho$ is the power allocation factor. Then $P_1=(1-\varrho)P$ and from \eqref{eq:A}, $A=\sqrt{P(1-\varrho)/(\varrho P \sigma_1^2+N_0)}$. By substituting $P_0$ and $A$ into \eqref{eq:Pb}, the optimization problem is to find the value of $\varrho$ for which $P_{\mathrm{b}}(E)$ is minimized. Since the best performance is obtained in slow-fading channels, $\alpha=1$ would be set in \eqref{eq:Pb}. Then, the minimization would be performed by a numerical exhaustive search.
%For the analytical approach, it is easy to see that the minimization does not depend on $\theta$. Thus minimizing $P_{\mathrm{b}}(E)$ is equivalent to minimizing $J(\theta)$, for any value of $\theta,$ say $\theta=\pi/2$. Therefore, the optimized power allocation is obtained by minimizing
%\begin{equation}
%\label{eq:Jtilde}
%\tilde{J}=\tilde{b}_3+\frac{\tilde{b}_3}{\sigma_2^2} (\tilde{b}_1-\tilde{b}_2) e^{\frac{\tilde{b}_2}{\sigma_2^2}} E_1\left( \frac{\tilde{b}_2}{\sigma_2^2}\right)
%\end{equation}
%where $\tilde{b}_3=\tilde{b}_2/\tilde{b}_1$, $\tilde{b}_1=1/A^2$ and $\tilde{b}_2=2/(2A^2+q(\pi/2)A^2\rho_1)$. It can be seen that for large $P$ or $P_0$, $\tb_1-\tb_2\approx \tb_1 $ and then $\tb_3(\tb_1-\tb_2)\approx \tb_2$. On the other hand, for small $x=\tb_2/\sigma_2^2$ one has the following approximation \cite[eq.5.1.20]{integral-tables}
%$$
%\label{eq:exE1x}
%e^x E_1(x)\approx \log\left( \frac{1}{x}\right).
%$$
%Thus, $\tilde{J}$ can be approximated as
%\begin{equation}
%\label{eq:Jtilde_approx}
%\tilde{J} \approx \tb_3 + \frac{\tb_2}{\sigma_2^2} \log\left( \frac{\sigma_2^2}{\tb_2}\right)
%\end{equation}
%Then, finding the minimum of \eqref{eq:Jtilde_approx} over the single variable $\varrho$. can be easily done by solving the derivative of \eqref{eq:Jtilde_approx} respect to $\varrho.$
The obtained optimum power allocations are listed in Table~\ref{table:sig12} for $P/N_0=35$ dB. Note that $P$ is the total power in the network and is divided between Source and Relay. It will be seen that, to achieve a low BER around $10^{-3}$ or $10^{-4}$, such a high value would be needed. In addition, the BER curves obtained from the exhaustive search are plotted versus $\varrho$ in Figure~\ref{fig:pw_m4_all} for $P/N_0=30,\;35,\;40$ dB and when DQPSK is employed. Similar results would be obtained when DBPSK is employed.  The results in Table~\ref{table:sig12} and Figure~\ref{fig:pw_m4_all} show that for symmetric and strong SR channels, more power should be allocated to Relay than Source and the BER is minimized at $\varrho_{\mathrm{opt}}\approx 0.3$ and $\varrho_{\mathrm{opt}}\approx 0.1$, for symmetric and strong SR channels, respectively. When the RD channel becomes stronger than the SR channel, the system would benefit more from an equal power allocation and the BER is minimized at $\varrho_{\mathrm{opt}}\approx 0.5$.

In all simulations, the channel coefficients, $h_1[k]$ and $h_2[k]$, are generated based on the simulation method of \cite{ch-sim}. This simulation method was developed to generate channel coefficients that are correlated in time. The amount of time-correlation is determined by the normalized Doppler frequency of the underlying channel, which is a function of the speed of the vehicle, carrier frequency and symbol duration. For fixed carrier frequency and symbol duration, a higher vehicle speed leads to a larger Doppler frequency and less time-correlation.

\begin{figure}[htb!]
\psfrag {f1 changes} [t][] [.8]{$f_1$ changes}
\psfrag {f12 change} [b][] [.8]{$f_1\&f_2$ change}
\psfrag {f} [t][] [1]{fade rate}
\psfrag {BER} [] [] [1] {Error Floor}
\psfrag {Analysis} [] [r] [.8] {Analysis \eqref{eq:PEP-floor}}
\psfrag {correlation} [] [] [1] {Autocorrelation}
\centerline{\epsfig{figure={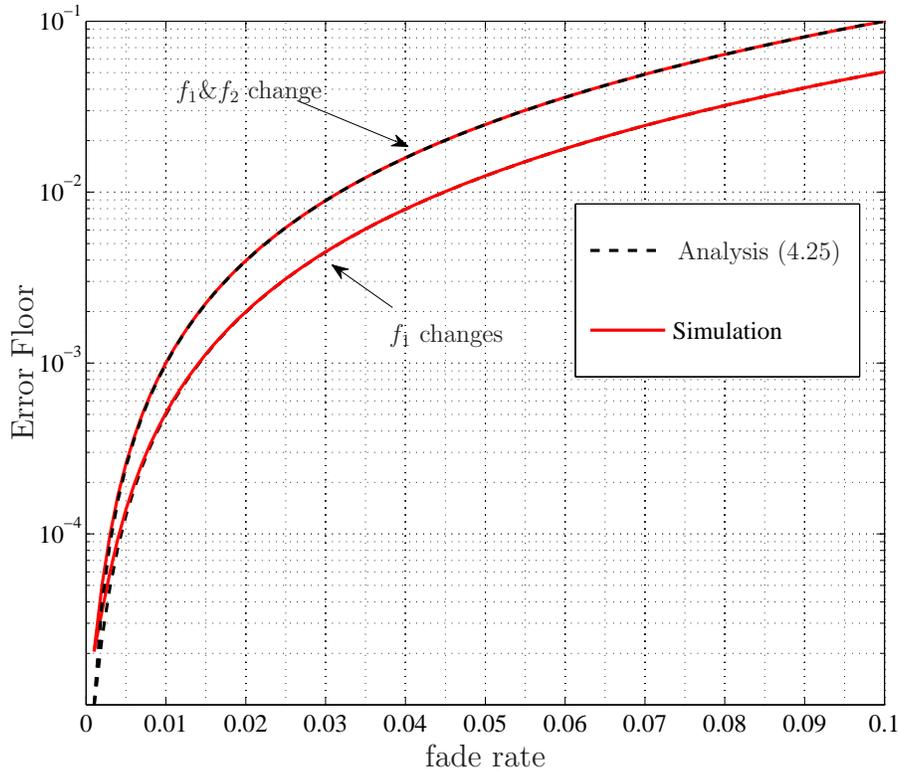},width=12cm}}
\caption{Theoretical and simulation values of error floor vs. fading rate, $M=2, \varrho=0.3$, $[\sigma_1^2,\sigma_2^2]=[1,1]$. }
\label{fig:ef_fd}
\end{figure}
To get a better understanding about the fading rate values and also verifying the analysis, the error floor values of CDD are obtained from both expression \eqref{eq:PEP-floor} and simulation at a high transmit power for a wide range of fading rates and plotted in Figure~\ref{fig:ef_fd}.
%Other parameters are $M=2$, with symmetric channels and $\varrho=0.25, P/N_0=60$dB.
In the figure, the lower graph corresponds to the case that only $f_1$ changes and $f_2=0.001$. In the upper graph both $f_1$ and $f_2$ change. Clearly, the error floors from both analysis and simulation tight together. For small values of fading rate around 0.001, the error floor is very low, around $10^{-5}$, but increases toward $10^{-3 }$ with increasing the fading rate to $0.01$. The fading rate around $0.01$ can be regarded as the margin beyond which the channels become fast-fading. As also seen, the error floor values are higher when both Doppler values change.

Based on the previous observation and the normalized Doppler frequencies of the two channels, different cases can be considered. In Case I, it is assumed that all nodes are fixed or slowly moving so that both channels are slow-fading with the normalized Doppler values of $f_1=.001$ and $f_2=.001$. In Case II, it is assumed that Source is moving so that the SR channel is fast-fading with $f_1=.01$. On the other hand, Relay and Destination are fixed and the RD channel is slow-fading with $f_2=.001$. In Case III, it is assumed that both Source and Relay are moving so that both the SR and RD channels are fast-fading with $f_1=.02$ and $f_2=.01$, respectively.
The normalized Doppler values are shown in Table \ref{table:f1f2}. %Also, a snapshot of realizations of the cascaded channel and their auto-correlation values in the three cases are plotted in Figure~\ref{fig:doub-ch}.
\begin{table}[!ht]
\begin{center}
\caption{Three fading scenarios.}
\vspace*{.1in}
\label{table:f1f2}
\begin{tabular}{ |c | c| c|c|  }
    \hline
				& $f_1$ & $f_2$ & Channels status   \\ \hline\hline
{Case I }& 0.001 & 0.001  & both are slow-fading \\ \hline
{Case II } & 0.01 & 0.001& SR is fast-fading  \\ \hline
{Case III } & 0.02 & 0.01 & both are fast-fading  \\
    \hline
  \end{tabular}
\end{center}
\end{table}
%\vspace*{-0.5cm}

%\begin{figure}[htb!]
%\psfrag {time} [][] [1]{$k$}
%\psfrag {h} [] [] [1] {$|h[k]|$}
%\psfrag {correlation} [] [] [1] {Autocorrelation}
%\centerline{\epsfig{figure={figure/dch_scs.eps},width=8.5cm}}
%\caption{Snapshot of realizations of the cascaded channel and corresponding autocorrelations in different cases.}
%\label{fig:doub-ch}
%\end{figure}

\begin{figure}[htb!]
\psfrag {P(dB)} [t][b] [1]{$(P/N_0)$ (dB)}
\psfrag {BER} [] [] [1] {BER}
\centerline{\epsfig{figure={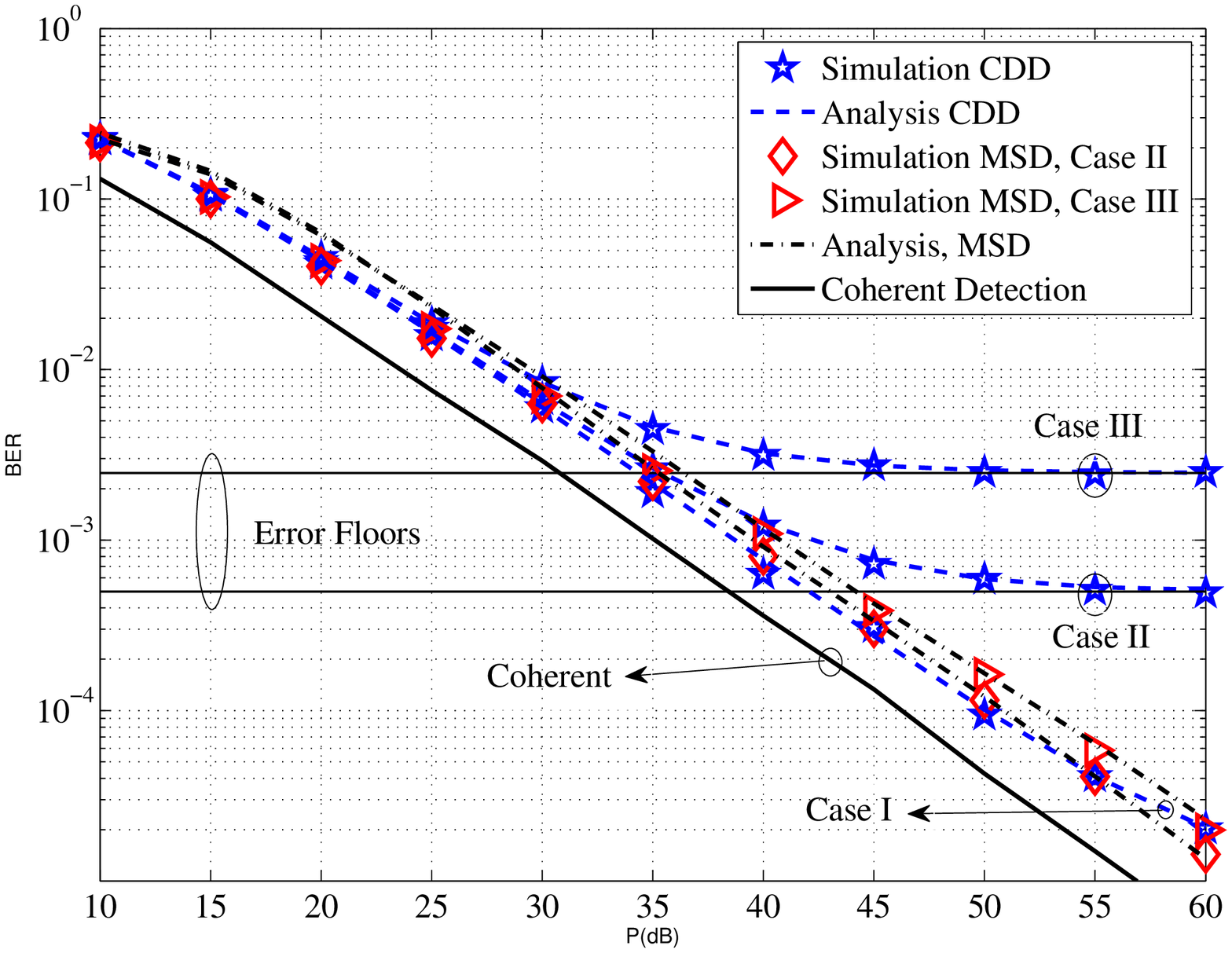},height=8.5cm,width=12cm}}
\caption{Theoretical and simulation BER of a D-DH relaying in different fading cases and $[\sigma_1^2,\sigma_2^2]=[1,1]$ using DBPSK and CDD ($N=2$) and MSD ($N=10$).}
\label{fig:ber_m2_sg11}
\end{figure}

\begin{figure}[htb!]
\psfrag {P(dB)} [t][b] [1]{$(P/N_0)$ (dB)}
\psfrag {BER} [] [] [1] {BER}
\centerline{\epsfig{figure={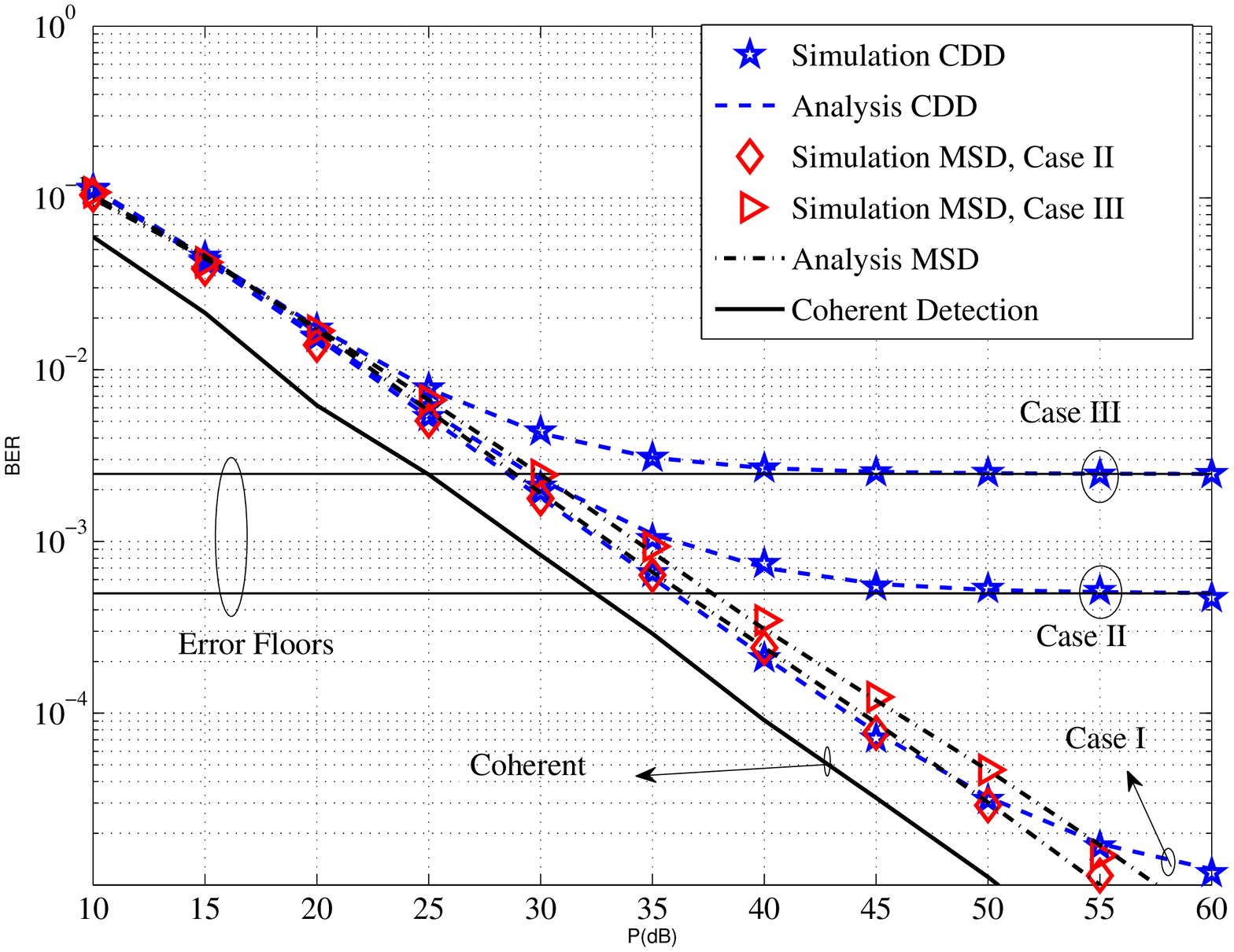},height=8.5cm,width=12cm}}
\caption{Theoretical and simulation BER of a D-DH relaying in different fading cases and $[\sigma_1^2,\sigma_2^2]=[1,10]$ using DBPSK and CDD ($N=2$) and MSD ($N=10$).}
\label{fig:ber_m2_sg110}
\end{figure}

\begin{figure}[htb!]
\psfrag {P(dB)} [t][b] [1]{$(P/N_0)$ (dB)}
\psfrag {BER} [] [] [1] {BER}
\centerline{\epsfig{figure={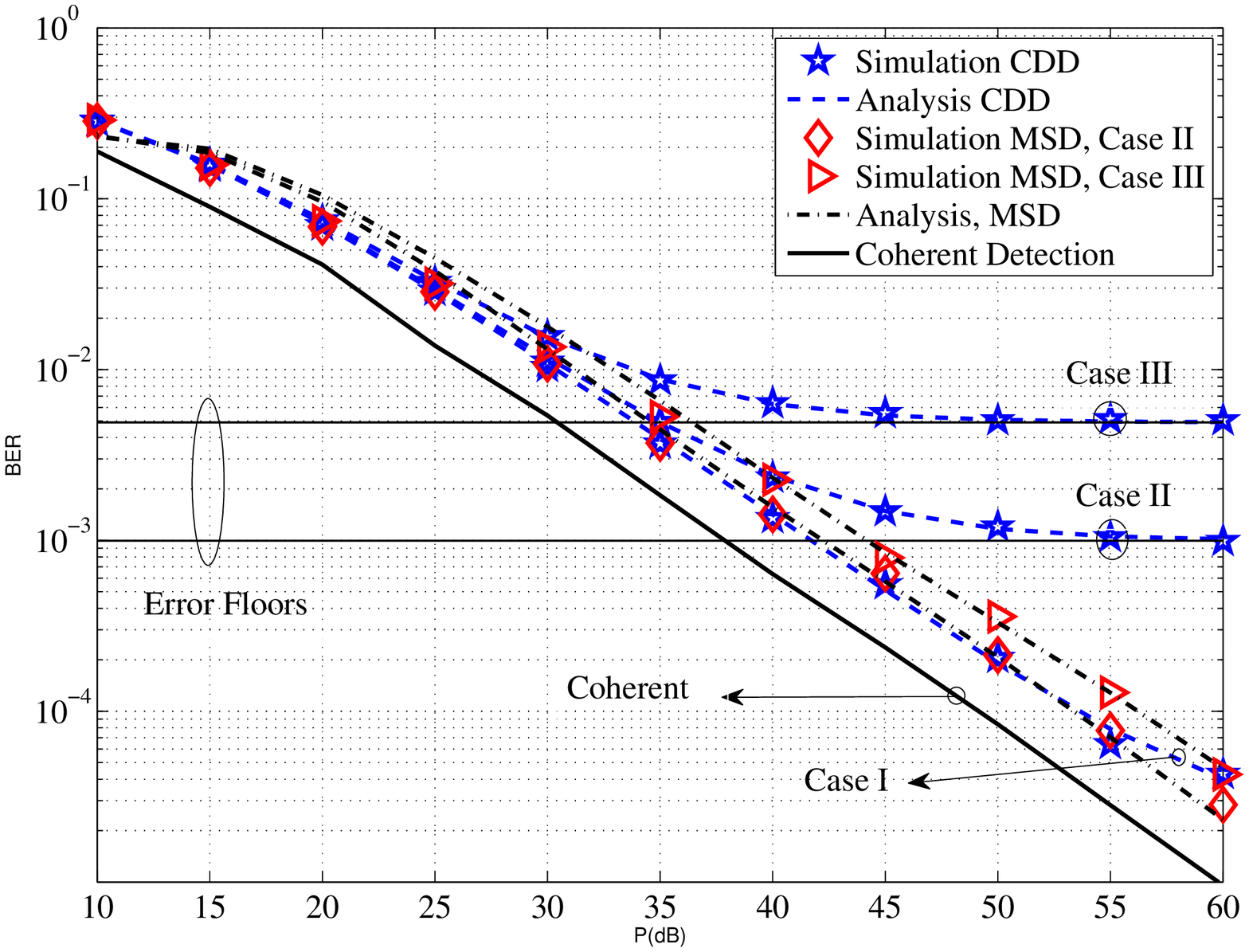},height=8.5cm,width=12cm}}
\caption{Theoretical and simulation BER of a D-DH network in different fading cases and $[\sigma_1^2,\sigma_2^2]=[1,1]$ using DQPSK and CDD ($N=2$) and MSD ($N=10$).}
\label{fig:ber_m4_sg11}
\end{figure}

\begin{figure}[htb!]
\psfrag {P(dB)} [t][b] [1]{$(P/N_0)$ (dB)}
\psfrag {BER} [] [] [1] {BER}
\centerline{\epsfig{figure={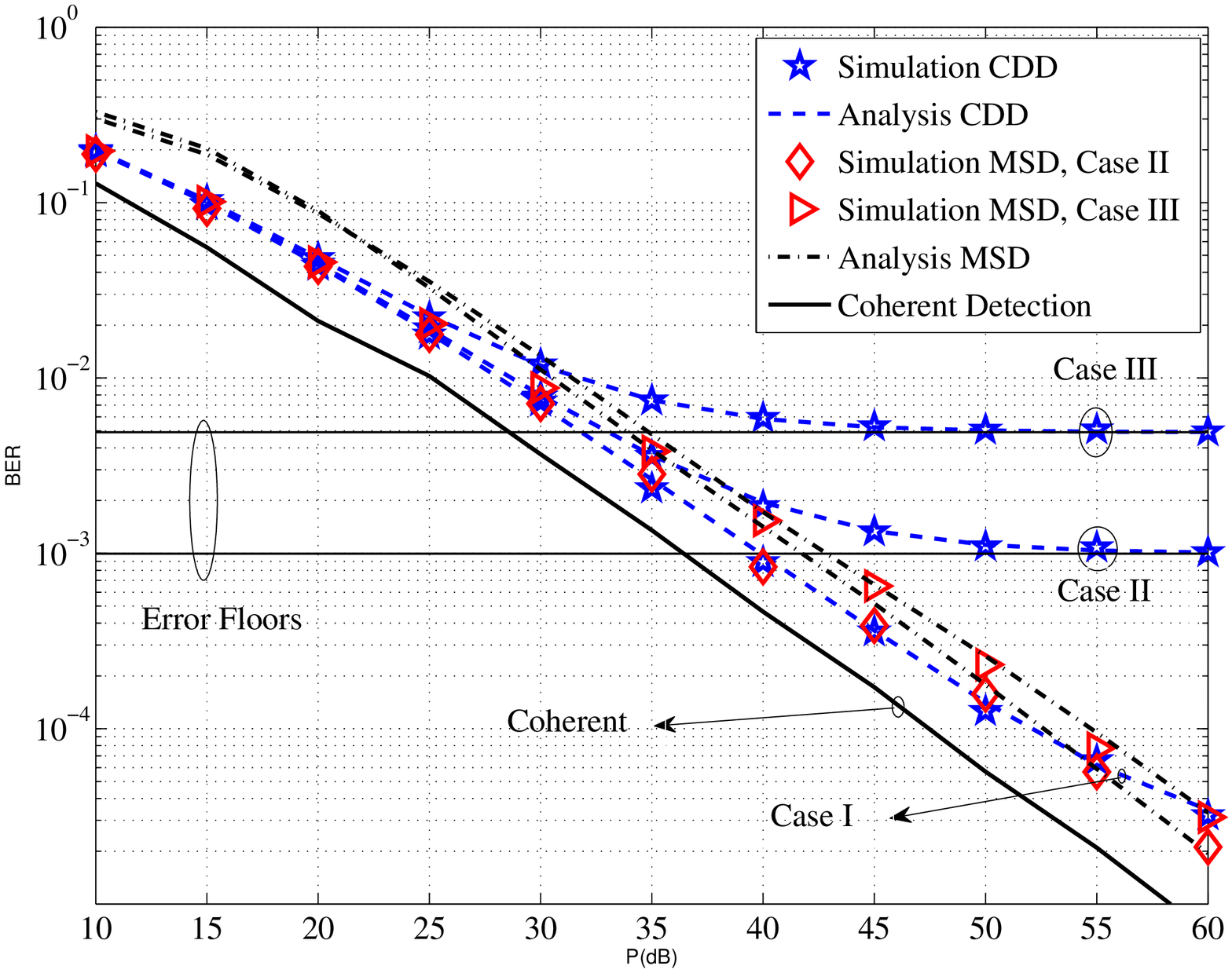},height=8.5cm,width=12cm}}
\caption{Theoretical and simulation BER of a D-DH relaying in different fading cases and $[\sigma_1^2,\sigma_2^2]=[10,1]$ using DQPSK and CDD ($N=2$) and MSD ($N=10$).}
\label{fig:ber_m4_sg101}
\end{figure}

Based on the obtained optimum power allocation values in Table~\ref{table:sig12}, the simulated BER values using the CDD are computed for all cases and different channel variances and are plotted versus $P/N_0$ in Figs.~\ref{fig:ber_m2_sg11}-\ref{fig:ber_m4_sg101}, for DBPSK and DQPSK modulations. Also, the simulated BER values using coherent detection are computed for Case I (slow-fading) and plotted in Figs.~\ref{fig:ber_m2_sg11}-\ref{fig:ber_m4_sg101}. Figures~\ref{fig:ber_m2_sg11} and \ref{fig:ber_m2_sg110} correspond to symmetric channels and strong RD channel, respectively, while Figures~\ref{fig:ber_m4_sg11} and \ref{fig:ber_m4_sg101} correspond to symmetric channels and strong SR channel, respectively. %In addition, the simulated outage probability for $\gth=0$ and 5 dB using the CDD are also computed for all cases and different channel variances. They are plotted versus $P/N_0$ in Figs.~\ref{fig:pout_m2_scs_sg101_gths}-\ref{fig:pout_m4_scs_sg110_gths}, for DBPSK with strong SR channel and DQPSK with strong RD channel, respectively.
On the other hand, the corresponding theoretical BER values (for all cases) and the error floors (cases II\&III) for the CDD are computed from (\ref{eq:Pb}) and \eqref{eq:PEP-floor} (with the corresponding variances) and plotted in Figs.~\ref{fig:ber_m2_sg11}-\ref{fig:ber_m4_sg101}.

As can be seen in Fig.~\ref{fig:ber_m2_sg11}, in Case I with symmetric channels, the BER is monotonically decreasing with $(P/N_0)$ and it is consistent with the theoretical values in (\ref{eq:Pb}). Approximately, 3 dB performance loss is seen between coherent and non-coherent detection in this case. The error floor in this case does not practically exist. In Case II, which involves one fast-fading channel, this phenomenon starts earlier, around $35$ dB, and leads to an error floor at $5\times 10^{-4}$, which can also be predicted from (\ref{eq:PEP-floor}). The performance degradation is much more severe after $25$ dB in Case III since both channels are fast-fading, which leads to an error floor at $3\times 10^{-3}$. Similar behaviours can be seen in Figs.~\ref{fig:ber_m2_sg110}-\ref{fig:ber_m4_sg101} under other channel variances or DQPSK modulation. As is clearly seen from Figs.~\ref{fig:ber_m2_sg11}-\ref{fig:ber_m4_sg101}, the simulation results verify our theoretical evaluations. Specifically, the error floors in Case II and Case III do not depend on the channel variances.

Given the poor performance of the two-symbol detection in Cases II and III, MSD-DH algorithm is applied which takes a window of $N=10$ symbols for detection.
The BER results of the MSD-DH algorithm are also plotted in Figures~\ref{fig:ber_m2_sg11} -\ref{fig:ber_m4_sg101} with points (different legends).
Also, theoretical BER of MSD in Cases II and III are obtained from \eqref{eq:msd-Pb} and plotted in the figures with dash-dot lines. Since the best performance is achieved in the slow-fading environment, the performance plot of Case I can be used as a benchmark to see the effectiveness of MSD-DH. It can be seen that the MSD-DH is able to bring the performance of the system in Case II and Case III very close to that of Case I. Moreover, the analytical results of MSD are consistent with the simulation results and they match well in high SNR region. %Furthermore, the MSD-DH algorithm in Case II even outperforms the performance in Case I after 50 dB in Fig.~\ref{fig:ber_m2_sg110} and after 55 dB in Figs.~\ref{fig:ber_m2_sg11},\ref{fig:ber_m4_sg11}-\ref{fig:ber_m4_sg101}, which the CDD results tend to flat out for high transmit power.

\begin{figure}[htb!]
\psfrag {N} [t][b] [1]{$N$}
\psfrag {BER} [] [] [1] {BER}
\psfrag {f=0.01} [][] [.8]{$\qquad f_1=0.01$}
\psfrag {f=0.02} [][] [.8]{$\qquad f_1=0.02$}
\psfrag {f=0.03} [][] [.8]{$\qquad f_1=0.03$}
\psfrag {f=0.04} [][] [.8]{$\qquad f_1=0.04$}
\psfrag {f=0.05} [][] [.8]{$\qquad f_1=0.05$}
\psfrag {f=0.06} [][] [.8]{$\qquad f_1=0.06$}
\psfrag {f=0.07} [][] [.8]{$\qquad f_1=0.07$}
\psfrag {f=0.08} [][] [.8]{$\qquad f_1=0.08$}
\centerline{\epsfig{figure={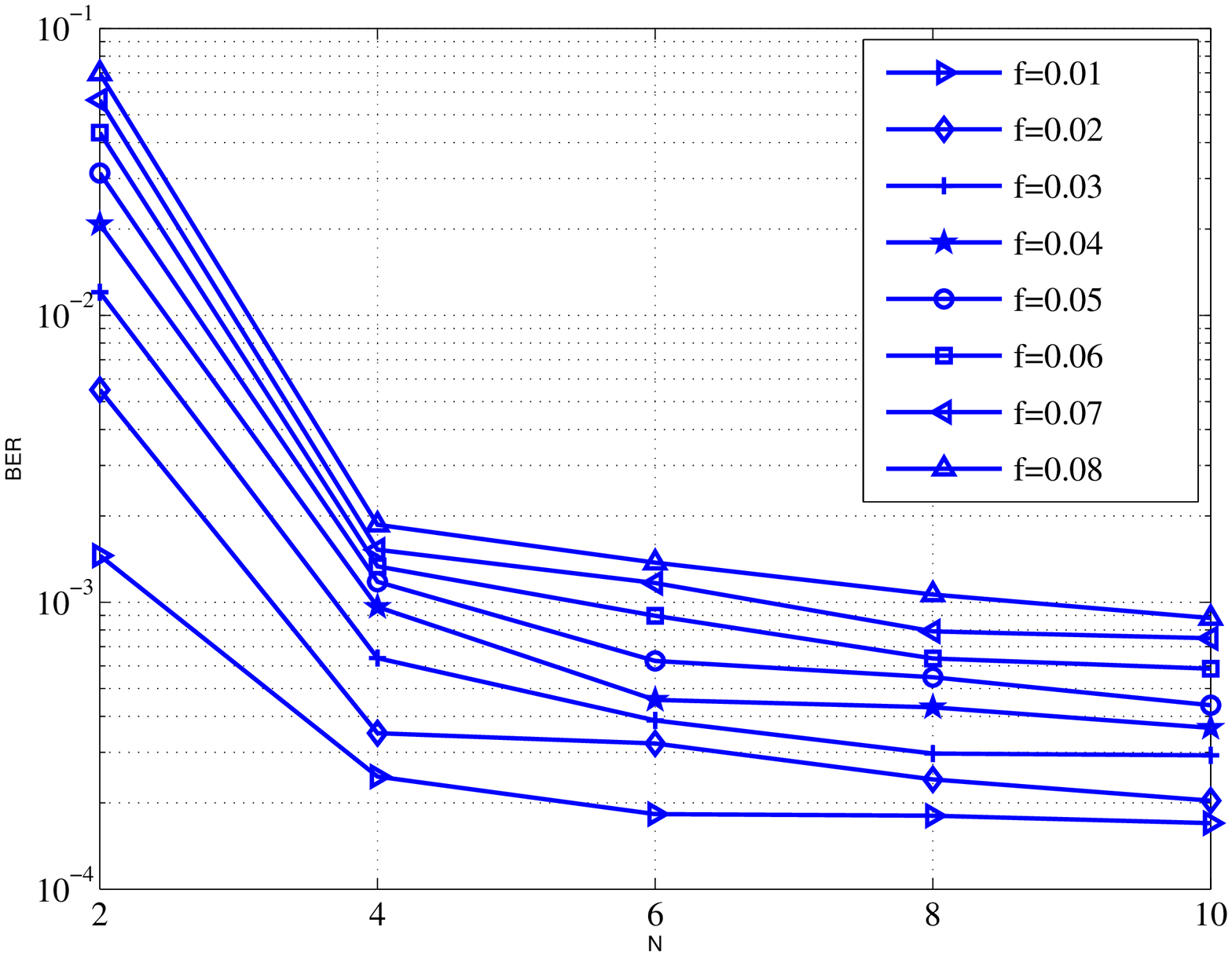},height=8.5cm,width=12cm}}
\caption{BER of MSDD as a function of block length at $P/N_0=50$ dB for different fade rates.}
\label{fig:ber_vs_N_fs}
\end{figure}

{
It is pointed out that the performance of MSDD is improved with increasing $N$. However, there would be a point beyond which the performance improvement is not significant. Here, the BER expression of MSDD can be used to determine the required block length for a desired BER at a certain fade rate. To see the effect of increasing the block length on the performance of MSDD at a specific fade rate and power, the BER expression derived in \eqref{eq:msd-Pb} is evaluated for a range of $N$ at $P/N_0=50$ dB and for various fade rates. The results are plotted in Figure.~\ref{fig:ber_vs_N_fs}. As it is seen from the figure, much of the performance improvement is achieved by increasing the block length from two to four, for all the fade rates. After this point, the BER is slowly decreasing with increasing $N$, and the improvement might not be significant. This observation suggests to keep $N$ small to reduce the complexity, albeit with some performance penalty.
}

%\begin{figure}[htb!]
%\psfrag {P(dB)} [][b] [1]{$(P/N_0)$ (dB)}
%\psfrag {Pout} [] [] [1] {$P_{\mathrm{out}}$}
%\centerline{\epsfig{figure={figure/pout_m2_scs_sg101_gths.eps},width=8.5cm}}
%\caption{Outage probability of a D-DH network in different fading cases and $\sigma_1^2=10,\sigma_2^2=1$ using DBPSK and $\gth=$ 0 and 5 dB .}
%\label{fig:pout_m2_scs_sg101_gths}
%\end{figure}
%
%\begin{figure}[htb!]
%\psfrag {P(dB)} [][b] [1]{$(P/N_0)$ (dB)}
%\psfrag {Pout} [] [] [1] {$P_{\mathrm{out}}$}
%\centerline{\epsfig{figure={figure/pout_m4_scs_sg110_gths.eps},width=8.5cm}}
%\caption{Outage probability of a D-DH network in different fading cases and $\sigma_1^2=1,\sigma_2^2=10$ using DQPSK and $\gth=$ 0 and 5 dB .}
%\label{fig:pout_m4_scs_sg110_gths}
%\end{figure}

%\vspace*{-0.35cm}

\section{Conclusion}
\label{sec:con}
Differential AF relaying for a dual-hop transmission has been studied for time-varying Rayleigh fading channels. The analysis showed that, while simple, two-symbol differential detection (CDD) suffers from an error floor in fast-fading channels. This analysis would be useful to predict the error floor phenomenon and control its value with changing other system parameters such as the symbol duration. On the other hand, it was seen that multiple-symbol detection (MSD) can overcome the limitations of CDD in fast-fading channels, although the improvement is obtained at the price of higher complexity. %Future work would consider developing a MSD scheme for blind statistics channels.

\begin{subappendices}
\section{Optimum Power Allocation}
\label{ddh:Appen1}
The optimization problem aims to minimize the BER for a given total power $P=P_0+P_1$. Let $P_0=\varrho P$, where $\varrho$ is the power allocation factor. Then $P_1=(1-\varrho)P$ and from \eqref{eq:A}, $A=\sqrt{P(1-\varrho)/(\varrho P \sigma_1^2+N_0)}$. By substituting $P_0$ and $A$ into \eqref{eq:Pb}, the optimization problem is to find the value of $\varrho$ for which $P_{\mathrm{b}}(E)$ is minimized. Since the best performance is obtained in slow-fading channels, $\alpha=1$ would be set in \eqref{eq:Pb}. Then, the minimization would be performed either by a numerical exhaustive search or an analytical approach. For the analytical approach, it is easy to see that the minimization does not depend on $\theta$. Thus minimizing $P_{\mathrm{b}}(E)$ is equivalent to minimizing $J(\theta)$, for any value of $\theta,$ say $\theta=\pi/2$. Therefore, the optimized power allocation is obtained by minimizing
\begin{equation}
\label{eq:Jtilde}
\tilde{J}=\tilde{b}_3+\frac{\tilde{b}_3}{\sigma_2^2} (\tilde{b}_1-\tilde{b}_2) e^{\frac{\tilde{b}_2}{\sigma_2^2}} E_1\left( \frac{\tilde{b}_2}{\sigma_2^2}\right)
\end{equation}
where $\tilde{b}_3=\tilde{b}_2/\tilde{b}_1$, $\tilde{b}_1=1/A^2$ and $\tilde{b}_2=2/(2A^2+q(\pi/2)A^2(P_0/N_0)\sigma_1^2)$. It can be seen that for large $P$ or $P_0$, $\tb_1-\tb_2\approx \tb_1 $ and then $\tb_3(\tb_1-\tb_2)\approx \tb_2$. On the other hand, for small $x=\tb_2/\sigma_2^2$ one has the following approximation \cite[eq.5.1.20]{integral-tables}
$$
\label{eq:exE1x}
e^x E_1(x)\approx \log\left( \frac{1}{x}\right).
$$
Thus, $\tilde{J}$ can be approximated as
\begin{equation}
\label{eq:Jtilde_approx}
\tilde{J} \approx \tb_3 + \frac{\tb_2}{\sigma_2^2} \log\left( \frac{\sigma_2^2}{\tb_2}\right)
\end{equation}
Then, finding the minimum of \eqref{eq:Jtilde_approx} over the single variable $\varrho$ can be easily done by solving the derivative of \eqref{eq:Jtilde_approx} with respect to $\varrho.$
\end{subappendices}

\chapter{Differential Amplify-and-Forward Relaying Using Semi-MRC Combining}
\label{ch:mnode}
The previous chapter considers single-branch dual-hop relaying for scenarios that the direct link between the source and the destination is not sufficiently strong to facilitate data transmission. However, single-branch dual-hop relaying only extends the coverage area and does not improve the diversity order of a wireless system. If the direct link between the source and the destination is also available, a single-branch dual-hop relaying with a direct link, see Figure~\ref{fig:dhpd}, can be considered.
This topology is useful to improve the diversity order to two, yet simple to implement. There is only one relay involved in the communication and no synchronization with the source or other users is required. Also, less computational burden is imposed to the destination in the combining and detection process. Moreover, studying this topology provides a background for studying multi-branch relay networks.

Under the scenario that more users are willing to cooperate, a multi-branch relay network with a direct link, depicted in Figure~\ref{fig:mbpd}, can be constructed. Thus the achievable diversity would increase to $R+1$, where $R$ is the number of relays. When more than one relay are employed in the network, it is important to decide on the cooperative strategy among the relays. Here, repetition-based strategy is chosen due to its simplicity of implementation. In the re-broadcasting phase, the relays send their data sequentially in time and only frame synchronization is needed in the network. Obviously, more signal processing would be needed at the destination for combining and detection, which is the price of getting a better diversity with using multiple relays.

In the multi-branch dual-hop relaying with a direct link considered in this chapter, differential $M$-PSK is used at the source, amplify-and-forward is performed at the relays and two-symbol non-coherent detection at the destination is employed. No channel estimation is required at the relays or destination. An important processing part at the destination of a D-AF relaying system is how to combine the received signals to achieve the cooperative diversity. Here, a semi maximum-ratio-combining (semi-MRC) scheme is considered. The prefix semi is added to the MRC acronym because the conventional optimum combining weights are not available in the system under consideration. The research focus is to examine the performance and limitations of the D-AF relaying system using semi-MRC scheme in practical time-varying channels. Moreover, new combining weights are obtained to improve the performance. It will be seen that, similar to the system considered in the previous chapter, the performance of the system degrades with channel variation and the existence of an error floor is inevitable. However, the effect of channel variation can be mitigated by increasing the number of relays in the network.

The results of our study are reported in manuscripts [Ch5-1] and [Ch5-2]. Manuscript [Ch5-1] considers a three-node relay network (one relay). Manuscript [Ch5-2] studies a general case with $R$ relays and is presented in this chapter.

[Ch5-1] M. R. Avendi, Ha H. Nguyen,``Differential Amplify-and-Forward
Relaying in Time-Varying Rayleigh Fading Channels", \textit{IEEE Wireless Communications and Networking Conference, Shanghai, China, Apr. 7-10, 2013.}

[Ch5-2] M. R. Avendi, Ha H. Nguyen,``Performance of Differential Amplify-and-Forward
Relaying in Multi-Node Wireless Communications", \textit{IEEE Transactions on Vehicular Technology}, vol. 62, no. 8, pp. 3603--3613, Oct. 2013.

\begin{center}
{\bf{\Large
Performance of Differential Amplify-and-Forward Relaying in Multi-Node Wireless Communications
}}
\end{center}
\begin{center}
M. R. Avendi, Ha H. Nguyen
\end{center}
\blfootnote{Manuscript received November 1, 2012; revised February 21, 2013; accepted March 20, 2013. Date of publication May 10, 2013.
This work was supported in part by the Dean's Scholarship from the University of Saskatchewan and a Natural Sciences and Engineering Research Council of Canada Discovery Grant. The authors are with the Department of Electrical and Computer Engineering, University of Saskatchewan, Saskatoon, Canada, S7N5A9.
Email: m.avendi@usask.ca, ha.nguyen@usask.ca.}

\begin{center}
\bf Abstract
\end{center}
This paper is concerned with the performance of differential amplify-and-forward (D-AF) relaying for multi-node wireless communications over time-varying Rayleigh fading channels. A first-order auto-regressive model is utilized to characterize the time-varying nature of the channels. Based on the second-order statistical properties of the wireless channels, a new set of combining weights is proposed for signal detection at the destination. Expression of pair-wise error probability (PEP) is provided and used to obtain the approximated total average bit error probability (BER). It is shown that the performance of the system is related to the auto-correlation of the direct and cascaded channels and an irreducible error floor exists at high signal-to-noise ratio (SNR). The new weights lead to a better performance when compared to the conventional combining scheme. Computer simulation is carried out in different scenarios to support the analysis.

\begin{center}
\bf Index terms
\end{center}
Differential amplify-and-forward relaying, non-coherent detection, time-varying channels, performance analysis, channel auto-correlation, auto-regressive models.

%\IEEEpeerreviewmaketitle

\section{Introduction}
\label{mnode:sec:intro}
The increasing demand for better quality and higher data rate in wireless communication systems motivated the use of multiple transmit/receive antennas, resulting in the so-called multiple-input multiple-output (MIMO) systems. However, using multiple antennas is not practical for mobile units due to insufficient space to make wireless channels corresponding to multiple antennas uncorrelated. This limitation was however addressed by the technique of cooperative communications \cite{uplink-Aazgang,user-coop1}, which has been shown to be applicable in many wireless systems and applications such as 3GPP LTE-Advances, WiMAX, WLANs, vehicle-to-vehicle communications and wireless sensor networks \cite{coop-deploy,coop-V2V,coop-WiMAX,coop-dohler,coop-LTE}.

Cooperative communication exploits the fact that, since other users in a network can also listen to a source during the source's transmission phase, they would be able to re-broadcast the received data to the destination in another phase to help the source. Therefore, the overall diversity and performance of the system would benefit from the virtual MIMO system that is constructed using the help of other users. Depending on the strategy that relays utilize to cooperate, the relay networks have been classified as decode-and-forward or amplify-and-forward \cite{coop-laneman}.

Among these strategies, amplify-and-forward (AF) has been the focus of many studies because of its simplicity in the relay's function. Specifically, the relay's function is to multiply the received signal with a fixed or variable gain and forward the result to the receiver. For convenience, the overall channel of source-relay-destination is called the cascaded, the equivalent or double-Rayleigh channel. Depending on the type of modulation, the relays may need full or partial channel state information (CSI) for determining the amplification factor. Also, the destination would need the CSI of both the direct and the cascaded channels in order to combine the received signals for coherent detection.

To avoid channel estimation at the relays and destination, differential AF (D-AF) scheme has been considered in \cite{DAF-Liu,DAF-MN-Himsoon,DAF-DDF-QZ,DAF-General} which only needs the second-order statistics of the channels at the relays. In the absence of instantaneous CSI, a set of fixed weights, based on the second-order statistics, has been used to combine the received signals over the relay-destination and the source-destination links. Then, the standard differential detection is applied to recover the transmitted symbol. However, all the previous works assume a slow-fading situation and show that the performance of D-AF is abound 3-4 dB worse than the performance of its coherent version. For future reference, we call such a scheme ``conventional differential detection'' (CDD). In practice, the increasing speed of mobile users leads to fast time-varying channels (also referred to as time-selective channels). Thus, the typical assumption made in the development of conventional differential detection, namely the approximate equality of two consecutive channel uses, is violated. Therefore, it is important to consider performance of D-AF relaying systems and its robustness under more practical and general channel variation scenarios. It should also be mentioned that the effect of time-varying channels on the performance of coherent AF relay networks has been investigated in \cite{AF-suraweera,AF-Diomidis-2012}.

In this paper, the performance of D-AF for a multi-relay network in \emph{fast} time-varying Rayleigh fading channels is studied. We call the detection scheme developed for such fast time-varying channels ``time-varying differential detection'' (TVD). The channels from the source to the relays (SR channels), the source to the destination (SD channel) and from the relays to the destination (RD channels) are changing continuously according to the Jakes' model \cite{microwave-jake}. Depending on the mobility of nodes with respect to each other, different cases are considered. The direct channel is modelled with a first-order auto-regressive model, AR(1) \cite{AR1-ch,AR2-ch}. Also, based on the AR(1) model of the individual Rayleigh-faded channels, a time-series model is proposed to characterize the time-varying nature of the cascaded channels. The statistical properties of this model are verified using theory and monte-carlo simulation. Taking into account the statistical properties of channel variations, new weights for combining the received signals over multiple channels are proposed. Since analyzing the performance of the proposed system using fixed combining weights is too complicated (if not impossible), the performance of the system using the optimum maximum ratio combining (MRC) weights is analyzed and the result is used as a lower bound for the system error performance. Specifically, the pair-wise error probability (PEP) is obtained and used to approximate the average bit error rate (BER) using nearest neighbour approximation. It is shown that an error floor exits at high signal-to-noise ratio (SNR) region. Such an error floor can be approximately determined and it is related to the auto-correlation values of both the direct and the cascaded channels. Simulation results are presented to support the analysis in various scenarios of fading channels and show that the TVD with the proposed weights always outperforms the CDD in time-selective channels.

The outline of the paper is as follows. Section \ref{mnode:sec:system} describes the system model. In Section III the channel model and the differential detection of D-AF relaying with MRC technique over fast time-varying channels is developed. The performance of the system is considered in Section \ref{mn:sec:symbol_error_probability}.
Simulation results are given in Section \ref{mn:sec:sim}. Section \ref{mn:sec:con} concludes the paper.

\emph{Notations:} $(\cdot)^*$, $|\cdot|$ and $\mbox{Re}\{\cdot\}$ denote conjugate, absolute value and the real part of a complex number, respectively. $\mathcal{CN}(0,\sigma^2)$ and $\chi_2^2$ stand for complex Gaussian distribution with mean zero and variance $\sigma^2$ and chi-squared distribution with two degrees of freedom, respectively. $\mbox{E}\{\cdot\}$, $\mbox{Var}\{\cdot\}$ denote expectation and variance operations, respectively. Both ${\mathrm{e}}^{(\cdot)}$ and $\exp(\cdot)$ show the exponential function.

\section{System Model}
\label{mnode:sec:system}
\begin{figure}[t]
\psfrag {Source} [] [] [1.0] {Source}
\psfrag {Relay1} [] [] [1.0] {Relay 1}
\psfrag {Relay2} [] [] [1.0] {Relay 2}
\psfrag {RelayR} [] [] [1.0] {Relay R}
\psfrag {Destination} [] [] [1.0] {Destination}
\psfrag {f1} [r] [] [1.0] {$h_{\mathrm{sr}_1}[k]$}
\psfrag {g1} [l] [] [1.0] {$h_{\mathrm{r}_1 \mathrm{d}}[k]$}
\psfrag {f2} [bl] [] [1.0] {$h_{\mathrm{sr}_2}[k]$}
\psfrag {g2} [] [] [1.0] {\;\;$h_{\mathrm{r}_2 \mathrm{d}}[k]$}
\psfrag {fR} [] [] [1.0] {$h_{\mathrm{sr}_R}[k]$\;\;\;}
\psfrag {gR} [l] [] [1.0] {\;\;$h_{\mathrm{r}_R \mathrm{d}}[k]$}
\psfrag {hsd} [] [] [1.0] {\;\;$h_0[k]$}
\centerline{\epsfig{figure={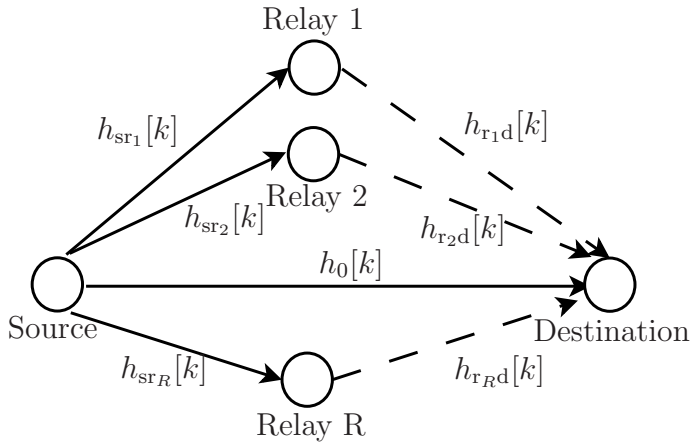},width=8.5cm}}
\caption{The wireless relay model under consideration.}
\label{fig:daf}
\end{figure}

The wireless relay model under consideration is shown in Figure~\ref{fig:daf}. It has one source, $R$ relays and one destination. The source communicates with the destination directly and also via the relays. Each node has a single antenna, and the communication between nodes is half duplex, i.e., each node can either send or receive in any given time. The channels from the source to the destination (SD), from the source to the $i$th relay ($\mathrm{SR}_i$) and from the $i$th relay, $i=1,\cdots,R$, to the destination ($\mathrm{R}_i$D) are shown with $h_0[k]$, $h_{\mathrm{sr}_i}[k]$ and $h_{\mathrm{r}_i\mathrm{d}}[k]$, respectively, where $k$ is the symbol time. A Rayleigh flat-fading model is assumed for each channel. The channels are spatially uncorrelated and changing continuously in time. The auto-correlation value between two channel coefficients, which are $n$ symbols apart, follows the Jakes' fading model \cite{microwave-jake}:
\begin{equation}
\label{eq:jakes-auto}
\mbox{E} \{h[k]h^*[k+n]\}=J_0(2\pi f n),
\end{equation}
where $J_0(\cdot)$ is the zeroth-order Bessel function of the first kind, $f$ is the maximum normalized Doppler frequency of the channel and $h$ is either $h_0$, $h_{\mathrm{sr}_i}$ or $h_{\mathrm{r}_i\mathrm{d}}$. The maximum Doppler frequency of the SD, $\mathrm{SR}_i$ and $\mathrm{R}_i$D channels are shown with $f_{\mathrm{sd}}$, $f_{\mathrm{sr}_i}$ and $f_{\mathrm{r}_i\mathrm{d}}$, respectively. Also, it is assumed that the carrier frequency is the same for all links.

Let $\mathcal{V}=\{{\mathrm{e}}^{j2\pi m/M},\; m=0,\dots, M-1\}$ denote the set of $M$-PSK symbols. At time $k$, a group of  $\log_2M$ information bits is transformed to $v[k]\in \mathcal{V}$. Before transmission, the symbols are encoded differentially as
\begin{equation}
\label{eq:s[k]}
s[k]=v[k] s[k-1],\quad s[0]=1.
\end{equation}
The transmission process is divided into two phases. Technically, either symbol-by-symbol or block-by-block dual-phase transmission protocol can be considered. In symbol-by-symbol protocol, first the source sends one symbol to the relays and then the relays re-broadcast the amplified versions of the corresponding received signals to the destination, in a time division manner. Hence, two channel uses are $R+1$ symbols apart. However, this protocol is not practical as it causes frequent switching between reception and transmission. Instead, in block-by-block protocol, a frame of information data is broadcasted in each phase and then two channel uses are one symbol apart. Therefore, block-by-block transmission is considered in this paper. However, the analysis is basically the same for both cases as only the channel auto-correlation values are different.

In phase I, the symbol $\sqrt{P_0}s[k]$ is transmitted by the source to the relays and the destination, where $P_0$ is the average source power. The received signal at the destination and the $i$th relay are
\begin{equation}
\label{eq:source_destination_rx}
y_0[k]=\sqrt{P_0}h_0[k]s[k]+w_0[k]
\end{equation}
\begin{equation}
\label{eq:ysri}
y_{\mathrm{sr}_i}[k]=\sqrt{P_0}h_{\mathrm{sr}_i}[k]s[k]+w_{\mathrm{sr}_i}[k]
\end{equation}
where $w_0[k],w_{\mathrm{sr}_i}[k]\sim \mathcal{CN}(0,1)$ are the noise components at the destination and the $i$th relay, respectively.

The received signal at the $i$th relay is then multiplied by an amplification factor $A_i$, and forwarded to the destination. The amplification factor can be either fixed or variable. A variable $A_i$ needs the instantaneous CSI. For D-AF, in the absence of the instantaneous CSI, the variance of the SR channels (here equals to one) is utilized to define the fixed amplification factor as \cite{DAF-Liu,DAF-MN-Himsoon,DAF-DDF-QZ,DAF-General}:
\begin{equation}
\label{eq:A_i}
A_i=\sqrt{\frac{P_i}{P_0+1}}
\end{equation}
where $P_i$ is the average transmitted power of the $i$th relay.\footnote{{Typically, the total power $P=P_0+\sum \limits_{i=1}^{R}P_i,$ is divided among the source and relays to minimize the average BER. This can be done by either exhaustive search or analytical approaches. The study of power allocation for multi-node differential amplify-and-forward relay networks can be found in \cite{DAF-Liu,DAF-MN-Himsoon,DAF-DDF-QZ,DAF-General}.}}

The corresponding received signal at the destination is
\begin{equation}
\label{eq:yi[k]}
y_i[k]=A_i h_{\mathrm{r}_i\mathrm{d}}[k]y_{\mathrm{sr}_i}[k]+w_{\mathrm{r}_i\mathrm{d}}[k],
\end{equation}
where $w_{\mathrm{r}_i\mathrm{d}}[k]\sim \mathcal{CN}(0,1)$ is the noise component at the destination. Substituting (\ref{eq:ysri}) into (\ref{eq:yi[k]}) yields
\begin{equation}
\label{eq:destination-rx}
y_i[k]= A_i \sqrt{P_0}h_i[k]s[k]+w_i[k],
\end{equation}
where the random variable $h_i[k]=h_{\mathrm{sr}_i}[k]h_{\mathrm{r}_i\mathrm{d}}[k]$ represents the gain of the equivalent double-Rayleigh channel, whose mean and variance equal zero and one, respectively. Furthermore,
$$
%\label{eq:noise at destination}
w_i[k]=A_i h_{\mathrm{r}_i\mathrm{d}}[k]w_{\mathrm{sr}_i}[k]+w_{\mathrm{r}_i\mathrm{d}}[k]
$$
is the equivalent noise component. It should be noted that for a given $h_{\mathrm{r}_i\mathrm{d}}[k]$, $w_i[k]$ and $y_i[k]$ are complex Gaussian random variables with mean zero and variances $\sigma_i^2=A_i^2 |h_{\mathrm{r}_i\mathrm{d}}[k]|^2+1$ and $\sigma_i^2(\rho_i+1)$, respectively, where $\rho_i$ is the average received SNR conditioned on $h_{\mathrm{r}_i\mathrm{d}}[k]$, defined as
\begin{equation}
\label{eq:rhoi}
\rho_i=\frac{A_i^2 P_0 |h_{\mathrm{r}_i\mathrm{d}}[k]|^2}{\sigma_i^2}.
\end{equation}
%Later, in the performance analysis, we take the average over the distribution of $|h_2[k]|^2$.

In the following section, we consider the differential detection of the combined received signals at the destination and evaluate its performance.

\section{Channel Models and Differential Detection}
\label{mnode:sec:CDD}
The CDD was developed under the assumption that two consecutive channel uses are approximately equal. However, such an assumption is not valid for fast time-varying channels. To find the performance of differential detection in fast time-varying channels, we need to model both the direct and the cascaded channels with time-series models. Depending on the mobility of the nodes with respect to each other, three cases are considered. The first case applies when a mobile user is communicating with a base station both directly and via other fixed users (or fixed relays) in the network. The second case can happen when the communication between two mobile users are conducted directly and also via other fixed relays. The last case is a situation that a mobile user communicates with another mobile user in the network both directly and with the help of other mobile users. The channel models in these three cases are detailed as follows.

\subsection{Mobile Source, Fixed Relays and Destination}
\label{mnode:subsec:mobile-source}
When the source is moving but the relays and the destination are fixed, the SD and all SR channels become time-varying and their statistical properties follow the fixed-to-mobile 2-D isotropic scattering channels \cite{microwave-jake}. However, all RD channels remain static.

First, the direct link is modelled with an AR(1) model \cite{AR1-ch,AR2-ch} as follows
\begin{equation}
\label{eq:AR1-model}
h_0[k]=\alpha_0 h_0[k-1]+\sqrt{1-\alpha_0^2}e_0[k]
\end{equation}
where $\alpha_0=J_0(2\pi f_{\mathrm{sd}}n)\leq 1$ is the auto-correlation of the SD channel and $e_0[k]\sim \mathcal{CN}(0,1)$ is independent of $h_0[k-1]$. Note also that $n=1$ for block-by-block transmission and $n=R+1$ for symbol-by-symbol transmission. The auto-correlation value is equal one for static channels and decreases with higher fade rates. Obviously, this value will be smaller for symbol-by-symbol transmission than for block-by-block transmission, which is another drawback of using symbol-by-symbol transmission in addition to its practical implementation issue.

Similarly, the $\SRi$ channel can be described as
\begin{gather}
\label{eq:AR_hsri[k]}
h_{\mathrm{sr}_i}[k]=\alpha_{\mathrm{sr}_i} h_{\mathrm{sr}_i}[k-1]+\sqrt{1-\alpha_{\mathrm{sr}_i}^2}e_{\mathrm{sr}_i}[k]
\end{gather}
where $\alpha_{\mathrm{sr}_i}=J_0(2\pi f_{\mathrm{sr}_i} n)\leq 1$ is the auto-correlations of the $\mathrm{SR}_i$ channel and $e_{\mathrm{sr}_i}[k]\sim \mathcal{CN}(0,1)$ is independent of $h_{\mathrm{sr}_i}[k-1]$.  Also, under the scenario of fixed relays and destination, two consecutive $\RDi$ channel uses are equal, i.e.,
\begin{equation}
\label{eq:hrdi-fixed1}
h_{\mathrm{r}_i\mathrm{d}}[k]=h_{\mathrm{r}_i\mathrm{d}}[k-1].
\end{equation}

Thus, for the $i$th cascaded channel, multiplying \eqref{eq:AR_hsri[k]} by \eqref{eq:hrdi-fixed1} gives
\begin{equation}
\label{eq:AR2-hi[k]}
h_i[k]=\alpha_{\sri} h_i[k-1] +\sqrt{1-\alpha_{\sri}^2} h_{\mathrm{r}_i\mathrm{d}}[k-1] e_{\mathrm{sr}_i}[k]
\end{equation}
which is an AR(1) model with the parameter $\alpha_{\sri}$ and $h_{\mathrm{r}_i\mathrm{d}}[k-1] e_{\mathrm{sr}_i}[k]$ as the input white noise.

\subsection{Mobile Source and Destination, Fixed Relays}
When both the source and the destination are moving, but the relays are fixed, all the SR and RD channels become time-varying and again follow the fixed-to-mobile scattering model \cite{microwave-jake}. Also, the SD channel follows the mobile-to-mobile channel model \cite{m2m-Akki} which is still Rayleigh fading but with the auto-correlation value of the corresponding model.
Therefore, the AR(1) models in \eqref{eq:AR1-model} and \eqref{eq:AR_hsri[k]} are used for modelling the SD and SR channels, respectively. However, for the SD channel, the value of $\alpha_0$ is obtained from the mobile-to-mobile channel model \cite{m2m-Akki}.

For $\RDi$ channel, the AR(1) model is
\begin{equation}
\label{eq:AR_hrdi[k]}
h_{\mathrm{r}_i\mathrm{d}}[k]=\alpha_{\mathrm{r}_i\mathrm{d}} h_{\mathrm{r}_i\mathrm{d}}[k-1]+\sqrt{1-\alpha_{\mathrm{r}_i\mathrm{d}}^2}e_{\mathrm{r}_i\mathrm{d}}[k]
\end{equation}
where $\alpha_{\mathrm{r}_i\mathrm{d}}=J_0(2\pi f_{\mathrm{r}_i\mathrm{d}} n)\leq1$ is the auto-correlation of $\RDi$ channel and $e_{\mathrm{r}_i\mathrm{d}}[k]\sim \mathcal{CN}(0,1)$ is independent of $h_{\mathrm{r}_i\mathrm{d}}[k-1]$.

Then, for the cascaded channel, multiplying \eqref{eq:AR_hsri[k]} by \eqref{eq:AR_hrdi[k]} gives
\begin{equation}
\label{eq:hi[k]}
h_i [k]=\alpha_i  h_i [k-1]+\Delta_i[k],
\end{equation}
where $\alpha_i=\alpha_{\sri}\alpha_{\rdi}\leq 1$ is the equivalent auto-correlation of the cascaded channel and
\begin{multline}
\label{eq:Deltai[k]}
\Delta_i[k]  =\alpha_{\mathrm{sr}_i} \sqrt{1-\alpha_{\mathrm{r}_i\mathrm{d}}^2} h_{\mathrm{sr}_i}[k-1] e_{\mathrm{r}_i\mathrm{d}}[k]+\alpha_{\mathrm{r}_i\mathrm{d}} \sqrt{1-\alpha_{\mathrm{sr}_i}^2}\\
h_{\mathrm{r}_i\mathrm{d}}[k-1] e_{\mathrm{sr}_i}[k]
+\sqrt{(1-\alpha_{\mathrm{sr}_i}^2)(1-\alpha_{\mathrm{r}_i\mathrm{d}}^2)}e_{\mathrm{sr}_i}[k]e_{\mathrm{r}_i\mathrm{d}}[k]
\end{multline}
represents the time-varying part of the equivalent channel, which is a combination of three uncorrelated complex-double Gaussian distributions \cite{DGC-M} and uncorrelated to $h_i[k-1]$. Since $\Delta_i[k]$ has a zero mean, its auto-correlation function is computed as
\begin{equation}
\label{eq:var_Deltai}
E\{\Delta_i[k]  \Delta_i^*[k+m]\}=
\begin{cases}
1-\alpha_i^2, & \text{if}\;\;  m=0,\\
0, & \text{if} \;\; m\neq 0
\end{cases}
\end{equation}
Therefore, $\Delta_i[k]$ is a white noise process with variance $\mbox{E}\{\Delta_i[k]\Delta_i^*[k]\}=1-\alpha_i^2$.

However, using $\Delta_i[k]$ in the way defined in \eqref{eq:Deltai[k]} is not feasible for the performance analysis. Thus, to make the analysis feasible, $\Delta_i[k]$ shall be approximated with an adjusted version of one of its terms as
\begin{equation}
\label{eq:delta_hi_hat}
\hat{\Delta}_i[k]= \sqrt{1-\alpha_i^2} {h}_{\mathrm{r}_i\mathrm{d}}[k-1] {e}_{\mathrm{sr}_i}[k]
\end{equation}
which is also a white noise process with first and second order statistical properties identical to that of $\Delta_i[k]$ and uncorrelated to $h_i[k-1]$.

By substituting \eqref{eq:delta_hi_hat} into \eqref{eq:hi[k]}, the time-series model of the equivalent channel can be described as
\begin{equation}
\label{eq:AR2-hi[k]-approx}
h_i[k]=\alpha_i h_i[k-1]+\sqrt{1-\alpha_i^2} h_{\mathrm{r}_i\mathrm{d}}[k-1]e_{\mathrm{sr}_i}[k]
\end{equation}
which is again an AR(1) with parameter $\alpha_i$ and $h_{\mathrm{r}_i\mathrm{d}}[k-1]e_{\mathrm{sr}_i}[k]$ as the input white noise.

Comparing the AR(1) models in \eqref{eq:AR2-hi[k]} and \eqref{eq:AR2-hi[k]-approx} shows that, in essence, they are only different in the model parameters: the parameter contains the effect of the $\SRi$ channel in the former model, while the effects of both the $\SRi$ and $\RDi$ channels are included in the later model. This means that the model in \eqref{eq:AR2-hi[k]-approx} can be used as the time-series model of the cascaded channel for the analysis in both cases. Specifically, for static $\RDi$ channels  $\alpha_{\rdi}=1$ and hence \eqref{eq:AR2-hi[k]-approx} turns to \eqref{eq:AR2-hi[k]}.

\begin{figure*}[t]
\psfrag {h} [] [] [.8] {$|h_i[k]|$}
\psfrag {delta} [] [] [.8] {\eqref{eq:Deltai[k]}}
\psfrag {mdelta} [] [] [.8] {$|\Delta_i[k]|$ or $|\hat{\Delta}_i[k]|$}
\psfrag {deltahat} [] [] [.8] {\eqref{eq:delta_hi_hat}}
\psfrag {PDF} [] [] [1] {pdf}
\psfrag {exact} [] [c] [.8] {\eqref{eq:hi[k]}}
\psfrag {approx} [] [c] [.8] {\eqref{eq:AR2-hi[k]-approx}}
\psfrag {theory} [] [c] [.8] {\eqref{eq:f_lambda}}
\centerline{\epsfig{figure={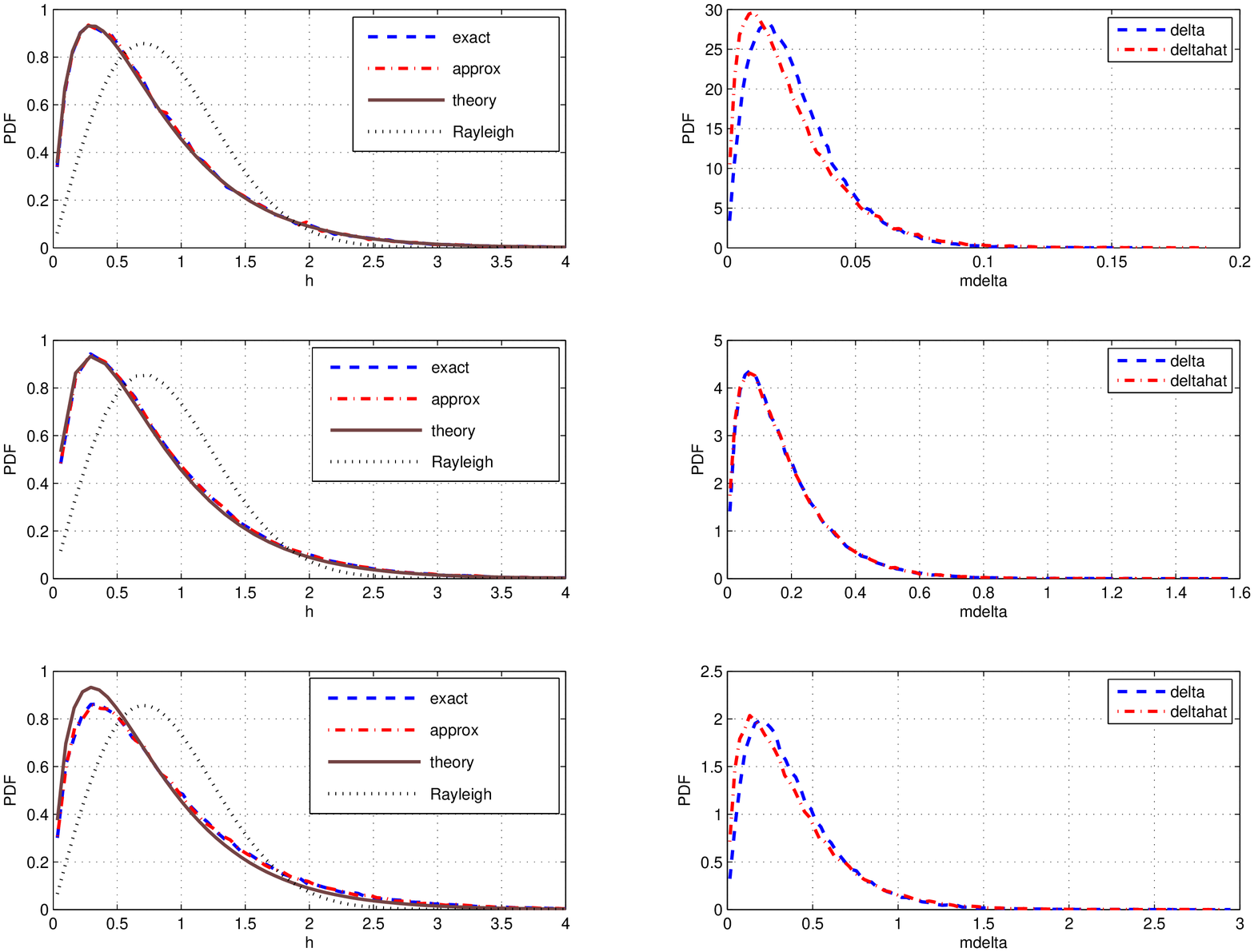},height=20cm,width=17cm}}
\caption{Theoretical pdf of $|h_i[k]|$ and obtained distributions of $|\Delta_i[k]|$, $|\hat{\Delta}_i[k]|$ and $|h_i[k]|$  in Scenario I (upper), Scenario II (middle) and Scenario III (lower). These scenarios are listed in Table \ref{table:scenarios}.}
\label{fig:pdfs}
\end{figure*}

To validate the model in \eqref{eq:AR2-hi[k]-approx}, its statistical properties are verified with the theoretical counterparts. Theoretical mean and variance of $h_i[k]$ are shown to be equal to zero and one, respectively \cite{SPAF-P,DGC-M}. This can be seen by taking expectation and variance operations over \eqref{eq:AR2-hi[k]-approx} so that $\mbox{E}\{h_i[k]\}=0$, $\mbox{Var}\{h_i[k]\}=1$. Also, the theoretical auto-correlation of $h_i[k]$ is obtained as the product of the auto-correlation of the $\SRi$ and $\RDi$ channels in \cite{SPAF-P}. By multiplying both sides of \eqref{eq:AR2-hi[k]-approx} with $h_i^*[k-1]$ and taking expectation, one has
\begin{equation}
\label{eq:Ehihi}
\mbox{E}\{h_i[k]h_i^*[k-1]\}=\alpha_i\mbox{E} \{h_i[k-1]h_i^*[k-1]\}+\mbox{E}\{\hat{\Delta}_i[k]h_i^*[k-1]\}
\end{equation}
Since $\hat{\Delta}_i[k]$ is uncorrelated to $h_i[k-1]$ then $\mbox{E}\{\hat{\Delta}_i[k]h_i^*[k-1] \}=0$ and it can be seen that
\begin{equation}
\label{eq:Ehh[k]}
\mbox{E}\{h_i[k]h_i^*[k-1]\}=\alpha_i=\alpha_{\mathrm{sr}_i} \alpha_{\mathrm{r}_i\mathrm{d}}.
\end{equation}
In addition, the theoretical pdf of the envelope $\lambda=|h_i[k]|$ is
\begin{equation}
\label{eq:f_lambda}
f_{\lambda}(\lambda)=4\lambda K_0\left( 2 \lambda \right)
\end{equation}
where $K_0(\cdot)$ is the zero-order modified Bessel function of the second kind \cite{SPAF-P}, \cite{DGC-M}.
To verify this, using Monte-Carlo simulation the histograms of $|h_i[k]|$, $|\Delta_i[k]|$ and $|\hat{\Delta}_i[k]|$, for different values of $\alpha_i$, are obtained for both models in \eqref{eq:hi[k]} and \eqref{eq:AR2-hi[k]-approx}. The values of $\alpha_i$ are computed from the normalized Doppler frequencies given in Table \ref{table:scenarios}, which as discussed in Section \ref{sec:sim} covers a variety of practical situations. These histograms along with the theoretical pdf of $|h_i[k]|$ are illustrated in Figure~\ref{fig:pdfs}.
Although, theoretically, the distributions of $\Delta_i[k]$ and $\hat{\Delta}_i[k]$ are not exactly the same, we see that for practical values of $\alpha_i$ they are very close. Moreover, the resultant distributions of $h_i[k]$, regardless of $\Delta_i[k]$ or $\hat{\Delta}_i[k]$, are similar and close to the theoretical distribution.
The Rayleigh pdf is depicted in the figure only to show the difference between the distributions of an individual and the cascaded channels.

\subsection{All Nodes are Mobile}
In this case, all links follow the mobile-to-mobile channel model \cite{m2m-Akki}. However, they are all individually Rayleigh faded and the only difference is that the auto-correlation of the channel should be replaced according to this model. Thus, the channel model in \eqref{eq:AR1-model} and \eqref{eq:AR2-hi[k]-approx} again can be used as the time-series model of the direct and cascaded channels in this case, albeit with appropriate auto-correlation values. We refer the reader to the discussion in \cite{SPAF-P} and \cite{m2m-patel} for more details on computing these auto-correlations as well as the tutorial survey on various fading models for mobile-to-mobile cooperative communication systems in \cite{M2M-Talha}. For our analysis, it is assumed that the equivalent maximum Doppler frequency of each link, regardless of fixed-to-mobile or mobile-to-mobile, is given and then the auto-correlation of each link is computed based on \eqref{eq:jakes-auto}.

\subsection{Combining Weights and Differential Detection}
By substituting the time-series models in \eqref{eq:AR1-model} and \eqref{eq:AR2-hi[k]-approx} for the direct and the cascaded channels into (\ref{eq:source_destination_rx}) and (\ref{eq:destination-rx}), respectively, one has
\begin{equation}
\label{eq:cddfast-source-destination}
y_0[k]=\alpha_0 v[k] y_0[k-1]+n_0[k],\\
\end{equation}
%where
\begin{equation}
\label{eq:n0}
%\begin{split}
n_0[k]=w_0[k]- \alpha_0 v[k] w_0[k-1]
+ \sqrt{1-\alpha_0^2} \sqrt{P_0} s[k]e_0[k],
%\end{split}
\end{equation}
and
\begin{equation}
\label{eq:cddfast-relay-destination}
y_i[k]=\alpha_i v[k] y_i[k-1]+n_i[k],
\end{equation}
%where
\begin{equation}
\label{eq:ni}
%\begin{split}
n_i[k]=w_i[k]- \alpha_i v[k] w_i[k-1]
+ \sqrt{1-\alpha_i^2}A_i\sqrt{P_0}h_{\mathrm{r}_i\mathrm{d}}[k-1]s[k]e_{\mathrm{sr}_i}[k].
%\end{split}
\end{equation}
Note that, the equivalent noise $n_0[k]$ and also $n_i[k]$ for a given $h_{\mathrm{r}_i\mathrm{d}}[k]$, are combinations of complex Gaussian random variables, and hence they are also complex Gaussian with variances
\begin{gather}
\label{eq:variance_eq_nosie1}
\sigma_{n_0}^2=1+\alpha_0^2+(1-\alpha_0^2)P_0\\
\label{eq:variance_eq_nosie2}
\sigma_{n_i}^2=\sigma_i^2 \left(1+\alpha_i^2+(1-\alpha_i^2)\rho_i \right)
\end{gather}

It can be seen that, compared with the CDD scheme, an additional term appears in the noise expression of \eqref{eq:n0} and \eqref{eq:ni} and their variances%%. This term is caused by the channel variation and can be significant in the high SNR region.

To achieve the cooperative diversity, the received signals from the two phases are combined as
\begin{equation}
\label{eq:combined}
\zeta=b_0 y_0^*[k-1]y_0[k]+\sum \limits_{i=1}^R b_i y_i^*[k-1]y_i[k]
\end{equation}
where $b_0$ and $b_i$ are the combining weights. Using the MRC technique \cite{Linear-Diversity}, the optimum combining weights, which takes into account the noise variance of each link, would be
\begin{equation}
\label{eq:optimum_weights}
\begin{split}
b_0^{\mathrm{opt}}&=\frac{\alpha_0}{\sigma_{n_0}^2}\\
b_i^{\mathrm{opt}}&=\frac{\alpha_i}{\sigma_{n_i}^2}, \;\; i=1,\cdots,R
\end{split}
\end{equation}
However, as can be see from \eqref{eq:variance_eq_nosie2}, even for slow-fading channels with $\alpha_i=1$, the noise variance depends on the channel coefficients $h_{\mathrm{r}_i\mathrm{d}}[k]$, which is not known in the system under consideration. To overcome this problem, for slow-fading channels, the average values of the noise variances, $\mbox{E}\{\sigma_{n_0}^2\}=2$ and $\mbox{E}\{\sigma_{n_i}^2\}=2(1+A_i^2)$, were utilized to define the  weights for the CDD scheme as
\begin{equation}
\label{eq:cdd_weights}
\begin{split}
b_0^{\mathrm{cdd}}&=\frac{1}{2}\\
b_i^{\mathrm{cdd}}&=\frac{1}{2(1+A_i^2)}, \;\; i=1,\cdots,R
\end{split}
\end{equation}
It is also shown in \cite{DAF-Liu,DAF-MN-Himsoon,DAF-DDF-QZ,DAF-General} that these weights give a performance close to the optimum combining in slow-fading channels.

For fast time-varying channels, the average variances of the equivalent noise terms in the direct and the cascaded links are $\mbox{E}\{\sigma_{n_0}^2\}=1+\alpha_0^2+(1-\alpha_0^2)P_0$ and $\mbox{E}\{\sigma_{n_i}^2\}=(1+\alpha_i^2)(1+A_i^2)+(1-\alpha_i^2)A^2_i P_0 $, respectively. Therefore, the new combining weights for fast time-varying channels are proposed as
\begin{equation}
\label{eq:b0_b1}
\begin{split}
&b_0=\frac{\alpha_0}{1+\alpha_0^2+(1-\alpha_0^2)P_0}\\
&b_i=\frac{\alpha_i}{(1+\alpha_i^2)(1+A_i^2)+(1-\alpha_i^2)A_i^2P_0}
\end{split}
\end{equation}
It can be seen that for slow-fading, $\alpha_0=1$ and $\alpha_i=1$, which gives $b_0=b_0^{\mathrm{cdd}}$ and $b_i=b_i^{\mathrm{cdd}}$ as expected. However, for fast-fading channels, the weights change with the channel auto-correlation and the source power. In essence, the new weights provide a dynamic combining of the received signals based on the fade rate of each link. The faster the channel changes in a communication link, the smaller portion of the received signal in that link is taken into account for detection. In terms of complexity, the proposed combining weights need the auto-correlation values of the channels which can be computed based on the Jakes' model once the corresponding Doppler frequencies are determined.

Finally, the well known minimum Euclidean distance (ED) detection is expressed as \cite{Dig-ITC-porakis}
\begin{equation}
\label{eq:vhat}
\hat{v}[k]= \arg \min \limits_{v[k] \in \mathcal{V}} |\zeta-v[k]|^2
\end{equation}
%which is also equivalent to maximum-likelihood detection with the assumption of given $h_2[k]$ and $n[k]$ Gaussian.
In the next section, we analyse the error performance of this detector.

\section{Error Performance Analysis}
\label{mn:sec:symbol_error_probability}
This section evaluates performance of the D-AF system over time-varying fading channels. Although, the practical combining weights given in \eqref{eq:b0_b1} are used in the detection process, finding the performance of the system with these weights appears infeasible. Instead, performance of the TVD scheme based on the optimum combining weights is carried out and used as a benchmark for the performance of the TVD approach with the proposed weights. It is noted that such an approach in performance analysis is also adopted for the CDD scheme as in \cite{DAF-DDF-QZ,DAF-Liu,DAF-MN-Himsoon}.

Without loss of generality, assume that symbol $v_1$ is transmitted and it is decoded erroneously as $v_2$, the nearest neighbour symbol, by the decoder. The corresponding PEP is defined as $
\label{eq:PEij}
P_s(E_{12})=P_s(v_1\rightarrow v_2).
$
An error occurs when
\begin{equation}
\label{eq:eulidian-distance}
|\zeta-v_1 |^2>|\zeta-v_2|^2
\end{equation}
which can be simplified to
\begin{equation}
\label{eq:pep-cond1}
\text{Re} \left\lbrace (v_1-v_2)^*\zeta  \right\rbrace < 0.
\end{equation}
By substituting $\zeta$ from \eqref{eq:combined} into the above inequality and using $b_0=b_0^{\mathrm{opt}}$ and $b_i=b_i^{\mathrm{opt}}$, the error event can be further simplified as $z>a$ where
\begin{multline}
\label{eq:z_gr_a}
a= |d_{\mathrm{min}}|^2 \left(\alpha_0 b_0^{\mathrm{opt}} |y_0[k-1]|^2+ \sum \limits_{i=1}^{R} \alpha_i b_i^{\mathrm{opt}} |y_i[k-1]|^2\right) \\
z=-2 \text{Re} \bigg\lbrace d_{\mathrm{min}}^* ( b_0^{\mathrm{opt}} y_0^*[k-1] n_0[k]
+ \sum \limits_{i=1}^{R} b_i^{\mathrm{opt}} y_i^*[k-1] n_i[k]) \bigg\rbrace
\end{multline}
and $d_{\mathrm{min}}=v_1-v_2$. Note that $n_0[k]$ is Gaussian, while, conditioned on $h_{\mathrm{r}_i\mathrm{d}}[k]$, $n_i[k]$ is also Gaussian. Thus, conditioned on $y_0[k-1]$, $\{y_i[k-1]\}_{i=1}^{R}$ and $\{h_{\mathrm{r}_i\mathrm{d}}[k]\}_{i=1}^R$, the variable $z$ is Gaussian as well. Its mean, $\mu_z$, and variance, $\sigma^2_z$, conditioned on the above variables, are given as (see proof in Appendix \ref{mn:Appen1}):
\begin{equation}
\label{eq:mean-z}
\mu_z=|d_{\mathrm{min}}|^2 \left( \frac{\alpha_0b_0^{\mathtt{opt}}}{P_0+1} |y_0[k-1]|^2 \right.
\left. +\sum \limits_{i=1}^{R}\frac{\alpha_i b_i^{\mathrm{opt}}}{\rho_i+1} |y_i[k-1]|^2\right)
\end{equation}

\begin{equation}
\label{eq:var-z}
\sigma_z^2= 2|d_{\mathrm{min}}|^2\left( \alpha_0 b_0^{\mathrm{opt}} |y_0[k-1]|^2\right.
\left.+ \sum \limits_{i=1}^{R}\alpha_i b_i^{\mathrm{opt}} |y_i[k-1]|^2 \right).
\end{equation}

Therefore, the conditional PEP can be written as
\begin{multline}
\label{eq:PEP_given_y_h }
P_s(E_{12}|y_0,\{y_i\}_{i=1}^R,\{h_{\mathrm{r}_i\mathrm{d}}\}_{i=1}^R)
=\text{Pr}(z> a|y_0,\{y_i\}_{i=1}^R,\{h_{\mathrm{r}_i\mathrm{d}}\}_{i=1}^R) \\
=Q\left( \frac{a-\mu_z}{\sigma_z}\right)
=Q\left(
\sqrt{\Gamma_0+
\sum \limits_{i=1}^{R}\Gamma_i}
\right)
\end{multline}
where $Q(x)=\int \limits_x^{\infty}\frac{1}{\sqrt{2\pi}}\exp\left(\frac{-t^2}{2}\right)\dd t$ and
\begin{gather}
\label{eq:Gamma0}
\Gamma_0=\frac{\gamma_0 |d_{\mathrm{min}}|^2}{P_0+1} |y_0[k-1]|^2\\
\label{eq:Gammai}
\Gamma_i=\frac{\gamma_i |d_{\mathrm{min}}|^2}{\sigma_i^2(\rho_i+1)} |y_i[k-1]|^2
\end{gather}
with
$\gamma_0$ and $\gamma_i$ defined as
\begin{gather}
\label{eq:gamma0}
\gamma_0=\frac{\alpha_0^2 P_0}{2P_0(1-\alpha_0^2)+4+\frac{2}{P_0}}\\
\label{eq:gammai}
\gamma_i=\frac{\alpha_i^2 \rho_i}{2\rho_i(1-\alpha_i^2)+4+\frac{2}{\rho_i}}.
\end{gather}

Now, take the average over the distribution of $|y_0[k-1]|^2$ and $|y_i[k-1]|^2$ by using the moment-generating function (MGF) technique \cite{DigComFad-Simon}, the conditional PEP can be written as
\begin{equation}
\label{eq:PEP-h2-MGF}
P_s(E_{12}|\{h_{\mathrm{r}_i\mathrm{d}}\}_{i=1}^R)=
\frac{1}{\pi} \int \limits_0^{\pi/2} M_{\Gamma_0} \left( - \frac{1}{2 \sin^2 \theta} \right) \prod \limits_{i=1}^{R} M_{\Gamma_i} \left( - \frac{1}{2 \sin^2 \theta} \right) \dd\theta
\end{equation}
where $M_{\Gamma_0}(\cdot)$ and $M_{\Gamma_i}(\cdot)$ are the MGFs of
$\Gamma_0$ and $\Gamma_i$, respectively.
Since $y_0[k-1]$ and $y_i[k-1]$, conditioned on $h_{\mathrm{r}_i\mathrm{d}}[k]$, are $\mathcal{CN}(0,P_0+1)$ and $\mathcal{CN}(0,\sigma_i^2(\rho_i+1))$, respectively, it follows that $|y_0[k-1]|^2\sim (P_0+1)/2 \chi_2^2$ and $|y_i[k-1]|^2\sim \sigma_i^2(\rho_i+1)/2 \chi_2^2$, respectively.
Hence, the MGFs of $\Gamma_0$ and $\Gamma_i$ can be shown to be \cite{probab-Miller}
\begin{equation}
\label{eq:MGF_Gamma}
\begin{split}
M_{\Gamma_0}(s)=\frac{1}{1-s \gamma_0 |d_{\mathrm{min}}|^2}\\
M_{\Gamma_i}(s)=\frac{1}{1-s \gamma_i |d_{\mathrm{min}}|^2}.
\end{split}
\end{equation}

Therefore, by substituting (\ref{eq:MGF_Gamma}) into (\ref{eq:PEP-h2-MGF}), one obtains
\begin{equation}
\label{eq:PEP-h2-MGF2}
P_s(E_{12}|\{h_{\mathrm{r}_i\mathrm{d}}\}_{i=1}^R)=\\
\frac{1}{\pi} \int \limits_0^{\pi/2} \frac{1}{1+\frac{1}{2\sin^2 \theta}\gamma_0 |d_{\mathrm{min}}|^2} \prod \limits_{i=1}^R\frac{1}{1+\frac{1}{2\sin^2 \theta}\gamma_i |d_{\mathrm{min}}|^2} \dd\theta.
\end{equation}
The above integral can be solved by partial fraction technique and then averaged over the distributions of $|h_{\mathrm{r}_i\mathrm{d}}[k]|^2$. However, this leads to a complicated expression without much insight. Instead, we take the average over the distributions of $|h_{\mathrm{r}_i\mathrm{d}}[k]|^2$, $f(\eta_i)=\exp(-\eta_i),\hspace{.1 in} \eta_i>0$, and the unconditioned PEP is given as
\begin{equation}
\label{eq:PEP-integral}
P_s(E_{12})=\\
\frac{1}{\pi} \int \limits_0^{\pi/2} \frac{\prod \limits_{i=1}^R I_i(\theta)}{1+\frac{1}{2\sin^2 \theta}\gamma_0 |d_{\mathrm{min}}|^2}  \dd\theta
\end{equation}
where %$I_i(\theta)$ is given as
\begin{equation}
\label{eq:I1}
%\begin{split}
I_i(\theta)=\int \limits_0^{\infty} \frac{e^{-\eta_i}}{1+\frac{1}{2\sin^2 \theta}\gamma_i |d_{\mathrm{min}}|^2} \dd\eta_i
=\varepsilon_{i}(\theta) \left[ 1+ (\beta_{i}-\epsilon_{i}(\theta)) e^{\epsilon_{i}(\theta)} E_1(\epsilon_{i}(\theta)) \right]
%\end{split}
\end{equation}
with $\varepsilon_i(\theta)$, $\beta_i$ and $\epsilon_i(\theta)$ defined as
\begin{multline}
\label{eq: c1_beta1_beta2}
\varepsilon_{i}(\theta)=\frac{4(1-\alpha_i^2)A_i^2P_0+8A_i^2}{\frac{1}{\sin^2(\theta)}\alpha_i^2A_i^2P_0|d_{\mathrm{min}}|^2+4(1-\alpha_i^2)A_i^2P_0+8A_i^2}\\
\beta_i=\frac{4}{2(1-\alpha_i^2)A_i^2P_0+4A_i^2}\\
\epsilon_i(\theta)=\frac{8}{\frac{1}{\sin^2(\theta)}\alpha_i^2A_i^2P_0|d_{\mathrm{min}}|^2+4(1-\alpha_i^2)A_i^2P_0+8A_i^2}
\end{multline}
and $E_1(x)=\int \limits_x^{\infty} ({\mathrm{e}}^{-t}/{t})\dd t$ is the exponential integral function.
The integral in \eqref{eq:PEP-integral}, then can be computed numerically to find the PEP.

It can be verified that, for DBPSK, the expression in \eqref{eq:PEP-integral} gives the exact bit-error rate (BER). On the other hand, for higher-order $M$-PSK constellations, the nearest-neighbour approximation \cite{Dig-ITC-porakis} shall be applied to obtain the overall symbol-error rate (SER) as
$
\label{eq:symbol-error-P}
P_s(E)\approx 2 P_s(E_{12}),
$
and the average BER for Gray-mapping as
\begin{equation}
\label{eq:BER}
P_b(E)\approx \frac{2}{\log_2 M} P_s(E_{12}).
\end{equation}

Finding an upper bound for the PEP expression can help to get more insights about the system performance. For $\theta=\frac{\pi}{2}$, \eqref{eq:PEP-integral} is bounded as
\begin{equation}
\label{eq:upper_bound}
P_s(E_{12})\leq \frac{\prod \limits_{i=1}^RI_i(\frac{\pi}{2})}{2+\gamma_0|d_{\mathrm{min}}|^2}.
\end{equation}
Based on the definition of $\gamma_0$ and $I_i({\pi}/{2})$, in \eqref{eq:gamma0} and \eqref{eq:I1}, it can be seen that, the error probability depends on the fading rates, $\alpha_0$ and $\alpha_i$, of both the direct and the cascaded channels. If all channels are very slow-fading, $\alpha_0= 1$ and $\alpha_i= 1$ for $i=1,\ldots, R$, and it can be verified that $\gamma_0 \propto P_0$ and $I_i({\pi}/{2})\propto ({1}/{P_0})$. Thus the diversity order of $R+1$ is achieved. On the other hand, if the channels are fast time-varying, the terms $(1-\alpha_0^2)P_0$ and $(1-\alpha_i^2)P_0$ in the denominator of $\gamma_0$ and $I_i({\pi}/{2})$ become significant in high SNR. This decreases the effective values of $\gamma_0$ and $\gamma_i$ and consequently the overall performance as well as the achieved diversity order of the system will be affected.

It is also informative to examine the expression of PEP at high SNR values. In this case,
\begin{equation}
\label{eq:gama_bar_0}
\bar{\gamma}_0=\lim \limits_{P_0\rightarrow \infty} \gamma_0= \frac{\alpha_0^2}{2(1-\alpha_0^2)}
\end{equation}
and (see proof in Appendix \ref{mn:app:B})
\begin{equation}
\label{eq:gama_bar_i}
\bar{\gamma}_i=\lim \limits_{P_0\rightarrow \infty} E[\gamma_i]= \frac{\alpha_i^2}{2(1-\alpha_i^2)}
\end{equation}
which is independent of $|h_{\mathrm{r}_i\mathrm{d}}[k]|^2$. Therefore, by substituting the above converged values into \eqref{eq:PEP-h2-MGF2} or \eqref{eq:PEP-integral}, it can be seen that the error floor appears as (see proof in Appendix \ref{mn:Appen3}),
\begin{equation}
\label{eq:Pef}
\lim \limits_{P_0 \rightarrow \infty}P_s(E_{12})=
\frac{1}{2}\sum\limits_{k=0}^R
\frac{\bar{\gamma}_k^{R}}{\prod \limits_{\substack{j=0 \\ j\neq k}}^R (\bar{\gamma}_k-\bar{\gamma}_j)}
\left\lbrace 1-\sqrt{\frac{\bar{\gamma}_k |d_{\mathrm{min}}|^2}{2+\bar{\gamma}_k |d_{\mathrm{min}}|^2}}
\right\rbrace
\end{equation}
when $\bar{\gamma}_k\neq \bar{\gamma}_j, \forall \; k,j \geq 0$

\begin{equation}
\label{eq:PEP-floor2}
\lim \limits_{P_0 \rightarrow \infty}P_s(E_{12})=\\
\frac{1}{2} \left\lbrace
1- \sqrt{\frac{\bar{\gamma}|d_{\mathrm{min}}|^2}{\bar{\gamma}|d_{\mathrm{min}}|^2+2}} \sum \limits_{l=0}^{R} \binom{2l}{l} \left( \frac{1}{4+2\bar{\gamma}|d_{\mathrm{min}}|^2} \right)^l
\right\rbrace
\end{equation}
when $\; \bar{\gamma}_0=\bar{\gamma}_i=\bar{\gamma}, \forall \; i>0$

\begin{multline}
\label{eq:PEP-floor3}
\lim \limits_{P_0 \rightarrow \infty}P_s(E_{12})=
\frac{\bar{\gamma}_0^R}{2(\bar{\gamma}_0-\bar{\gamma})^R}\left\lbrace 1- \sqrt{\frac{\bar{\gamma}_0|d_{\mathrm{min}}|^2}{\bar{\gamma}_0|d_{\mathrm{min}}|^2+2}} \right\rbrace\\-
\sum \limits_{k=1}^{R} \frac{\bar{\gamma}_0^{R-k}\bar{\gamma}}{2(\bar{\gamma}_0-\bar{\gamma})^{R-k+1}}
\left\lbrace
1- \sqrt{\frac{\bar{\gamma} |d_{\mathrm{min}}|^2}{\bar{\gamma} |d_{\mathrm{min}}|^2+2}} \sum \limits_{l=0}^{k-1} \binom{2l}{l} \left( \frac{1}{4+2\bar{\gamma}|d_{\mathrm{min}}|^2}\right)^l
\right\rbrace\\
\end{multline}
when $\; \bar{\gamma}_0\neq \bar{\gamma}_i=\bar{\gamma}, \forall \; i>0$

%Another case is when the nodes are divided into several groups such that the nodes in each group have a similar converged value but different from other groups. The expression of this case is ignored here because of its complexity, although it could be obtained using partial fraction technique.

It should be noted that the PEP and the error floor expressions are obtained based on the optimum combining weights and hence, as will be observed in the simulation results, they give a lower bound for the PEP and error floor of the system using the proposed weights. The superior performance of the proposed TVD scheme over the CDD scheme as illustrated in the next section comes with the price of requiring the channel auto-correlations for determining the new combining weights. The accurate determination of these auto-correlations is important since it would affect both the actual system performance and the performance analysis.

\section{Simulation Results}
\label{mn:sec:sim}
In this section a typical multi-node D-AF relay network is simulated in different channel scenarios and for the case that all nodes are mobile (the general case). In all simulations, the channels $h_0[k]$, $\{h_{\mathrm{sr}_i}[k]\}_{i=1}^{R}$ and $\{h_{\mathrm{r}_i \mathrm{d}}[k]\}_{i=1}^{R}$ are generated individually according to the simulation method of \cite{ch-sim}. Based on the normalized Doppler frequencies of the channels, three different scenarios are considered: (I) all the channels are fairly slow fading, (II) the SD and SR channels are fairly fast, while the RD channels are fairly slow, (III) the SD and SR channels are very fast and the RD channels are fairly-fast fading. The normalized Doppler frequencies of the three scenarios are shown in Table \ref{table:scenarios}. The values in the table can be translated to different vehicle speeds of communication nodes in typical wireless systems. For example, in a system with carrier frequency $f_c=2$ GHz and symbol duration $T_s=0.1$ ms, the corresponding Doppler shifts for the SD channel would be around $f_D={f_{\sd}}/{T_s}=50,\; 500,\; 1000$ Hz, which would correspond to the speeds of $v={cf_D}/{f_c}=25,\; 270,\; 540$ km/hr, respectively, where $c=3\times 10^8$ m/s is the speed of light. Usually, the value of 75 km/hr is assumed for a typical vehicle speed in the literature but much faster speeds are common in vehicles such as hi-speed trains. Thus, Table \ref{table:scenarios} covers a wide range of practical situations, from very slow to very fast fading, and these situations can be applicable in present and future wireless applications. In fact, Scenario I is practically equivalent to the case of static channels.
%\vspace*{-0.25cm}
\begin{table}[!ht]
\begin{center}
\caption{Three simulation scenarios.}
\vspace*{.1in}
\label{table:scenarios}
  \begin{tabular}{ |c | c| c| c | }
    \hline
				& $f_{\mathrm{sd}}$ & $f_{\mathrm{sr}_i}$ & $f_{\mathrm{r}_i\mathrm{d}}$  \\ \hline\hline
{Scenario I}    & .005              & .005              & .005  			 \\ \hline
{Scenario II}   & .05               & .05 			    & .005   			  \\ \hline
{Scenario III}  & .1               & .1 			    & .05   			  \\
    \hline
  \end{tabular}
\end{center}
\end{table}
%\vspace*{-0.5cm}

\begin{figure}[tb]
\psfrag {P(dB)} [][] [.8]{$P$ (dB)}
\psfrag {BER} [] [] [.8] {BER}
\psfrag {Scenario I} [] [] [.8] {Scenario I}
\psfrag {Scenario II} [] [] [.8] {Scenario II}
\psfrag {Scenario III} [b] [] [.8] {Scenario III}
%\psfrag {Slow-Fading} [] [cr] [.8] {Slow-fading}
\psfrag {Analysis} [l] [l] [.75] {\small Lower Bound}
\centerline{\epsfig{figure={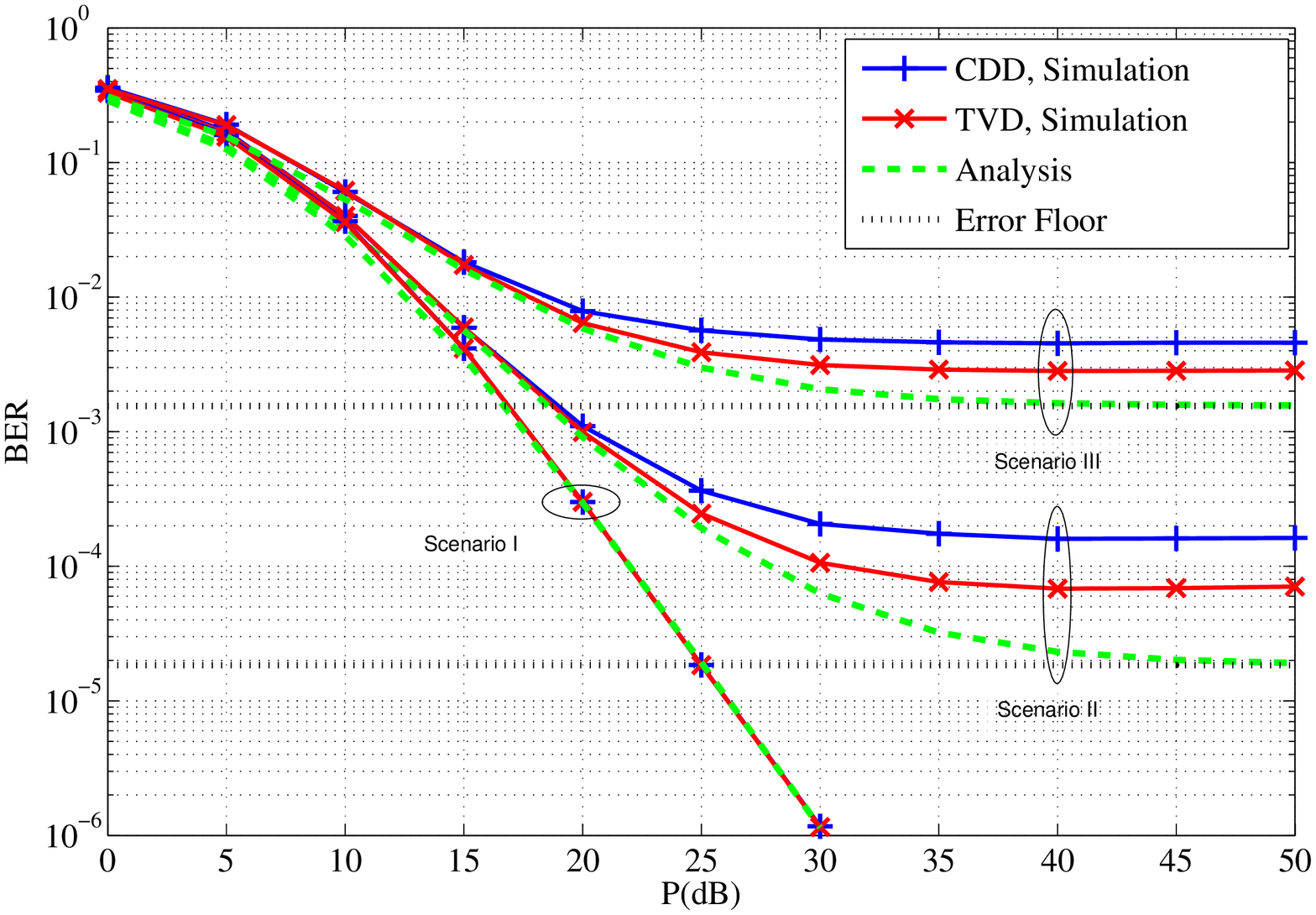},height=8.5cm,width=12cm}}
\caption{Theoretical and simulation results of D-AF relaying with two relays in Scenario I (lower plots), Scenario II (middle plots) and Scenario III (upper plots) using DBPSK.}
\label{fig:r2_m2_scs}
\end{figure}

\begin{figure}[tb]
\psfrag {P(dB)} [][] [.8]{$P$ (dB)}
\psfrag {BER} [] [] [.8] {BER}
\psfrag {Scenario I} [] [] [.8] {Scenario I}
\psfrag {Scenario II} [] [] [.8] {Scenario II}
\psfrag {Scenario III} [] [] [.8] {Scenario III}
\psfrag {Analysis} [l] [l] [.75] {\small Lower Bound }
\centerline{\epsfig{figure={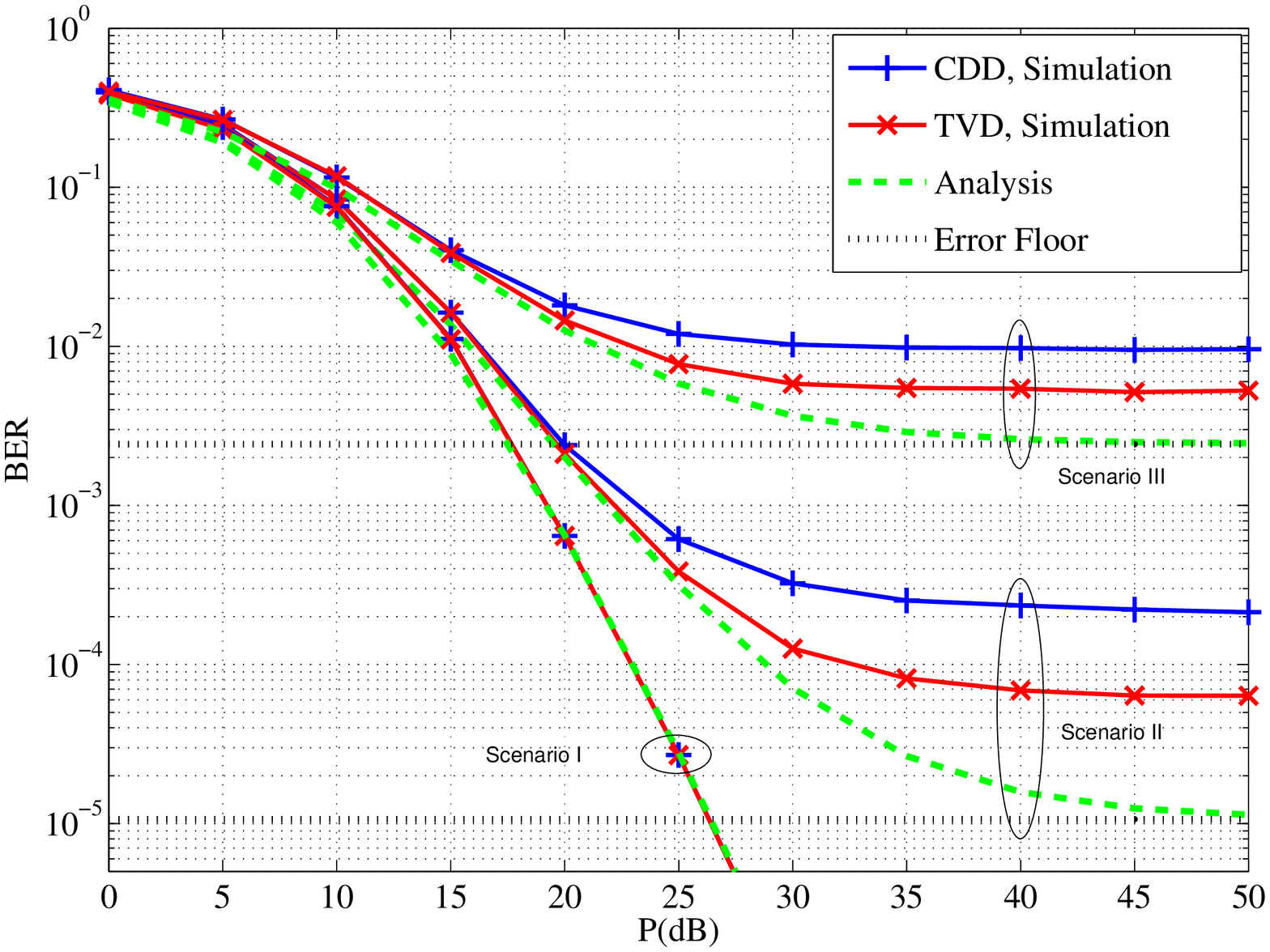},height=8.5cm,width=12cm}}
\caption{ Theoretical and simulation results of D-AF relaying with three relays in Scenario I (lower plots), Scenario II (middle plots) and Scenario III (upper plots) using DQPSK.}
\label{fig:r3_m4_scs}
\end{figure}

In each scenario, binary data is differentially encoded for DBPSK ($M=2$) or DQPSK ($M=4$) constellations. Block-by-block transmission is conducted in all scenarios. The amplification factor at the relay is fixed to $A_i=\sqrt{{P_i}/{(P_0+1)}}$ to normalize the average relay power to $P_i$. The power allocation among the source and relays is such that $P_0={P}/{2}$ and $P_i={P}/{(2R)}$, where $P$ is the total power consumed in the network. Note that, due to the way the variance of all AWGN components and channel gains is normalized to unity, the total power $P$ also has the meaning of a signal-to-noise ratio (SNR). At the destination, the received signals are first combined with the proposed weights so that the minimum Euclidean-distance detection can then be carried out. The simulation is run for various values of the total power in the network. For comparison, the same simulation process but with the combining weights given in \eqref{eq:cdd_weights} is repeated for the CDD system. The practical BER values obtained with the CDD and TVD schemes are plotted versus $P$ in Fig.~\ref{fig:r2_m2_scs} (solid lines but different markers) for DBPSK and a two-relay network. Fig. \ref{fig:r3_m4_scs} shows similar BER plots but for DQPSK and a three-relay network.

On the other hand, for computing the theoretical BER values, first the values of $\alpha_i$ and $\alpha_0$ are computed for each scenario. Also, $|d_{\min}|^2=4   \sin^2({\pi}/{M})$ for $M$-PSK symbols is computed to give $|d_{\min}|^2=4$ for $M=2$, and $|d_{\min}|^2=2$ for $M=4$. Then, the corresponding theoretical BER values from \eqref{eq:BER} are plotted in the two figures with dashed lines.

As can be seen from Figs.~\ref{fig:r2_m2_scs} and ~\ref{fig:r3_m4_scs}, in Scenario I of very slow-fading (practically the scenario of static channels), the desired cooperative diversity is achieved with both the CDD and TVD schemes. The BER curves for both schemes monotonically decrease with increasing $P$ and are consistent with the theoretical values. Since in this scenario, all the channels are fairly slow, the combining weights are approximately equal in both CDD and TVD systems and the BER results are very tight to the theoretical values which are determined using the optimum combining weights. Also, the error floor is very low and does not practically exist in this slow-fading situation and it is not plotted.

In Scenario II, which involves two fast-fading channels, the BER plots gradually deviate from the BER results obtained in Scenario I, at around 15 dB, and reach an error floor for $P\geq 30$ dB. The error floor is also calculated theoretically from (\ref{eq:PEP-floor3}) and plotted in the figures with dotted lines. The error floor is around $6\times 10^{-5}$ for TVD scheme, while it is around $2\times 10^{-4}$ for the CDD scheme in both figures. The significantly-lower error floor of the TVD scheme clearly shows its performance improvement over the CDD scheme.  The ``deviating'' phenomenon starts earlier, around 10 dB in Scenario III, and the performance degradation is much more severe since all the channels are fast fading in this scenario. Although the existence of the error floor is inevitable in both detection approaches, the TVD scheme with the proposed weights always outperforms the CDD scheme and it performs closer to the theoretical results using the optimum weights. %Generally speaking, for a medium value of $P$, around 5 dB can be saved with the TVD scheme as compared to the CDD scheme in both Scenarios II and III.
As expected, for both Scenarios II and III, the theoretical BER plots corresponding to the optimum combining weights give lower bounds for the actual performance. Another important observation is that the achieved diversity is severely affected by the high fade rates of time-varying fading channels, although the multiple fading channels are still independent. %The information about the error floor would help the system designer to prevent the system from operating in such an error floor region. One possibility is to mitigate the effect of channel variations by deploying more relays in the network.

%\begin{figure}[tb]
%\psfrag {P(dB)} [][] [.8]{$P$ (dB)}
%\psfrag {BER} [] [] [.8] {BER}
%%\psfrag {Scenario I} [] [cr] [.8] {Scenario I}
%%\psfrag {Scenario II} [] [cr] [.8] {Scenario II}
%%\psfrag {Scenario III} [] [cr] [.8] {Scenario III}
%\psfrag {Theoritical Optimum} [] [c] [.6] {Theoretical Optimum}
%\centerline{\epsfig{figure={figure/r2_m4_scs.eps},width=8.5cm}}
%\caption{ Theoretical and simulation results of D-AF relaying with two relays in Scenario I (lower plots), Scenario II (middle plots) and Scenario III (upper plots) using DQPSK.}
%\label{fig:r2_m4_scs}
%\end{figure}

%\begin{figure}[tb]
%\psfrag {P(dB)} [][] [.8]{$P$ (dB)}
%\psfrag {BER} [] [] [.8] {BER}
%%\psfrag {Scenario I} [] [cr] [.8] {Scenario I}
%%\psfrag {Scenario II} [] [cr] [.8] {Scenario II}
%%\psfrag {Scenario III} [] [cr] [.8] {Scenario III}
%%\psfrag {Theoritical Optimum} [] [c] [.6] {Theoretical Optimum}
%\centerline{\epsfig{figure={figure/r3_m2_scs.eps},width=8.5cm}}
%\caption{ Theoretical and simulation results of D-AF relaying with three relays in Scenario I (lower plots), Scenario II (middle plots) and Scenario III (upper plots) using DBPSK.}
%\label{fig:r3_m2_scs}
%\end{figure}

%In another experiment we compare the symbol-by-symbol and block-by-block transmission protocols. The simulation is run for the two relay network in Scenario II and the average BER for both transmissions are plotted in Figure~\ref{fig:r2_m2_sc2_sym_vs_blc}. As we mentioned before, in symbol-by-symbol transmission, two channel uses are $R+1$ symbols apart in compare with one symbol apart for block-by-block transmission. Hence, the overall auto-correlations are smaller for symbol-by-symbol transmission at similar channel fade rate, which leads to higher degradation in the performance. This effect can be clearly seen in the figure. In the figure, the upper plots correspond to symbol-by-symbol transmission and the lower plots correspond to block-by-block transmission protocol. %As we can see, an average BER of $10^{-2}$ is the best performance that we can achieve for symbol-by-symbol transmission in compare with $6 \times 10^{-5}$ in block-by-block transmission.

%Now, we repeat the above experiment for a three relays network in Scenario II. The results can be seen in Figure~\ref{fig:r3_m2_sc2_sym_vs_blc}. Despite our expectation, adding one relay to the network is more disruptive than helpful for the performance of symbol-by-symbol transmission. This comes from the fact that the two channel uses become more apart and hence the overall auto-correlations reduce. The experiments suggest that using symbol-by-symbol transmission should be carefully considered in D-AF relay networks, specially when channels are time-selective.

\section{Conclusion}
\label{mn:sec:con}
Performance of multi-node relay networks has been analyzed when differential $M$-PSK modulation along with the amplify-and-forward strategy are used over fast time-varying channels. The time-varying nature of the channels was related to their auto-correlation values. Using the auto-correlation values, the new combining weights at the destination were provided. The obtained error probability expression serves as a lower bound of the actual BER. It was shown that the error performance depends on the fading rates of the direct and the cascaded channels. For fast fading channels, a large fading rate can lead to a severe degradation in the error probability. It was also shown that there exists an error floor at high SNR in time-varying channels and such an error floor was determined in terms of the channel auto-correlations. The analysis is supported with simulation in different scenarios and depicts that the proposed combining gains lead to a better performance over that achieved with the conventional combining weights.

%\begin{figure}[b]
%\psfrag {P(dB)} [][] [.8]{$P$ (dB)}
%\psfrag {BER} [] [] [.8] {BER}
%\centerline{\epsfig{figure={figure/r2_m2_sc2_sym_vs_blc.eps},width=8.5cm}}
%\caption{ Comparing symbol-by-symbol (upper plots) and block-by-block (lower plots) transmission protocols for two relays network using DBPSK}
%\label{fig:r2_m2_sc2_sym_vs_blc}
%\end{figure}
%
%\begin{figure}[b]
%\psfrag {P(dB)} [][] [.8]{$P$ (dB)}
%\psfrag {BER} [] [] [.8] {BER}
%\centerline{\epsfig{figure={figure/r3_m2_sc2_sym_vs_blc.eps},width=8.5cm}}
%\caption{ Comparing symbol-by-symbol (upper plots) and block-by-block (lower plots) transmission protocols for three relays network using DBPSK}
%\label{fig:r3_m2_sc2_sym_vs_blc}
%\end{figure}

\begin{subappendices}
%\addappheadtotoc

\section{Proof of \eqref{eq:mean-z} and \eqref{eq:var-z}}
\label{mn:Appen1}

\begin{multline}
\label{eq:proof of mu_z}
\mu_z=\mbox{E}\{z|y_0[k-1],\{y_i[k-1]\}_{i=1}^R,\{h_{\rdi}[k-1]\}_{i=1}^R\}=\\
-2 \text{Re} \left\lbrace d_{\mathrm{min}}^* ( b_0^{\mathrm{opt}} y_0^*[k-1] \mbox{E}\{n_0[k]|y_0[k-1]\} \right.\\ \left.
+ \sum \limits_{i=1}^{R} b_i^{\mathrm{opt}} y_i^*[k-1] \mbox{E}\{n_i[k]|y_i[k-1],h_{\rdi}[k-1]\} ) \right\rbrace\\
=2\alpha_0 \text{Re} \left\lbrace d_{\mathrm{min}}^* ( b_0^{\mathrm{opt}} y_0^*[k-1] \mbox{E}\{w_0[k-1]|y_0[k-1]\} \right.\\ \left.
+ \sum \limits_{i=1}^{R} b_i^{\mathrm{opt}} y_i^*[k-1] \mbox{E}\{w_i[k-1]|y_i[k-1],h_{\rdi}[k-1]\} ) \right\rbrace
\end{multline}
The conditional means of Gaussian noise components $w_0[k-1]$ and $w_i[k-1]$ are obtained as \cite{prob-papo}
\begin{gather}
\label{eq:Ew0-y0}
\mbox{E}\{w_0[k-1]|y_0[k-1]\}=\frac{1}{P_0+1} d_{\min} y_0[k-1]\\
\label{eq:Ewi-yi}
\mbox{E}\{w_i[k-1]|y_i[k-1],h_{\rdi}[k-1]\}=\frac{1}{\rho_i+1} d_{\min} y_i[k-1].
\end{gather}
Substituting \eqref{eq:Ew0-y0} and \eqref{eq:Ewi-yi} into \eqref{eq:proof of mu_z} gives \eqref{eq:mean-z}.

\begin{multline}
\label{eq:proof of var_z}
\sigma^2_z=\Var \{z|y_0[k-1],\{y_i[k-1]\}_{i=1}^R,\{h_{\rdi}[k-1]\}_{i=1}^R\}\\
=2  |d_{\min}|^2 \left( \left(b_0^{\mathrm{opt}}\right)^2 |y_0[k-1]|^2 \mbox{Var}\{n_0[k]|y_0[k-1]\} \right. + \\  \left.
\sum \limits_{i=1}^{R} \left(b_i^{\mathrm{opt}}\right)^2 |y_i[k-1]|^2 \mbox{Var}\{n_i[k]|y_i[k-1],h_{\rdi}[k-1]\} \right).
\end{multline}
The conditional variances of $n_0[k-1]$ and $n_i[k-1]$ are obtained as
\begin{equation}
\label{eq:Var-n0-y0}
\mbox{Var}\{n_0[k-1]|y_0[k-1]\}=1+\alpha_0^2+(1-\alpha_0^2)P_0=\frac{\alpha_0}{b_{0}^{\opt}}
\end{equation}
\vspace{-.2in}
\begin{equation}
\label{eq:Var-ni-yi}
\mbox{Var}\{n_i[k-1]|y_i[k-1],h_{\rdi}[k-1]\}=1+\alpha_i^2+(1-\alpha_i^2)\rho_i
=\frac{\alpha_i}{b_{i}^{\opt}}.
\end{equation}
Substituting \eqref{eq:Var-n0-y0} and \eqref{eq:Var-ni-yi} into \eqref{eq:proof of var_z} gives \eqref{eq:var-z}.
It should be noted that since $z$ is proportional to the real part of $n_0[k]$ and $n_i[k]$, its variance is proportional to half of the total variance.

\section{Proof of \eqref{eq:gama_bar_i}}
\label{mn:app:B}
By substituting \eqref{eq:rhoi} into \eqref{eq:gammai} we have
\begin{equation}
\begin{split}
\lim \limits_{P_0\rightarrow \infty} \mbox{E}\{\gamma_i\}&= \mbox{E} \{ \lim \limits_{P_0 \rightarrow \infty} \gamma_i \}= \\
& \mbox{E} \left\lbrace
\lim \limits_{P_0\rightarrow \infty}
\frac{\alpha_i^2 A_i^2 P_0 \eta_i}
{
\left(
2A_i^2P_0(1-\alpha_i^2) +4A_i^2
\right)\eta_i+4
}
\right\rbrace \\
&= \mbox{E} \left\lbrace
\frac{\alpha_i^2}{2(1-\alpha_i^2)}
\right\rbrace =\frac{\alpha_i^2}{2(1-\alpha_i^2)}
\end{split}
\end{equation}

\section{Proof of \eqref{eq:Pef}-\eqref{eq:PEP-floor3}}
\label{mn:Appen3}
\begin{multline}
%\begin{split}
\lim \limits_{P_0 \rightarrow \infty} P_s(E_{12})=
\lim \limits_{P_0 \rightarrow \infty} \frac{1}{\pi} \int \limits_{0}^{\frac{\pi}{2}}
\frac{\prod \limits_{i=1}^{R} I_i(\theta)}{1+\frac{1}{2sin^2(\theta)}\gamma_0|d_{\mathrm{min}}|^2} \dd \theta=
 \frac{1}{\pi} \int \limits_{0}^{\frac{\pi}{2}}
\frac{\lim \limits_{P_0 \rightarrow \infty}\prod \limits_{i=1}^{R} I_i(\theta)}{\lim \limits_{P_0 \rightarrow \infty}\left( 1+\frac{1}{2sin^2(\theta)}\gamma_0|d_{\mathrm{min}}|^2 \right)} \dd \theta\\=
\frac{1}{\pi} \int \limits_{0}^{\frac{\pi}{2}}
\frac{1}{ 1+\frac{1}{2\sin^2(\theta)}\bar{\gamma}_0|d_{\mathrm{min}}|^2 }
\prod \limits_{i=1}^{R} \frac{1}{ 1+\frac{1}{2\sin^2(\theta)}\bar{\gamma}_i|d_{\mathrm{min}}|^2 }
\dd \theta\\
%\end{split}
\end{multline}

Now, for the first case that $\bar{\gamma}_k \neq \bar{\gamma}_j, \forall \; k,j\geq 0$, using the partial fraction technique gives
\begin{multline}
\label{eq:partial-fraction}
\frac{1}{ 1+\frac{1}{2\sin^2(\theta)}\bar{\gamma}_0|d_{\mathrm{min}}|^2 }
\prod \limits_{i=1}^{R} \frac{1}{ 1+\frac{1}{2\sin^2(\theta)}\bar{\gamma}_i|d_{\mathrm{min}}|^2 }=
\prod \limits_{k=0}^{R} \frac{1}{ 1+\frac{1}{2\sin^2(\theta)}\bar{\gamma}_k|d_{\mathrm{min}}|^2 }\\=
\sum \limits_{k=0}^{R} c_k \bar{\gamma}_k \frac{1}{1+\frac{1}{2\sin^2(\theta)} \bar{\gamma}_k|d_{\min}|^2}
\end{multline}
where
$c_k=\frac{\bar{\gamma}_k^{R-1}}{\prod \limits_{\substack{j=0 \\j\neq k}}^{R}(\bar{\gamma}_k-\bar{\gamma}_j)}$.
Then,
\begin{multline}
\label{eq:err-floor-proof1}
\frac{1}{\pi} \int \limits_{0}^{\frac{\pi}{2}}
\sum \limits_{k=0}^{R} c_k \bar{\gamma}_k \frac{1}{1+\frac{1}{2\sin^2(\theta)} \bar{\gamma}_k|d_{\min}|^2} \dd \theta=
\sum \limits_{k=0}^{R} c_k \bar{\gamma}_k \int \limits_{0}^{\frac{\pi}{2}} \frac{1}{1+\frac{1}{2\sin^2(\theta)} \bar{\gamma}_k |d_{\min}|^2} \dd \theta\\=
\frac{1}{2} \sum \limits_{k=0}^{R}  \frac{\bar{\gamma}_k^R}{\prod \limits_{\substack{j=0 \\j\neq k}}^{R}(\bar{\gamma}_k-\bar{\gamma}_j)}
\left\lbrace
1-\sqrt{\frac{\bar{\gamma}_k |d_{\mathrm{min}}|^2}{2+\bar{\gamma}_k |d_{\mathrm{min}}|^2}}
\right\rbrace
\end{multline}

Now, for the second case that $\bar{\gamma}_0 = \bar{\gamma}_i=\bar{\gamma}, \forall \; i>0$, again using the partial fraction technique gives
\begin{equation}
\label{eq:par-frac-case2}
\frac{1}{ 1+\frac{1}{2\sin^2(\theta)}\bar{\gamma}_0|d_{\mathrm{min}}|^2 }
\prod \limits_{i=1}^{R} \frac{1}{ 1+\frac{1}{2\sin^2(\theta)}\bar{\gamma}_i|d_{\mathrm{min}}|^2 }=\\
\left(
\frac{1}{ 1+\frac{1}{2\sin^2(\theta)}\bar{\gamma}|d_{\mathrm{min}}|^2 }
\right)^{R+1}.
\end{equation}

Hence, using the integral techniques in \cite{integral-tables}, one obtains
\begin{multline}
\label{eq:err-floor-proof2}
\frac{1}{\pi} \int \limits_{0}^{\frac{\pi}{2}}
\left(
\frac{1}{ 1+\frac{1}{2\sin^2(\theta)}\bar{\gamma}|d_{\mathrm{min}}|^2 }
\right)^{R+1}\dd \theta=\\
\frac{1}{2} \left\lbrace
1- \sqrt{\frac{\bar{\gamma}|d_{\mathrm{min}}|^2}{\bar{\gamma}|d_{\mathrm{min}}|^2+2}} \sum \limits_{l=0}^{R} \binom{2l}{l} \left( \frac{1}{4+2\bar{\gamma}|d_{\mathrm{min}}|^2} \right)^l
\right\rbrace
\end{multline}

For the last case $\; \bar{\gamma}_0\neq \bar{\gamma}_i=\bar{\gamma}, \forall \; i>0$, one has
\begin{multline}
\label{eq:par-frac-case3}
\frac{1}{ 1+\frac{1}{2\sin^2(\theta)}\bar{\gamma}_0|d_{\mathrm{min}}|^2 }
\prod \limits_{i=1}^{R} \frac{1}{ 1+\frac{1}{2\sin^2(\theta)}\bar{\gamma}_i|d_{\mathrm{min}}|^2 }=\\
\frac{1}{ 1+\frac{1}{2\sin^2(\theta)}\bar{\gamma}_0|d_{\mathrm{min}}|^2 }
\left(
\frac{1}{ 1+\frac{1}{2\sin^2(\theta)}\bar{\gamma}|d_{\mathrm{min}}|^2 }
\right)^{R}=\\
\frac{b_0}{1+\frac{1}{2\sin^2(\theta)} \bar{\gamma}_0 |d_{\min}|^2}+
\sum \limits_{k=1}^{R} \frac{b_k}{\left(1+\frac{1}{2\sin^2(\theta)}\bar{\gamma} |d_{\min}|^2\right)^k}
\end{multline}
where $b_0=\left(\frac{\bar{\gamma}_0}{\bar{\gamma}_0-\bar{\gamma}}\right)^R$ and $b_k=\frac{-\bar{\gamma}_0^{R-k}\bar{\gamma}}{(\bar{\gamma}_0-\bar{\gamma})^{R-k+1}}$. Then taking the integration from \eqref{eq:par-frac-case3} gives the error floor expression in \eqref{eq:PEP-floor3}.

\vspace*{1cm}

\end{subappendices}
%{\bf M. R. Avendi} [Photo and bio are not available.]\\
%
%\vspace*{0.5cm}
%
%{\bf Ha H. Nguyen} (M'01, SM'05) [Photo and bio are not available.]

\chapter{ Selection Combining for Differential Amplify-and-Forward Relaying: Slow-Fading Case}
\label{ch:sc}
In the previous chapter, multi-branch relaying systems with a direct link were considered as a mean to improve the overall diversity. Take for instance the single-branch dual-hop relaying with direct channel depicted in Figure~\ref{fig:dhpd}. After Source and Relay send their signals to Destination, an important task at Destination  is to combine the received signals from Source and Relay to achieve cooperative diversity. A semi Maximum-Ratio-Combining (semi-MRC) scheme was studied in the previous chapter. The semi-MRC method requires the second-order statistics of the transmission links to determine the combining weights. However, to avoid this requirement, the use of selection combining is investigated in this chapter. The goal is to examine the possibility of using selection combining instead of semi-MRC scheme. It would be very useful to simplify the detection process at the destination while not sacrificing much of the performance.

The manuscript in this chapter considers a single-branch dual-hop relaying system with a direct link employing selection combining at Destination. The selection combining method computes the decision variable for each link and then chooses the link with the maximum magnitude for detection. The advantage of this method is that no combining weights are necessary. The exact BER and outage probability of this combiner using differential $M$-PSK in symmetric slow-fading channels are obtained and verified with simulation results. Interestingly, the performance of both SC method and semi-MRC method are very close to each other. It is pointed out that the study in this chapter is limited to single-branch dual-hop relaying with a direct link. Extension to multi-branch relaying systems is worthwhile but the analysis appears to be very complicated.

The results of our study are reported in the following manuscript.

[Ch6-1] M. R. Avendi, Ha H. Nguyen,``Selection Combining for Differential Amplify-and-Forward Relaying Over Rayleigh-Fading Channels", \textit{IEEE Signal Processing Letters}, vol. 13, pp. 277-280, Mar. 2013.

\begin{center}
{\bf{\Large
Selection Combining for Differential Amplify-and-Forward Relaying Over Rayleigh-Fading Channels
}}
\end{center}
\begin{center}
M. R. Avendi, Ha H. Nguyen
\end{center}
\blfootnote{
Manuscript received November 12, 2012; revised January 10, 2013; accepted
January 26, 2013. Date of publication February 01, 2013; date of current version
February 07, 2013. The associate editor coordinating the review of this manuscript
and approving it for publication was Prof. Azadeh Vosoughi.
The authors are with the Department of Electrical and Computer Engineering,
University of Saskatchewan, Saskatoon, SK S7N 5A9 Canada (e-mail:
m.avendi@usask.ca; ha.nguyen@usask.ca).}

\begin{center}
\bf Abstract
\end{center}
This paper proposes and analyses selection combining (SC) at the destination for differential amplify-and-forward (D-AF) relaying over slow Rayleigh-fading channels. The selection combiner chooses the link with the maximum magnitude of the decision variable to be used for non-coherent detection of the transmitted symbols. Therefore, in contrast to the maximum ratio combining (MRC), no channel information is needed at the destination. The exact average bit-error-rate (BER) of the proposed SC is derived and verified with simulation results. It is also shown that the performance of the SC method is very close to that of the MRC method, albeit with lower complexity.

\begin{center}
\bf Index terms
\end{center}
Differential amplify-and-forward relaying, differential modulation, selection combining.

\section{Introduction}
\label{sc1:sec:intro}
The idea of employing other wireless users as relays in a communication network was proposed more than a decade ago \cite{uplink-Aazgang}. Cooperative communication exploits the fact that, since other users can also listen to a source, they would be able to receive, process and re-broadcast the received data to the destination. Depending on the strategy that relays utilize for cooperation, the relay networks are generally classified as decode-and-forward (DF) and amplify-and-forward (AF)\cite{coop-laneman}.

Among these two strategies, AF is very attractive in terms of having less computational burden at the relays. Specifically, the relay's function is simply to multiply the received signal with a fixed or variable amplification factor, depending on the availability of the channels state information (CSI). In the case of having no CSI at the relays, the second-order statistics of source-relay channels can be used to determine a fixed amplification factor. Also, using differential encoding, differential AF (D-AF) scheme has been considered in \cite{DAF-Liu,DAF-DDF-QZ,DAF-General} to avoid channel estimation at the destination. In the absence of CSI at the destination, a set of fixed weights, based on the second-order statistics of all channels, have been used to combine the received signals from the relay-destination and the source-destination links \cite{DAF-Liu,DAF-DDF-QZ,DAF-General}. For future reference, this combiner is called semi-maximum ratio combining (semi-MRC). Since the exact performance analysis of semi-MRC appears to be too complicated (if not impossible), the performance of a system using instantaneous combining weights (i.e., the instantaneous MRC) is usually conducted for benchmarking the performance of a semi-MRC system \cite{DAF-Liu,DAF-DDF-QZ,DAF-General}. It was shown that the performance of D-AF using semi-MRC is close to the performance of an instantaneous MRC and about 3-4 dB worse than its coherent version \cite{DAF-Liu,DAF-DDF-QZ,DAF-General}.

While obtaining the second-order statistics of  all channels at the destination for combining the received signals could be an issue, the need for a simpler combiner, without sacrificing much of the performance, that can also be analysed exactly, is the motivation of this paper.

In particular, this paper studies D-AF relaying over slow Rayleigh-fading channels using post-detection selection combining (SC) which can be seen as a counterpart to MDPSK for point-to-point communications with reception diversity \cite{SC-DPSK}. At the destination, the decision variable is computed for each link and the one with the maximum magnitude is chosen for non-coherent detection. Hence, different from the semi-MRC, the selection combiner does not need the second-order statistic of any of the channels, which simplifies the destination's detection task. The probability density function (pdf) and commutative density function (cdf) of the instantaneous signal to noise ratio (SNR) in each link and the combiner's output are derived and used to obtain the exact average bit-error-rate (BER) and the outage probability of the system. The analysis is verified with simulation. Comparison of SC and semi-MRC systems shows that the performance of SC is very close to that of the semi-MRC, of course with a lower complexity.

The outline of the paper is as follows. Section \ref{sc1:sec:system} describes the system model. In Section III the non-coherent detection of D-AF relaying using SC technique is developed. The performance of the system is considered in Section \ref{sc1:sec:symbol_error_probability}. Simulation results are given in Section \ref{sc1:sec:sim}. Section \ref{sc1:sec:con} concludes the paper.

\emph{Notation}: $(\cdot)^*$, $|\cdot|$ denote conjugate and absolute values of a complex number, respectively. $\mathcal{CN}(0,\sigma^2)$ stand for complex Gaussian distribution with mean zero and variance $\sigma^2$.

\section{System Model}
\label{sc1:sec:system}
The wireless relay model under consideration
%is shown in Figure~\ref{fig:sysmodel}. It
has one source, one relay and one destination. The source communicates with the destination both directly and via the relay. Each node has a single antenna, and the communication between nodes is half duplex (i.e., each node is able to only send or receive in any given time). The channel coefficients at time $k$, from the source to the destination (SD), from the source to the relay (SR) and from the relay to the destination (RD) are shown with $h_{\mathrm{sd}}[k]$, $h_{\mathrm{sr}}[k]$ and $h_{\mathrm{rd}}[k]$, respectively. All channels are $\mathcal{CN}(0,1)$ (i.e., Rayleigh flat-fading) and follow Jakes' correlation model \cite{microwave-jake}. Also, the channels are spatially uncorrelated and are approximately constant for two consecutive channel uses.

Let $\mathcal{V}=\{{\mathrm{e}}^{j2\pi m/M},\; m=0,\cdots, M-1\}$ be the set of $M$-PSK symbols. A group of $\log_2M$ information bits at time $k$ are transformed to an $M$-PSK symbol $v[k]\in \mathcal{V}$. Before transmission, the symbols are encoded differentially as
\begin{equation}
\label{eq:skvk}
s[k]=v[k] s[k-1],\quad s[0]=1.
\end{equation}
The transmission process is divided into two phases. Block-by-block transmission protocol is utilized to transmit a frame of symbols in each phase as symbol-by-symbol transmission causes frequent switching between reception and transmission, which is not practical.
%Instead, in block-by-block transmission, a block of data is broadcast in each phase. Therefore, we utilize block-by-block transmission in this paper. However, the analysis is the same for both cases.

In phase I,  symbol $s[k]$ is transmitted by the source to the relay and the destination. Let $P_0$ be the average source power per symbol. The received signal at the destination and the relay are
\begin{equation}
\label{eq:ysd[k]}
y_{\mathrm{sd}}[k]=\sqrt{P_0}h_{\mathrm{sd}}[k]s[k]+w_{\mathrm{sd}}[k]
\end{equation}
\begin{equation}
\label{eq:ysr[k]}
y_{\mathrm{sr}}[k]=\sqrt{P_0}h_{\mathrm{sr}}[k]s[k]+w_{\mathrm{sr}}[k]
\end{equation}
where $w_{\mathrm{sd}}[k],w_{\mathrm{sr}}[k]\sim \mathcal{CN}(0,1)$ are noise components at the destination and the relay, respectively.

The received signal at the relay is then multiplied by an amplification factor, and re-transmitted to the destination. The common amplification factor, based on the variance of SR channel, is commonly used in the literature as $A=\sqrt{P_1/(P_0+1)}$, where $P_1$ is the average power per symbol at the relay. However, $A$ can be any arbitrarily fixed value.\footnote{{Typically, the total power $P=P_0+P_1$ is divided between the source and rely to minimize the average BER of the system.}} The corresponding received signal at the destination is
\begin{equation}
\label{eq:yrd[k]}
y_{\mathrm{rd}}[k]=A \; h_{\mathrm{rd}}[k]y_{\mathrm{sr}}[k]+w_{\mathrm{rd}}[k],
\end{equation}
where $w_{\mathrm{rd}}[k]\sim \mathcal{CN}(0,1)$ is the noise at the destination. Substituting (\ref{eq:ysr[k]}) into (\ref{eq:yrd[k]}) yields
\begin{equation}
\label{eq:yrd[k]2}
y_{\mathrm{rd}}[k]= A\; \sqrt{P_0}h[k]s[k]+w[k],
\end{equation}
where $h[k]=h_{\mathrm{sr}}[k]h_{\mathrm{rd}}[k]$ is the equivalent double-Rayleigh channel with zero mean and variance one \cite{SPAF-P} and
$
%\label{eq:noise at destination}
w[k]=A\; h_{\mathrm{rd}}[k]w_{\mathrm{sr}}[k]+w_{\mathrm{rd}}[k]
$
is the equivalent noise. It should be noted that for a given $h_{\mathrm{rd}}[k]$, $w[k]$ is complex Gaussian random variable with zero mean and variance $A^2 \; |h_{\mathrm{rd}}[k]|^2+1$.

The following section presents the selection combining of the received signals at the destination and its differential detection.

%\begin{figure}[t]
%\psfrag {Source} [] [] [1.0] {Source}
%\psfrag {Relay} [] [] [1.0] {Relay}
%\psfrag {Destination} [] [] [1.0] {Destination}
%\psfrag {h1} [] [] [1.0] {$h_{\mathrm{sr}}[k]$\;\;\;}
%\psfrag {h2} [] [] [1.0] {\;\;$h_{\mathrm{rd}}[k]$}
%\psfrag {hsd} [] [] [1.0] {\;\;$h_{\mathrm{sd}}[k]$}
%\centerline{\epsfig{figure={figure/daf.eps},width=8.5cm}}
%\caption{The wireless relay model under consideration.}
%\label{fig:sysmodel}
%\end{figure}

\section{Selection Combining and Differential Detection}
\label{sc1:sec:ch-model}
By substituting \eqref{eq:skvk} into (\ref{eq:ysd[k]}) and (\ref{eq:yrd[k]2}), and using the slow-fading assumption, $h_{\sd}[k]\approx h_{\sd}[k-1]$ and $h[k]\approx h[k-1]$, one has
\begin{equation}
\label{eq:y_sd_k}
y_{\mathrm{sd}}[k]= v[k] y_{\mathrm{sd}}[k-1]+n_{\mathrm{sd}}[k]
\end{equation}
\begin{equation}
\label{eq:n_sd}
n_{\mathrm{sd}}[k]=w_{\mathrm{sd}}[k]- v[k] w_{\mathrm{sd}}[k-1]
\end{equation}
\begin{equation}
\label{eq:y_rd}
y_{\mathrm{rd}}[k]=v[k] y_{\mathrm{rd}}[k-1]+n_{\mathrm{rd}}[k],
\end{equation}
\begin{equation}
\label{eq:n_rd}
n_{\mathrm{rd}}[k]=w[k]-  v[k] w[k-1].
\end{equation}

Note that, the equivalent noise components $n_{\mathrm{sd}}[k]$ and $n_{\mathrm{rd}}[k]$ (for a given $h_{\mathrm{rd}}[k]$) are  combinations of complex Gaussian random variables, and hence they are also complex Gaussian with variances equal 2 and $2(1+A^2|h_{\rd}[k]|^2)$, respectively.
%$
%\label{eq:variance_eq_nosie1}
%\mbox{E}\{\sigma_{n_{\mathrm{sd}}}^2\}=2
%$
%and
%$
%\label{eq:variance_eq_nosie2}
%\mbox{E}\{\sigma_{n_{\mathrm{rd}}}^2\}=2(1+A^2).
%$

To achieve the cooperative diversity, the received signals from the two phases should be combined using some combining technique \cite{Linear-Diversity}. For the semi-MRC, the variance of $n_{\sd}$ and the expected value of the variance of $n_{\rd}$ were utilized to combine the signals as \cite{DAF-Liu,DAF-DDF-QZ,DAF-General}
\begin{equation}
\label{eq:y_combined}
\zeta=\frac{1}{2} y_{\mathrm{sd}}^*[k-1]y_{\mathrm{sd}}[k]+\frac{1}{2(1+A^2)} y_{\mathrm{rd}}^*[k-1]y_{\mathrm{rd}}[k]
\end{equation}

\begin{figure}[t]
\psfrag {y1} [] [] [1.0] {$y_{\sd}[k]$}
\psfrag {y2} [] [] [1.0] {$y_{\rd}[k]$}
\psfrag {Delay} [] [] [1.0] {Delay}
\psfrag {Decision} [] [] [1.0] {Selection}
\psfrag {y1k} [l] [] [1.0] {$y_{\sd}^*[k-1]$}
\psfrag {y2k} [l] [] [1.0] {$y_{\rd}^*[k-1]$}
\psfrag {zeta1} [l] [] [1.0] {$\zeta_{\sd}$}
\psfrag {zeta2} [l] [] [1.0] {$\zeta_{\rd}$}
\psfrag {zeta} [l] [] [1.0] {$\zeta$}
\psfrag {*} [] [] [1.0] {*}
\centerline{\epsfig{figure={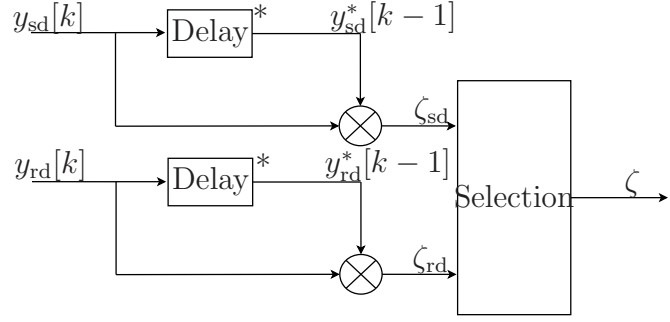},width=8.5cm}}
\caption{Block diagram of the post-detection selection combiner at the destination.}
\label{fig:sc-block}
\end{figure}

However, instead of the semi-MRC which needs the second-order statistics of all channels, we propose to combine the received signals using a selection combiner as illustrated in Figure~\ref{fig:sc-block}. As it is seen, the decision statistics for the direct link, $\zeta_{\sd}=y_{\sd}^*[k-1]y_{\sd}[k]$, and the cascaded link, $\zeta_{\rd}=y_{\rd}^*[k-1]y_{\rd}[k]$, are computed and compared to choose the link with a higher magnitude. The output of the combiner is therefore
\begin{equation}
\label{eq:zeta-sc}
\zeta =
\begin{cases}
\zeta_{\sd} & \mbox{if} \;\; |\zeta_{\sd}|>|\zeta_{\rd}|\\
\zeta_{\rd} & \mbox{if} \;\; |\zeta_{\rd}|>|\zeta_{\sd}|
\end{cases}
\end{equation}
Obviously, no channel information is needed at the destination.

Finally, the well known minimum Euclidean distance (ED) detection is applied to detect the transmitted signal as \cite{Dig-ITC-porakis}
\begin{equation}
\label{eq:vh[k]2}
\hat{v}[k]= \arg \min \limits_{x\in \mathcal{V}} |\zeta-x|^2,
\end{equation}
where the minimization is taken over all symbols $x$ of the constellation $\mathcal{V}$.

In the next section, the performance of the above selection combining detector is analysed.

\section{Error Performance Analysis}
\label{sc1:sec:symbol_error_probability}
In order to evaluate the performance of the system, the distribution of the instantaneous SNR at the output of the selection combiner is derived and used in the unified approach \cite{unified-app} to obtain the BER. To simplify the notation, the time index of the channels is omitted in this section.

The instantaneous received SNRs of two links are given as \cite{DAF-Liu,DAF-DDF-QZ,DAF-General}
\begin{gather}
\label{eq:gama_sd}
\gamma_{\sd}=P_0 |h_{\sd}|^2\\
\label{eq:gama_rd}
\gamma_{\rd}= c |h_{\sr}|^2
\end{gather}
where $c=A^2 P_0 |h_{\rd}|^2/(1+A^2 |h_{\rd}|^2)$. Since, $|h_{\sd}|^2$ has an exponential distribution, $\gamma_{\sd}$ is also exponentially distributed with the following pdf and cdf:
$
\label{eq:pdf-gama-sd}
f_{\gamma_{\sd}}(\gamma)=(1/P_0) e^{-\frac{\gamma}{P_0}}
$,
$
\label{eq:cdf-gamma-sd}
F_{\gamma_{\sd}}(\gamma)=1-e^{-\frac{\gamma}{P_0}}.
$

Since, the quantity $c$ conditioned on $h_{\rd}$ is a constant, the conditional pdf and cdf of $\gamma_{\rd}$ are given as
%\begin{gather}
$
\label{eq:pdf-gama-rd}
f_{\gamma_{\rd}|h_{\rd}}(\gamma)=(1/c) e^{-\frac{\gamma}{c}}
$,
$
\label{eq:cdf-gama-rd}
F_{\gamma_{\rd}|h_{\rd}}(\gamma)=1-e^{-\frac{\gamma}{c}}.
%\end{gather}
$

The instantaneous SNR at the output of the combiner is defined as
$
%\begin{equation}
\label{eq:gama-max}
\gamma_{\max}=\max (\gamma_{\sd},\gamma_{\rd})
%\end{equation}
$
Thus, its cdf, conditioned on $h_{\rd}$, can be expressed as
\begin{equation}
\label{eq:cdf-gama-max}
\begin{split}
F_{\gamma_{\max}|h_{\rd}}(\gamma)&= \mbox{Pr}(\max(\gamma_{\sd},\gamma_{\rd}) \leq \gamma |h_{\rd} )\\
&=\mbox{Pr}(\gamma_{\sd}\leq \gamma,\gamma_{\rd} \leq \gamma | h_{\rd} )\\
&= F_{\gamma_{\sd}}(\gamma) F_{\gamma_{\rd}|h_{\rd}}(\gamma)\\
&=\left( 1-e^{-\frac{\gamma}{P_0}}\right) \left(1-e^{-\frac{\gamma}{c}}\right)
\end{split}
\end{equation}
By taking the derivative of \eqref{eq:cdf-gama-max}, the conditional pdf of $\gamma_{\max}$ is
\begin{equation}
\label{eq:pdf-gama-max}
f_{\gamma_{\max}|h_{\rd}}(\gamma)=\frac{1}{P_0} e^{-\frac{\gamma}{P_0}}+ \frac{1}{c} e^{-\frac{\gamma}{c}}-\frac{1}{c'} e^{-\frac{\gamma}{c'}}
\end{equation}
where $c'=c P_0/(c+P_0)$ conditioned on $h_{\rd}$ is a constant.

Using the unified approach \cite[eq.25]{unified-app}, it follows that the conditional BER can be written as
\begin{equation}
\label{eq:Pb|g,hrd}
P_b(E|\gamma_{\max},h_{\rd})=\frac{1}{4\pi} \int \limits_{-\pi}^{\pi} g(\theta) e^{-\alpha(\theta)\gamma_{\max}} \dd \theta
\end{equation}
where $g(\theta)=(1-\beta^2)/(1+2\beta\sin(\theta)+\beta^2)$, $\alpha(\theta)=(b^2/(2\log_2 M))(1+\beta^2+2\beta\sin(\theta))$, and $\beta=a/b$. The values of $a$ and $b$ depend on the modulation size \cite{unified-app}.

Next, the average over the distribution of $\gamma_{\max}$ is taken to give
\begin{equation}
\label{eq:Pb|hrd}
P_b(E|h_{\rd})=\frac{1}{4\pi} \int \limits_{-\pi}^{\pi} \int \limits_{0}^{\infty} g(\theta) e^{-\alpha(\theta)\gamma} f_{\gamma_{\max}|h_{\rd}}(\gamma) \dd \gamma \dd \theta
\end{equation}
By substituting \eqref{eq:pdf-gama-max} into \eqref{eq:Pb|hrd}, one obtains
\begin{equation}
\label{eq:Pb-hrd-sub}
P_b(E|h_{\rd})=\frac{1}{4\pi} \int \limits_{-\pi}^{\pi} g(\theta) [I_1(\theta)+I_2(\theta)-I_3(\theta)] \; \dd \theta
\end{equation}
where
\begin{equation}
\label{eq:I1-theta}
I_1(\theta)=\int \limits_{0}^{\infty} e^{-\alpha(\theta)\gamma} \frac{1}{P_0} e^{-\frac{\gamma}{P_0}} \dd \gamma
= \frac{1}{P_0 \alpha(\theta)+1}
\end{equation}

\begin{equation}
\label{eq:I2-theta}
I_2(\theta)=\int \limits_{0}^{\infty} e^{-\alpha(\theta)\gamma} \frac{1}{c} e^{-\frac{\gamma}{c}} \dd \gamma
= \frac{1}{c \alpha(\theta)+1}
\end{equation}
\begin{equation}
\label{eq:I3-theta}
I_3(\theta)=\int \limits_{0}^{\infty} e^{-\alpha(\theta)\gamma} \frac{1}{c'} e^{-\frac{\gamma}{c'}} \dd \gamma
= \frac{1}{c' \alpha(\theta)+1}
\end{equation}

Finally, substituting $c$ and $c'$ and taking the average over the distribution of $|h_{\mathrm{rd}}|^2$, $f_{\lambda}(\lambda)=e^{-\lambda},\hspace{.1 in} \lambda>0$, the unconditioned BER is given as
\begin{equation}
\label{eq:BER-integral}
P_b(E)=\frac{1}{4\pi} \int \limits_{-\pi}^{\pi} g(\theta)[J_1(\theta+J_2(\theta)-J_3(\theta)] \; \dd \theta
\end{equation}
where
\begin{equation}
\label{eq:J1-theta}
J_1(\theta)=\int \limits_{0}^{\infty} I_1(\theta) e^{-\lambda} \dd \lambda\\
= \frac{1}{P_0 \alpha(\theta)+1}
\end{equation}
\begin{equation}
\label{eq:J2-theta}
%\begin{split}
J_2(\theta)=\int \limits_{0}^{\infty} I_2(\theta) e^{-\lambda} \dd \lambda
= b_3(\theta) [1+(b_1-b_2(\theta))e^{b_2(\theta)}E_1(b_2(\theta))]
%\end{split}
\end{equation}
with $b_1=1/A^2$, $b_2(\theta)=1/(A^2(1+P_0\alpha(\theta))$ and $b_3(\theta)=1/(P_0\alpha(\theta)+1)$.
\begin{equation}
\label{eq:J3-theta}
%\begin{split}
J_3(\theta)=\int \limits_{0}^{\infty} I_3(\theta) e^{-\lambda} \dd \lambda
= d_3(\theta) [1+(d_1-d_2(\theta))e^{d_2(\theta)}E_1(d_2(\theta))]
%\end{split}
\end{equation}
with $d_1=1/(2A^2)$, $d_2(\theta)=1/(A^2(2+P_0\alpha(\theta))$ and $d_3(\theta)=2/(P_0\alpha(\theta)+2)$.
Also, $E_1(x)=\int \limits_x^{\infty} (e^{-t}/t)\dd t$ is the exponential integral function.
The integral in \eqref{eq:BER-integral} can be computed numerically to find the exact BER.

To get more insights about the achieved diversity, approximating $J_2(\theta)$ and $J_3(\theta)$ with $b_3(\theta)$ and $d_3(\theta)$ as
\begin{equation}
\label{eq:upper-bounds}
\begin{split}
%J_1(\theta) & \leq \frac{1}{\alpha(\theta)P_0+1},\\
J_2(\theta) & \gtrsim \frac{1}{\alpha(\theta)P_0+1},\\
J_3(\theta) & \gtrsim \frac{2}{\alpha(\theta)P_0+2}
\end{split}
\end{equation}
Using the above values in \eqref{eq:BER-integral}, it can be seen that
\begin{equation}
\label{eq:approx-ub}
P_b(E)\propto \frac{2}{(1+\alpha(\theta)P_0)(2+\alpha(\theta)P_0)} \propto \frac{1}{P_0^2}
\end{equation}
which shows that the diversity order of two can be achieved in high SNR region.

Before closing this section, it is pointed out that the outage probability can be straightforwardly obtained from \eqref{eq:pdf-gama-max}. Specifically, the probability that the instantaneous SNR at the output of the SC combiner drops below a SNR threshold $\gth$ is
\begin{multline}
\label{eq:Poutage}
P_{\mathrm{out}}=\mathrm{Pr}(\gamma_{\max}\leq \gth)%=F_{\gamma_{\max}}(\gth)
= \int \limits_{0}^{\infty} F_{\gamma_{\max}|h_{\rd}} (\gth) e^{-\lambda} \dd \lambda \\=
%1-e^{-\frac{\gth}{P_0}}-(1-e^{-\frac{\gth}{P_0}})  \int \limits_{0}^{\infty} e^{-\frac{\gth}{c}} e^{-\lambda}\dd %\lambda =\\
%1-e^{-\frac{\gth}{P_0}}-(1-e^{-\frac{\gth}{P_0}}) e^{-\frac{\gth}{P_0}} \int \limits_{0}^{\infty} e^{-\frac{\gth}%{A^2P_0}\frac{1}{\lambda} -\lambda}  \dd \lambda =\\
\left(1-e^{-\frac{\gth}{P_0}}\right)\left[1- e^{-\frac{\gth}{P_0}} \sqrt{\frac{4\gth}{A^2P_0}}K_1\left(\sqrt{\frac{4\gth}{A^2P_0}}\right)\right]
\end{multline}
where $K_1(\cdot)$ is the first-order modified Bessel function of the second kind.

\section{Simulation Results}
\label{sc1:sec:sim}
To verify the BER performance analysis, computer simulation was carried out.\footnote{Due to space limitation, simulation results that verify the outage probability analysis are not included.} In the simulation, the channels $h_{\mathrm{sd}}[k]$, $h_{\mathrm{sr}}[k]$ and $h_{\mathrm{rd}}[k]$ are generated individually according to the simulation method of \cite{ch-sim}. The normalized Doppler frequency of all channels is set to $0.001$, so that the channels are slow-fading. Binary data is differentially encoded for $M=2,\;4$ constellations. At the destination, the received signals are combined using the SC technique and the decision variable is used to recover the transmitted signal using the minimum Euclidean-distance detection. The simulation is run for various values of the total power in the network, whereas the amplification factor at the relay is fixed to $A=\sqrt{P_1/(P_0+1)}$ to normalized the average relay power to $P_1$.

\begin{figure}[t!]
\psfrag {BER} [] [] [1] {BER}
\psfrag {c} [t] [b] [1] {$q=\frac{P_0}{P}$}
\centerline{\epsfig{figure={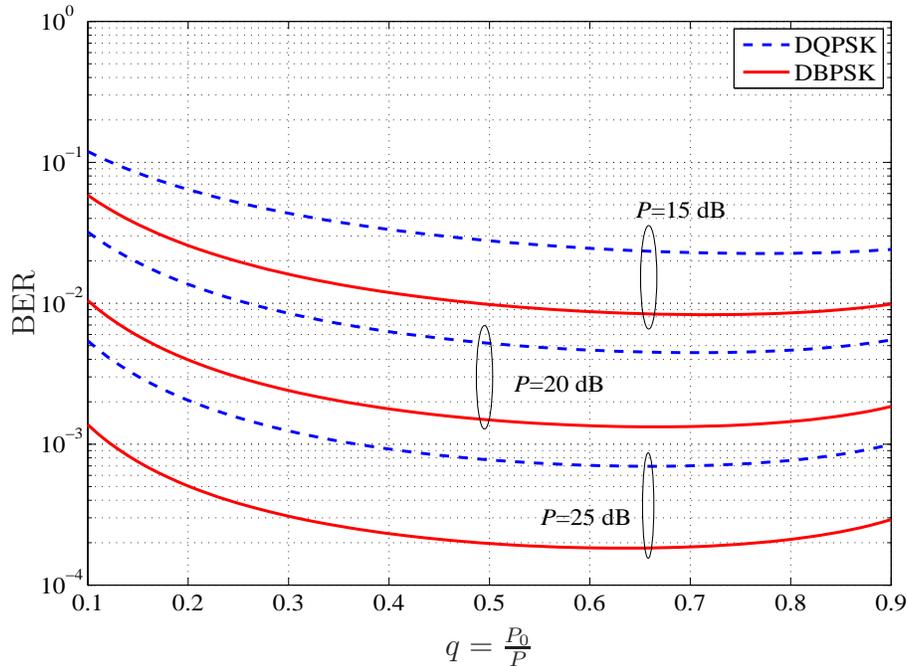},height=8.5cm,width=12cm}}
\caption{BER as a function of power allocation factor $q$ for $P=15,\;20,\;25$ dB.}
\label{fig:pw_all}
\end{figure}

First, to find the optimum power allocation between the source and the relay, the expression of BER is examined for different values of power allocation factor $q=P_0/P$, where $P=P_0+P_1$ is the total power in the system. The BER curves are plotted versus $q$ in Figure~\ref{fig:pw_all} for P=15,\;20,\;25 dB and when DBPSK and DQPSK are employed. Note that, for computing the theoretical BER in \eqref{eq:BER-integral}, $\left\lbrace a=0,\; b=\sqrt{2}\right\rbrace$ and $\left\lbrace a=\sqrt{2-\sqrt{2}},\; b=\sqrt{2+\sqrt{2}}\right\rbrace$ are obtained for DBPSK and DQPSK, respectively \cite{unified-app}.
The figure shows that more power should be allocated to the source than the relay and the BER is minimized at $q\approx 0.7$. This observation is similar to what reported in \cite{DAF-Liu} for the semi-MRC technique. Based on Figure~\ref{fig:pw_all} the power allocation factor $q=0.7$ is used in all the simulations.

\begin{figure}[th]
\psfrag {P(dB)} [][] [1]{$P$ (dB)}
\psfrag {BER} [] [] [1] {BER}
%\psfrag {MRC} [] [] [1] {\quad \quad \; semi-MRC}
%\psfrag {SC} [] [] [1] {\quad SC}
%\psfrag {Theory SC} [] [] [1] {\quad Theory SC}
\centerline{\epsfig{figure={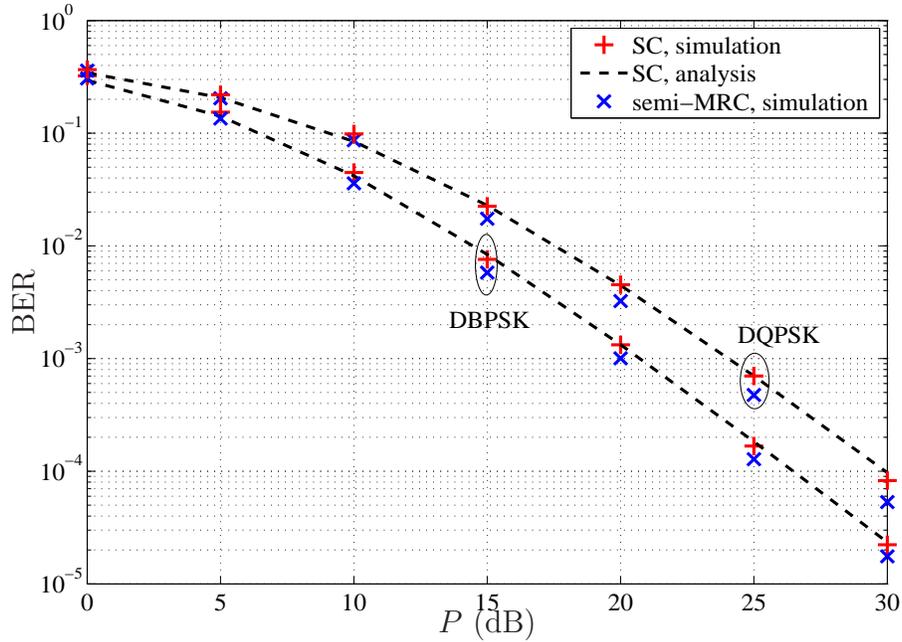},height=8.5cm,width=12cm}}
\caption{Theoretical and simulation BER of the D-AF system with semi-MRC and SC methods using DBPSK (lower) and DQPSK (upper).}
\label{fig:sc_m2m4}
\end{figure}

Figure~\ref{fig:sc_m2m4} plots the BER curves versus the total power $P$ that are obtained with the SC technique (both theoretical and simulation results) and the semi-MRC technique, and for both DBPSK (lower plots) and DQPSK (upper plots). As can be seen, the simulation results of SC technique are very close to the theoretical values. Moreover, the diversity order of two is achieved for both SC and semi-MRC methods and their results are also very close to each other. The small difference between the two methods can be accepted in many practical applications which seek a trade-off between simplicity and performance.

\balance

\section{Conclusion}
\label{sc1:sec:con}
A selection combining of the received signals at the destination of a D-AF relay network was studied. Thanks to the differential encoding and selection combiner, no channel state information is needed at the destination for detection of the transmitted symbols. The distribution of the instantaneous SNR at the output of the combiner was derived and the exact bit error rate and the outage probability of the system have been obtained. It was shown that the desired diversity order of two can be achieved by the SC system. Simulation results verified the analysis and show that the selection combiner performs very close to the more-complicated semi-MRC technique (which needs the second-order statistics of all channels).

%\bibliographystyle{IEEEbib}
%\bibliography{H:/latex/references}

\chapter{Selection Combining for Differential Amplify-and-Forward Relaying: General Time-Varying Case}
\label{ch:sc_tv}
In the previous chapter, the use of selection combining (SC) at the destination in a differential amplify-and-forward (D-AF) relaying system was studied. It was shown that, while being simpler, the SC method is able to deliver performance very close to that of the semi- maximum-ratio-combining (semi-MRC) method studied in Chapter~\ref{ch:mnode}. However, the analysis in the previous chapter was limited to slow-fading channels with symmetric fading powers.

The study in this chapter on selection combining is more comprehensive and useful than that of Chapter~\ref{ch:sc} as the nodes in the networks can be in different locations or with different mobility and therefore they would experience different fading powers and fading rates. Here, only DBPSK modulation is considered. Using DBPSK allows one to take only the real part of the decision variable for the selection and detection process. The exact average BER of the SC scheme is derived and thoroughly verified with simulation results in various fading and channel scenarios. Similar to the semi-MRC scheme considered in Chapter~\ref{ch:mnode}, the existence of an error floor in fast-fading channels is inevitable in the SC method as well. It should also be mentioned that due to the complexity of analysis, the study is limited to single-branch dual-hop relaying with a direct link and DBPSK modulation.

The results of our study are reported in the following manuscript.

[Ch7-1] M. R. Avendi, Ha H. Nguyen,``Performance of Selection Combining for Differential Amplify-and-Forward Relaying Over Time-Varying Channels", \textit{submitted to IEEE Transactions on Wireless Communications}.
\vspace*{-.5in} 
\begin{center}
{\bf{\Large
Performance of Selection Combining for Differential Amplify-and-Forward Relaying Over Time-Varying Channels
}}
\end{center}
\begin{center}
M. R. Avendi, Ha H. Nguyen
\end{center}

\begin{center}
\bf Abstract
\end{center}
Selection combining (SC) at the destination for differential amplify-and-forward (AF) relaying is attractive as it does not require channel state information as compared to the maximum-ratio-combining (MRC) while delivering close performance. Performance analysis of the SC scheme was recently reported but only for the case of slow-fading channels. This paper provides an exact average bit-error-rate (BER) of the SC scheme over a general case of time-varying Rayleigh fading channels and when the DBPSK modulation is used together with the non-coherent detection at the destination. The presented analysis is thoroughly verified with simulation results in various fading scenarios. It is shown that the performance of the system is related to the auto-correlation values of the channels. It is also shown that the performance of the SC method is very close to that of the MRC method and the existence of an error floor at high signal-to-noise ratio region is inevitable in both methods. The obtained BER analysis for the SC method can also be used to approximate the BER performance of the MRC method, whose exact analytical evaluation in time-varying channels appears to be difficult.

%\begin{keywords}
%Differential amplify-and-forward relaying, differential modulation, non-coherent detection, selection combining, time-varying channels, auto-regressive models.
%\end{keywords}

\section{Introduction}
\label{sc2:sec:intro}

Cooperative communications has now become a mature research topic. Currently, a special type of cooperative communications (with the help of one relay) has been standardized in the 3 GPP LTE technology to leverage the coverage problem of cellular networks and it is envisaged that LTE-advanced version will include cooperative relay features to overcome other limitations such as capacity and interference \cite{coop-dohler}. There are also applications for cooperative relay networks in wireless LAN, vehicle-to-vehicle communications and wireless sensor networks that have been discussed in \cite{coop-LTE,coop-WiMAX,coop-deploy,coop-sensor} and references therein.

In cooperative communications, a user in the network act as a relay to receive signals from a source, processes and re-broadcasts to a destination. In this way, additional links, other than the direct link from a source to a destination, can be constructed via relays and hence the overall spatial diversity of the system would be increased. Depending on the signal processing strategy that a relay utilizes, relay networks are generally classified as decode-and-forward (DF) and amplify-and-forward (AF) \cite{coop-laneman}.

Among these two strategies, AF or its non-coherent version, differential AF (D-AF) is very attractive as it requires less computational burden at the relays and destination. In D-AF, data symbols are differentially encoded at the source. The relay's function is simply to multiply the received signal with a fixed amplification factor. At the destination, the received signals from multi-links are combined to achieve the diversity, and used for non-coherent detection of the transmitted signals without the need of instantaneous channel state information (CSI). In \cite{DAF-Liu,DAF-DDF-QZ,DAF-General}, a maximum-ratio combiner using a set of fixed weights, based on the second-order statistics of all channels, has been used to combine the received signals from the relay-destination and source-destination links. For future reference, this combiner is called semi-maximum ratio combiner (semi-MRC).

With the motivation of reducing the detection complexity at the destination, selection combining for differential AF relay networks was recently investigated and analyzed in \cite{DAF-SC}. This combiner can be seen as a counterpart of selection combining of DPSK in point-to-point communications with receive diversity studied in \cite{SC-Kam1,SC-Norman,SC-Kam2,SC-DPSK}. However, the analysis reported in our previous work \cite{DAF-SC} only apply for symmetric \emph{slow-fading} channels. The slow-fading assumption requires approximate equality of two consecutive channel uses, which would be violated in practice under high mobility of users. %Moreover, depending on the quality of the channels, they would have different variances.

This paper studies D-AF relaying over general \emph{time-varying} Rayleigh-fading channels using post-detection selection combining (SC). The DBPSK modulation is used and the AF strategy with fixed gain at the relay is employed. Two links are involved in the communication: the direct link from the source to the destination (SD) and the cascaded link from the source to the destination via the relay. The decision variable is computed for each link and the one with the maximum magnitude is chosen for non-coherent detection. Hence, different from the semi-MRC, the selection combiner does not need the second-order statistic of any of the channels, which simplifies the detection at the destination. To characterize the time-varying nature of the channels, first-order auto-regressive models \cite{AR1-ch,DAF-ITVT} are employed for the direct and cascaded channels. The probability density function (pdf) and cumulative density function (cdf) of the decision variable in each link are derived and used to obtain the exact average bit-error-rate (BER). The analysis is verified with simulation results in different fading and channel scenarios. Comparison of the SC and semi-MRC systems shows that the performance of the SC method is very close to that of the semi-MRC. For fast-fading channels, it is seen that the performance of both SC and semi-MRC systems degrades and reaches an error floor. The expression of the error floor is also derived for the SC method. On the other hand, the close performance of both the SC and semi-MRC schemes implies that one can use the exact BER analysis of the SC method to closely approximate the performance of the semi-MRC method in time-varying channels. This is useful since the exact BER of the semi-MRC method in time-varying channels appears to be difficult \cite{DAF-ITVT} and only a loose lower bound was derived for this system in \cite{DAF-ITVT}.

The outline of the paper is as follows. Section \ref{sc2:sec:system} describes the system model. In Section \ref{sc2:sec:combinig} the non-coherent detection of D-AF relaying using SC technique is developed. The performance of the system is considered in Section \ref{sc2:sec:symbol_error_probability}. Simulation results are given in Section \ref{sc2:sec:sim}. Section \ref{sc2:sec:con} concludes the paper.

\emph{Notation}: $(\cdot)^*$, $|\cdot|$ denote conjugate and absolute values of a complex number, respectively. $\mathcal{CN}(0,\sigma^2)$ stands for a complex Gaussian distribution with mean zero and variance $\sigma^2$, while $\chi_2^2$ stands for chi-squared
distribution with two degrees of freedom. $\Et\{\cdot\}$ and $\Var\{\cdot\}$ are expectation and variance operations, respectively. Both $\exp(\cdot)$ and ${\mathrm{e}}^{(\cdot)}$ indicate exponential function and $E_1(x)=\int \limits_{x}^{\infty} ({\mathrm{e}}^{-t}/t)\dd t$ is the exponential integral function.

\section{System Model}
\label{sc2:sec:system}
\begin{figure}[t]
\psfrag {Source} [] [] [1.0] {Source}
\psfrag {Relay} [] [] [1.0] {Relay}
\psfrag {Destination} [] [] [1.0] {Destination}
\psfrag {h1} [] [] [1.0] {$h_1[k]$\;\;\;}
\psfrag {h2} [] [] [1.0] {\;\;$h_2[k]$}
\psfrag {h0} [] [] [1.0] {\;\;$h_0[k]$}
\centerline{\epsfig{figure={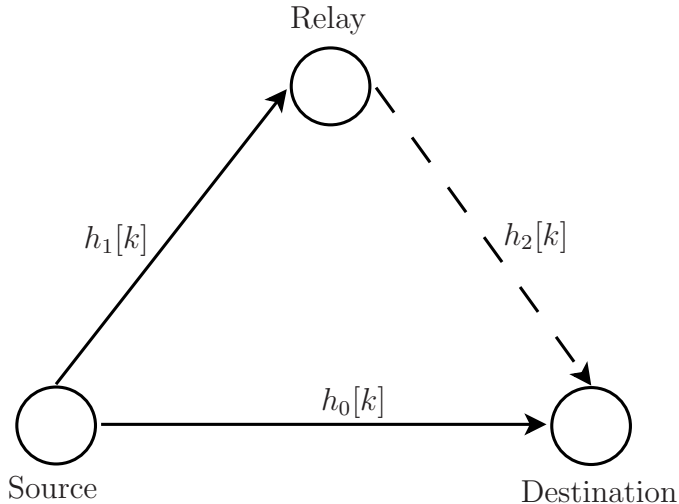},width=8.5cm}}
\caption{The wireless relay model under consideration.}
\label{fig:sc_tv}
\end{figure}

The system model in this article is very similar to that of \cite{DAF-WCNC,DAF-SC,DAF-ITVT}. As such, the formulation and description of the system model are similar to those in \cite{DAF-WCNC,DAF-SC,DAF-ITVT}. Figure~\ref{fig:sc_tv} depicts the wireless relay model under consideration, which has three nodes: one Source, one Relay and one Destination. There are a direct link and a cascaded link, via Relay, from Source to Destination. The inherent diversity order of the system is therefore two. A common half-duplex communication between the nodes is assumed, i.e., each node employs a single antenna and able to only send or receive in any given time.

The channel coefficients at time $k$, from Source to Destination (SD), from Source to Relay (SR) and from Relay to Destination (RD) are shown with $h_0[k]$, $h_1[k]$ and $h_2[k]$, respectively. A Rayleigh flat-fading model is assumed for each channel, i.e., $h_i\sim \CN(0,\sigma_i^2),\; i=0,1,2$. The channels are spatially uncorrelated and changing continuously in time. The time correlation between two channel coefficients, $n$ symbols apart, follows the Jakes' model \cite{microwave-jake}:
\begin{equation}
\label{eq:phi}
\varphi_i(n)=\Et\{h_i[k] h_i^*[k+n]\}=\sigma_i^2 J_0(2\pi f_i n), \quad i=0,1,2
\end{equation}
where $J_0(\cdot)$ is the zeroth-order Bessel function of the first kind and $f_i$ is the maximum normalized Doppler frequency of the $i$th channel. The normalized Doppler frequency is a function of the velocity of the nodes. A higher velocity leads to a higher Doppler value and hence a lower time-correlation between the channel coefficients.

Let $\mathcal{V}=\{{\mathrm{e}}^{j2\pi m/M},\; m=0,\cdots, M-1\}$ be the set of $M$-PSK symbols. A group of $\log_2M$ information bits at time $k$ is transformed to an $M$-PSK symbol $v[k]\in \mathcal{V}$. Before transmission, the symbols are encoded differentially as
\begin{equation}
\label{eq:s[k]3}
s[k]=v[k] s[k-1],\quad s[0]=1.
\end{equation}
Here it is assumed that $M=2$, i.e., DBPSK modulation is employed.

The transmission process is divided into two phases. A symbol or a frame of symbols could be transmitted in each phase. Symbol-by-symbol transmission is not practical as it causes frequent switching between reception and transmission. Hence, frame-by-frame transmission protocol is utilized here. However, the analysis is the same for both cases and only the channels auto-correlation values are different. In symbol-by-symbol transmission, two channel uses are two symbols apart $(n=2),$ while in frame-by-frame transmission two channel uses are one symbol apart $(n=1)$.

In the first phase, symbol $s[k]$ is transmitted by Source to Relay and Destination. Let $P_0$ be the average Source's power per symbol. The received signals at Destination and Relay are
\begin{equation}
\label{eq:y0}
y_0[k]=\sqrt{P_0}h_0[k]s[k]+w_0[k]
\end{equation}
\begin{equation}
\label{eq:y1[k]}
y_1[k]=\sqrt{P_0}h_1[k]s[k]+w_1[k]
\end{equation}
where $w_0[k],\; w_1[k]\sim \CN(0,N_0)$ are noise components at Destination and Relay, respectively. It is easy to see that, for given $s[k]$, $y_0[k]\sim \CN(0,N_0(\rho_0+1))$, where $\rho_0$ is the average received SNR per symbol from the direct link, defined as
\begin{equation}
\label{eq:rho0}
\rho_0=\frac{P_0 \sigma_0^2}{N_0}.
\end{equation}
Also, the average received SNR per symbol at Relay is defined as
\begin{equation}
\label{eq:rho_1}
\rho_1=\frac{P_0\sigma_1^2}{N_0}.
\end{equation}
The received signal at Relay is then multiplied by an amplification factor, and re-transmitted to Destination. Based on the variance of SR channel, the amplification factor commonly used in the literature is
\begin{equation}
\label{eq:AmpFactor}
A=\sqrt{\frac{P_1}{P_0 \sigma_1^2+N_0}},
\end{equation}
where $P_1$ is the average transmitted power per symbol at Relay. In general, $A$ can be any arbitrarily fixed value. If the total power in the network, $P$, is divided between Source and Relay such that $P_0=qP,\;P_1=(1-q)P$, where $q$ is the power amplification factor, then $A=\sqrt{(1-q)P/(qP\sigma_1^2+N_0)}$.

The corresponding received signal at Destination is
\begin{equation}
\label{eq:y}
y_2[k]=A \; h_2[k]y_1[k]+w_2[k],
\end{equation}
where $w_2[k]\sim \mathcal{CN}(0,N_0)$ is the noise component at Destination in the second phase. Substituting (\ref{eq:y1[k]}) into (\ref{eq:y}) yields
\begin{equation}
\label{eq:y2[k]}
y_2[k]= A\; \sqrt{P_0}h[k]s[k]+w[k],
\end{equation}
where $h[k]=h_1[k]h_2[k]$ is the equivalent double-Rayleigh channel with zero mean and variance $\sigma_1^2 \sigma_2^2$ \cite{SPAF-P} and $w[k]=A\; h_2[k]w_1[k]+w_2[k]$ is the equivalent noise component. It should be noted that for a given $h_2[k]$, $w[k]$ is a complex Gaussian random variable with zero mean and variance
\begin{equation}
\label{eq:sig2wk}
\sigma_{w}^2=N_0(1+A^2 \; |h_2[k]|^2)
\end{equation}
and hence $y_2[k]$, conditioned on $s[k]$ and $h_2[k]$, is a complex Gaussian random variable with zero mean and variance $(\rho_2+1)\sigma_{w}^2$. Here, $\rho_2$ is the average received SNR per symbol from the cascaded link at Destination, conditioned on $h_2[k]$. It is given as
\begin{equation}
\label{eq:rhok}
\rho_2= \frac{A^2 \rho_1 |h_2[k]|^2}{1+A^2|h_2[k]|^2}.
\end{equation}

The next section presents the selection combining of the received signals at Destination and its non-coherent detection.

\section{Selection Combining and Non-Coherent Detection}
\label{sc2:sec:combinig}
Based on two consecutive received symbols, non-coherent detection of the transmitted symbols can be obtained. For DBPSK, the decision variables for the direct and cascaded links are computed from the two latest symbols as
\begin{align}
\label{eq:zeta_0}
\zeta_0= \Re\{y_0^*[k-1] y_0[k] \} \\
\label{eq:zeta_1}
\zeta_2= \Re\{ y_2^*[k-1] y_2[k] \}
\end{align}

To achieve the cooperative diversity, the decision variables from the two transmission phases should be combined using some combining technique \cite{Linear-Diversity}. For the semi-MRC method, over slow-fading channels, the decision variables were combined as \cite{DAF-Liu,DAF-DDF-QZ,DAF-General}
\begin{equation}
\label{eq:zeta}
\zeta=\frac{1}{N_0} \zeta_0+\frac{1}{N_0(1+A^2\sigma_2^2)} \zeta_2
\end{equation}

\begin{figure}[t]
\psfrag {y1} [] [] [1.0] {$y_0[k]$}
\psfrag {y2} [] [] [1.0] {$y_2[k]$}
\psfrag {Delay} [] [] [1.0] {Delay}
\psfrag {Decision} [] [] [1.0] {Selection}
\psfrag {y1k} [l] [] [1.0] {$y_0^*[k-1]$}
\psfrag {y2k} [l] [] [1.0] {$y_2^*[k-1]$}
\psfrag {zeta1} [l] [] [1.0] {$\zeta_0$}
\psfrag {zeta2} [l] [] [1.0] {$\zeta_2$}
\psfrag {zeta} [l] [] [1.0] {$\zeta$}
\psfrag {*} [] [] [1.0] {*}
\centerline{\epsfig{figure={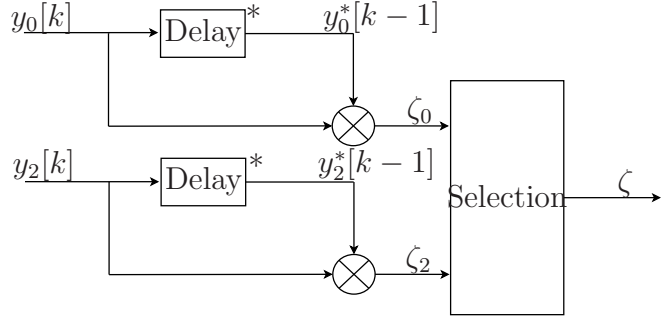},width=8.5cm}}
\caption{Block diagram of the selection combiner at Destination of a D-AF relay network.}
\label{fig:scblk}
\end{figure}

However, instead of the semi-MRC which needs the second-order statistics of all channels, it is proposed to combine the received signals using a selection combiner as illustrated in Fig.~\ref{fig:scblk} \cite{DAF-SC}. As it is seen, the decision statistics for the direct link, $\zeta_0$, and the cascaded link, $\zeta_2$, are computed and compared to choose the link with a higher magnitude. The output of the combiner is therefore
\begin{equation}
\label{eq:zeta02}
\zeta =
\begin{cases}
\zeta_0, & \mbox{if} \;\; |\zeta_0|>|\zeta_2|\\
\zeta_2, & \mbox{if} \;\; |\zeta_2|>|\zeta_0|
\end{cases}
\end{equation}
Obviously, using this scheme, no channel information is needed at Destination.

Finally, the output of the combiner is used to decode the transmitted signal as
\begin{equation}
\label{eq:vh3}
\hat{v}[k]=
\begin{cases}
-1, & \mbox{if} \;\; \zeta<0\\
+1, & \mbox{if} \;\; \zeta>0
\end{cases}.
\end{equation}

The next section analyzes the performance of the above selection combining detector.

\section{Error Performance Analysis}
\label{sc2:sec:symbol_error_probability}
As usual, the transmitted symbols are assumed to be equally likely. Without loss of generality, assume that symbol $v[k]=+1$ is transmitted and let $\hat{v}[k]$ denote the decoded symbol. The BER can be expressed as
\begin{equation}
\label{eq:PbE}
P_{\bt}(E)=\Pr(\zeta<0, v[k]=+1)=\Pr(\zeta_0<0,|\zeta_0|>|\zeta_2|)
+\Pr(\zeta_2<0,|\zeta_2|>|\zeta_0|).
\end{equation}
Since $\zeta_0$ and $\zeta_2$ have different distributions, the two terms in \eqref{eq:PbE} should be computed separately. The first term can be written as
\begin{multline}
\label{eq:Pb1}
P_{\bt}(E_1)=\Pr(\zeta_0<0,|\zeta_0|>|\zeta_2|)
=\Pr(|\zeta_2|+\zeta_0<0)
= \int \limits_{-\infty}^{0} \int \limits_{0}^{-\beta} f_{\zeta_0}(\beta) f_{|\zeta_2|}(r) \dd r \dd \beta\\
=\int \limits_{-\infty}^{0} f_{\zeta_0}(\beta) \left[F_{|\zeta_2|}(-\beta)-F_{|\zeta_2|}(0)\right] \dd \beta
\end{multline}
Likewise, the second term of \eqref{eq:PbE} can be expressed as
\begin{multline}
\label{eq:Pb2}
P_{\bt}(E_2)=\Pr(\zeta_2<0,|\zeta_2|>|\zeta_0|)
=\Pr(|\zeta_0|+\zeta_2<0)
= \int \limits_{-\infty}^{0} \int \limits_{0}^{-\beta} f_{\zeta_2}(\beta) f_{|\zeta_0|}(r) \dd r \dd \beta\\
=\int \limits_{-\infty}^{0} f_{\zeta_2}(\beta) \left[F_{|\zeta_0|}(-\beta)-F_{|\zeta_0|}(0)\right] \dd \beta.
\end{multline}

In \eqref{eq:Pb1} and \eqref{eq:Pb2}, $f_{\zeta_0}(\cdot)$ and $f_{\zeta_2}(\cdot)$ are the pdfs of $\zeta_0<0$ and $\zeta_2<0$, respectively. Also, $F_{|\zeta_0|}(\cdot)$ and $F_{|\zeta_2|}(\cdot)$ are the cdfs of $|\zeta_0|>0$ and $|\zeta_2|>0$, respectively. They can be written as
\begin{equation}
\label{eq:F_z1}
\begin{split}
F_{|\zeta_2|}(\beta)=\Pr(|\zeta_2|\leq \beta)=\Pr(-\beta\leq \zeta_2 \leq \beta)\\
= F_{\zeta_2}(\beta)-F_{\zeta_2}(-\beta).
\end{split}
\end{equation}
\begin{equation}
\label{eq:F_|z0|}
\begin{split}
F_{|\zeta_0|}(\beta)=\Pr(|\zeta_0|\leq \beta)=\Pr(-\beta \leq \zeta_0 \leq \beta)\\
= F_{\zeta_0}(\beta)-F_{\zeta_0}(-\beta).
\end{split}
\end{equation}
where $F_{\zeta_0}(\cdot)$ and $F_{\zeta_2}(\cdot)$ are the cdfs of $\zeta_0$ and $\zeta_2$, respectively.

To proceed with the computation of \eqref{eq:Pb1} and \eqref{eq:Pb2}, the pdfs and cdfs of $\zeta_0$ and $\zeta_2$ are required. To obtain these functions, the relationship between two consecutive channel uses is required. The conventional assumption is that two consecutive channel uses are approximately equal, i.e., $h_0[k]\approx h_0[k-1]$ and $h[k]\approx h[k-1]$. However, such an assumption is not valid for fast-fading channels.

For time-varying channels, individual channels are expressed by an AR(1) model as
\begin{gather}
\label{eq:ARi-hi}
h_i[k]=\alpha h_i[k-1]+\sqrt{1-\alpha^2} e_i[k],\quad i=0,1, 2
\end{gather}
where $\alpha=\varphi_i(1)/\sigma_i^2$ is the auto-correlation of the $i$th channel and $e_i[k]\sim \mathcal{CN}(0,\sigma_i^2)$ is independent of $h_i[k-1]$. Based on these expressions, a first-order time-series model was derived in \cite{DAF-ITVT} to characterize the evolution of the cascaded channel in time. The time-series model of the cascaded channel is given as (see \cite{DAF-ITVT} for the detailed derivation and verification):
\begin{equation}
\label{eq:AR_h[k]}
h[k]=\alpha h[k-1]+\sqrt{1-\alpha^2}\ h_2[k-1]e_1[k]
\end{equation}
where $\alpha=\alpha_1 \alpha_2 \leq 1$ is the equivalent auto-correlation of the cascaded channel, which is equal to the product of the auto-correlations of individual channels, and $e_1[k]\sim \mathcal{CN}(0,\sigma_1^2)$ is independent of $h[k-1]$.

By substituting \eqref{eq:s[k]3}, \eqref{eq:ARi-hi} and \eqref{eq:AR_h[k]} into \eqref{eq:y0} and \eqref{eq:y}, one has
\begin{equation}
\label{eq:y0k}
y_0[k]=\alpha_0 v[k] y_0[k-1]+\widetilde{w}_0[k],\\
\end{equation}
where
\begin{equation}
\label{eq:wt0[k]}
\widetilde{w}_0[k]=w_0[k]- \alpha_0 v[k] w_0[k-1]
+ \sqrt{1-\alpha_0^2} \sqrt{P_0} s[k]e_0[k]
\end{equation}
and
\begin{equation}
\label{eq:yk}
y_2[k]=\alpha v[k] y_2[k-1]+\widetilde{w}[k],
\end{equation}
where
\begin{equation}
\label{eq:wt[k]}
\widetilde{w}[k]=w[k]- \alpha v[k] w[k-1]
+ \sqrt{1-\alpha^2} A\sqrt{P_0}h_2[k-1]s[k]e_1[k].
\end{equation}
It should be pointed out that, compared with slow-fading channels (see \cite[Eqs. (7) and 9]{DAF-SC}), additional terms appear in the noise expressions, which are functions of the channel auto-correlations and transmit power.

Then by substituting \eqref{eq:y0k} and \eqref{eq:yk} into \eqref{eq:zeta_0} and \eqref{eq:zeta_1}, one has
\begin{align}
\label{eq:z0_simp1}
\zeta_0= \Re \left\lbrace \alpha_0 v[k] |y_0[k-1]|^2+y_0^*[k-1]\widetilde{w}_0[k] \right\rbrace \\
\label{eq:z1_simp1}
\zeta_2=\Re \left\lbrace \alpha v[k] |y_2[k-1]|^2+y_2^*[k-1]\widetilde{w}[k]\right\rbrace.
\end{align}
It is seen that, for given $y_0[k-1]$, $\zeta_0$ is a combination of complex Gaussian random variables, whose  conditional mean and variance are computed as
\begin{multline}
\label{eq:mu_z0}
\mu_{\zeta_0}=\Et \{ \zeta_0 | y_0[k-1],v[k]=+1\}= \alpha_0 |y_0[k-1]|^2
+\Et\{\Re\{y_0^*[k-1] \widetilde{w}_0[k] \}\}\\=
\alpha_0 |y_0[k-1]|^2 -\alpha_0 \Et\{\Re\{w_0[k-1]|y_0[k-1]\} \} \\
= \alpha_0 |y_0[k-1]|^2 -\frac{\alpha_0 }{\rho_0+1} |y_0[k-1]|^2= \frac{\alpha_0 \rho_0}{\rho_0+1} |y_0[k-1]|^2
\end{multline}
\begin{multline}
\label{eq:var_z0}
\Sigma_{\zeta_0}=\Var\{\zeta_0|y_0[k-1],v[k]=+1\}=
\Var\{\Re\{ \alpha_0 |y_0[k-1]|^2+y_0^*[k-1] \widetilde{w}_0[k] \} \}\\
= \Var\{\Re\{ y_0^*[k-1] \widetilde{w}_0[k] \} \}=
\frac{1}{2} \left[ N_0 + \alpha_0^2 \Var\{w[k-1]|y_0[k-1]\} +
(1-\alpha_0^2) P_0 \sigma_0^2
 \right] |y_0[k-1]|^2\\=
\frac{1}{2}N_0 \left( 1+ \frac{\alpha_0^2\rho_0}{\rho_0+1} +
(1-\alpha_0^2) \rho_0
 \right)|y_0[k-1]|^2
\end{multline}

Furthermore, for given $y_2[k-1]$ and $h_2[k-1]$, $\zeta_2$ is a combination of complex Gaussian random variables and hence it is Gaussian as well. Its conditional mean and variance are computed as
\begin{multline}
\label{eq:mu_z1}
\mu_{\zeta_2}=\Et \{ \zeta_2 | y_2[k-1],h_2[k-1],v[k]=+1\}\\= \alpha |y_2[k-1]|^2
+\Et\{\Re\{y_2^*[k-1] \widetilde{w}[k]|y_2[k-1],h_2[k-1] \}\}\\=
\alpha |y_2[k-1]|^2 -\alpha \Et\{\Re\{w[k-1]|y_2[k-1],h_2[k-1] \}\} \\
= \alpha |y_2[k-1]|^2 -\frac{\alpha }{\rho_2+1} |y_2[k-1]|^2= \frac{\alpha \rho_2}{\rho_2+1} |y_2[k-1]|^2
\end{multline}
\begin{multline}
\label{eq:var_z1}
\Sigma_{\zeta_2}=\Var\{\zeta_2|y_2[k-1],h_2[k-1],v[k]=1\}\\=
\Var\{\Re\{ \alpha |y_2[k-1]|^2+y_2^*[k-1] \widetilde{w}[k] |y_2[k-1],h_2[k-1] \}\}\\
= \Var\{\Re\{ y_2^*[k-1] \widetilde{w}[k] |y_2[k-1],h_2[k-1] \}\}\\=
\frac{1}{2} \left( \sigma_w^2 + \alpha^2 \Var\{w[k-1]|y_2[k-1],h_2[k-1]\}\right. \left.  +(1-\alpha^2) P_0 \sigma_1^2 |h_2[k-1]|^2 \right) |y_2[k-1]|^2 \\=
\frac{1}{2}\sigma_w^2 \left[ 1+ \frac{\alpha^2\rho_2}{\rho_2+1} +
(1-\alpha^2) \rho_2 \right]|y_2[k-1]|^2
\end{multline}
%\begin{equation}
%\label{eq:var_z1}
%\Sigma_{\zeta_2}=\frac{1}{2} \sigma_{w}^2 \left( 1+(1-\alpha^2)\rho+\frac{\alpha^2 \rho}{\rho+1}\right) |y[k-1]|^2.
%\end{equation}

In the remaining of the paper, the time index $[k-1]$ is omitted for notational simplicity. From \eqref{eq:mu_z0} and \eqref{eq:var_z0}, the conditional pdf of $\zeta_0$ is given as
\begin{equation}
\label{eq:fz0_y0}
f_{\zeta_0}(\beta|y_0)=\frac{1}{\sqrt{2\pi \Sigma_{\zeta_0}}} \exp \left( - \frac{(\beta-\mu_{\zeta_0})^2}{2\Sigma_{\zeta_0}}\right).
\end{equation}
Since $y_0\sim \CN(0,N_0(\rho_0+1))$, $|y_0|^2\sim 0.5{N_0(\rho_0+1)}\chi_2^2$, i.e.,
\begin{equation}
\label{eq:f_y0}
f_{|y_0|^2}(\eta)=\frac{1}{N_0(\rho_0+1)} \exp\left( -\frac{\eta}{N_0(\rho_0+1)} \right)
\end{equation}
By taking the expectation of \eqref{eq:fz0_y0} over the distribution of $|y_0|^2$, the pdf of $\zeta_0$ is obtained as \cite[Eq. 3.471.15]{integral-tables} %\cite[eq. 36]{SC-Kam2}
\begin{equation}
\label{eq:f_z0}
f_{\zeta_0}(\beta)=
\begin{cases}
b_0 \exp\left(c_0 \beta\right) & , \;\; \beta\leq 0 \\
b_0 \exp\left(d_0 \beta\right) & , \;\; \beta\geq 0
\end{cases}
\end{equation}
where
\begin{gather}
\nonumber
b_0=\frac{1}{N_0(1+\rho_0)} \\
\label{eq:b0c0d0}
c_0=\frac{2}{N_0(1+(1-\alpha_0)\rho_0)}\\\nonumber
d_0=\frac{-2}{N_0(1+(1+\alpha_0)\rho_0)}.
\end{gather}
%$b_0={1}/{(1+\rho_0)},$ $c_0=\frac{2}{1+(1-\alpha_0)\rho_0},$ $d_0=\frac{-2}{1+(1+\alpha_0)\rho_0}$.

Thus, the cdf of $\zeta_0$ is expressed as
\begin{equation}
\label{eq:F_z0}
F_{\zeta_0}(\beta)=
\begin{cases}
\frac{b_0}{c_0} \exp\left(c_0 \beta\right) & , \;\; \beta\leq 0\\
1+\frac{b_0}{d_0} \exp\left(d_0 \beta\right) & , \;\; \beta\geq 0
\end{cases}.
\end{equation}
By substituting \eqref{eq:F_z0} into \eqref{eq:F_|z0|}, the cdf of $|\zeta_0|$ is obtained as
\begin{equation}
\label{eq:F_|z0|_sim1}
F_{|\zeta_0|}(\beta)=1+\frac{b_0}{d_0} \exp\left(d_0 \beta\right)- \frac{b_0}{c_0} \exp\left(-c_0 \beta\right).
\end{equation}

On the other hand, it follows from \eqref{eq:mu_z1} and \eqref{eq:var_z1} that the conditional pdf of $\zeta_2$ is
\begin{equation}
\label{eq:fz1_y2_h2}
f_{\zeta_2}(\beta|y_2,h_2)=\frac{1}{\sqrt{2\pi \Sigma_{\zeta_2}}} \exp \left( - \frac{(\beta-\mu_{\zeta_2})^2}{2\Sigma_{\zeta_2}}\right).
\end{equation}
Since, conditioned on $h_2$, $y_2\sim \CN(0,\sigma_{w}^2(\rho_2+1))$. Therefore, $|y_2|^2\sim 0.5{\sigma_w^2 (\rho_2+1)}\chi_2^2$, i.e.,
\begin{equation}
\label{eq:f_|y|}
f_{|y_2|^2}(\eta|h_2)=\frac{1}{\sigma_w^2(\rho_2+1)} \exp\left( -\frac{\eta}{\sigma_w^2(\rho_2+1)} \right)
\end{equation}
And by taking the expectation of \eqref{eq:fz1_y2_h2} over the distribution of $|y_2|^2$, one has \cite[Eq. 3.471.15]{integral-tables}
\begin{equation}
\label{eq:f_z1_h2}
f_{\zeta_2}(\beta|h_2)=
\begin{cases}
b_2 \exp\left(c_2 \beta\right) & , \;\; \beta\leq 0 \\
b_2 \exp\left(d_2 \beta\right) & , \;\; \beta\geq 0
\end{cases}
\end{equation}
where
\begin{gather}
\nonumber
b_2=\frac{1}{\sigma_w^2(\rho_2+1)} \\
\label{eq:b2c2d2}
c_2=\frac{2}{\sigma_w^2(1+(1-\alpha)\rho_2)}\\ \nonumber
d_2=\frac{-2}{\sigma_w^2(1+(1+\alpha)\rho_2)}.
\end{gather}
are functions of random variable $\lambda=|h_2|^2$, whose pdf is $f_{\lambda}(\lambda)=(1/\sigma_2^2)\exp(\lambda/\sigma_2^2)$.
%$b_2=\frac{1}{\rho+1},$ $c_2=\frac{2}{1+(1-\alpha)\rho},$ $d_2=\frac{-2}{1+(1+\alpha)\rho}.$

Thus, the cdf of $\zeta_2$ conditioned on $h_2$ is given as
\begin{equation}
\label{eq:F_z1_h2}
F_{\zeta_2}(\beta|h_2)=
\begin{cases}
\frac{b_2}{c_2} \exp\left(c_2 \beta\right) & , \;\; \beta\leq 0\\
1+\frac{b_2}{d_2} \exp\left(d_2 \beta\right) & , \;\; \beta\geq 0
\end{cases}.
\end{equation}
By substituting \eqref{eq:F_z1_h2} into \eqref{eq:F_z1}, the cdf of $|\zeta_2|$, conditioned on $h_2$, is
\begin{equation}
\label{eq:F_|z1|}
F_{|\zeta_2|}(\beta|h_2)=1+\frac{b_2}{d_2} \exp\left(d_2 \beta\right)- \frac{b_2}{c_2} \exp\left(-c_2 \beta\right).
\end{equation}
Using \eqref{eq:f_z0} and \eqref{eq:F_|z1|}, \eqref{eq:Pb1} can be evaluated as follows:
\begin{multline}
\label{eq:Pb1_sim1}
P_{\bt}(E_1|h_2)= \int \limits_{-\infty}^{0} f_{\zeta_0}(\beta) \left[ F_{|\zeta_2|}(-\beta|h_2)-F_{|\zeta_2|}(0|h_2) \right] \dd \beta \\
=\int \limits_{-\infty}^{0} b_0 {\mathrm{e}}^{c_0 \beta} \left( \frac{b_2}{d_2} {\mathrm{e}}^{-d_2 \beta}-\frac{b_2}{c_2}{\mathrm{e}}^{c_2 \beta} -\frac{b_2}{d_2} +\frac{b_2}{c_2} \right) \dd \beta
= \frac{b_0b_2}{c_0} \left( \frac{1}{c_0-d_2} + \frac{1}{c_0+c_2}\right).
\end{multline}

Also, using \eqref{eq:F_|z0|_sim1} and \eqref{eq:f_z1_h2}, \eqref{eq:Pb2} is computed as
\begin{multline}
\label{eq:Pb2_sim1}
P_{\bt}(E_2|h_2)= \int \limits_{-\infty}^{0} f_{\zeta_2}(\beta|h_2) \left[ F_{|\zeta_0|}(-\beta)-F_{|\zeta_0|}(0) \right] \dd \beta \\
=\int \limits_{-\infty}^{0} b_2 {\mathrm{e}}^{c_2\beta} \left( \frac{b_0}{d_0} {\mathrm{e}}^{-d_0 \beta}-\frac{b_0}{c_0}{\mathrm{e}}^{c_0 \beta} -\frac{b_0}{d_0} +\frac{b_0}{c_0} \right) \dd \beta
= \frac{b_0b_2}{c_2} \left( \frac{1}{c_2-d_0} + \frac{1}{c_0+c_2}\right).
\end{multline}

Therefore, from \eqref{eq:Pb1_sim1} and \eqref{eq:Pb2_sim1}, the conditional BER can be expressed as
\begin{equation}
\label{eq:Pb_sim1}
P_{\bt}(E|h_2)=P_{\bt}(E_1|h_2)+P_{\bt}(E_2|h_2)=
\frac{b_0 b_2}{c_0c_2}+ \frac{b_0b_2}{c_0(c_0-d_2)}+\frac{b_0 b_2}{c_2(c_2-d_0)}.
\end{equation}

Finally, by substituting $b_2,c_2,d_2$ from \eqref{eq:b2c2d2} and taking the average over the distribution of $\lambda=|h_2|^2$, one has
\begin{equation}
\label{eq:Pb_sim2}
P_{\bt}(E)= I_1+I_2+I_3,
\end{equation}
where the terms $I_1$, $I_2$ and $I_3$ are determined in the following.

First, $I_1$ is computed as
\begin{multline}
\label{eq:I1:sc2}
I_1=\frac{b_0}{c_0} \int \limits_{0}^{\infty} \frac{b_2}{c_2}f_{\lambda}(\lambda) \dd \lambda=\frac{b_0}{c_0} \int \limits_{0}^{\infty} \frac{B_2}{2B_1}\frac{\lambda+B_1}{\lambda+B_2}\frac{1}{\sigma_2^2} {\mathrm{e}}^{-\frac{\lambda}{\sigma_2^2}} \dd \lambda\\=
\frac{b_0B_2}{2c_0B_1} \left(1+\frac{B_1-B_2}{\sigma_2^2}\exp\left(\frac{B_2}{\sigma_2^2}\right)E_1\left(\frac{B_2}{\sigma_2^2}\right) \right),
\end{multline}
where $B_1,\; B_2$ are defined as
\begin{equation}
\label{eq:B1B2}
\begin{split}
B_1=&\frac{1}{A^2\left(1+(1-\alpha)\rho_1\right)}\\
B_2=&\frac{1}{A^2(1+\rho_1)}.
\end{split}
\end{equation}
Second, $I_2$ is obtained as
\begin{multline}
\label{eq:I2}
I_2=\frac{b_0}{c_0} \int \limits_{0}^{\infty} \frac{b_2}{c_0-d_2}f_{\lambda}(\lambda) \dd \lambda
=\frac{b_0}{c_0} \int \limits_{0}^{\infty} \tilde{B}_3\frac{1}{\lambda+\tilde{B}_2}\frac{1}{\sigma_2^2} {\mathrm{e}}^{-\frac{\lambda}{\sigma_2^2}} \dd \lambda\\=
\frac{b_0}{c_0}\tilde{B}_3 \left(\frac{1}{\sigma_2^2}\exp\left(\frac{\tilde{B}_2}{\sigma_2^2}\right)E_1\left(\frac{\tilde{B}_2}{\sigma_2^2}\right) \right),
\end{multline}
where
\begin{gather}
%\nonumber
%\tilde{B}_1=\frac{2\tilde{a}_0+\tilde{a}_1 P_0/N_0\sigma_1^2}{A^2(\tilde{a}_0+\tilde{a}%_1P_0/N_0\sigma_1^2)}\\
\label{eq:B1B2B3t}
\tilde{B}_2= \frac{(3+\alpha+(1-\alpha_0)\rho_0)\rho_1+3+(1-\alpha_0)\rho_0}{A^2(1+(2+\alpha)\rho_1+(1+\alpha)\rho_1^2)}
\\ \nonumber
\tilde{B}_3= \frac{(1+(1+\alpha)\rho_1)(1+(1-\alpha_0)\rho_0)}{2A^2(1+(2+\alpha)\rho_1+(1+\alpha)\rho_1^2)}.
\end{gather}

Third, $I_3$ is determined as
\begin{multline}
\label{eq:I3}
I_3=b_0 \int \limits_{0}^{\infty} \frac{b_2}{c_2(c_2-d_0)}f_{\lambda}(\lambda) \dd \lambda
= \int \limits_{0}^{\infty} \breve{B}_3\frac{\lambda+\breve{B}_1}{\lambda+\breve{B}_2}\frac{1}{\sigma_2^2} {\mathrm{e}}^{-\frac{\lambda}{\sigma_2^2}} \dd \lambda\\=
\breve{B}_3 \left(1+\frac{\breve{B}_1-\breve{B}_2}{\sigma_2^2}\exp\left(\frac{\breve{B}_2}{\sigma_2^2}\right)E_1\left(\frac{\breve{B}_2}{\sigma_2^2}\right) \right),
\end{multline}
where
\begin{gather}
\nonumber
\breve{B}_1=\frac{2}{A^2(1+(1-\alpha)\rho_1)}\\
\label{eq:B1bB2bB3b}
\breve{B}_2=\frac{(3-\alpha+(1+\alpha_0)\rho_0)\rho_1+3+(1+\alpha_0)\rho_0}{A^2(1+(2-\alpha)\rho_1)+(1-\alpha)\rho_1^2)}\\
\nonumber
\breve{B}_3=\frac{(1+(1-\alpha)\rho_1)^2(1+(1+\alpha_0)\rho_0)}{4(1+\rho_0)(1+(2-\alpha)\rho_1+(1-\alpha)\rho_1^2)}.
\end{gather}

In summary, the obtained BER expression in \eqref{eq:Pb_sim2} gives the exact BER of the D-AF relaying system using DBPSK and selection combining in general time-varying Rayleigh fading channels. For the special case of $\alpha_0=1,\;\alpha=1,\;\sigma_i^2=1,\;i=0,1,2,$ this expression yields the BER of the system considered in \cite{DAF-SC} for symmetric slow-fading channels. It should be mentioned that, although, the BER expression of \cite[Eq.23]{DAF-SC} looks different than \eqref{eq:Pb_sim2} in the special case, both expressions give the same results for slow-fading symmetric channels. However, the expression in \eqref{eq:Pb_sim2} only involves computing the exponential integral function, whereas the expression of \cite[Eq.23]{DAF-SC} was derived in the integral form which also involves the exponential integral function.

It is also seen that the obtained BER expression depends on the channel auto-correlations. This dependence is the reason for performance degradation in fast-fading channels and the fact that the BER reaches an error floor at high signal-to-noise ratio. The error floor can be obtained as (see Appendix for the proof):
\begin{equation}
\label{eq:PEf}
\lim \limits_{P_0/N_0 \rightarrow \infty} P_{\bt}(E)=\bar{I}_1+\bar{I}_2+\bar{I}_3,
\end{equation}
where
\begin{equation}
\label{eq:I1b}
\bar{I}_1=\lim \limits_{P_0/N_0\rightarrow \infty} I_1=\frac{1}{4}(1-\alpha_0)(1-\alpha)
\end{equation}
\begin{equation}
\label{eq:I2b}
\bar{I}_2=\lim \limits_{P_0/N_0\rightarrow \infty} I_2=
\frac{1-\alpha_0}{2\sigma_2^2}\bar{\tilde{B}}_3
\exp\left(\frac{\bar{\tilde{B}}_2}{\sigma_2^2}\right)E_1\left(\frac{\bar{\tilde{B}}_2}{\sigma_2^2}\right)
\end{equation}
with
\begin{gather}
\label{eq:lim_B2t}
\bar{\tilde{B}}_2=\lpn \tilde{B}_2= \frac{q \sigma_0^2}{1-q} \frac{1-\alpha_0}{1+\alpha}  \\
\label{eq:lim_B3t}
\bar{\tilde{B}}_3=\lpn \tilde{B}_3=\frac{q(1-\alpha_0)\sigma_0^2}{2(1-q)}
\end{gather}
and
\begin{equation}
\label{eq:I3b}
\bar{I}_3=\lpn I_3= \bar{\breve{B}}_3 \left(1-\frac{\bar{\breve{B}}_2}{\sigma_2^2} \exp\left(\frac{\bar{\breve{B}}_2}{\sigma_2^2} \right) E_1\left(\frac{\bar{\breve{B}}_2}{\sigma_2^2} \right) \right)
\end{equation}
with
\begin{gather}
\label{eq:lim_B2b}
\bar{\breve{B}}_2=\lpn \breve{B}_2 =\frac{q \sigma_0^2}{1-q} \frac{1+\alpha_0}{1-\alpha}  \\
\label{eq:lim_B3b}
\bar{\breve{B}}_3=\lpn \breve{B}_3 =\frac{1}{4} (1-\alpha)(1+\alpha_0).
\end{gather}
The above expressions show that the error floor is a function of the second-order statistics of the channels (auto-correlation and variance) and the power amplification factor and it is independent of the (high) transmitted power.

\section{Simulation Results}
\label{sc2:sec:sim}
In this section the D-AF relay network under consideration is simulated for various channel qualities using both the SC and semi-MRC methods. The obtained theoretical BER and error floor of the SC method are verified by simulation results.

The channel coefficients are assumed to be Rayleigh flat-fading, i.e., $h_0[k]\sim \CN(0,\sigma_0^2),\; h_1[k]\sim \CN(0,\sigma_1^2),\; h_2[k]\sim \CN(0,\sigma_2^2).$ Based on the location of the nodes with respect to each other and channel qualities, variances of the channels would be different. Here, three scenarios are considered: (i) symmetric channels with $\sigma_0^2=1,\;\sigma_1^2=1,\; \sigma_2^2=1$, (ii) non-symmetric channels with strong SR channel $\sigma_0^2=1,\;\sigma_1^2=10,\;\sigma_2^2=1$, and (iii) non-symmetric channels with strong RD channel $\sigma_0^2=1,\;\sigma_1^2=1,\;\sigma_2^2=10$. The channel scenarios are summarized in Table~\ref{table:sig012}.
\begin{table}[!ht]
\begin{center}
\caption{Channel variances and corresponding optimum power allocation factors}
\vspace*{.1in}
\label{table:sig012}
  \begin{tabular}{ |c | c| c|  }
    \hline
				& {$[\sigma_0^2,\sigma_1^2, \sigma_2^2]$}& $q_{\opt}$  \\ \hline\hline
{Symmetric}		& $[1,1,1]$    & 0.67   \\ \hline
{Strong SR } 		& $[1,10,1]$   & 0.58    \\ \hline
{Strong RD} 	& $[1,1,10]$   & 0.85    \\
    \hline
  \end{tabular}
\end{center}
\end{table}

The simulation method of \cite{ch-sim} is utilized to generate the time-correlated channel coefficients $h_0[k],h_1[k],h_2[k].$ The amount of time-correlation is determined by the normalized Doppler frequency of the underlying channel, which is a function of the mobility, carrier frequency and symbol duration. Obviously, for fixed carrier frequency and symbol duration, a higher mobility leads to a larger Doppler frequency and less time-correlation.

\begin{figure}[tb!]
\psfrag {f} [t][] [.9]{$f_0=f_1$}
\psfrag {BER} [][] [.9]{Error Floor}
\centerline{\epsfig{figure={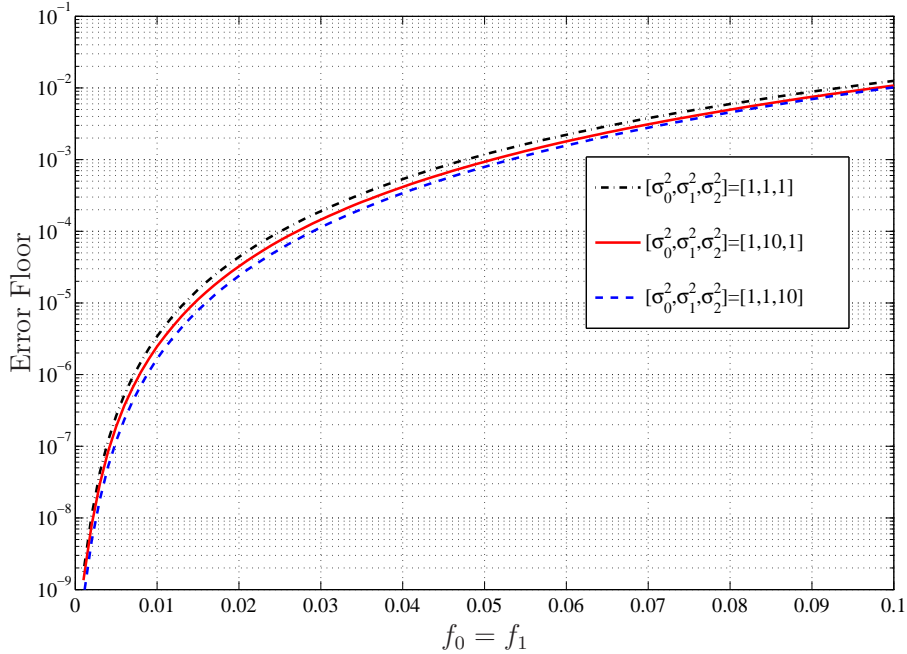},height=8.5cm,width=12cm}}
\caption{Error floors as functions of channel fade rates for different channel qualities, $f_0=f_1,\;f_2=0.001$.}
\label{fig:err_floor}
\end{figure}

Based on the normalized Doppler frequency of the three channels, different cases are considered. To get an understanding about choosing the normalized Doppler frequency values, the obtained error floor expression is examined for a large range of Doppler values. Fig.~\ref{fig:err_floor} plots of error floors versus channel fade rates for the three scenarios of Table~\ref{table:sig012}. It is assumed that the SD and SR channels have similar normalized Doppler frequencies, i.e., $f_0=f_1$, which changes from 0.001 to 0.1 and $f_2=0.001$. The plots in Fig.~\ref{fig:err_floor} can be divided into three regions. For fade rates less than 0.01, the error floor is very small and this region would be regarded as the slow-fading region. The fade rate of 0.01 is an approximate threshold beyond which the channels become relatively fast-fading. For fade rates between 0.01 to 0.05, the error floor increases asymptotically in a linear manner with a relatively sharp slope from $10^{-6}$ to $10^{-3}$. When the fade rate is larger than 0.05, the error floors continue to increase from $10^{-3}$ to $10^{-2}$. A BER value of around $10^{-2}$ is obtained in this region. This is very high for reliable communication and therefore the fade rate of 0.05 would be regarded as the threshold beyond which the channels become very fast-fading.

\begin{figure}[thb!]
\psfrag {time1} [][] [1]{$k$}
\psfrag {time2} [][] [1]{$n$}
\psfrag {h} [] [] [.8] {$|h[k]|$}
\psfrag {h0} [] [] [.8] {$|h_0[k]|$}
\psfrag {corr0} [] [] [.8] {$\varphi_0(n)$}
\psfrag {corr2} [] [] [.7] {$\varphi_1(n)\varphi_2(n)$}
\centerline{\epsfig{figure={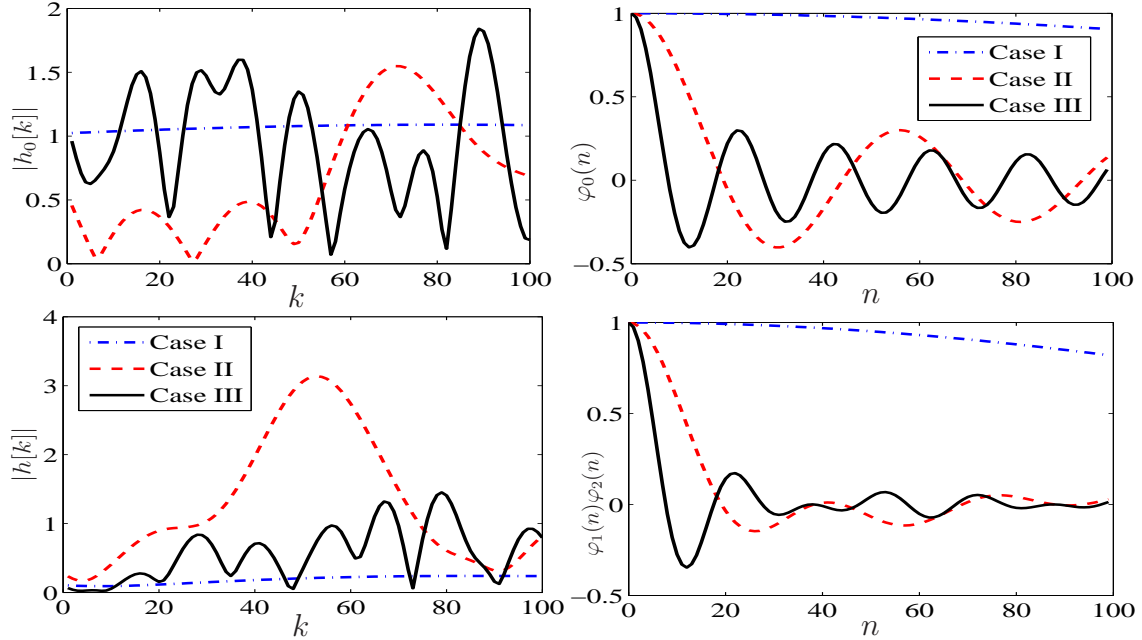},height=8.5cm,width=15cm}}
\caption{Snapshot of realizations of the direct and cascaded channels and the corresponding autocorrelations in different cases. Here $k$ and $n$ are defined as in Eq. \eqref{eq:phi}.}
\label{fig:doub-ch}
\end{figure}

From the discussion concerning Fig.~\ref{fig:err_floor}, three fading cases are considered. In Case I, it is assumed that all nodes are fixed or slowly moving so that all channels are slow-fading with the normalized Doppler values of $f_0=0.001,\;f_1=0.001,\;f_2=0.001$. In Case II, it is assumed that Source is moving so that the SD and SR channels are fast-fading with $f_0=0.02,\;f_1=0.02$ but Relay and Destination are fixed and the RD channel is slow-fading with $f_2=0.001$. In Case III, it is assumed that both Source and Destination are moving so that all the channels are fast-fading with $f_0=0.05,\;f_1=0.01$ and $f_2=0.05$, respectively. The normalized Doppler values are listed in Table \ref{table:f0f1f2}. Also, a snapshot of realizations of the direct and cascaded channels and their auto-correlation values in the three cases are plotted in Fig.~\ref{fig:doub-ch}. The plots show that when the normalized Doppler frequency values get larger, the channel coefficients fluctuate wider and the corresponding auto-correlation values decline faster over time. The auto-correlation of the cascaded channel declines faster than that of the direct channel as it involves the effects of two channels.
\begin{table}[!ht]
\begin{center}
\caption{Three fading cases.}
\vspace*{.1in}
\label{table:f0f1f2}
  \begin{tabular}{ |c | c| c|c|c|  }
    \hline
				& $f_0$ & $f_1$ & $f_2$ & Channels status   \\ \hline\hline
{Case I }		& 0.001  & 0.001  & 0.001  & all channels are slow-fading \\ \hline
{Case II } 		& 0.02   & 0.02   &  0.001	& SD and SR are fast-fading \\ \hline
{Case III } 	& 0.05   & 0.01   &   0.05 & all channels are fast-fading \\
    \hline
  \end{tabular}
\end{center}
\end{table}
%\vspace*{-0.5cm}

%Information bits are differentially encoded for $M=2$ to obtain DBPSK symbols at Source. The amplification factor at Relay is fixed to $A=\sqrt{(1-q)P/(qP\sigma_1^2+N_0)}$ to normalize the average relay power to $P_1=(1-q)P$. Also, $N_0=1$ is assumed.

\begin{figure}[thb!]
\psfrag {BER} [] [] [1] {BER}
\psfrag {PAF} [t] [b] [1] {$q={P_0}/{P}$}
\psfrag {P/N_0} [t] [b] [1] {$P/N_0$}
\centerline{\epsfig{figure={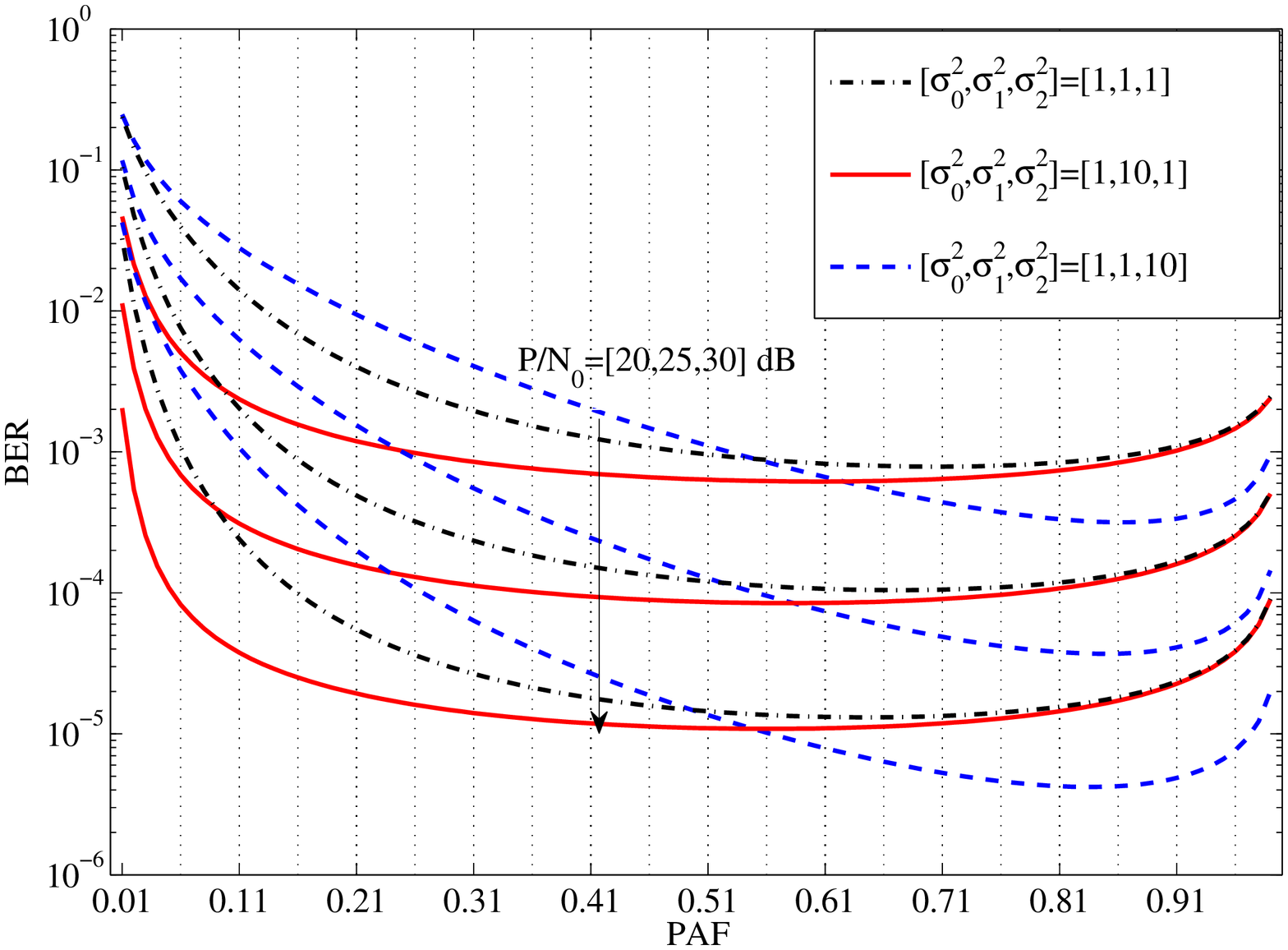},height=9cm,width=12cm}}
\caption{BER as a function of power allocation factor $q$ for $P/N_0=20,\;25,\;30$ dB.}
\label{fig:pw_sigs}
\end{figure}

First, the optimum power allocation between Source and Relay to minimize the BER is investigated in the three scenarios of Table~\ref{table:sig012} and fading rates of Case I (slow-fading). For each scenario listed in Table~\ref{table:sig012}, the BER expression is examined for different values of power allocation factor $q=P_0/P$, where $P=P_0+P_1$ is the total power. The BER curves are plotted versus $q$ in Fig.~\ref{fig:pw_sigs} for $P/N_0$=20,\;25,\;30 dB. Also, the optimum values obtained for the SC method are listed in Table~\ref{table:sig012} when $P/N_0=25$ dB. The optimum values in Table~\ref{table:sig012} and Fig.~\ref{fig:pw_sigs} show that in general more power should be allocated to Source than Relay. The BER is minimized at $q\approx 0.67$ and $\approx 0.58$ for symmetric and strong SR channels, respectively. When the RD channel becomes stronger than the SR channel, even more power should be allocated to Source and the BER is minimized at $q\approx 0.85$. This observation is similar to what reported in \cite[Table I]{DAF-MN-Himsoon} for the semi-MC technique.

\begin{figure}[thb!]
\psfrag {P(dB)} [t][] [1]{$P/N_0$ (dB)}
\psfrag {BER} [] [] [1] {BER}
\centerline{\epsfig{figure={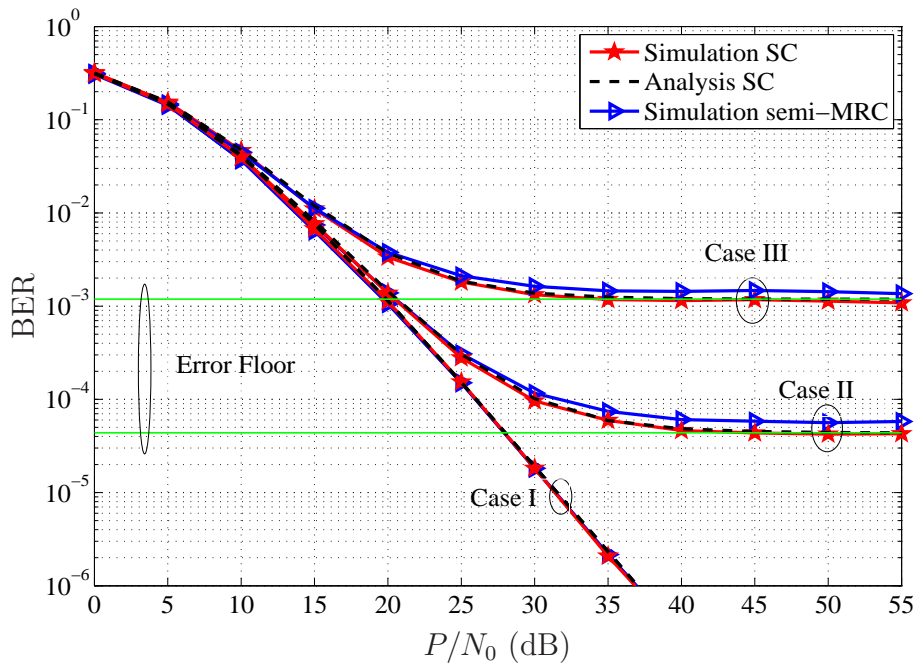},height=8.5cm,width=12cm}}
\caption{Theoretical and simulation BER of the D-AF system with SC and semi-MRC methods using DBPSK in different fading rates and symmetric channels: $\sigma_0^2=1,\sigma_1^2=1,\sigma_2^2=1$.}
\label{fig:scs_m2_sig11}
\end{figure}

\begin{figure}[thb!]
\psfrag {P(dB)} [t][] [1]{$P/N_0$ (dB)}
\psfrag {BER} [] [] [1] {BER}
\centerline{\epsfig{figure={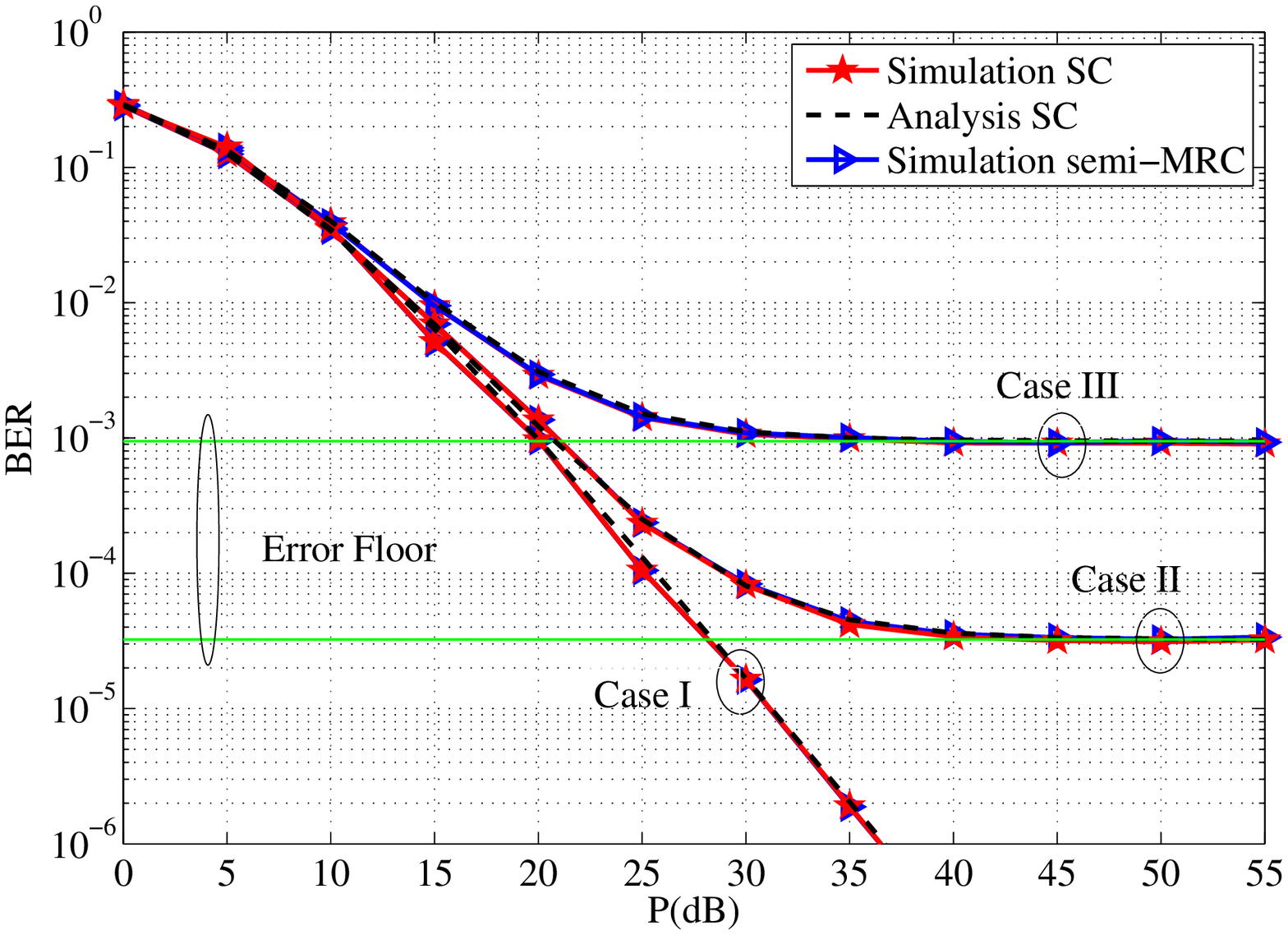},height=8.5cm,width=12cm}}
\caption{Theoretical and simulation BER of the D-AF system with SC and semi-MRC methods using DBPSK in different fading rates and strong SR channel: $\sigma_0^2=1,\sigma_1^2=10,\sigma_2^2=1$.}
\label{fig:scs_m2_sig101}
\end{figure}

\begin{figure}[thb!]
\psfrag {P(dB)} [t][] [1]{$P/N_0$ (dB)}
\psfrag {BER} [] [] [1] {BER}
\centerline{\epsfig{figure={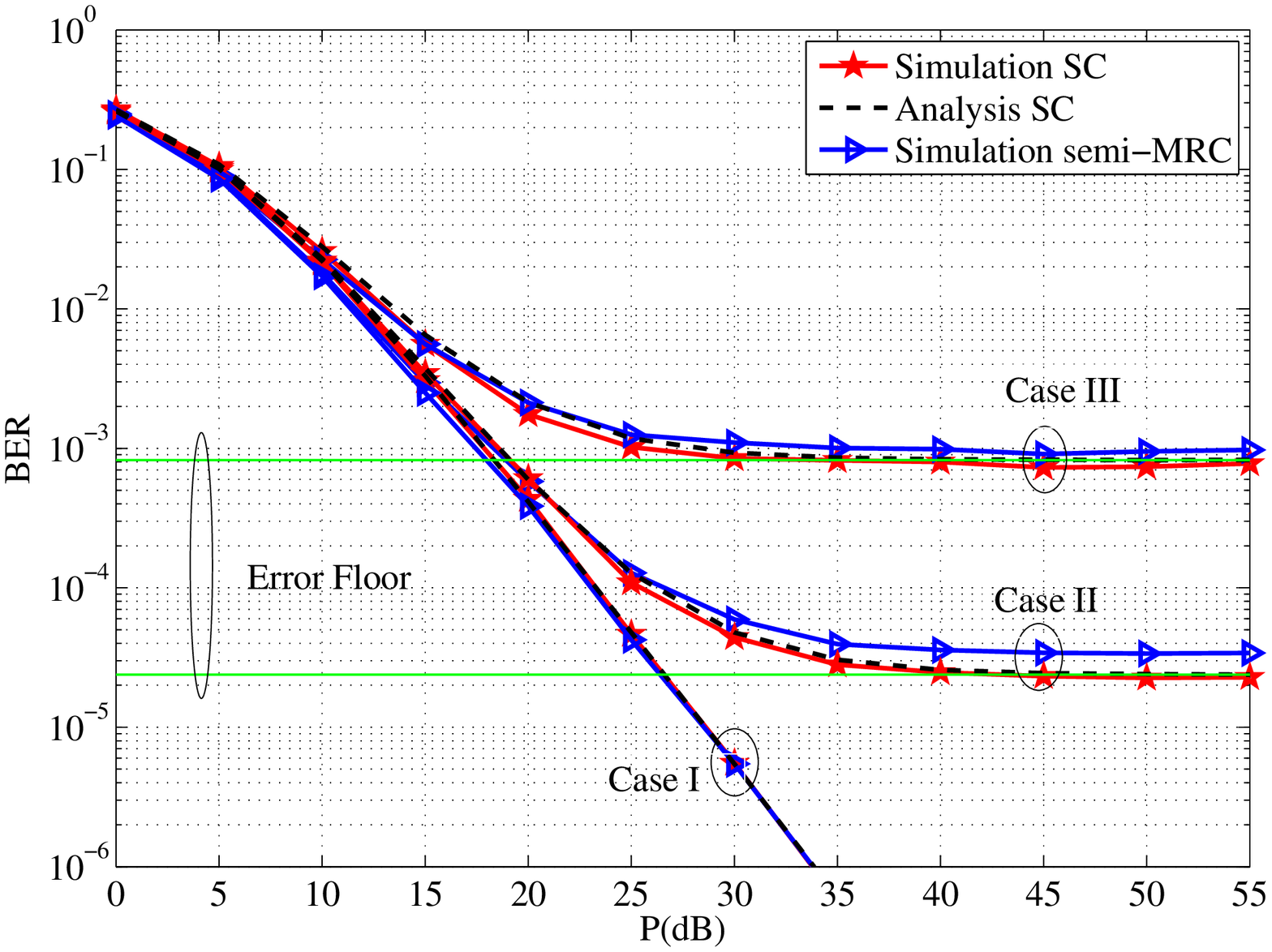},height=8.5cm,width=12cm}}
\caption{Theoretical and simulation BER of the D-AF system with SC and semi-MRC methods using DBPSK in different fading rates and strong RD channel: $\sigma_0^2=1,\sigma_2^2=1,\sigma_3^2=10$.}
\label{fig:scs_m2_sig110}
\end{figure}

The obtained power allocation factors are used in the simulation and the simulated BER using the SC and semi-MRC methods are computed for the three fading cases and channel variances. The BER results are plotted versus $P/N_0$ in Figs.~\ref{fig:scs_m2_sig11}--\ref{fig:scs_m2_sig110} with solid lines (different legends). In each figure, three fading cases are shown. It should be mentioned that the results of Case I in Figure~\ref{fig:scs_m2_sig11} correspond to the results of \cite{DAF-SC} considered for symmetric slow-fading channels. On the other hand, the corresponding theoretical BER values of the SC method (for all cases) are computed from \eqref{eq:Pb_sim2} and plotted in Figs.~\ref{fig:scs_m2_sig11}--\ref{fig:scs_m2_sig110} with dashed lines. The horizontal plotted lines show the theoretical values of the error floor for Case II and III, computed from \eqref{eq:PEf}.

As it seen in all the figures, the simulation results of the SC method are tight to the  analytical results for all fading cases and channel variances. This observation verifies our analysis. Specifically, in Case I (slow-fading) of all the figures, the BER of the SC method (and semi-MRC) is consistently decreasing with $P/N_0$ and a diversity of two is achieved. The error floor in this case is very small ($\approx 10^{-9}$) and practically does not exist.

However, in Case II of all the figures, the situation is different from Case I. Since this case involves two fast-fading channels, the effect of channels variation is clearly observed in the obtained BER of both methods for $P/N_0>20$ dB. The BER gradually deviates from the results in Case I and eventually reaches to an error floor between $10^{-5}$ to $10^{-4}$. The exact amount of the error floors can be read from the horizontal lines in the figures.

Similarly, in Case III of all the figures, the obtained BER is degraded for $P/N_0>10$ dB. A severe degradation is seen in this case as all channels are fast-fading, specially two channels are around the threshold of very fast-fading region. There is no benefit in increasing the transmit power since an error floor around $10^{-3}$ appears for $P/N_0>30$ dB.

Moreover, the results of both SC and semi-MRC methods are very close to each other in all the figures. In Cases II and III of Figs.~\ref{fig:scs_m2_sig11} and \ref{fig:scs_m2_sig110}, the results of the SC method are slightly better than that of the semi-MRC method at high $P/N_0$. This is due to the fact that the fixed combining weights (used in \eqref{eq:zeta}) of the semi-MRC method are not optimum and determined based on the second-order statistics of the channels and not their instantaneous CSI. Note that, in MRC method, the optimum combing  weights should be computed based on the noise variance of each link. The noise variance of the cascaded link is a function of the instantaneous CSI of the RD channel, which is not available in the considered system.

On the other hand, the close performance of both methods allows one to use the BER analysis of the SC method to tightly approximate the performance of the semi-MRC method in time-varying channels. It should be mentioned that the exact performance evaluation of the semi-MRC in time-varying channels appears difficult and only a loose lower bound was reported in \cite{DAF-ITVT}.

\balance

\section{Conclusion}
\label{sc2:sec:con}
Selection combining of the received signals at Destination in a D-AF relay network employing DBPSK was studied in general time-varying Rayleigh fading channels. Thanks to the differential encoding and selection combiner, no channel state information is needed at Destination for information detection. The exact bit-error-rate of the system was derived. Simulation results in various fading rates and channel variances verified the analysis and show that the selection combiner performs very close to the more-complicated semi-MRC method (which needs the second-order statistics of all channels at Destination). The analytical results show that the error performance depends on the fading rates of the equivalent channel and direct channel and an error floor exists at high signal-to-noise ratio region.

\begin{subappendices}

\section{Proof of \eqref{eq:PEf}}

\begin{equation}
\label{eq:prf:I1b}
\bar{I}_1=\lim \limits_{P_0/N_0 \rightarrow \infty} I_1
= \lpn \frac{b_0B_2}{2c_0B_1}+
\lpn \frac{b_0B_2(B_1-B_2)}{2c_0B_1\sigma_2^2}\exp\left(\frac{B_2}{\sigma_2^2}\right)E_1\left(\frac{B_2}{\sigma_2^2}\right).
\end{equation}
From \eqref{eq:b0c0d0}, one has
\begin{equation}
\label{eq:b0/c0}
\lpn \frac{b_0}{c_0}=\lpn \frac{1+(1-\alpha_0)\rho_0}{2(1+\rho_0)}=\frac{1-\alpha_0}{2}.
\end{equation}
Likewise, taking the limit in \eqref{eq:B1B2} gives
\begin{equation}
\label{eq:B2/B1}
\lpn \frac{B_2}{2B_1}=\lpn \frac{A^2\left(1+(1-\alpha)\rho_1\right)}{2A^2\left(1+\rho_1\right)}=
\frac{1-\alpha}{2}.
\end{equation}
Again from \eqref{eq:B1B2}, one has
\begin{multline}
\label{eq:B1-B2}
B_1-B_2=\frac{1}{A^2\left(1+(1-\alpha)\rho_1\right)}-\frac{1}{A^2\left(1+\rho_1\right)}
= \frac{\rho_1}{\left(1+\rho_1\right)} \frac{\alpha}{A^2\left(1+(1-\alpha)\rho_1\right)}\\ \approx \frac{\alpha}{A^2\left(1+(1-\alpha)\rho_1\right)}=\alpha B_1,
\end{multline}
where the approximation has been made for large $P_0/N_0$.
Hence,
\begin{equation}
\label{eq:B2/B1(B1-B2)}
\frac{B_2(B_1-B_2)}{B_1}\approx \alpha B_2.
\end{equation}
On the other hand, for large $P_0/N_0$, $x=B_2/\sigma_2^2 \rightarrow 0$ and using the following approximation \cite[Eq. 5.1.20]{abro72ha}, one has
\begin{equation}
\label{eq:exE1x}
{\mathrm{e}}^x E_1(x)\approx \log \left(\frac{1}{x}\right)
\end{equation}
and
\begin{equation}
\label{eq:limexpE1x}
\lim \limits_{x \rightarrow 0} x {\mathrm{e}}^x E_1(x)=\lim \limits_{x \rightarrow 0} x \log(\frac{1}{x})=\lim \limits_{y \rightarrow \infty} \frac{\log(y)}{y}=0.
\end{equation}
Therefore, by substituting \eqref{eq:b0/c0}, \eqref{eq:B2/B1} and \eqref{eq:limexpE1x} into \eqref{eq:prf:I1b}, \eqref{eq:I1b} is obtained.

Finding $\bar{I}_2$ and $\bar{I}_3$ by taking the limit of $I_2$ and $I_3$ is straightforward. Note that, in deriving $\bar{I}_2$ and $\bar{I}_3$, by substituting $P_1=(1-q)P_0/q$ into \eqref{eq:AmpFactor}, one has
\begin{equation}
\label{eq:A2q}
A^2=\frac{P_1}{P_0\sigma_1^2+N_0}=\frac{(1-q)P_0/N_0}{q(P_0/N_0\sigma_1^2+1)},
\end{equation}
\begin{equation}
\lpn \frac{1}{A^2}=\frac{q\sigma_1^2}{1-q}.
\end{equation}

\end{subappendices}

\chapter{Performance of Differential Distributed Space-Time Coding}
\label{ch:dstc}
As discussed in Chapter~\ref{ch:coop}, in cooperative relay networks, relays can utilize either repetition-based strategy or distributed space-time coding (DSTC) strategy to cooperate and deliver their signals to the destination. In the repetition-based strategy, each relay re-broadcasts its signals sequentially in time, while the other relays are silent (see Figure~\ref{fig:tx_rep}). Repetition based strategy is simple to implement but has a low spectral efficiency. The repetition-based strategy has been considered in Chapters~\ref{ch:mnode}-\ref{ch:sc_tv}.
Another way to take advantage of relay resources in cooperative networks is to use the DSTC strategy. In the DSTC strategy, relays transmit their signals simultaneously in time to the destination (see Figure~\ref{fig:dstc_tx}). The simultaneous transmission using space-time coding structure improves the spectral efficiency of the system at the price of higher complexity in both encoding and decoding.

In this chapter, a multi-branch dual-hop relaying without a direct link employing DSTC strategy is considered. This topology, depicted in Figure~\ref{fig:mbdh}, can be used for both coverage extension and diversity improvement. To avoid channel estimation at the destination, differential encoding is applied at the source.
Similar to the repetition-based strategy, the first goal in this manuscript is to study the performance of two-codeword differential detection and its robustness in time-varying channels. When using the DSTC strategy, the required coherence time for two-codeword detection is larger than that of the repetition-based strategy. This makes differential DSTC (D-DSTC) more vulnerable against the mobility of users in the network. Thus, it is important to analyse the performance of D-DSTC and its achievable diversity in time-varying channels. It will be seen through the analysis and simulation that the pair-wise error probability of two-codeword differential detection in fast-fading channels hits an error floor and the achieved diversity goes to zero.

In the second part of this chapter, a multiple-codeword differential detection is designed to improve the performance of D-DSTC in fast-fading channels. Since the multiple-codeword detection is complicated, the decision metric is simplified such that the available sphere-decoding techniques used in point-to-point MIMO systems can be easily adapted for D-DSTC system. It is very important to carefully determine the parameters of the detection process based on the system and channels information to achieve the optimal performance expected by multiple-codeword detection. The performance of the developed multiple-codeword differential detection algorithm is also shown by simulation results in various fading scenarios.

The results of our study is reported in the following manuscript.

[Ch8-1] M. R. Avendi, Ha H. Nguyen,``Effect of Mobility on the Performance of Differential Distributed Space-Time Coding", {\it submitted to ComManTel 2014}.

\begin{center}
{\bf{\Large
Effect of Mobility on the Performance of Differential Distributed Space-Time Coding
}}
\end{center}
\begin{center}
M. R. Avendi, Ha H. Nguyen
\end{center}

\begin{center}
\bf Abstract
\end{center}
This article studies the behaviour of Differential distributed space-time coding (D-DSTC) using two-codeword and multiple-codeword differential detection over time-varying Rayleigh fading channels. Instead of the conventional slow-fading assumption, a time-series model is utilized and the performance of two-codeword differential detection is analysed by deriving an upper bound for the pair-wise error probability. The derivation reveals that the two-codeword differential detection performs poorly in fast-fading channels and the obtained diversity approaches to zero regardless of the number of relays. On the other hand, to overcome the error floor experienced with fast-fading channels, a nearly optimal ``multiple-codeword'' differential detection is developed. The multiple-codeword detection algorithm jointly processes a larger window of received signals for differential detection and significantly improves the performance of D-DSTC in fast-fading channels. Computer simulation is carried out in different fading scenarios to support the analysis and the effectiveness of the multiple-codeword detection algorithm.

\begin{center}
\bf Index terms
\end{center}
Distributed relaying,differential modulation, time-varying fading channels, channel auto-correlation, two-codeword detection, multiple-codeword differential detection.

\section{Introduction}
\label{dstc:sec:intro}
It has been well-known that using antenna arrays can offer a substantial diversity gain, which helps to overcome the effect of channel fading and achieve a better quality and/or higher data rate in wireless networks. However, using multiple antennas is impractical for mobile applications since there is insufficient space to mount multiple antennas on a mobile unit in order to create uncorrelated wireless fading channels. This limitation can be addressed by using the technique of cooperative communications, which has been shown to be applicable in many wireless systems and applications such as 3GPP LTE-Advances, WiMAX, WLANs, vehicle-to-vehicle communications and wireless sensor networks \cite{coop-LTE,coop-V2V,coop-WiMAX,coop-deploy,coop-dohler}.

The technique of cooperative communications makes use of the fact that, since users in a network can listen to a source during its transmission phase, they would be able to re-broadcast the received data to the destination in another phase. Therefore, the overall diversity and performance of the network would benefit from the virtual antenna array that is constructed cooperatively by multiple users.

Depending on the protocol that relays utilize to process and re-transmit the received signal to the destination, the relay networks have been generally classified as decode-and-forward or amplify-and-forward \cite{coop-laneman}. Among these two protocols, amplify-and-forward (AF) has been the focus of many studies because of its simplicity in the relay's operation. Moreover, depending on the strategy that relays utilize to cooperate, relay networks are categorized as repetition-based and distributed space-time coding (DSTC)-based \cite{DSTC-Laneman}. The later strategy yields a higher spectral efficiency than the former but it is more complicated to design and build \cite{DSTC-Laneman}.

In DSTC networks \cite{DSTC-Laneman,DSTC-Y,DSTC-Kaveh}, the relays cooperate to combine the received symbols by multiplying them with a fixed or variable factor and forward the resulted signals to the destination. The cooperation is such that a space-time code is effectively constructed at the destination. Coherent detection of transmitted symbols can be achieved by providing the instantaneous channels state information (CSI) of all transmission links at the destination. Although this requirement can be accomplished by sending pilot (training) signals and using channel estimation techniques in slow-fading environments, it is not feasible in fast-fading channels. Moreover, collecting the CSI of SR channels at the destination is questionable due to noise amplification at relays. Furthermore, the computational complexity and overhead of channel estimation increase proportionally with the number of relays. In addition, in fast-fading channels a more frequent channel estimation is needed, which reduces the effective transmission rate and spectral efficiency. Also, all channel estimation techniques are subject to impairments that would directly translate to performance degradation.

When no CSI is available at the relays and destination, differential DSTC (D-DSTC) scheme has been studied in \cite{D-DSTC-Y,PC-DSTC-Kiran,D-DSTC-Amin,D-DSTC-Giannakis} which only needs the second-order statistics of the channels at the relays. Also, the constructed unitary space-time code at the destination provides the opportunity to apply two-codeword non-coherent detection without any CSI.
Although the simplicity of two-codeword differential detection makes it very appealing, there has not been any study on its capability in practical time-varying channels. Most of research works on D-DSTC schemes assume a \emph{slow-fading} situation and show, via simulation, that the performance of the D-DSTC scheme is about 3-4 dB worse than the performance of its coherent version. In reality, due to the mobility of users, channels become time-varying. Thus, the common assumption used in two-codeword differential detection, namely approximate equality of two consecutive channel uses, is violated. Based on this motivation, the first goal of this article is to examine the performance of D-DSTC using two-codeword differential detection and its robustness under practical channel variation scenarios.

Considered is the case that a source communicates with a destination via multiple relays and all the channels change over time depending on the mobility of users. Instead of the slow-fading assumption, a time-series model originally developed in \cite{DAF-ITVT} is utilized to characterize the time-varying nature of the channels. Based on this model, an upper bound for the pair-wise error probability (PEP) is derived and used to analyse the obtained diversity under general time-varying Rayleigh fading channels. It is seen that the full diversity is achieved in slow-fading channels. However, the diversity diminishes quickly and approaches to zero in fast-fading channels. In other words, the performance of the system hits an error floor. Simulation results are provided to support the analysis in various scenarios of fading channels.

To overcome the limitations of two-codeword detection in fast-fading channels, the second goal of this article is to design a robust differential detection for the D-DSTC system. It is noted that a similar problem was addressed in point-to-point communications using multiple-symbol detection. Multiple-symbol differential detection, first proposed for point-to-point communications in \cite{msdd-div2}, jointly processes a larger window of the received symbols for detection. As the complexity increases exponentially with the window size, Lampe et. al \cite{MSDSD-L} developed a multiple-symbol differential sphere detection algorithm to reduce the complexity of multiple-symbol detection. Later, Pauli et al. \cite{MSDUSTC-P} developed a multiple-codeword detection algorithm for unitary space-time codes in MIMO systems. In the context of relay networks, due to the complexity of the distribution of the received signals at the destination, the optimal decision rule of multiple-codeword detection does not have a closed-form solution. For simple implementation, an alternative decision rule is proposed and further simplified so that the multiple-codeword sphere detection algorithm of \cite{MSDUSTC-P} can be applied. It is also important to accurately determine the parameters of the detection process based on the channel and system information to achieve the optimal results. Finally, the near optimal performance of the proposed detection algorithm is illustrated with simulation results in different fading scenarios. It is seen that the proposed multiple-codeword detection technique, using a window of $N=10$ codewords, is able to significantly improve the performance of the D-DSTC system in fast-fading channels.

The outline of the paper is as follows. Section \ref{dstc:sec:system} describes the system model. In Section \ref{dstc:sec:two-CW}, the two-codeword differential detection of D-DSTC and its performance over time-varying Rayleigh-fading channels are studied. In Section \ref{dstc:sec:MSDSD}, multiple-codeword differential detection is developed. Simulation results are given in Section \ref{dstc:sec:sim}. Section \ref{dstc:sec:con} concludes the paper.

Notations: $(\cdot)^t$, $(\cdot)^*$, $(\cdot)^H$, $|\cdot|$, $\Re\{\cdot\}$ and $\Im\{\cdot\}$ denote transpose, complex conjugate, transpose conjugate, absolute value, real part and imaginary part of a complex vector or matrix, respectively. $\I_R$ and $\0_R$ are $R \times R$ identity matrix and zero matrix, respectively. $\CN(\0,\sigma^2 \I_R)$ and $\chi_{2R}^2$ stand for circular symmetric Gaussian random vector with zero mean and covariance $\sigma^2 \I_R$ and chi-squared distribution with $2R$ degrees of freedom, respectively. $\mbox{E}\{\cdot\}$, $\mbox{Var}\{\cdot\}$ denote expectation and variance operations, respectively. Both ${\mathrm{e}}^{(\cdot)}$ and $\exp(\cdot)$ show the exponential function. $\| \cdot \|$ denotes the Euclidean norm of a vector. $\diag\{x_1,\cdots,x_R\}$ is the diagonal matrix with $x_1,\cdots,x_R$ as its diagonal entries and $\diag\{\X_1,\cdots,\X_N\}$ is $RN\times RN$ block diagonal matrix with the $R \times R$ matrices $\X_l$ on its main diagonal. $\otimes$ is Kronecker product. A symmetric $N\times N$ Toeplitz matrix is defined by $\toep\{x_1,\cdots,x_N\}$. $\det\{\cdot\}$ denotes determinant of a matrix.

\section{System Model}
\label{dstc:sec:system}
\begin{figure}[t]
\psfrag {Source} [] [] [1.0] {Source}
\psfrag {Relay1} [] [] [1.0] {Relay 1}
\psfrag {Relay2} [] [] [1.0] {Relay 2}
\psfrag {RelayR} [] [] [1.0] {Relay R}
\psfrag {Destination} [] [] [1.0] {Destination}
\psfrag {f1} [r] [] [1.0] {$q_1[k]$}
\psfrag {g1} [l] [] [1.0] {$g_1[k]$}
\psfrag {f2} [bl] [] [1.0] {$q_2[k]$}
\psfrag {g2} [] [] [1.0] {\;\;$g_2[k]$}
\psfrag {fR} [] [] [1.0] {$q_R[k]$\;\;\;}
\psfrag {gR} [l] [] [1.0] {\;\;$g_R[k]$}
\centerline{\epsfig{figure={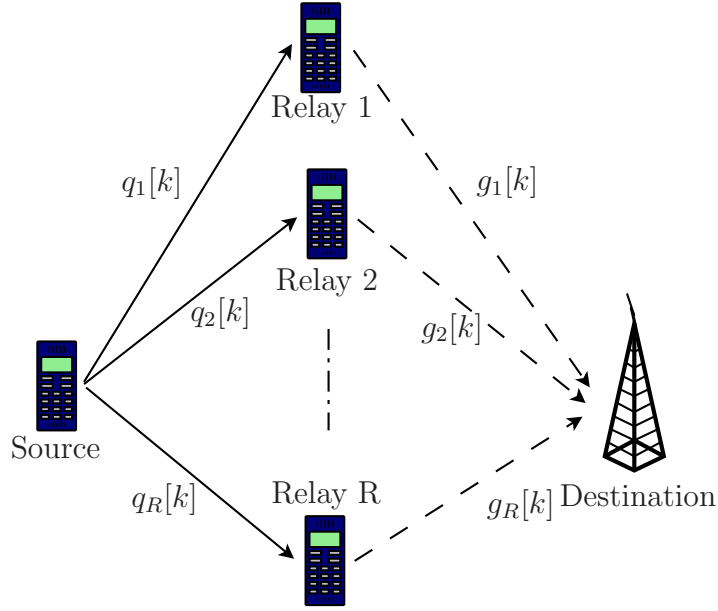},width=8.5cm}}
\caption{The wireless relay network under consideration.}
\label{fig:dstc}
\end{figure}

The wireless relay network under consideration, shown in Fig.~\ref{fig:dstc}, is similar to what considered in \cite{DSTC-Y,D-DSTC-Y}. It has one source, $R$ relays and one destination. Source communicates with Destination via the relays. Each node has a single antenna, and the communication between nodes is half duplex (i.e., each node is able to only send or receive in any given time). Individual channels from Source to the $i$th relay ($\mathrm{SR}_i$) and from the $i$th relay to Destination ($\mathrm{R}_i$D) are Rayleigh flat-fading and spatially uncorrelated. Moreover, due to the mobility of nodes, the channels would change over time. The amount of channel variation is quantified by the channel auto-correlation function, which is a function of fading rate. The fading rate is related to the Doppler shift which is also a function of the velocity of the user. Hence, there is a direct relation between the mobility, channel variation and Doppler frequency.
%with Jakes' fading model \cite{microwave-jake}:
%\begin{equation}
%\label{eq:jakes-auto}
%\mathcal{E} \{h[k]h^*[k+n]\}=J_0(2\pi f n),
%\end{equation}
%where $J_0(\cdot)$ is the zeroth-order Bessel function of the first kind, $f$ is the maximum normalized Doppler frequency of the channel and $h$ is either $h_{\mathrm{sr}_i}[k]$ or $h_{\mathrm{r}_i\mathrm{d}}[k]$. The maximum Doppler frequency of the $\SRi$ and $\RDi$ channels are shown with $f_{\mathrm{sr}_i}$ and $f_{\mathrm{r}_i\mathrm{d}}$, respectively. Also, it is assumed that the carrier frequency is the same for all links.

Information bits are converted to symbols using a modulation technique such as PSK, QAM at Source.
Depending on the number of relays and the type of constellation, appropriate $R\times R$ unitary matrices $\mathcal{V}=\{\V_l| \V^H_l \V_l=\V_l\V^H_l=\I_R,\; l=1,\cdots,L \}$ are used, where $L$ is the total number of codewords. We refer the reader to \cite{DSTC-Y,D-DSTC-Y,PC-DSTC-Kiran} for more details on selecting these matrices.
The transmission process is divided into two phases and sending a codeword (or matrix) from Source to Destination in two phases is referred to as one transmission block, indexed by $k$. Either codeword-by-codeword or frame-by-frame transmission protocol can be used. However, codeword-by-codeword is not practical, due to frequent switching of transmitter and receiver, and instead a frame of codewords is to be sent in each phase. The analysis is the same for both cases and only the channel auto-correlation value is different. In codeword-by-codeword transmission, the distance between two channel uses is $2R$ symbols apart, whereas in frame-by-frame transmission it is $R$ symbols apart.

Information symbols are encoded into codeword $\V[k]\in \mathcal{V}$. Before transmission, the codeword is differentially encoded as
\begin{equation}
\label{eq:s[k]V[k]}
\s[k]=\V[k] \s[k-1],\quad \s[0]=[1 \quad 0 \quad \cdots \quad 0]^t.
\end{equation}
Obviously, the length of vector $\s[k]$ is $R$.

In phase I, vector $\sqrt{P_0R}\s[k]$ is transmitted by Source to all the relays, where $P_0$ is the average source power per transmission. The fading channels between Source and relays are assumed to be quasi-static during each block but change continuously from block to block. Also, for simplicity of notations and feasibility of analysis, it is assumed that all SR links have the same variance, fade-rate and carrier frequency. Let the coefficient of the $\SRi$ channel, $i=1,\cdots,R$, during the $k$th block be represented by $q_i[k]\sim \mathcal{CN}(0,\sigma_{\sr}^2)$. The auto-correlation between two channel coefficients, which are $n$ blocks apart, follows the Jakes' fading model \cite{microwave-jake}:
\begin{equation}
\label{eq:phi_sr}
\varphi_{\sr}(n)=\E\{q_i[k] q_j^*[k+n]\}=
\left\lbrace
\begin{matrix}
\sigma_{\sr}^2 J_0(2 \pi f_{\sr} n R) & i=j \\
0 & i \neq j
\end{matrix} \right.
\end{equation}
where $J_0(\cdot)$ is the zeroth-order Bessel function of the first kind and $f_{\sr}$ is the maximum Doppler frequency of $\SRi$ channel. Note, the auto-correlation value equals one for static channels and decreases with higher fade-rates.

The received vector at the $i$th relay is
\begin{equation}
\label{eq:rbi}
\rb_i[k]=\sqrt{P_0R}\; q_i[k]\s[k]+\ub_i[k],
\end{equation}
where $\ub_i[k]\sim \mathcal{CN}(\0,N_0\I_R)$ is the noise vector at the $i$th relay.

The received vector at the $i$th relay is linearly combined with its conjugate as
\begin{equation}
\label{eq:ti}
\xb_i[k]=c \left(\A_i \rb_i[k]+\B_i {\rb}_i^*[k]\right),
\end{equation}
where $\A_i$ and $\B_i$ are the combining matrices and determined based on the space-time code that is used for the network. Usually, one matrix is chosen as an unitary matrix and the other is set to zero. Again, we refer the reader to \cite{DSTC-Y,D-DSTC-Y} for more details on determining the combining matrices. Also, $c$ is the amplification factor at the relay that can be either fixed or varying. A variable $c$ needs the instantaneous CSI. For D-DSTC, in the absence of the instantaneous CSI, the variance of $\SRi$ channels is utilized to define a fixed amplification factor as
\begin{equation}
\label{eq:C_i}
c=\sqrt{\frac{P_{\rt}}{P_0\sigma_{\sr}^2+N_0}},
\end{equation}
where $P_{\rt}$ is the average power per symbol of the $i$th relay.
{It was shown in \cite{D-DSTC-Y,DSTC_Ha} that for a given total power in the network, $P$, for symmetric Source-Relay (SR) and Relay-Destination (RD) channels, $P_0={P}/{2}$ and $P_{\rt}={P}/{(2R)}$ form the optimum power allocation between Source and relays to minimize the PEP. Hence, the amplification factor $c=\sqrt{{P}/{R(P+2N_0)}},\quad i=1,\cdots,R$, is chosen for all the relays.}

Then, in phase II, the relays send their data simultaneously to Destination. Again, under the quasi-static fading assumption, the coefficients of $\RDi$ channels during the $k$th block are represented by $g_i[k]\sim \mathcal{CN}(0,\sigma_{\rd}^2)$. The auto-correlation between the channel coefficients, $n$ blocks apart, follows the Jakes' model \cite{microwave-jake}
\begin{equation}
\label{eq:phi_rd}
\varphi_{\rd}(n)=\E \{ g_i[k] g_j^*[k+n] \}=
\left\lbrace
\begin{matrix}
\sigma_{\rd}^2 J_0(2\pi f_{\rd}n R) & i=j \\
0 & i \neq j
\end{matrix}
\right.,
\end{equation}
where $f_{\rd}$ is the maximum normalized Doppler frequency of $\RDi$ channel.

The corresponding received vector at Destination is
\begin{equation}
\label{eq:yb-dstc}
\yb[k]= \sum \limits_{i=1}^{R} g_i[k] \xb_i[k]+\z_i[k],
\end{equation}
where $\z_i[k]\sim \mathcal{CN}(\0,N_0\I_R)$ is the noise vector at Destination. Substituting (\ref{eq:ti}) into (\ref{eq:yb-dstc}) yields \cite{D-DSTC-Y}:
\begin{equation}
\label{eq:yii}
\yb [k]= c \sqrt{P_0 R} \mathbf{S}[k] \h[k]+\w[k],
\end{equation}
where
\begin{equation}
\label{eq:Sk-hk-wk}
\begin{split}
\mathbf{S}[k]&=[\hat{\A}_1 \hat{\s}_1 \; \cdots \; \hat{\A}_R \hat{\s}_R]=\V[k] \mathbf{S}[k-1]\\
\h[k]&=[\; h_1[k]\; \cdots \; h_R[k]\;]^t\\
\w[k]&=c \sum \limits_{i=1}^{R} g_i[k] \hat{\A}_i \hat{\ub}_i[k]+\z_i[k]
\end{split}
\end{equation}
and
\begin{equation*}
\left.
\begin{aligned}
&\hat{\A}_i=\A_i,\; h_i[k]=q_i[k] g_i[k],\\
&\hat{\ub}_i[k]=\ub_i[k],\; \hat{\s}_i[k]=\s[k]
\end{aligned}
\right\}
\quad  \mbox{if}\quad \B_i=\0
\end{equation*}
\begin{equation*}
\left.
\begin{aligned}
&\hat{\A}_i=\B_i,\; h_i[k]={q}_i^*[k] g_i[k],\\
&\hat{\ub}_i[k]={\ub}_i^*[k],\; \hat{\s}_i[k]={\s}^*[k]
\end{aligned}
\right\}
\quad \mbox{if}\quad \A_i=\0
\end{equation*}
are the distributed space-time code, the equivalent cascaded channel vector and the equivalent noise vector, respectively.

It should be noted that for given $\{g_i[k]\}_{i=1}^{R}$, $\w[k]$ is $\mathcal{CN}(\0,\sigma_{\w}^2 \I_R)$ where
\begin{equation}
\label{eq:sig_w}
\sigma_{\w}^2=N_0\left(1+c^2 \sum \limits_{i=1}^{R} |g_i[k]|^2\right).
\end{equation}
It follows that condition on $\Sb[k]$ and $\{g_i[k]\}_{i=1}^{R}$, $\yb[k]$ is $\mathcal{CN}(\0,\sigma_{\w}^2(\rho+1) \I_R)$, where $\rho$ is the average received SNR per symbol, conditioned on $\{g_i[k]\}_{i=1}^R$, defined as
\begin{equation}
\label{eq:rho}
\rho=\frac{ P_0 \sigma_{\sr}^2 c^2  \sum \limits_{i=1}^R |g_i[k]|^2}{N_0\left(1+ c^2 \sum \limits_{i=1}^{R} |g_i[k]|^2\right)}.
\end{equation}

In the following sections, the two-codeword and multiple-codeword differential detections of the received signals at Destination are considered.

\section{Two-Codeword Differential Detection}
\label{dstc:sec:two-CW}

\subsection{Time-Series Model and Differential Detection}
Coherent detection of transmitted codeword is possible only with the knowledge of both SR and RD channels based on the model in \eqref{eq:yii}. In the absence of channel information, in conventional D-DSTC that was considered in references \cite{D-DSTC-Y,D-DSTC-Amin,D-DSTC-Giannakis}, it is assumed that the channels are fixed for two consecutive block uses, i.e.,
\begin{equation}
\label{eq:h_slow}
\h[k]\approx \h[k-1].
\end{equation}
By substituting \eqref{eq:s[k]V[k]} and \eqref{eq:h_slow} into \eqref{eq:yii}, one has
\begin{equation}
\label{eq:yk_k-1_slow}
\yb[k]=\V[k]\yb[k-1]+ \w[k]-\V[k] \w[k-1].
\end{equation}
Given $\yb[k]$ and $\yb[k-1]$, differential non-coherent detection is applied to detect the transmitted codeword as \cite{D-DSTC-Y}
\begin{equation}
\label{eq:diff_detect}
\hat{\V}[k]= \arg \min \limits_{\V[k] \in \mathcal{V}} \|\yb[k]-\V[k] \yb[k-1] \|.
\end{equation}
Comparing \eqref{eq:yii} and \eqref{eq:yk_k-1_slow} reveals that the equivalent noise power is enhanced by a factor of two, which explains why the differential non-coherent detection performs approximately $10\log_{10} 2 \approx 3$ dB worse than the coherent detection in slow-fading channels.

However, slow-fading assumption requires a coherence interval of $3R$ for both SR and RD channels \cite{D-DSTC-Y}, which would be easily violated for fast-fading channels. To determine the performance of D-DSTC detection in time-varying channels, we need to model the cascaded channels with a time-series model.

Based on the first-order auto-regressive AR(1) model of individual Rayleigh channels \cite{AR2-ch}, reference \cite{DAF-ITVT} developed a time-series model to characterize the evolution of the cascaded channels in time. In particular, the first-order time-series model of the $i$th cascaded channel is given as (the reader is referred to \cite{DAF-ITVT} for the details in deriving this model):
\begin{equation}
\label{eq:AR-h-hat}
h_i[k]=\alpha h_i[k-1]+\sqrt{1-\alpha^2} g_i[k-1]e_i[k],
\end{equation}
where $\alpha=\varphi_{\sr}(1)\varphi_{\rd}(1)/(\sigma_{\sr}^2\sigma_{\rd}^2)$ is the equivalent auto-correlation of the cascaded channel and $e_i[k]\sim \mathcal{CN}(0,\sigma_{\sr}^2)$ is independent of $h_i[k-1]$. Please note that the time-series model given in \eqref{eq:AR-h-hat} is used only for the performance analysis and not for generating channel coefficients in the simulation.

The vector form of the cascaded channels is expressed as
\begin{equation}
\label{eq:AR-h-hat-vector}
\h[k]=\alpha \; \h[k-1] + \sqrt{1-\alpha^2} \; \G[k-1] \e[k],
\end{equation}
where $\G[k-1]=\diag \left\lbrace \; g_1 [k-1],\cdots,g_R[k-1]\; \right\rbrace$ and $\e[k]=[\; e_1[k],\cdots,e_R[k]\;]^t \sim \CN(\0,\sigma_{\sr}^2\I_R)$.

By substituting the time-series model \eqref{eq:AR-h-hat-vector} into (\ref{eq:yii}), one has
\begin{equation}
\label{eq:yk-yk-1}
\yb[k]=\alpha \V[k] \yb[k-1]+\widetilde{\w}[k],
\end{equation}
where
\begin{equation}
\label{eq:n}
%\begin{split}
\widetilde{\w}[k]=\w[k]- \alpha \V[k]\w[k-1]
+ \sqrt{1-\alpha^2} c\sqrt{P_0R}\; \mathbf{S}[k] \G[k-1] \e[k].
%\end{split}
\end{equation}
Note that, for given $\G[k-1]$, the equivalent noise $\widetilde{\w}[k]$ is a combination of complex Gaussian random vectors, and hence it is also a complex Gaussian random vector.

As it can be seen, the equivalent noise power is enhanced by an additional factor which is related to the equivalent channel's auto-correlation, the transmitted power and RD channel coefficients. Due to this term, compared to the case of differential detection under slow-fading assumption as in \cite{D-DSTC-Y,D-DSTC-Amin,D-DSTC-Giannakis}, a larger performance loss can be seen between coherent and non-coherent detections in fast-fading channels.
%By computing the average equivalent noise power as
%\begin{equation}
%\label{eq:ave_noise_power}
%\E\{ \widetilde{\w}^H[k] \widetilde{\w}[k] \}= (1+\alpha^2)\E\{\sigma_{\w}^2\}+(1-\alpha^2)\E\{\rho \sigma_{\w}^2\},
%\end{equation}
%the performance loss between coherent and non-coherent detections in time-varying channels can be approximately quantified as
%\begin{equation}
%\label{eq:loss}
%10\log_{10}\left(\frac{1+\alpha^2}{\alpha^2}+\frac{1-\alpha^2}{\alpha^2}\frac{c^2P_0R }{N_0\left(1+c^2R\right)} \right).
%\end{equation}
%It should be noted that the dependence of the equivalent noise power to $\alpha$ and $P_0/N_0$ leads to a dynamic performance loss between coherent and non-coherent detections in time-varying channels. Clearly, by substituting $\alpha \approx 1$ for slow-fading channels into \eqref{eq:loss}, $10 \log_{10} 2 \approx 3$ dB is obtained.

In the next section, the error performance of the system under consideration is analysed.
\subsection{Performance Analysis}
\label{subsec:PEP}
Assume that codeword $\V_i\in \Vc$ is transmitted and it is decoded erroneously to other codeword $\V_j\in \Vc$, by the decoder. The corresponding PEP is defined as
$
\label{eq:PEij-dstc}
P_e(E_{ij})=P_e(\V_i\rightarrow \V_j).
$
An error occurs when
\begin{equation}
\label{eq:PEP-err1}
\|\yb[k]-\V_i \yb[k-1] \|^2> \|\yb[k]-\V_j \yb[k-1] \|^2,
\end{equation}
which can be simplified to
\begin{equation}
\label{eq:pep-err2}
\Re \left\lbrace \yb^H[k-1] (\V_i-\V_j)^H \yb[k]  \right\rbrace < 0.
\end{equation}
By substituting $\yb[k]$ from \eqref{eq:yk-yk-1} into the above inequality, the error event can be further simplified as $\zeta>b$, where (see proof in Appendix \ref{app1})
\begin{equation}
\label{eq:z_gr_b}
\begin{split}
\zeta=& -2 \Re \left\lbrace \yb^H[k-1] \bDelta^H \widetilde{\w}[k] \right\rbrace,\\
b=&  \alpha \delta \yb^H[k-1] \yb[k-1],
\end{split}
\end{equation}
and $\bDelta=\V_i-\V_j$, $\bDelta^H\bDelta=\delta\; \I_R$. Since $\widetilde{\w}[k]$, conditioned on $\G[k-1]$, is a Gaussian random vector, the variable $\zeta$ conditioned on $\yb[k-1]$ and $\G[k-1]$ is a Gaussian random variable. From now on, for simplicity of notation, time index $[k-1]$ is omitted. The conditional mean and variance of $\zeta$ are obtained as (see proof in Appendix \ref{app2})
\begin{equation}
\label{eq:mu-z}
\mu_\zeta= 2 \alpha \sigma_{\w}^2 \Re\{\yb^H \bDelta^H \V_i \Sb(\Omg+\sigma_{\w}^2 \I_R)^{-1} \Sb^H \yb\}
%\mu_\zeta=\frac{\alpha \delta}{\rho+1} \yb^H[k-1] \yb[k-1]
%{\mu}_\zeta \approx 0
\end{equation}
\begin{equation}
\label{eq:sig-z}
\sigma_\zeta^2= 2\yb^H \bDelta^H ( \sigma_{\w}^2\I_R+(1-\alpha^2)\Sb \Omg \Sb^H+
\alpha^2 \sigma_{\w}^2 \V_i \Sb \Omg (\Omg+\sigma_{\w}^2\I_R)^{-1} \Sb^H \V_i^H ) \bDelta \; \yb
\end{equation}
%{\sigma}_\zeta^2\approx 2\left( 1+\alpha^2+(1-\alpha^2)\rho \right) \delta \sigma_{\w}^2 %\yb^H[k-1] \yb[k-1],
where $$\Omg=c^2P_0 \sigma_{\sr}^2 R \G \G^H=(c^2P_0  \sigma_{\sr}^2 R)\diag\{|g_1|^2,\cdots,|g_R|^2\}.$$

Therefore, the conditional PEP can be upper bound as
\begin{equation}
\label{eq:PEPgyh}
P_e(E_{ij}|\yb,\G)
=\text{Pr}(\zeta> b|\yb,\G)
=Q\left( \frac{b-\mu_\zeta}{\sigma_\zeta}\right) = Q\left(\sqrt{\Lambda} \right) \leq \frac{1}{2} \exp \left( - \frac{\Lambda}{2} \right)
\end{equation}
where $Q(x)=\int \limits_x^{\infty} (1/\sqrt{2\pi}) \exp\left({-t^2}/{2}\right)\dd t$. Since the diagonal elements of matrix $\Omg$ depend on RD channel coefficients and are not equal, using the exact expressions of \eqref{eq:mu-z} and \eqref{eq:sig-z} into \eqref{eq:PEPgyh} leads to a complicated fractional expression that cannot be further simplified. Hence, to facilitate the analysis an approximation has to be made. Here, we replace matrix $\Omg$ with $\left(c^2 P_0 \sigma_{\sr}^2 \sum \limits_{i=1}^{R} |g_i|^2\right) \I_R$, which is a diagonal matrix whose entries are the average of diagonal entries of $\Omg$. With this replacement, $\mu_\zeta$ and $\sigma_\zeta^2$ can be approximated as (see proof in Appendix \ref{app3})
\begin{equation}
\label{eq:mu-z-approx}
\mu_\zeta\approx \frac{\alpha \delta}{\rho+1} \yb^H \yb,
\end{equation}
\begin{equation}
\label{eq:sig-z-approx}
\sigma_\zeta^2\approx 2\delta \left( 1+(1-\alpha^2)\rho+\frac{\alpha^2 \rho}{\rho+1} \right)  \sigma_{\w}^2 \yb^H \yb.
\end{equation}
It then follows that
\begin{gather}
\label{eq:Gamma}
\Lambda \approx \frac{\gamma \delta}{2\sigma_{\w}^2 (\rho+1)} \yb^H \yb,
\end{gather}
with
$\gamma$, defined as
\begin{gather}
\label{eq:gamma}
\gamma=\frac{\alpha^2 \rho}{1+\alpha^2+\rho(1-\alpha^2)}.
\end{gather}
The parameter $\gamma$ can be interpreted as the effective SNR at the output of two-codeword differential decoder of the D-DSTC system in time-varying channels. For slow-fading channels, $\alpha=1$ and $\gamma=\rho/2$ which shows that the effective SNR is reduced by half, giving the well-known 3 dB performance loss between coherent and non-coherent detection. For fast-fading channels, $\alpha<1$ and $\gamma<\rho/2$, which leads to a higher degradation in the system performance.

To proceed with computing \eqref{eq:PEPgyh}, the distribution of $\yb$ or equivalently $\Lambda$ is required. Since, conditioned on $\G$, $\yb$ is $\mathcal{CN}(\0,\sigma_{\w}^2(\rho+1)\I_R)$, one has \begin{equation}
\label{eq:pdf_yy}
\frac{2}{\sigma_{\w}^2(\rho+1)} \yb^H\yb \sim \chi_{2R}^2.
\end{equation}
Also, it is known that if $ X \sim \chi_{2R}^2$ then $t X \sim \Gamma(R,2t),$ where $\Gamma(\cdot,\cdot)$ is the gamma distribution. Hence, $\Lambda \sim \Gamma (R,0.5 \gamma \delta),$ i.e.,
\begin{equation}
\label{eq:f_Lambda}
f_{\Lambda}(\lambda)=\frac{\lambda^{R-1}}{(R-1)! (0.5\gamma \delta)^R} \exp\left( \frac{-\lambda}{0.5\gamma \delta}\right).
\end{equation}
Now by taking the average over the distribution of $\Lambda$, the conditional upper bound can be written as
\begin{equation}
\label{eq:Pu_G}
P_e(E_{ij}|\G) \leq
\int \limits_0^{\infty} \frac{1}{2} \et^{-\frac{\lambda}{2}} f_{\Lambda}(\lambda) \dd \lambda=\frac{1}{2} \left( \frac{4}{4+\gamma \delta}\right)^R.
\end{equation}
As it seen, the main parameter to achieve the full diversity is the value of $\gamma,$ which is a function of $\rho$ and $\alpha$. For slow-fading channels, $\gamma\approx \rho/2$ and then in high SNR region, the upper bound decays with $\rho^{R}$ and the full diversity can be achieved. Nevertheless, due to the inherent randomness of $\rho$ caused by RD channels, the performance of D-DSTC is always inferior to that of MIMO systems with the same number of antennas. For fast-fading channels, $\alpha < 1$ and the average value of $\rho (1-\alpha^2)$ in the denominator of $\gamma$ becomes significant and leads to a degradation in the overall performance and the achievable diversity. For large SNR, one has
\begin{multline}
\label{eq:gbi}
%\bar{\gamma}=
\lim \limits_{P_0/N_0\rightarrow \infty} \mbox{E}\{\gamma\}= \mbox{E} \{ \lim \limits_{P_0/N_0 \rightarrow \infty} \gamma \}
=\\ \mbox{E} \left\lbrace
\lim \limits_{P_0/N_0\rightarrow \infty} \right.  \left.
\frac{\alpha^2 c^2 \sigma_{\sr}^2 P_0/N_0 \eta}
{(c^2 \sigma_{\sr}^2 P_0/N_0(1-\alpha^2) +(1+\alpha^2)c^2)\eta+(1+\alpha^2)}
\right\rbrace \\
= \mbox{E} \left\lbrace
\frac{\alpha^2}{1-\alpha^2}
\right\rbrace =\frac{\alpha^2}{1-\alpha^2},
\end{multline}
where $\eta=\sum \limits_{i=1}^{R}|g_i|^2$. It shows that an SNR ceiling beyond which no performance benefit can be achieved by increasing the transmitted power exists. As can be seen, the value of the SNR ceiling is independent of $R$ and only depends on the equivalent auto-correlation. %However, the starting point that the ceiling appears is related to the number of relays.
By substituting \eqref{eq:gbi} into \eqref{eq:Pu_G}, it can be seen that the upper bound on the error floor appears as
\begin{equation}
\label{eq:Pef-dstc}
\lim \limits_{P_0/N_0 \rightarrow \infty}P_{e}(E_{ij})
\leq \frac{1}{2}
\left(  \frac{4(1-\alpha^2)}{4(1-\alpha^2)+\alpha^2 \delta} \right)^R.
\end{equation}
In other words, the obtained diversity, regardless of the number of relays, approaches to zero for fast-fading channels. It is also pointed out that for slow-fading channels ($\alpha \approx 1$), the error floor practically does not exist (close to zero). On the other hand, for completely random channels with no correlation between the coefficients $(\alpha=0)$, the error floor is equal to $0.5$ (as expected).

Now, by substituting $\gamma$ from \eqref{eq:gamma} into \eqref{eq:Pu_G} one has
\begin{equation}
\label{eq:Pe-eta1}
P_{e}(E_{ij}|\eta) \leq \frac{1}{2}
\left(\frac{\beta_2}{\beta_1}\right)^R \left( \frac{\eta+\beta_1}{\eta+\beta_2} \right)^R=
\frac{1}{2}
\left(\frac{\beta_2}{\beta_1} \right)^R \sum \limits_{k=0}^{R} \binom{R}{k} \frac{(\beta_1-\beta_2)^k}{(\eta+\beta_2)^k},
\end{equation}
where the later expression is obtained using power expansion $(1+x)^n=\sum \limits_{k=0}^{n} \binom{n}{k} x^k$ \cite[page 25]{integral-tables}. Also,
\begin{equation}
\label{eq:eta-beta1-beta2}
\begin{split}
\eta &=\sum \limits_{i=1}^{R} |g_i|^2\\
\beta_1 &=\frac{8}{4(1-\alpha^2)c^2 P_0 \sigma_{\sr}^2 /N_0 +8c^2}\\
\beta_2 &= \frac{8}{\alpha^2 c^2 P_0 \sigma_{\sr}^2 /N_0 \delta+4(1-\alpha^2)c^2P_0 \sigma_{\sr}^2/N_0+8c^2}
\end{split}.
\end{equation}

The final step is to take the average over the distribution of $\eta$, which is $f_{\eta}(\eta)=\eta^{R-1} \et^{-\eta}/(R-1)!,\; \eta>0$ \cite{DSTC-Y}. The (unconditioned) upper bound for the PEP is given as
\begin{equation}
\label{eq:Pe}
P_e(E_{ij})\leq \frac{1}{2} \left(\frac{\beta_2}{\beta_1}\right)^R
\sum \limits_{k=0}^{R} \binom{R}{k} \left({\beta_1}-{\beta_2}\right)^k
I_1(\beta_2,k),
\end{equation}
where
\begin{multline}
\label{eq:I}
I_1(\beta_2,k)= \frac{1}{(R-1)!}\int \limits_{0}^{\infty}
\frac{\eta^{R-1}}{(\eta+{\beta_2})^k} \et^{-\eta} \dd \eta= \\
\frac{\et^{\beta_2}}{(R-1)!} \sum \limits_{j=0}^{R-1} \binom{R-1}{j} (-{\beta_2})^{R-1-j}
I_2(j-k,\beta_2),
\end{multline}
and \cite[page 339]{integral-tables}
\begin{equation}
\label{eq:int-eta-exp1}
I_2(j-k,\beta_2)=\int \limits_{\beta_2}^{\infty} \eta^{j-k} \et^{-\eta} \dd \eta
= \et^{-\beta_2} \sum \limits_{m=0}^{j-k} \frac{(j-k)!}{m!} \beta_2^m, \quad \mbox{if} \quad j-k \geq 0,
\end{equation}
\begin{multline}
\label{eq:int-eta-exp2}
I_2(j-k,\beta_2)=\int \limits_{\beta_2}^{\infty} \frac{1}{\eta^{k-j}} \et^{-\eta} \dd \eta=\\
(-1)^{k-j-1} \frac{E_1(\beta_2)}{(k-j-1)!}+\frac{\et^{-\beta_2}}{\beta_2^{k-j-1}}
\sum \limits_{m=0}^{k-j-2}
\frac{(-1)^m \beta_2^m (k-j-2-m)!}{(k-j-1)!}, \quad \mbox{if} \; \; j-k<0,
\end{multline}
and $E_1(x)=\int \limits_x^{\infty} \left(\et^{-t}/t\right) \dd t$ is the exponential integral function.

The expression in \eqref{eq:Pe} provides an upper bound for the PEP of the two-codeword differential detection of D-DSTC system in time-varying Rayleigh fading channels. The total block-error-rate (BLER) can be obtained using the union bound.

%Moreover, the obtained upper bound on the PEP can be used to determine an upper bound on the BER for a network with two relays using the differential space-time encoding of Tarokh-Jafakhani \cite{TJ_DSTC} which is actually based on the well-known Alamouti scheme \cite{STC-Alam}. As it was shown in \cite{dstc-vitetta}, for such a system using BPSK constellation, the PEP of the two nearest codewords
%$$\V_0=\frac{1}{\sqrt{2}}
%\begin{bmatrix}
%1 & -1 \\
%1 & 1
%\end{bmatrix},
%\V_2=\frac{1}{\sqrt{2}}
%\begin{bmatrix}
%1 & 1 \\
%-1 & 1
%\end{bmatrix},
%$$
%can be used to find the BER of the system as \cite[eq.40]{dstc-vitetta}
%\begin{equation}
%\label{eq:BER}
%P_b(E) =P_e\{\V_0 \rightarrow \V_2\}.
%\end{equation}
%The eigenvalue of matrix $\bDelta^H \bDelta$ in this case is $\delta=2$.
%In addition, as it was shown in \cite[eq. (13) and eq. (16)]{ustc-ber}, the BER of the differential Alamouti space-time code using BPSK and QPSK is obtained from the same expression but with a different parameter ($\delta=2$ for BPSK and $\delta=1$ for QPSK). %Therefore, the obtained PEP expression with $\delta=2$ and $\delta=1$ can be used to find the upper bound BER of the D-DSTC system with two relays and Alamouti scheme using BPSK and QPSK modulations, respectively, in time-varying channels. This will be verified in Section \ref{sec:sim} using computer simulations.

\section{Multiple-Codeword Differential Detection}
\label{dstc:sec:MSDSD}
As discussed in the previous section, two-codeword differential detection suffers from a large performance degradation in fast-fading channels. To overcome such a limitation, this section develops a multiple-codeword differential detection scheme that takes a window of the received symbols at the destination for detecting the transmitted signals.

Rewrite \eqref{eq:yii} as
\begin{equation}
\label{eq:yii_recall}
\yb [k]= c \sqrt{P_0 R} \mathbf{S}[k] \h[k]+\w[k]
= c \sqrt{P_0 R} \Sb[k] \G[k] \q[k]+\w[k]
\end{equation}
with
\begin{gather*}
\G[k]=\diag\{g_1[k],\cdots,g_R[k]\}\\
\q[k]=[\; q_1[k],\cdots,q_R[k]\;]^t.
\end{gather*}

Let the $N$ received symbols be collected in vector
\begin{equation}
\label{eq:ybar}
\ovy=\left[\; \yb^t[1],\yb^t[2],\dots, \yb^t[N]\; \right]^t,
\end{equation}
which can be written as
\begin{equation}
\label{eq:ovy}
\ovy=c\sqrt{P_0R} \; \ovS \; \ovh +\ovw
=c\sqrt{P_0R}\; \ovS \; \ovG \ovq +\ovw
\end{equation}
where
$$\ovS= \diag \left\lbrace\; \Sb[1],\cdots, \Sb[N]\; \right\rbrace,$$
$$\ovh=\left[\; \h^t[1],\cdots, \h^t[N]\; \right]^t,$$
$$\ovG=\diag\left\lbrace \; \G[1],\cdots, \G[N] \;\right\rbrace,$$
$$\ovq=\left[\; \q^t[1],\cdots, \q^t[N]\; \right]^t,$$
$$\ovw=\left[\; \w^t[1],\cdots, \w^t[N]\; \right]^t.$$
It should be mentioned that $\ovS$ is a unitary block diagonal matrix ($\ovS^H\ovS=\ovS\;\ovS^H=\I_{RN}$) and it contains $N$ transmitted codewords corresponding to $N-1$ data codewords collected in $\ovV=\diag\{\V[1],\cdots,\V[N-1]\}$ such that
\begin{equation}
\label{eq:SVS}
\Sb[n+1]=\V[n]\Sb[n],\quad n=1,\cdots,N-1
\end{equation}
and $\Sb[N]=\I_R$ is set as the reference symbol.

Therefore, conditioned on both $\ovV$ (or $\ovS$) and $\overline{\G}$, $\ovy$ is a circularly symmetric complex Gaussian vector with the following pdf:
\begin{equation}
\label{eq:Povy|VG}
P(\ovy|\ovV,\ovG)=\frac{1}{\pi^N \mathrm{det}\{\Sig_{\ovy}\}} \exp\left( -\ovy^H \Sig_{\ovy}^{-1} \ovy \right).
\end{equation}
In \eqref{eq:Povy|VG}, matrix $\Sig_{\ovy}$ is the conditional covariance matrix of $\ovy$, defined as
\begin{equation}
\label{eq:SigYb}
%\begin{split}
\Sig_{\ovy}=\E \left\lbrace \ovy\; \ovy^H | \ovV,\ovG \right\rbrace=
c^2 P_0R \ovS \; \ovG \Sig_{\ovq} \ovG^H \ovS^H +\Sgwb
%\end{split}
\end{equation}
with $\Sig_{\ovq}$ and $\Sig_{\ovw}$ as the covariance matrices of $\ovq$ and $\ovw$, respectively. They are given as follows (see proof in Appendix D):
\begin{equation}
\label{eq:Sigma_qbar}
\Sig_{\ovq}=\E\{ \ovq \;\ovq^H \}=\C_{\ovq}\otimes \I_R,
\end{equation}
$$ \C_{\ovq}=\mathrm{toeplitz}\{ \varphi_{\sr}(0),\varphi_{\sr}(1),\dots,\varphi_{\sr}(N-1)\} $$
and
\begin{equation}
\label{eq:Sigma_wbar1}
\Sgwb=\E\{\ovw\; \ovw^H \}=\C_{\ovw} \otimes \I_R
\end{equation}
\begin{equation}
\label{eq:C_wb}
\C_{\ovw}=N_0
 \diag\left\lbrace \left(1+c^2\sum\limits_{i=1}^{R}|g_i[1]|^2\right),\cdots \right.  \left. ,
 \left(1+c^2\sum\limits_{i=1}^{R}|g_i[N]|^2\right) \right\rbrace.
\end{equation}

Based on \eqref{eq:Povy|VG}, the maximum likelihood (ML) detection of $N$ transmitted codewords collected in $\ovS$ or the corresponding $N-1$ data codewords collected in $\ovV$ would be given as
\begin{equation}
\label{eq:ovVh}
\widehat{\ovV}=\arg \max \limits_{\ovV \in \Vcb^{N-1}} \left\lbrace \underset{\ovG}{\E} \left\lbrace
\frac{1}{\pi^N \mathrm{det}\{\Sig_{\ovy}\}} \exp\left( -\ovy^H \Sig_{\ovy}^{-1} \ovy \right)
\right\rbrace \right\rbrace,
\end{equation}
where $\widehat{\ovV}= \diag \left\lbrace\; \widehat{\V}[1],\cdots, \widehat{\V}[N] \; \right\rbrace$.
As it can be seen, the ML metric needs the expectation over the distribution of $\ovG$, which does not yield a closed-form expression. As an alternative, it is proposed to use the following modified decision metric:
\begin{equation}
\label{eq:ovV-Modified}
\widehat{\ovV}=\arg \max \limits_{\ovV \in \Vcb^{N-1}} \left\lbrace
\frac{1}{\pi^N \mathrm{det}\{\widehat{\Sig}_{\ovy}\}} \exp\left( -\ovy^H \widehat{\Sig}_{\ovy}^{-1} \ovy \right)
\right\rbrace
\end{equation}
where
\begin{equation}
%\label{eq:Ry_bar}
\widehat{\Sig}_{\ovy}=\underset{\ovG}{\E} \{ \Sig_{\ovy} \}=\\ c^2P_0 R \ovS  (\C_{\ovh}\otimes \I_R) \ovS^H +(1+c^2 \sigma_{\rd}^2 R)N_0 (\I_N \otimes \I_R)
=\ovS \; (\C \otimes \I_R) \; \ovS^H
\end{equation}
with
\begin{equation}
\label{eq:Cb}
\C=c^2 P_0 R  \C_{\ovh} +N_0(1+c^2 \sigma_{\rd}^2 R)\I_N
\end{equation}
%\begin{equation}
%\label{eq:R_h}
%\Rb_{\ovG\ovq}=\E\left\lbrace \ovG \C_{\ovq} \ovG^H \right\rbrace=\Rb_{\h}\otimes \I_R
%\end{equation}
\begin{equation}
\label{eq:Sig_h}
\C_{\ovh}=\mathrm{toeplitz} \{ \varphi_{\sr}(0)\varphi_{\rd}(0),\dots,\varphi_{\sr}(N-1)\varphi_{\rd}(N-1) \}.
\end{equation}
Although the alternative decision metric is not optimal in the ML sense, it will be shown by simulation results that nearly identical performance to that obtained with the optimal metric can be achieved.

Using the rule $\det \{\A\B\}=\det\{\B\A\}$, the determinant in \eqref{eq:ovV-Modified} is no longer dependent to $\ovS$ and the modified decision metric can be further simplified as
\begin{multline}
\label{eq:ovVh2}
\widehat{\ovV}=\arg \min \limits_{\ovV \in \Vcb^{N-1}} \left\lbrace \ovy^H \widehat{\Sig}_{\ovy}^{-1} \ovy \right\rbrace
=\arg \min \limits_{\ovV \in \Vcb^{N-1}} \{\ovy^H \ovS (\C^{-1}\otimes \I_R) \ovS^H \ovy \}\\
=\arg \min \limits_{\ovV \in \Vcb^{N-1}} \{\ovy^H \ovS (\U^H \otimes \I_R) (\U \otimes \I_R) \ovS^H \ovy \}
= \arg \min \limits_{\ovV \in \Vcb^{N-1}} \left\lbrace \norm{\bb}^2 \right\rbrace
\end{multline}
where $\U$ is an upper triangular matrix obtained by the Cholesky decomposition of $\C^{-1}=\U^H \U$ and
\begin{equation}
\label{eq:bb}
\bb=(\U \otimes \I_R) \ovS^H \ovy=
\begin{bmatrix}
\sum \limits_{j=1}^{N} u_{1,j} \Sb^H[j] \yb[j]\\
\sum \limits_{j=2}^{N} u_{2,j} \Sb^H[j] \yb[j]\\
\vdots \\
u_{N,N} \Sb^H[N] \yb[N]\\
\end{bmatrix}
\end{equation}
and $u_{i,j}$ is the element of $\U$ in row $i$ and column $j$.

Since $\Sb[N]=\I_R$, the last term of vector $\bb$ does not have any effect on the minimization and it can be ignored. Then by substituting $\Sb^H[n]=\Sb^H[n+1]\V[n]$  (obtained from \eqref{eq:SVS}) into \eqref{eq:bb}, it follows that
\begin{equation}
\label{eq:ovVh22}
\widehat{\ovV}=\arg \min \limits_{\ovV \in \Vcb^{N-1}} \left\lbrace \sum \limits_{n=1}^{N-1} \| u_{n,n} \V[n] \yb[n]  \right.   \left. +\Sb[n+1] \sum \limits_{j=n+1}^{N} u_{n,j} \Sb^H[j] \yb[j] \| ^2 \right\rbrace.
\end{equation}
The simplified alternative minimization in \eqref{eq:ovVh2} is a sum of $N-1$ non-negative scalar terms and similar to the decision metric of multiple-codeword detection of unitary space-time coding for MIMO systems given in \cite[eq.5]{MSDUSTC-P}. Therefore, this minimization can be solved using the sphere decoding algorithm described in \cite{MSDUSTC-P} to obtain $N-1$ data codewords with low complexity. The multiple-codeword differential sphere-decoding (MCDSD) algorithm adapted to the D-DSTC system under consideration is summarized in \emph{Algorithm I}. It should be mentioned that Steps 1 to 3 are performed once, whereas Step 4 will be repeated for every $N$ consecutive received symbols. Also, the processed blocks overlap by one vector symbol, i.e., the observation window of length $N$ moves forward by $N-1$ symbols at a time.

\begin{table}[thb!]
\renewcommand{\arraystretch}{2}
\label{tb:msdsd-dstc}
\centering
\begin{tabular}{l}
\hline
\bf Algorithm 1: MCDSD-DSTC \\
\hline
{\bf Input:} $f_{\sr}, f_{\rd}, c, P_0,N_0,N,R, \ovy$ \\
{\bf Output:} $\widehat{\V}[k],\quad k=1,\cdots,N-1$ \\
\hline
1: Find $\C_{\ovh}$ from \eqref{eq:Sig_h} \\
2: Find $\C$ from \eqref{eq:Cb} \\
3: Find $\U$ from $\C^{-1}=\U^H\U$  \\
4: Apply sphere decoding algorithm $\widehat{\ovV}$=MCDSD ($\U$,$\ovy$) \cite{MSDUSTC-P}\\
\hline
\end{tabular}
\end{table}

\section{Numerical Results}
\label{dstc:sec:sim}
To support our analysis and development, in this section a relay network with one source, $R = 2$ relays and one destination is simulated in different fading scenarios while both two-codeword and multiple-codeword detection schemes are applied. For the case of two-codeword detection, the simulation results are verified with the obtained upper bound. Effectiveness of multiple-codeword differential detection is shown when a window of length $N=10$ is processed at a time.

The Alamouti space-time code is chosen for the network.
The combining matrices at the relays are designed as \cite{D-DSTC-Y}
$$
\A_1=
\begin{bmatrix}
1 & 0 \\
0 & 1
\end{bmatrix},\;
\B_1=\0,\;
\A_2=\0,\;
\B_2=
\begin{bmatrix}
0 & -1 \\
1 & 0
\end{bmatrix}.
$$
Also, the set of unitary codewords are designed as \cite{D-DSTC-Y}
\begin{equation}
\label{eq:unitary-alamouti}
\mathcal{U}=\left\lbrace \frac{1}{\sqrt{2}}
\begin{bmatrix}
u_1 & -u_2^* \\
u_2 & u_1^*
\end{bmatrix} | u_i \in \mbox{PSK}, \; i=1,2.
\right\rbrace.
\end{equation}
The power allocation between Source and the relays is such that $P_0={P}/{2}$ and $P_{\rt}={P}/{4}$, where $P$ is the total power in the network. Also, $\sigma_{\sr}^2=1,\sigma_{\rd}^2=1,N_0=1.$

Please note that the time-series model given in \eqref{eq:AR-h-hat} has been utilized only for \emph{performance analysis} and not for generating the channel coefficients in simulations. In all simulations, channel coefficients $\{q_i[k]\}_{i=1}^{R}$ and $\{g_i[k]\}_{i=1}^{R}$ are generated independently according to the simulation algorithm of \cite{ch-sim}. This simulation algorithm utilizes a sum-of-sinusoids method to generate time-correlated Rayleigh-faded channel coefficients. For instance, to generate $q_i[k]$:
\begin{gather}
\label{eq:qi_sim}
\nonumber
q_i[k]=\Re\{q_i[k]\}+j \Im\{q_i[k]\}, \quad i=1,\cdots,R \\
\Re\{q_i[k]\}=\sqrt{\frac{2}{M}} \sum \limits_{m=1}^{M} \cos(2\pi f_{\sr} k \cos(a_m)+\phi_m) \\
\Im\{q_i[k]\}=\sqrt{\frac{2}{M}} \sum \limits_{m=1}^{M} \cos(2\pi f_{\sr} k \sin(a_m)+\psi_m)\\ \nonumber
a_m=\frac{2\pi m-\pi +\theta}{4M},\quad m=1,2,\cdots,M
\end{gather}
where $\phi_m,\psi_m,$ and $\theta$ are statistically independent and uniformly distributed on $[\pi,\pi)$ for all $m$ and $M=8$ is the number of multipaths chosen arbitrarily large enough for an accurate model \cite{ch-sim}. Similarly $g_i[k]$ channel coefficients are generated for $i=1,\cdots,R$, except that $f_{\sr}$ is replaced with $f_{\rd}$. The input to the simulation algorithm is the normalized Doppler frequency of the channels, which is a function of the velocity of users. A higher velocity causes a higher fade-rate and thus less correlation between channel coefficients. Therefore, by changing the Doppler values, various fading scenarios from slow-fading to fast-fading channels can be simulated.

\begin{figure}[tb]
\psfrag {fsr or frd} [l] [] [.8] {$f_{\sr}$ or $f_{\rd}$ changes}
\psfrag {fsr and frd} [] [] [.8] {\qquad \qquad \qquad \quad both $f_{\sr}$ and $f_{\rd}$ change}
\psfrag {fade} [t] [] [1] {normalized fade rate}
\psfrag {BER} [] [] [1] {Error Floor}
\centerline{\epsfig{figure={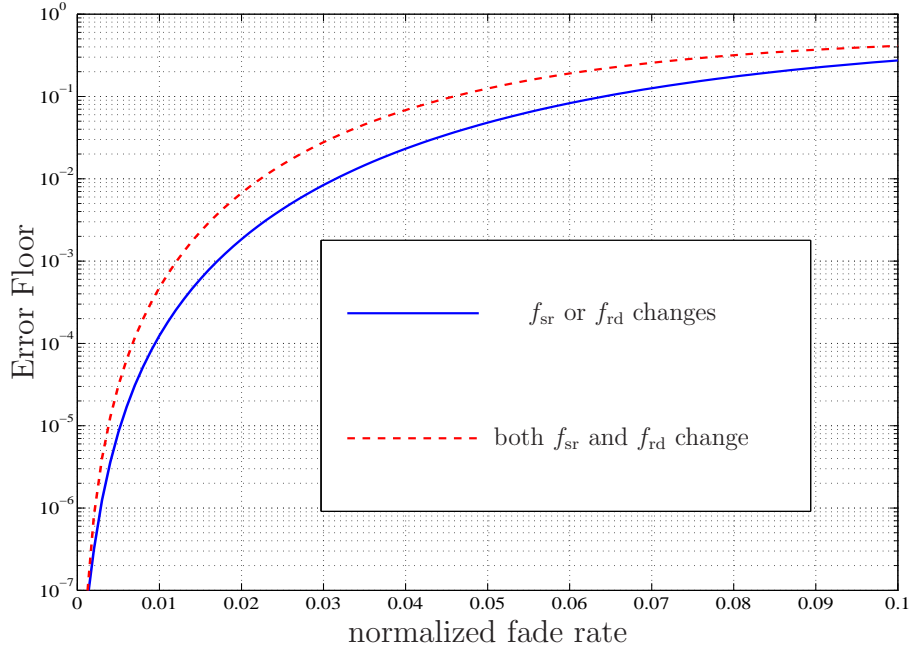},height=8.5cm,width=12cm}}
\caption{ Error floor vs. normalized channel fade rates using Alamouti space-time code and BPSK.}
\label{fig:efvsfd}
\end{figure}

To get a better understanding about the Doppler values, the error floor expression given in \eqref{eq:Pef-dstc} is examined for a wide range of fading values from $0.001$ to $0.1$ and the computed results are plotted in Figure~\ref{fig:efvsfd}. In the figure, the lower plot corresponds to the case that either $f_{\sr}$ or $f_{\rd}$ varies in a wide range and the other one fixed to $0.001$. In the upper plot both $f_{\sr}$ and $f_{\rd}$ vary in a wide range. Clearly, the error floor is higher when both SR and RD channels are changing. For small fade rates less than 0.005, the amount of error floor is small and channels would be regarded as slow-fading. For moderate fade rates around 0.01 the error floor increases quickly toward $10^{-3}$ and channels would be fairly fast-fading. Fade rates around 0.02 and higher lead to fast-fading channels with error floor of $10^{-2}$ or higher.

Based on the previous observations, next, three scenarios are considered. In Scenario I, it is assumed that all the channels are slow-fading with $f_{\sr}=0.002,f_{\rd}=0.002$. In Case II, both SR and RD channels are moderately fast-fading such that $f_{\sr}=0.012,\;f_{\rd}=0.008$. In Case III, both SR and RD channels are very fast-fading with $f_{\sr}=0.018,\;f_{\rd}=0.02$. The normalized Doppler frequencies of SR and RD channels are summarized in Table \ref{tb:scs} for different scenarios.
\begin{table}[!ht]
\begin{center}
\caption{Three simulation scenarios.}
\vspace*{.1in}
\label{tb:scs}
  \begin{tabular}{ |c | c| c| c | }
    \hline
				& $f_{\sr}$ & $f_{\rd}$  \\ \hline\hline
{Case I}    & 0.002              & 0.002  			 \\ \hline
{Case II}   & 0.012 			    & 0.008   			  \\ \hline
{Case III}  & 0.018 			    & 0.02   			  \\
    \hline
  \end{tabular}
\end{center}
\end{table}
%\vspace*{-0.5cm}

\begin{figure}[tb]
\psfrag {P(dB)} [t][] [1]{$P/N_0$ (dB)}
\psfrag {BER} [] [] [1] {BER}
\centerline{\epsfig{figure={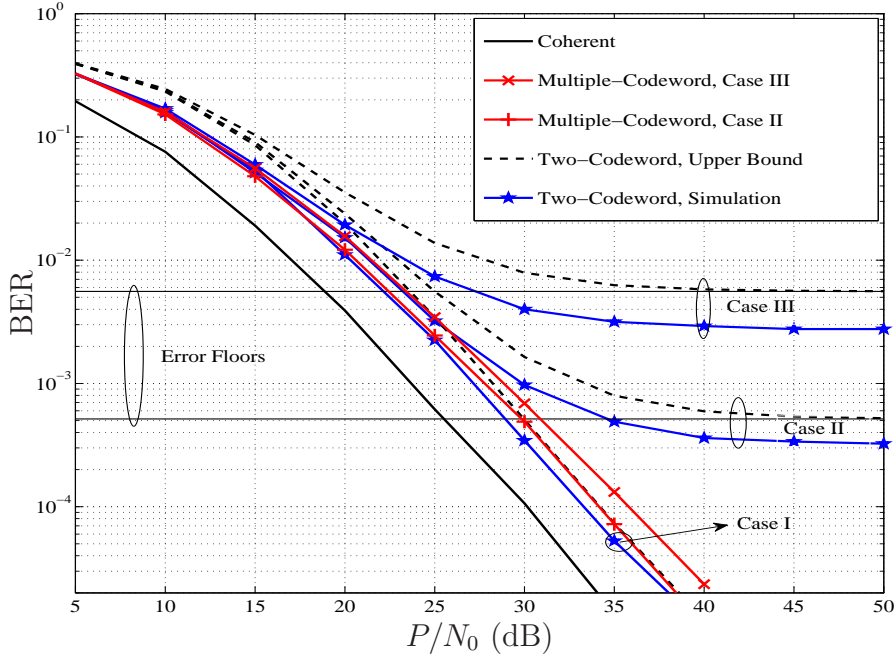},height=8.5cm,width=12cm}}
\caption{ BER results of two-codeword and multiple-codeword differential detection of D-DSTC relaying with two relays in different cases using Alamouti code and BPSK.}
\label{fig:ber_m2_all}
\end{figure}

\begin{figure}[tb]
\psfrag {P(dB)} [t][] [1]{$P/N_0$ (dB)}
\psfrag {BER} [] [] [1] {BER}
\centerline{\epsfig{figure={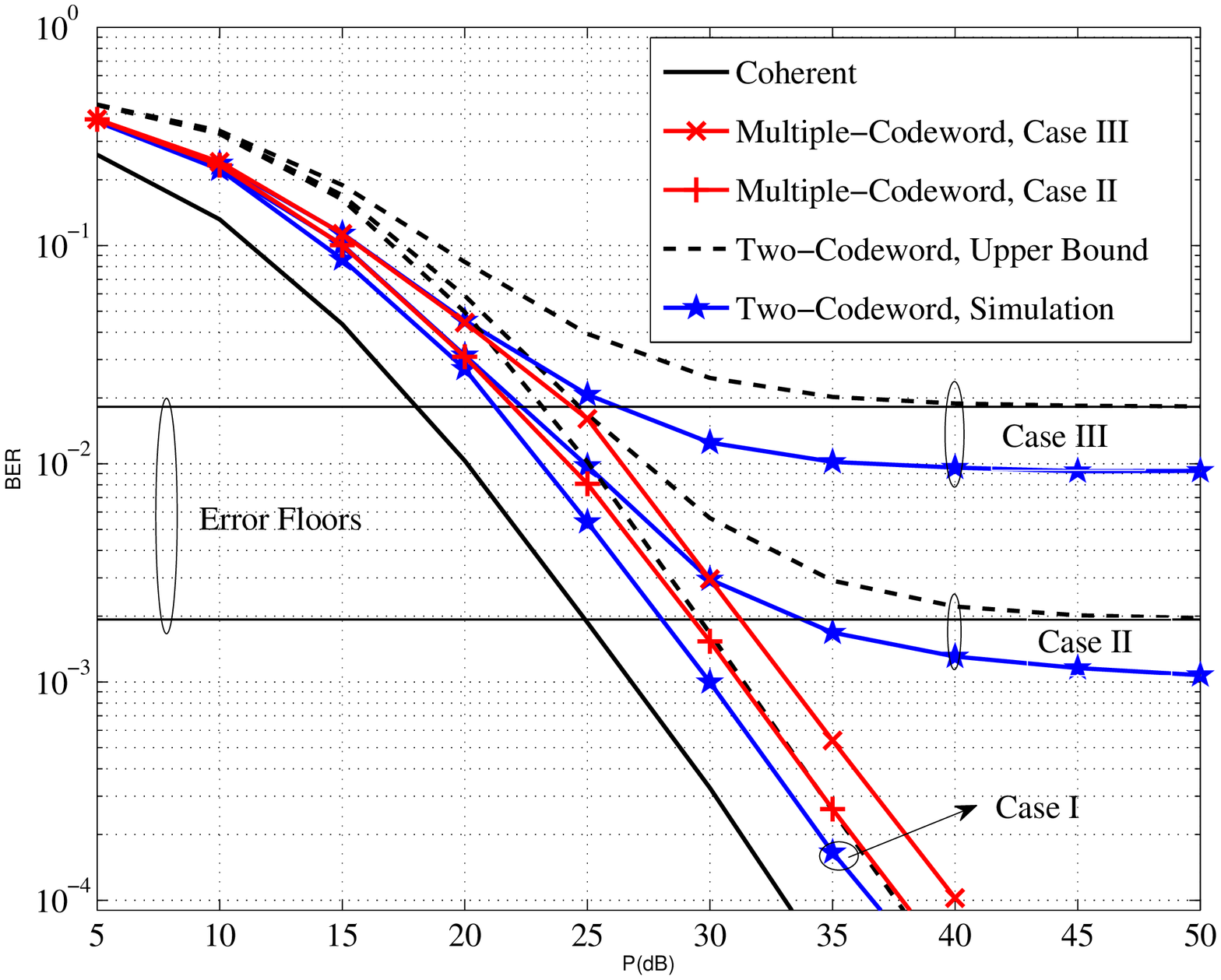},height=8.5cm,width=12cm}}
\caption{ BER results of two-codeword and multiple-codeword differential detection of D-DSTC relaying with two relays in different cases using Alamouti code and QPSK.}
\label{fig:ber_m4_all}
\end{figure}

To evaluate the BER of the system, in each case, binary data is converted to BPSK/QPSK constellation and then to unitary codewords based on \eqref{eq:unitary-alamouti}. Next, the codewords are encoded differentially according to \eqref{eq:s[k]V[k]}. At Destination, first the two-codeword differential detection \eqref{eq:diff_detect} is applied. The simulation is run for various values of the total power in the network. The practical values of the BER are computed for all cases and plotted versus $P/N_0$ in Figs.~\ref{fig:ber_m2_all}-\ref{fig:ber_m4_all}. For comparison purpose, performance of coherent detection of the received symbols for slow-fading channels is also evaluated and plotted in the figures.

On the other hand, the obtained upper bound on the PEP can be used to determine an upper bound on the BER for a network with two relays using Alamouti space-time code \cite{STC-Alam}. As it was shown in \cite{dstc-vitetta}, for such a system using BPSK constellation, the PEP of the two nearest codewords
$$\V_0=\frac{1}{\sqrt{2}}
\begin{bmatrix}
1 & -1 \\
1 & 1
\end{bmatrix},
\V_2=\frac{1}{\sqrt{2}}
\begin{bmatrix}
1 & 1 \\
-1 & 1
\end{bmatrix},
$$
can be used to find the BER of the system as \cite[eq.40]{dstc-vitetta}
\begin{equation}
\label{eq:Pb-dstc}
P_b(E) =P_e\{\V_0 \rightarrow \V_2\}.
\end{equation}
The eigenvalue of matrix $\bDelta^H \bDelta$ in this case is $\delta=2$.
In addition, as it was shown in \cite[eq. (13) and eq. (16)]{ustc-ber}, the BER of the differential Alamouti space-time code using BPSK and QPSK is obtained from the same expression but with a different parameter ($\delta=2$ for BPSK and $\delta=1$ for QPSK). Hence, the corresponding upper bound for the BER are computed based on \eqref{eq:Pb-dstc} and \eqref{eq:Pe} and plotted in Figs.~\ref{fig:ber_m2_all}-\ref{fig:ber_m4_all} with dashed lines. The horizontal lines show the error floors in Cases II and III.

It can be seen from Figs.~\ref{fig:ber_m2_all}-\ref{fig:ber_m4_all} that,
as expected, coherent detection gives the best performance, which is obtained at the price of providing the instantaneous CSI of all transmission links at the destination. For two-codeword differential detection, in Case I (slow-fading channels), the error probability is monotonically decreasing with $P/N_0$ and the desired cooperative diversity is achieved for the D-DSTC system. Approximately 3-4 dB performance degradation can be seen between coherent and non-coherent detections in this case. However, in Case II, with fairly fast-fading channels, the plot gradually deviates from the results in Case I, at $P/N_0>20$ dB, and reaches the error floor at $P/N_0>35$ dB. This phenomena starts earlier after $P/N_0>15$ dB in Case III (fast-fading channels). The performance degradation is much more severe and error floors of $3 \times 10^{-3}$ (BPSK) and $10^{-2}$ (QPSK) can be seen at $P/N_0>30$ dB.  It can also be seen that, in all cases, the upper bound values are consistent with the simulation results.

Given the poor performance of the two-codeword differential detection in cases II and III, the multiple-codeword algorithm with $N=10$ is applied to Case II and Case III. %Because of the orthogonality of Alamouti space-time code, the decision rule \eqref{eq:ML-simp2} can be divided into two separate decision rules, one for each $u_i, i=1,2$ symbol, and solved separately with low complexity using multiple-symbol differential detection algorithm of point-to-point systems given in \cite{MSDSD-L}.
The BER results of multiple-codeword detection algorithm are also plotted in Figures~\ref{fig:ber_m2_all} and \ref{fig:ber_m4_all}. Since the best performance is achieved in the slow-fading environment, the BER plot of Case I using two-codeword detection can be used as a benchmark to see the effectiveness of multiple-codeword detection algorithm. As can be seen, the MCDSD-DSTC algorithm is able to bring the performance of the system in Case II and Case III very close to that of Case I.

\section{Conclusion}
\label{dstc:sec:con}
This article aimed to provide an insight into the effect of mobility of users on the performance of differential distributed  space-time coding in cooperative relay networks. Following this goal, the conventional two-codeword differential detection, its achievable diversity and approximate performance are evaluated against channels variations. The observations revealed that two-codeword differential detection suffers from a huge performance loss in time-varying channels. Next, a near optimal multiple-codeword differential detection was developed and its effectiveness was shown through simulation results. Obviously, the improvement gained by the multiple-codeword detection is at the price of a higher complexity in the detection process. However, with the available sphere decoding algorithms, this complexity would be significantly reduced. The theoretical analysis and supporting simulations of this article would bring valuable design criteria for system engineers to design more robust systems against channel variation and users mobility. Future work can focus on theoretical performance evaluation of multiple-codeword detection and providing a trade-off between the number of required symbols and the desired performance.

\begin{subappendices}

\section{Proof of \eqref{eq:z_gr_b}}
\label{app1}
%\begin{proof}[Proof of \eqref{eq:z_gr_b}]

Substituting \eqref{eq:yk-yk-1} into \eqref{eq:pep-err2} gives
\begin{multline}
2\Re \{\yb^H[k-1] (\V_i-\V_j)^H (\alpha \V_i \yb[k-1]+\widetilde{\w}[k])\}=\\
{2\Re \{\alpha \yb^H[k-1] (\I_R-\V_j^H \V_i)\yb[k-1]\}} +
2\Re \{\yb^H[k-1](\V_i-\V_j)^H \widetilde{\w}[k]\} <0.
\end{multline}
The second part of the left hand side of the above inequality is the definition of $\zeta$. Using $x+x^*=2\Re\{x\}$, the first part of the inequality can be re-written as
\begin{multline}
\label{eq:proof-a}
2\Re \{\alpha \yb^H[k-1] (\I_R-\V_j^H \V_i)\yb[k-1]\}=\\
\alpha \yb^H[k-1] (2\I_R-\V_j^H \V_i-\V_i^H\V_j)\yb[k-1]\}=\\
 \alpha \yb^H[k-1] (\V_i-\V_j)^H (\V_i-\V_j) \yb[k-1]\}=\\
{\alpha \yb^H[k-1] \bDelta^H \bDelta \; \yb[k-1]}.
\end{multline}
Using the fact that the eigenvalues of unitary matrices are equal, one has $\bDelta^H \bDelta=\delta \;\I_R$. Thus \eqref{eq:proof-a} is the same as the definition of $b$.
%\end{proof}

\section{Proof of \eqref{eq:mu-z} and \eqref{eq:sig-z}}
\label{app2}
%\begin{proof}[Proof of \eqref{eq:mu-z}]

For notational simplicity, time index $[k-1]$ is omitted.
\begin{equation}
\label{eq:prf mu_z}
\mu_\zeta=\E\{\zeta|\yb,\G\}=
-2 \Re \left\lbrace
\yb^H \bDelta^H \E\{\widetilde{\w}[k]|\yb,\G \}
\right\rbrace
= 2\alpha \Re \left\lbrace
\yb^H \bDelta^H \V_i \mbox{E}\{\w|\yb,\G\}
\right\rbrace.
\end{equation}
To find the conditional mean of $\w$, first re-write \eqref{eq:yb-dstc} as
\begin{equation}
\yb=c\sqrt{P_0R}\Sb\h+\w=c\sqrt{P_0R}\Sb\G\q +\w
\end{equation}
where $\q=[q_1,\cdots,q_R]^t$. Hence, using the conditional mean technique \cite{prob-papo}
\begin{multline}
\label{eq:mu_w_yg}
\E\{\w|\yb,\G\}=\sigma_{\w}^2(c^2P_0\sigma_{\sr}^2 R\Sb \G \G^H \Sb^H+\sigma_{\w}^2\I)^{-1} \yb=\\
\sigma_{\w}^2 \Sb (c^2P_0 \sigma_{\sr}^2 R \G \G^H +\sigma_{\w}^2\I)^{-1} \Sb^H \yb
\end{multline}
By substituting \eqref{eq:mu_w_yg} into \eqref{eq:prf mu_z}, the expression in \eqref{eq:mu-z} is obtained.
%\end{proof}

%\begin{proof}[Proof of \eqref{eq:sig-z}]
\begin{multline}
\label{eq:prf var_z}
\sigma^2_{\zeta}=\Var \{\zeta|\yb,\G\}= 2 \yb^H \bDelta^H \mbox{Var}\{\widetilde{\w}[k]|\yb,\G\} \bDelta \yb=\\
2\yb^H \bDelta^H (\sigma_{\w}^2\I_R+(1-\alpha^2)  \Sb (c^2P_0\sigma_{\sr}^2 R\G\G^H) \Sb^H +\alpha^2 \V_i
\Var \{\w|\yb,\G\}\V_i^H) \bDelta \; \yb.
%2\yb^H[k-1]\bDelta^H \left(\sigma_{\w}^2\I_R+(1-\alpha^2)\rho \sigma_{\w}^2\I_R+ \right. \\ \left.
%\alpha^2 \frac{\sigma_{\w}^2\rho}{\rho+1} \I_R \right) \bDelta \yb[k-1]=\\
%2 \left(1+(1-\alpha^2)\rho +\frac{\alpha^2 \rho}{\rho+1} \right) \sigma_{\w}^2 \yb^H[k-1]\bDelta^H \bDelta \yb[k-1]
\end{multline}
Using the conditional variance technique \cite{prob-papo}, one has
\begin{multline}
\label{eq:Var_w_yg}
\Var\{\w|\yb,\G \}=
\sigma_{\w}^2\left(c^2P_0 \sigma_{\sr}^2 R \Sb\G\G^H\Sb^H\right)
\left(c^2 P_0 \sigma_{\sr}^2 R \Sb\G\G^H \Sb^H+\sigma_{\w}^2\I\right)^{-1} \\=
\sigma_{\w}^2\Sb (c^2P_0 \sigma_{\sr}^2 R\G\G^H) \left(c^2 P_0 \sigma_{\sr}^2 R \G\G^H+\sigma_{\w}^2 \right)^{-1} \Sb^H.
\end{multline}
Substituting \eqref{eq:Var_w_yg} into \eqref{eq:prf var_z} gives the expression in \eqref{eq:sig-z}.
%\end{proof}

\section{Proof of \eqref{eq:mu-z-approx} and \eqref{eq:sig-z-approx}}
\label{app3}
%\begin{proof}[Proof of \eqref{eq:mu-z-approx}]

Re-write \eqref{eq:mu-z} as
\begin{equation}
\label{eq:proof_mean-z_approx}
\mu_{\zeta}= 2 \alpha \sigma_{\w}^2 \Re\{\yb^H \bDelta^H \V_i \Sb(\Omg+\sigma_{\w}^2 \I_R)^{-1} \Sb^H \yb\}
\end{equation}
By replacing $\Omg$ with $c^2P_0 \sigma_{\sr}^2 \sum \limits_{i=1}^{R} |g_i|^2$, one has
\begin{multline}
\label{eq:SXS}
\Sb(\Omg+\sigma_{\w}^2\I_R)^{-1}\Sb^H \approx \Sb (c^2P_0 \sigma_{\sr}^2 \sum \limits_{i=1}^{R}|g_i|^2\I_R+\sigma_{\w}^2 \I_R)^{-1} \Sb^H \\
= \frac{1}{\sigma_{\w}^2} \Sb \left(\frac{c^2P_0 \sigma_{\sr}^2 \sum \limits_{i=1}^{R}|g_i|^2}{\sigma_{\w}^2}\I_R+ \I_R\right)^{-1} \Sb^H \\
= \frac{1}{\sigma_{\w}^2} \Sb (\rho+ 1)^{-1} \Sb^H = \frac{1}{\sigma_{\w}^2(\rho+1)} \I_R.
\end{multline}
Substituting \eqref{eq:SXS} into \eqref{eq:proof_mean-z_approx} gives
\begin{multline}
\label{eq:proof_mean-z_approx2}
\mu_{\zeta}\approx 2  \frac{\alpha}{\rho+1} \Re\{\yb^H \bDelta^H \V_i \yb\}= \frac{\alpha}{\rho+1} \yb^H (\bDelta^H \V_i+ \V_i^H \bDelta) \yb \\
= \frac{\alpha}{\rho+1} \yb^H (\bDelta^H \bDelta) \yb
=\frac{\alpha \delta}{\rho+1}  \yb^H \yb
\end{multline}
%\end{proof}

%\begin{proof}[Proof of \eqref{eq:sig-z-approx}]

Re-write \eqref{eq:sig-z} as
\begin{equation}
\label{eq:proof_var-z_approx}
\sigma_{\zeta}^2= 2\yb^H \bDelta^H ( \sigma_{\w}^2\I_R+(1-\alpha^2)\Sb^H \Omg \Sb+
\alpha^2 \sigma_{\w}^2 \V_i \Sb^H \Omg (\Omg+\sigma_{\w}^2\I_R)^{-1} \Sb \V_i^H ) \bDelta \yb
%{\sigma}_z^2\approx 2\left( 1+\alpha^2+(1-\alpha^2)\rho \right) \delta \sigma_{\w}^2 %\yb^H[k-1] \yb[k-1].
\end{equation}
Again if we replace $\Omg$ with the approximated scalar matrix, it is seen that
\begin{equation}
\label{eq:SXS_approx}
\Sb \Omg \Sb^H \approx \Sb (c^2P_0 \sigma_{\sr}^2) \sum \limits_{i=1}^{R} |g_i|^2 \Sb^H = c^2P_0 \sigma_{\sr}^2 \sum \limits_{i=1}^{R} |g_i|^2 \I_R= \rho \sigma_{\w}^2 \I_R,
\end{equation}
and similarly,
\begin{multline}
\label{eq:SXIS_approx}
\Sb \Omg (\Omg+\sigma_{\w}^2\I_R)^{-1} \Sb^H \approx
\Sb c^2P_0 \sigma_{\sr}^2 \sum\limits_{i=1}^{R} |g_i|^2 \left(c^2P_0 \sigma_{\sr}^2 \sum\limits_{i=1}^{R} |g_i|^2\I_R+\sigma_{\w}^2\I_R\right)^{-1} \Sb^H \\
=\Sb \left( \frac{c^2P_0 \sigma_{\sr}^2 \sum\limits_{i=1}^{R}|g_i|^2}{\sigma_{\w}^2} \right)   \left(\frac{c^2P_0 \sigma_{\sr}^2 \sum\limits_{i=1}^{R} |g_i|^2}{\sigma_{\w}^2}\I_R+\I_R\right)^{-1} \Sb^H \\
= \Sb \rho (\rho+1)^{-1} \Sb^H=\frac{\rho}{\rho+1} \I_R.
\end{multline}
Substituting \eqref{eq:SXS_approx} and \eqref{eq:SXIS_approx} into \eqref{eq:proof_var-z_approx} gives the expression in \eqref{eq:sig-z-approx}.
%\end{proof}

\section{Proof of \eqref{eq:Sigma_qbar} and \eqref{eq:Sigma_wbar1}}
\label{app4}
%\begin{proof}[Proof of \eqref{eq:Sigma_qbar}]

\begin{multline}
\label{eq:C_qbar}
\Sig_{\ovq}=
\E
\left\lbrace
\begin{bmatrix}
\q[1]\\
\vdots\\
\q[N]
\end{bmatrix}
\begin{bmatrix}
\q^H[1],\cdots,\q^H[N]
\end{bmatrix}
\right\rbrace
=
\begin{bmatrix}
\E\{\q[1]\q^H[1]\} & \cdots & \E\{\q[1]\q^H[N]\}\\
\vdots  &\ddots &\vdots\\
\E\{\q[N]\q^H[1]\} & \cdots & \E\{\q[N]\q^H[N]\}
\end{bmatrix},
\end{multline}
where
\begin{multline*}
\E\{\q[k]\q^H[k+n]\}=
\E\left\lbrace
\begin{bmatrix}
q_1[k]\\
\vdots \\
q_R[k]
\end{bmatrix}
\begin{bmatrix}
q_1^*[k+n],\cdots,q_R^*[k+n]
\end{bmatrix}
\right\rbrace\\=
\begin{bmatrix}
\E\{q_1[k]q_1^*[k+n]\} & \cdots & \E\{q_1[k]q_R^*[k+n]\}\\
\vdots  &\ddots &\vdots\\
\E\{q_R[k]q_1^*[k+n]\} & \cdots & \E\{q_R[k]q^*[k+n]\}
\end{bmatrix} \\
=\begin{bmatrix}
\varphi_{\sr}(n)  & \cdots & 0 \\
0  & \cdots &  \varphi_{\sr}(n)\}
\end{bmatrix}
= \varphi_{\sr}(n) \I_R.
\end{multline*}
Therefore,
\begin{multline}
\label{eq:C_qbar_proof}
\Sig_{\ovq}=
\begin{bmatrix}
\varphi_{\sr}(0)\I_R & \cdots & \varphi_{\sr}(N-1)\I_R \\
\vdots  &\ddots &\vdots\\
\varphi_{\sr}(N-1)\I_R & \cdots & \varphi_{\sr}(0)\I_R
\end{bmatrix}= \\
\begin{bmatrix}
\varphi_{\sr}(0) & \cdots & \varphi_{\sr}(N-1) \\
\vdots  &\ddots &\vdots\\
\varphi_{\sr}(N-1) & \cdots & \varphi_{\sr}(0)
\end{bmatrix}\otimes \I_R = \C_{\ovq} \otimes \I_R
\end{multline}
%\end{proof}

%\begin{proof}[Proof of \eqref{eq:Sigma_wbar1}]
\begin{multline}
\label{eq:Sigma_wbar}
\Sgwb=
\E
\left\lbrace
\begin{bmatrix}
\w[1]\\
\vdots\\
\w[N]
\end{bmatrix}
\begin{bmatrix}
\w^H[1],\cdots,\w^H[N]
\end{bmatrix}
\right\rbrace
=\\
\begin{bmatrix}
\E\{\w[1]\w^H[1]\} & \cdots & \E\{\w[1]\w^H[N]\}\\
\vdots  &\ddots &\vdots\\
\E\{\w[N]\w^H[1]\} & \cdots & \E\{\w[N]\w^H[N]\}
\end{bmatrix}.
\end{multline}
From \eqref{eq:sig_w} one has
$
\E \{\w[k]\w^H[k+n]\}
= \sigma^2_{\w} \I_R \delta[n],
$
where $\delta[n]$ is Kronecker delta function. Then,
\begin{multline}
\label{eq:Sigma_wbar2}
\Sgwb= N_0
\begin{bmatrix}
(1+c^2\sum\limits_{i=1}^{R}|g_i[1]|^2) \I_R & \cdots & \0_R \\
\vdots  &\ddots &\vdots\\
\0_R & \cdots & (1+c^2\sum\limits_{i=1}^{R}|g_i[N]|^2) \I_R
\end{bmatrix} \\ = N_0
\begin{bmatrix}
1+c^2\sum\limits_{i=1}^{R}|g_i[1]|^2 & \cdots & \0 \\
\vdots  &\ddots &\vdots \\
\0 & \cdots & 1+c^2\sum\limits_{i=1}^{R}|g_i[N]|^2
\end{bmatrix} \otimes \I_R =
\C_{\ovw} \otimes \I_R.
\end{multline}
%\end{proof}

\end{subappendices} 

\chapter{Conclusions and Suggestions for Further Studies}
\label{ch:sum}

\section{Conclusions}
This thesis focused mainly on developing and analyzing differential modulation and non-coherent detection techniques for wireless amplify-and-forward (AF) relay networks. Such techniques do not waste overhead information required for channel estimation and hence are very useful, especially as the number of channels grows linearly with the number of relays employed in a network. Moreover, new and emerging wireless communication technologies, such as LTE, are expected to support high mobility with high performance. As channel estimation becomes highly inaccurate in highly-mobile environments, employing differential modulation and non-coherent detection techniques would be necessary for these technologies. In general, the studies in this thesis provide various options for using differential modulation and non-coherent detection and demonstrate that they are effective and promising solutions for information transmission over wireless relay networks.

In particular, the main findings and contributions of this thesis are as follows:

\begin{itemize}
\item In Chapter~\ref{ch:dh}, a single-branch dual-hop relaying system without a direct link employing differential $M$-PSK and non-coherent detection was studied. First, two-symbol non-coherent detection was examined in time-varying channels and an exact BER expression was derived for the system performance. It was seen that the coverage service can be extended beyond the cell edge by employing a relay in the network and good performance with diversity order of one can be achieved. Although, dependency of the BER to the channels auto-correlations causes a severe degradation and an error floor in the performance over rapid time-varying channels. Next, a near optimal multiple-symbol differential detection was designed and theoretically analysed. Simulation and theoretical results showed that the multiple-symbol differential detection is able to significantly improve the system performance in fast-fading channels.

\item
In Chapter~\ref{ch:mnode}, a multi-branch dual-hop relaying with a direct link employing differential $M$-PSK and two-symbol non-coherent detection was considered. A linear combiner with fixed combining weights was chosen to achieve the spatial diversity. The fixed combining weights were determined based on the second-order statistics of the channels. The performance of the system in practical time-varying channels was studied and a lower bound of the BER was derived. The theoretical and simulation results showed that diversity can be improved by employing more relays in the network. Also, the performance of the system degrades with channel variation and the existence of an error floor is inevitable. However, this degradation can be mitigated by increasing the number of relays in the network. In other words, by increasing the diversity, the desired performance can be obtained in lower values of SNR before falling into the error floor region.

\item
In Chapter~\ref{ch:sc} and Chapter~\ref{ch:sc_tv}, a single-branch dual-hop relaying system with a direct link employing selection combining at the destination was studied. Compared to the linear combiner, the selection combiner does not need any channel or system information. For the case of slow-fading channels, the exact BER and outage probability of this combiner using differential $M$-PSK were obtained. For the case of general time-varying channels (including both scenarios of slow-fading and fast-fading channels), an exact expression of the BER using DBPSK was derived. The simulation and theoretical results showed that the selection combiner achieves diversity order of two and its performance is very close to that of the semi-MRC method, while it is simpler. Therefore, when the second-order statistics of the channels are not available, the SC method can be used instead of the semi-MRC method, without loosening the performance.

\item
In Chapter~\ref{ch:dstc}, a multi-branch dual-hop relaying without a direct link employing differential distributed space-time coding (D-DSTC) strategy was considered. The performance of the system using two-symbol detection over time-varying channels was evaluated in terms of diversity. It was shown that using this topology, the coverage area, diversity and data rate of the network can be improved. Two-symbol differential detection was effective in slow-fading channels. However, it fails to perform well in fast-fading channels and an error will be seen and diversity goes to zero. In contrast to its better spectral efficiency, D-DSTC is more vulnerable to channel variation than repetition based strategy. This is due to its requirement to a longer channel coherence time for two-symbol differential detection. In addition, a near optimal multiple-symbol differential detection (MSDS) was developed. The simulation results showed the effectiveness of the multiple-symbol detection to improve the system performance in fast-fading channels.

\end{itemize}

\section{Future Studies}
The main objective of this thesis was to examine several important issues in cooperative communications using differential encoding and decoding. On the path toward achieving this objective, other issues came up and worthwhile to be investigated further. These issues are elaborated next.

\begin{itemize}

\item Selection combining was considered and analysed for a single-branch dual-hop relaying with a direct link. This combiner is attractive as it reduces the requirement of channels information while delivering similar performance to that of the linear combiner. For a multi-branch relaying system, more channel information would be needed for a linear combining. Although it appears difficult, it would be interesting to consider and analyse the performance of the selection combiner for multi-branch dual-hop relaying with/without a direct link as well.

\item
The performance analysis of multiple-symbol differential detection is important to reach a trade-off between the desired performance and the complexity of the decoding process. In this thesis, the performance of multiple-symbol differential detection for a single-branch dual-hop relaying without a direct link was derived. It would be then interesting to analytically evaluate the performance of multiple-symbol differential detection for multi-branch dual-hop relaying system with/without a direct link.

\item The developed multiple-symbol differential detection scheme requires the second-order statistics of all transmission links and also several other system parameters such as the noise variance and the amplification factor of the relays. In case that such information is not available nor accurate, multiple-symbol detection is not effective or near optimal. An alternative solution, yet challenging, is to develop a blind multiple-symbol differential detection that can deliver near optimal results.

\item
{
In this thesis, it is assumed that no interference exists in the relay networks. In practice, due to frequency re-use and simultaneous transmission of multi users in the same frequency band, co-channel interference (CCI) would arise \cite{CCI-Fei,DH-CCI,AF-CCI}. In addition, the lack of perfect synchronization between multi relays would cause inter-symbol interference (ISI) \cite{Asynch-Valenti,DMT-Asynch,DSTC-Asynch}. The existing studies on CCI and ISI problems in relay networks are either limited to static channels or consider under some special scenarios. Therefore it is worthwhile to consider both CCI and ISI for general scenarios in time-varying channels and develop robust detection techniques against the effects of CCI and ISI.
}

\item
{
Channel capacity is defined as the upper bound of data rate transmission over a channel with arbitrary low error probability.
There are several studies on the capacity of wireless relay networks in \cite{Cap-Coop1,Cap-Coop2,user-coop1,Cap-Cover}. Also, there are studies on the capacity of time-varying channels for point-to-point communications \cite{Cap-TV1,Cap-Goldsmith}. It would be interesting to extend these studies to determine the capacity of wireless relay networks in time-varying environments.
}

\end{itemize}

\addcontentsline{toc}{chapter}{References}
\bibliographystyle{IEEEbib}
\bibliography{references/references}
%\bibliography{references}

\appendix
\addcontentsline{toc}{chapter}{Copyright Permissions}
\chapter*{Copyright Permissions}

This section contains the copyright permissions for papers included in this thesis that have already been published by the IEEE (Institute of Electrical and Electronics Engineers).

\begin{figure}[t]
\centerline{\epsfig{figure={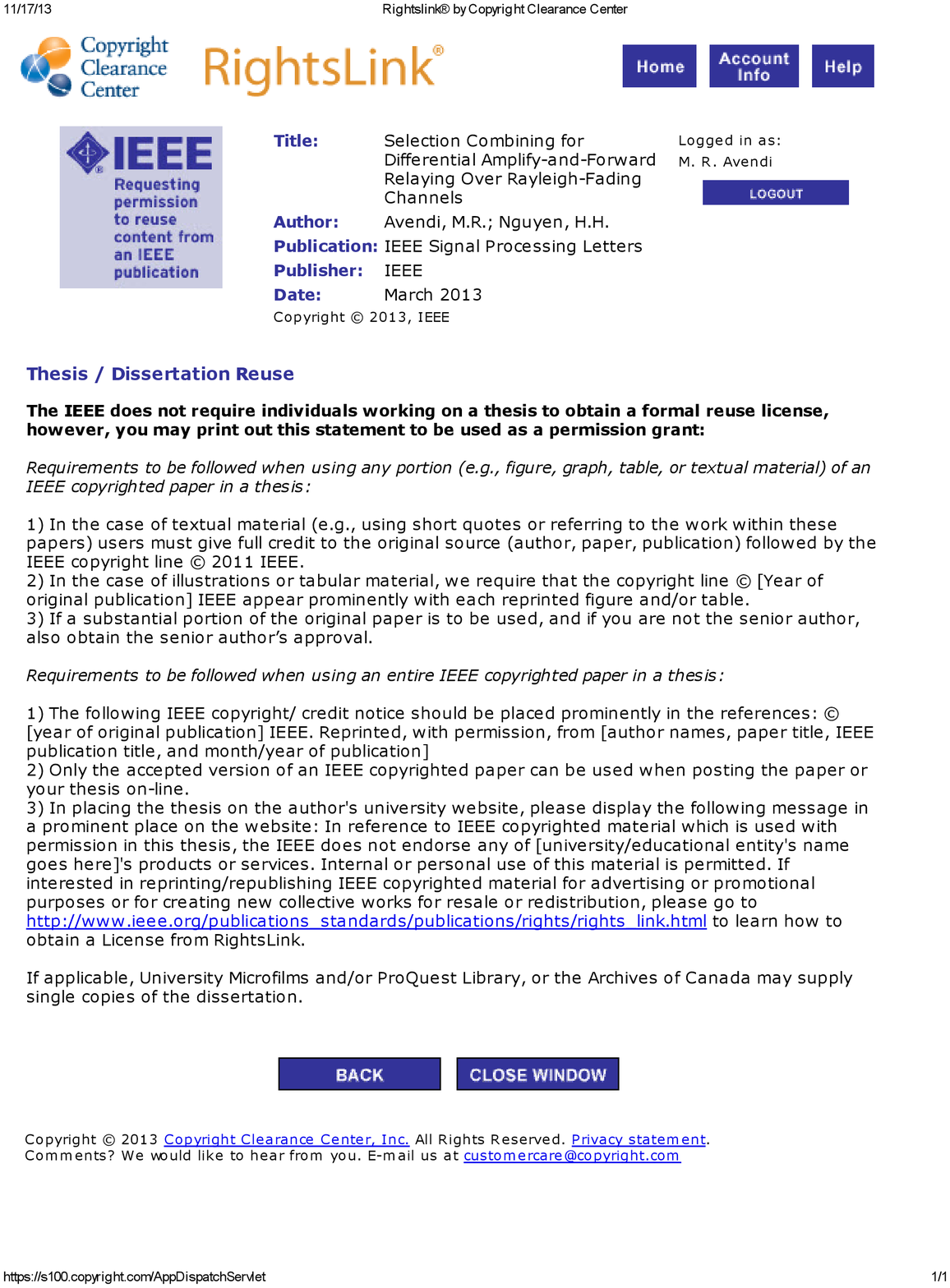},height=10in,width=8in}}
\end{figure}

\begin{figure}[t]
\centerline{\epsfig{figure={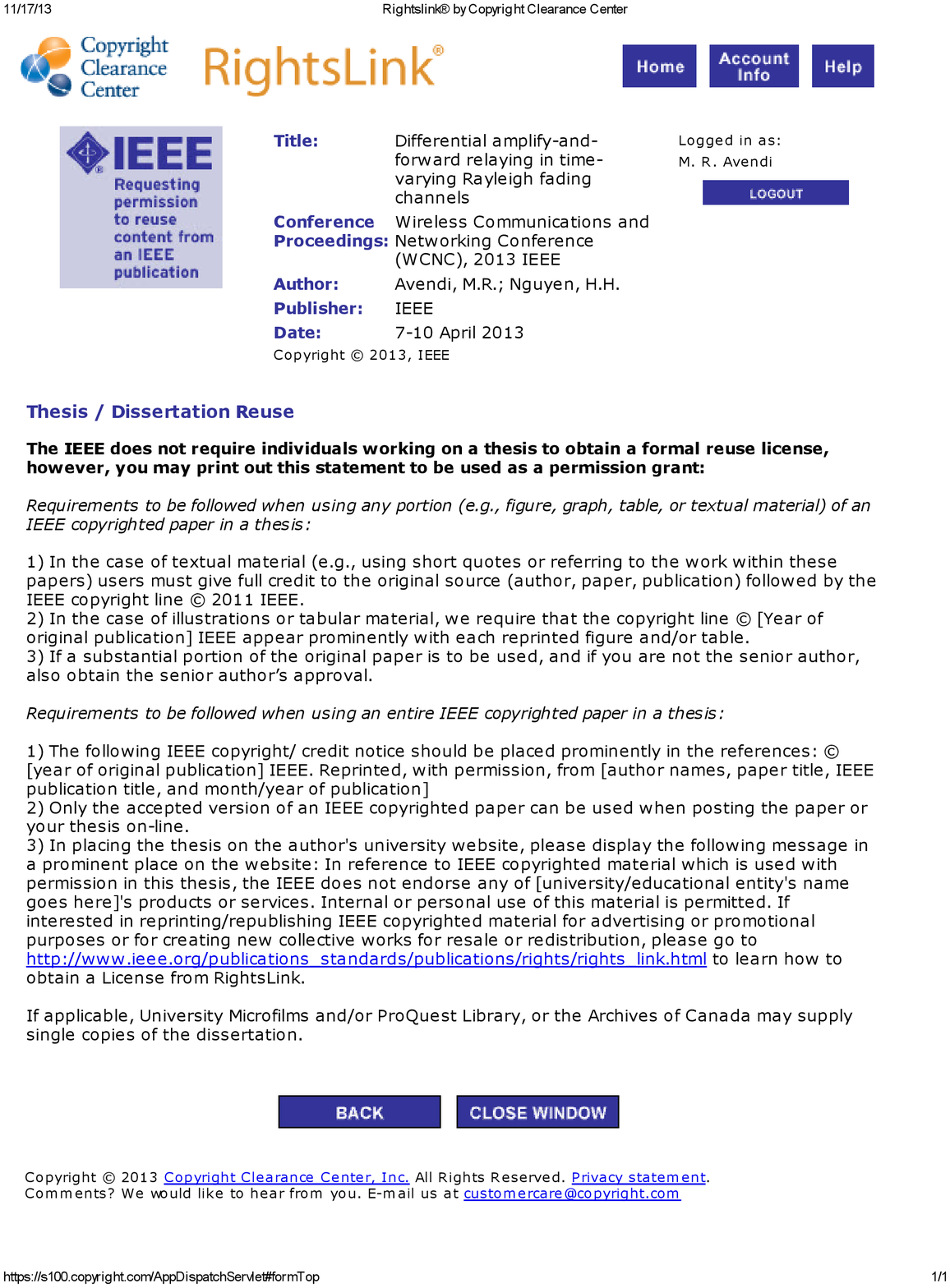},height=10in,width=8in}}
\end{figure}

\begin{figure}[t]
\centerline{\epsfig{figure={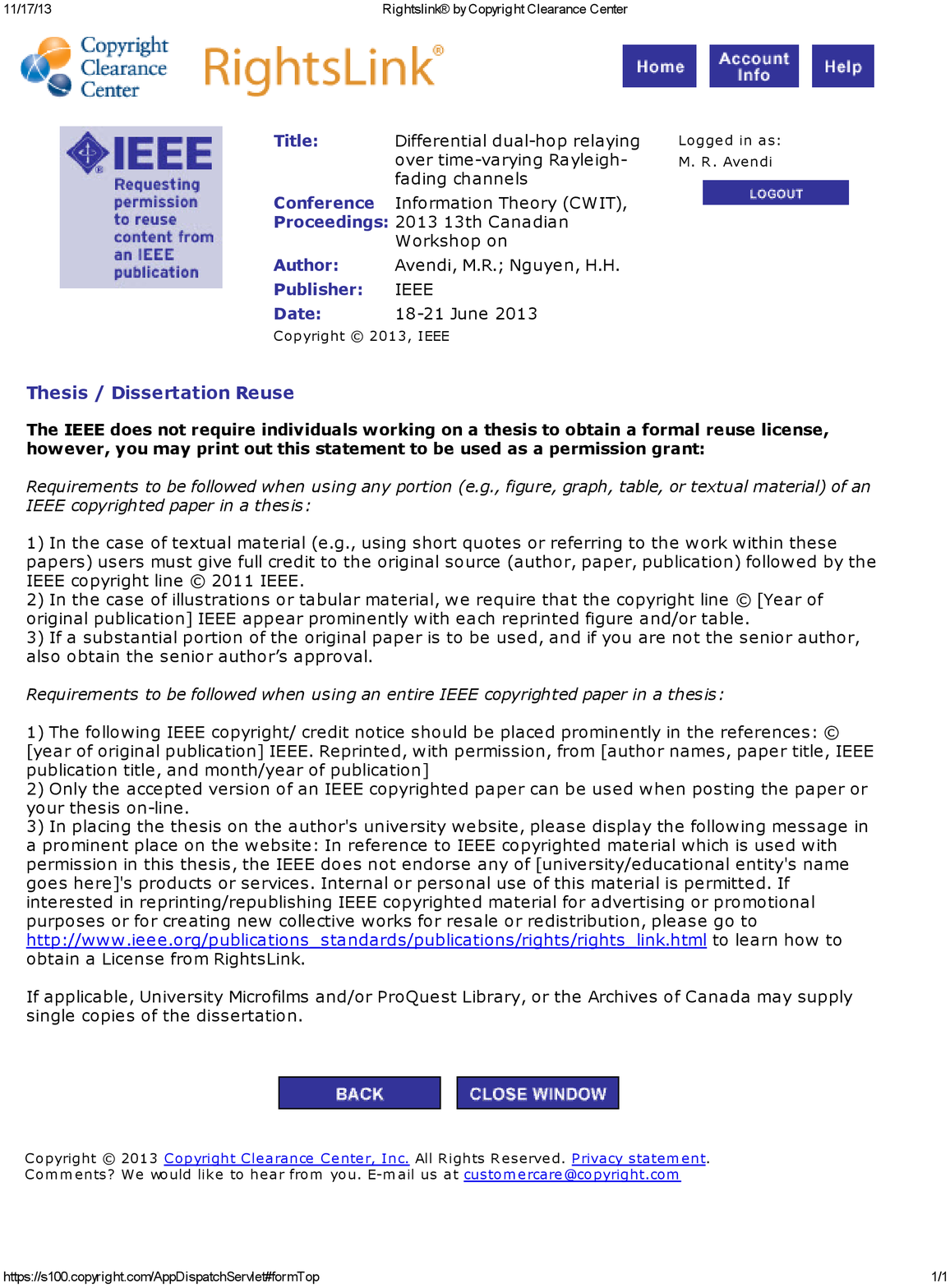},height=10in,width=8in}}
\end{figure}

\begin{figure}[t]
\centerline{\epsfig{figure={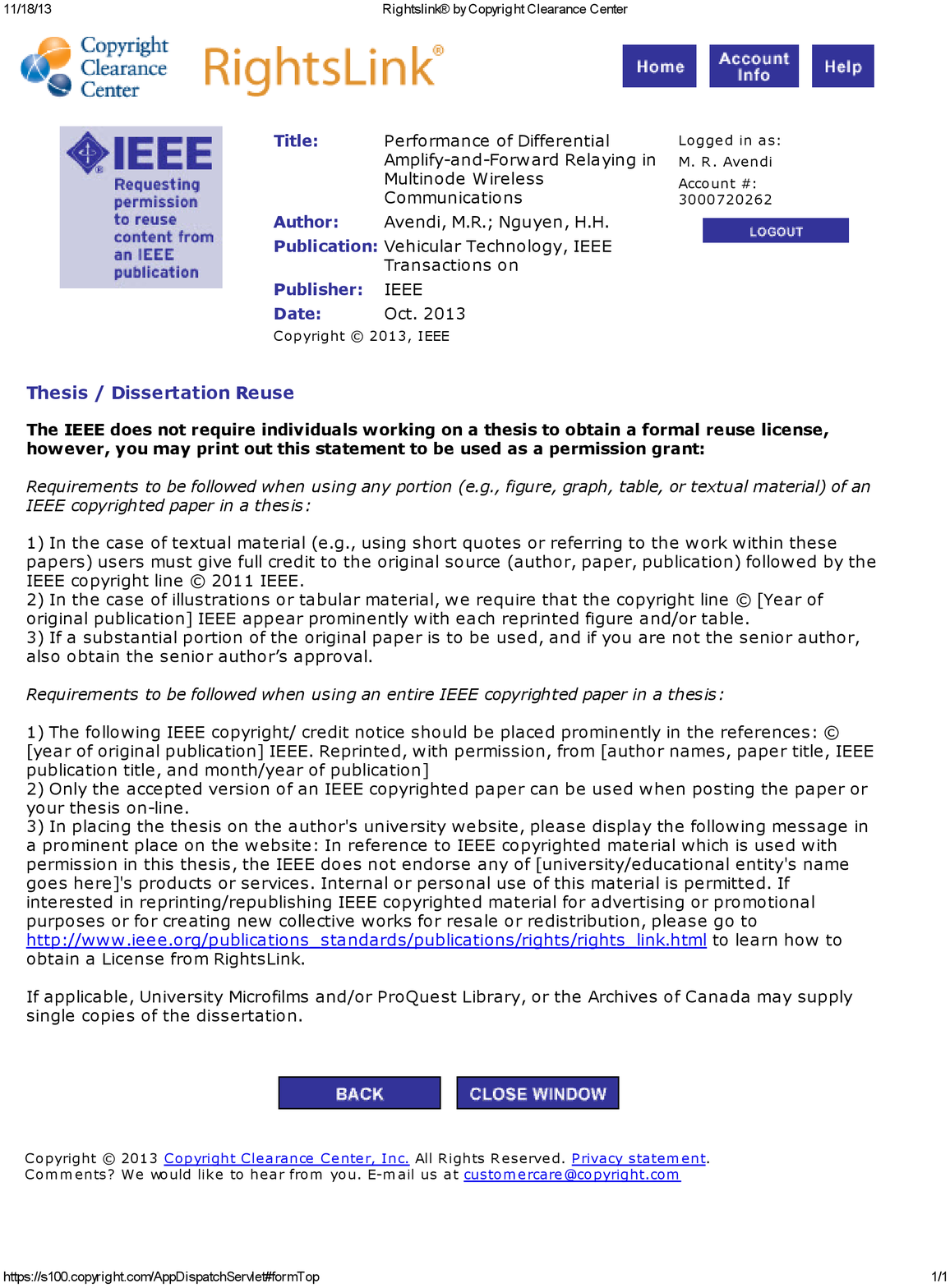},height=10in,width=8in}}
\end{figure} 

\end{document}